\documentclass{appolb}
\usepackage{amssymb}
\usepackage{amsmath}
\usepackage{epsfig}
\usepackage{amsfonts}
\usepackage{graphicx}%
\usepackage{xcolor}
\usepackage{xspace}
\providecommand{\U}[1]{\protect\rule{.1in}{.1in}}
\begin{document}

\title{Introductory visual lecture on QCD at large-$N_{c}$: \\
bound states, chiral models, and phase diagram}
\author{Francesco Giacosa
\address{Institute of Physics, Jan Kochanowski University, \\ ul. Uniwersytecka 7, 25-406, Kielce, Poland \\
Institute for Theoretical Physics, Johann Wolfgang Goethe University,
Max-von-Laue-Str.\ 1, 60438,  Frankfurt am Main, Germany} }
\maketitle

\begin{abstract}
In these lectures, we present the behavior of conventional $\bar{q}q$
mesons, glueballs, and hybrids in the large-$N_{c}$ limit of QCD. To this end,
we use an approach based on rather simple \textquotedblleft
NJL-like\textquotedblright\ bound-state equations. 
The obtained large-$N_{c}$
scaling laws are general and coincide with the known results. A series of
consequences, such as the narrowness of certain mesons and the smallness of some interaction types,
the behavior of chiral and dilaton models at  large-$N_{c},$ and the relation
to the compositeness condition and the standard derivation of large-$N_{c}$
results, are explained. The bound-state formalism shows also that mesonic
molecular and dynamically generated states do not form in the large-$N_{c}$
limit. The same fate seems to apply also for  tetraquark states, but here
further studies are needed. Next, following the same approach, baryons are
studied as bound states of a generalized diquark ($N_{c}-1$ antisymmetric
object) and a quark. Similarities and differences with regular mesons are discussed. All the
standard scaling laws for baryons and their interaction with mesons are
correctly reproduced. The behavior of chiral models involving baryons and
describing chirally invariant mass generation is investigated.
Finally,
properties of QCD in the medium at large-$N_{c}$ are studied: the
deconfinement phase transition is investigated along the temperature and the
chemical potential directions, respectively. While the critical temperature for
deconfinement $T_{dec}$ is $N_{c}$ independent$,$ the critical chemical potential is not and increases for growing $N_{c},$ thus for very large $N_{c}$ one has
confined matter below $T_{dec}$ and deconfined above. Yet,  in the confined
phase but for large densities, one has a `stiff-matter' phase whose pressure is
proportional to $N_{c}$ (just as a gas of quarks would do) in agreement with a
quarkyonic phase. Within the QCD phase diagrams, the features of different
models at large-$N_{c}$ are reviewed and the location of the critical endpoint
is discussed. In the end, the very existence of nuclei and the implications of
large-$N_{c}$ arguments for neutron stars are outlined.

\end{abstract}

\section{Introduction}
In Quantum Chromodynamics (QCD) each quark can appear in three charges
denoted as colors: red (R), green (G), and blue (B). This applies for any of the six quark flavors present in Nature (the light quarks flavors u, d, s and the 
heavy quark flavors c, b, t \cite{pdg}). The force carriers, the gluons, can
be thought as color-anticolor objects, for a total of $9-1=8$ combinations
\cite{weisebook,ratti}.

The origin of colors can be better understood by looking at the fundamental
properties of QCD, which is a gauge theory built under local invariance of the
color group $SU(3).$ The quarks transform under the fundamental representation
and the gluons under the adjoint representation.

Why QCD has 3 colors ($N_{c}=3$, where $N_{c}$ stands for the number of
colors) and not, e.g. 7? At present, there is no compelling answer for that,
at least not within the Standard Model (SM). One may eventually ask if the choice $N_{c}\neq3$ would
allow for stable nuclei, and if this is not the case \cite{bonanno}, resort to
a kind of anthropic argument.

Yet, here we are not interested in this type of questions, but rather in the study
of $N_{c}$ different from $3,$ and in particular the study of the limit in which $N_{c}$ is a `large' number, in order to understand better our word with
$N_{c}=3.$ 
This is indeed the so-called QCD in the large-$N_{c}$ limit,
initiated by 't Hooft \cite{thooft} and further investigated in 
several review papers and lectures
\cite{witten,thooftlect,lebedlect,colemanlect,lucinilect,luciniglueballsN,lucha} (and refs. therein).

At first, the idea to consider an expansion along $N_{c}$ may sound as strange:
how could \textit{anything} valid for, say, $N_{c}=101,$ be also somewhat relevant 
for the physical case $N_{c}=3$? 
In other words, how can $N_{c}=3$ be
considered a `large' number \cite{witten}? As we shall see, that depends. In
some (indeed the majority of) cases, the number $3$ turns out to be large.

In particular, in these lectures we intend to revisit the behavior of bound states of QCD in the large-$N_{c}$ limit. To this end, we recall that
quarks and gluons are confined in hadrons, further classified as mesons
(bosonic hadrons) and baryons (baryonic hadrons).

Mesons can be divided into conventional ones corresponding to quark-antiquark
states (quarkonia), and to exotic or non-conventional types, such as
glueballs, hybrids but also mesonic molecules, dynamically
generated states, and compact tetraquark states (bound states of diquarks). Quite interestingly, quarkonia, glueballs, and hybrids `survive'
in the large-$N_{c}$ limit: this means that their masses are $N_c$-independent, and their widths decrease with $N_c$, implying that these objects become stable in the large-$N_c$ limit. We shall revisit these well known results as well as the specific scaling laws in a novel
fashion, that involves the study of bound-state equations. For the
latter, we chose the easiest possible approach that describes bound-state equations similar to the ones found in the Nambu Jona-Lasinio (NJL) model \cite{njlorig,njl,volkov} (technically, the kernel is separable).  Indeed, these bound state equations are also similar to approaches involving
the compositeness conditions, e.g. \cite{compcond,compo,gutsche,pionsigma}. 
Yet, it should be stressed that our aim is not to actually solve these equations,
but just to discuss their large-$N_{c}$ behavior. The latter is (thought to be)
independent on the particular approach and applies also to more advanced
methods for bound states, such as Bethe-Salpeter equations in QCD
\cite{alkofer,fischer,eichmann}. Quite interestingly, the proposed
large-$N_{c}$ treatment can also help to understand, from a different
perspective, various large-$N_{c}$ features. Namely, we shall re-derive known
results in a different and quite simple way. The coupling of bound objects to their constituents is also an intermediate consequence of the chosen approach. 

As additional applications, we
shall present the large-$N_{c}$ study of the linear sigma model (LSM)
\cite{dick,cpnc}, the dilaton \cite{migdal,salo,ellis,stani}, and their
interconnection. Many other properties (weak decay constant, decay chains,
etc.) shall be discussed as well. The connection of our bound-state approach
to the commonly implemented one that uses correlators and currents is also shown. 

The fate of mesonic molecular and dynamically generated states is different:
they fade away in the large-$N_{c}$ limit. Indeed, we shall recover this result within the bound state approach. A peculiar case, not yet fully solved, is if all tetraquark types (among which molecules are only a specific example) fade out as well. It was long believed that this is the case, but this conclusion was revisited by Weinberg in 2013 \cite{weinberg}, in which he argued that certain tetraquarks could exist in the large-$N_c$ and, if that is the case, their mass scale as $N_c^0$ and their widh as $N_c^{-1}$, just as regular quarkonia. The work \cite{weinberg} was followed by a series of papers on the subject  \cite{lucha,lebed,cohenlebed1,cohenlebed2,estrada,knecht}.
Up to now, the existence of such peculiar tetraquarks in the large-$N_{c}$ limit is not settled. We shall discuss what the bound-state approach has to say for tetraquarks as well.

Baryons will be also briefly  discussed in this work. We shall concentrate on
conventional baryons, which for $N_{c}=3$ are made by 3 quarks and for an
arbitrary $N_{c}$ by $N_{c}$ quarks. We shall present some interesting
similarities between conventional baryons and conventional mesons. To this
end, we treat baryons in a way similar to conventional mesons: they shall be
seen as a bound state of a quark and of a generalized diquark, the latter being the
antisymmetric combination of $N_{c}-1$ quarks. Within this context, we will
re-derive the large-$N_{c}$ scaling for baryons. As an
application, we study chiral models implementing baryonic fields and
investigate possible ways to generate their mass in a chiral invariant manner
and in agreement with large-$N_{c}$ expectations.

In the end, we recall -concisely- some relevant facts concerning the large-$N_{c}$ behavior of QCD at finite temperature and density,
concentrating on the phase diagram and the quarkyonic phase
\cite{chiralspirals1,chiralspirals2,redlich,mclerranissues}, chiral models in
the medium \cite{cpnc,achimnc}, nuclear matter \cite{bonanno,dichotomous}, and
neutron stars \cite{quarkyonicneutronstars,pagliara}.

The style of these lectures is focused on the conceptual and qualitative
features of QCD at large-$N_c$. Moreover, many pictures shall be presented for a better
visualization of the scaling properties. These lectures on large-$N_{c}$
complete previous lectures on chiral models for mesons beyond the quark-antiquark picture
given few years ago \cite{beyond}.

The article is organized as follows. In Sec. 2 we introduce QCD for an arbitrary number of colors together with the double-line notation and the groups $SU(N)$ and $U(N)$, we discuss the QCD running coupling in the framework of the 't Hooft large-$N_c$ limit, and we qualitatively argue that the main features of the propagators of quarks and gluons are $N_c$-independent. Next, in Sec. 3 we study mesons in the large-$N_c$ limit, first the conventional ones (quarkonia) and related topics (chiral models,...), then the exotic glueballs (together with the dilaton), hybrids, and four-quark objects. In Sec. 4 we present conventional baryons and their implementation in chiral models. In Sec. 5 we discuss the main properties of QCD matter at nonzero temperature and chemical potential for large $N_c$. In Sec. 6 conclusions are summarized.

\section{QCD for arbitrary $N_{c}$}

\subsection{QCD Lagrangian of any $N_c$}
We present the Lagrangian of QCD for an arbitrary number of colors $N_{c}$ and
quark flavors $N_{f}$ (see, for instance, \cite{weisebook,ratti}):
\begin{align}
\mathcal{L}_{QCD}  &  =\mathrm{Tr}\left[  \overline{q}_{i}(i\gamma^{\mu}%
D_{\mu}-m_{i})q_{i}-\frac{1}{2}G_{\mu\nu}G^{\mu\nu}\right]  ,\text{ }D_{\mu
}=\partial_{\mu}-ig_{0}A_{\mu}\text{ ,}\nonumber\\
G_{\mu\nu}^{a}  &  =\partial_{\mu}A_{\nu}-\partial_{\nu}A_{\mu}-ig_{0}[A_{\mu
},A_{\nu}]\text{  .} \label{lqcd}%
\end{align}
Above,  $A_{\mu}$ is $N_{c}\times N_c$ 
Hermitian matrix, and $q_i(x)$ is a vector in color space for each value of the flavor index  $i=1,...,N_{f}$ (with $N_{f}$ being the number of quark flavors), see details below. 
Moreover, the
coupling constant $g_{0}$ is an adimensional parameter of the classical
Lagrangian and $m_{i}$ is the bare mass of the $i$-th quark flavor. 
Note, in the
chiral limit ($m_{i}=0$, for each $i$) the Lagrangian is invariant under dilatation
transformation, since no dimensionful parameter is present in the classical theory. This symmetry is broken by quantum fluctuations (trace anomaly, see Sec. 3.2).

\begin{figure}[h]
        \centering       \includegraphics[scale=0.45]{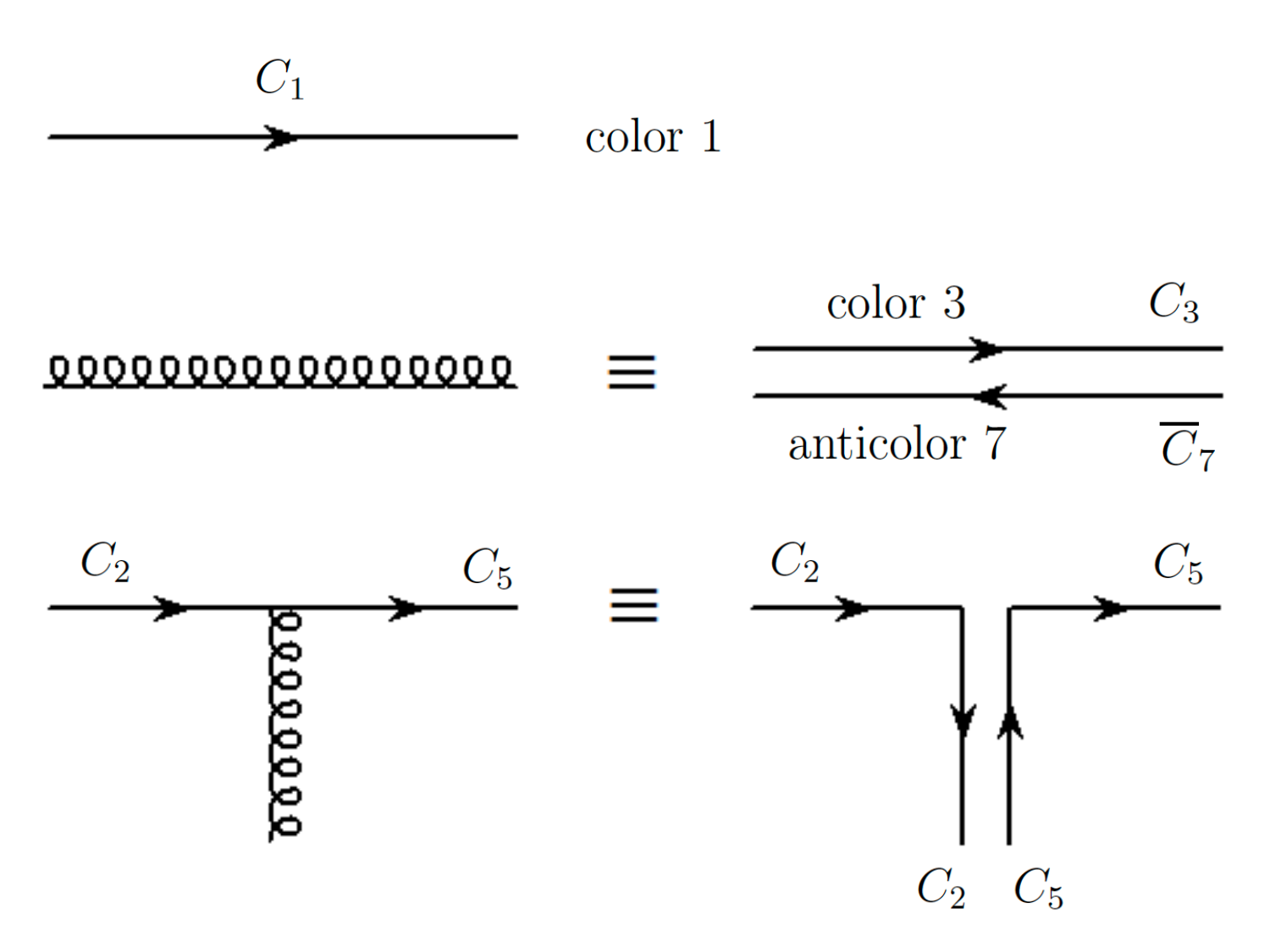}\\
        \caption{Free quark, free gluon, and quark-gluon vertex. The double-line notation for the gluons is also shown. The specific choice of colors refers to illustrative examples: the free quark is taken with color $C_1$, the free gluon carries $\bar{C}_7C_3$, and the vertex shows how a quark $C_2$ changes into $C_5$ via the interaction with a $\bar{C}_2C_5$ gluon. }
        \label{qcd}
   \end{figure}

   \begin{figure}[h]
        \centering       
        \includegraphics[scale=0.45]{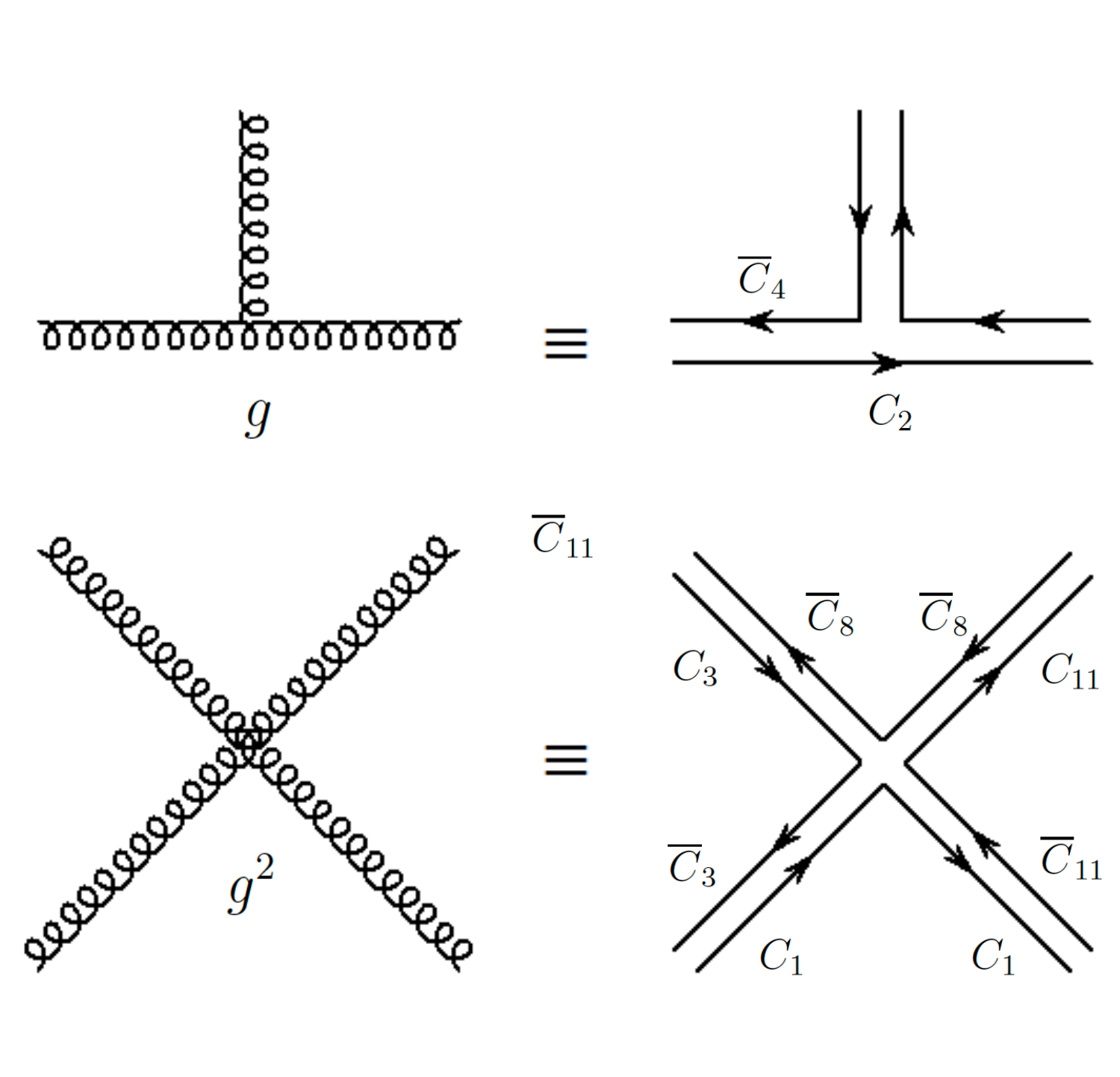}\\
        \caption{Three-leg and four-leg in the standard and double-line notation. The specific choice of colors on the right part refers to examples. }
        \label{qcd2}
   \end{figure}

The part of Eq. (\ref{lqcd}) containing only gluons is called the Yang-Mills (YM) Lagrangian:
\begin{equation}
\mathcal{L}_{YM}=\mathrm{Tr}\left[  -\frac{1}{2}G_{\mu\nu}G^{\mu\nu}\right]
\text{ .}
\end{equation}
For $N_{c}>1$, the YM Lagrangian contains 3-gluon and 4-gluon vertices. The
gluonic self-interactions are a fundamental property of nonabelian theories. In turn, this feature implies that gluonic bound states, called glueballs, are possible \cite{mainlattice,bag,vento}, see also Sec. 3.2.

In Nature, $N_{c}=3$ and $N_{f}=6.$ However, depending on the problem, one can
consider different values for $N_{c}$ and $N_{f}.$ For instance, low-energy
QCD is realized for $N_{c}=3$ and $N_{f}=3,$ i.e. only light quarks are
retained. Moreover, varying $N_{c}$ is the main goal of large-$N_c$ studies.

In Figs. \ref{qcd} and \ref{qcd2} we present the Feynman diagrams that follow form the QCD Lagrangian. In particular, in Fig. \ref{qcd} the fundamental quark-gluon vertex is depicted, while in Fig. \ref{qcd2} the gluonic self-interactions are shown. In both cases, gluons are also represented via the so-called double line notation, that `naively speaking' corresponds to a quark and an antiquark. In order to understand this point better, we need to have a closer look at the gluon field.

The $N_{c}\times N_{c}$ Hermitian matrix  Yang-Mills field $A_{\mu}(x)$ can be expressed as:
\begin{equation}
A_{\mu}(x)=\sum_{a=1}^{N_{c}^{2}-1}A_{\mu}^{a}(x)t^{a}\text{ ,}%
\label{amu}
\end{equation}
where $t^{a}$ is an appropriate set of $N_{c}^{2}-1$ matrices basis. Usually, they are taken as Hermitian and traceless, see the next subsection. In this way the coefficients $A_{\mu}^{a}(x)$ are real numbers.

The the quark field is a vector in color space with
\begin{equation}
q_{i}=\left(
\begin{array}
[c]{c}%
q_{1,i}\\
q_{1,i}\\
...\\
q_{N_{c},i}%
\end{array}
\right)  \text{ ,}%
\end{equation}
where $i=1,...,N_{f}$ is the flavor index.

Under $SU(N_{c})$ local gauge transformations these fields transform as:
\begin{equation}
\text{ }q_{i}\rightarrow U(x)q_{i}\text{ , }A_{\mu}(x)\rightarrow A_{\mu
}^{\prime}(x)=U(x)A_{\mu}(x)U^{\dagger}(x)-\frac{i}{g_{0}}U(x)\partial_{\mu
}U^{\dagger}(x)\text{ }, \label{transf}%
\end{equation}
where $U(x)$ is an arbitrary function of the space-time variable $x \equiv x^{\mu} \equiv (t,\bold{x}) $.

The Lagrangian $\mathcal{L}_{QCD}$ has been constructed to be  invariant under the local gauge transformations of Eq. (\ref{transf}). 

A particularly useful limit is the one in which $U(x)=U$ is a constant matrix, leading to:
\begin{equation}
\text{ }q_{i}\rightarrow Uq_{i}\text{ , }A_{\mu}(x)\rightarrow A_{\mu}%
^{\prime}(x)=UA_{\mu}(x)U^{\dagger}\text{ }.
\end{equation}

In Figs. \ref{qcd} and \ref{qcd2} a  double-line notation for the gluon is presented, according to which the
gluon field is described with the help of components carrying two indices
\begin{equation}
A_{\mu}^{(a,b)}(x)\equiv A_{\mu}^{(a,b)}(x)\text{ with }a,b=1,\text{...,
}N_{c} \text{ .}
\end{equation}
This point reflects the adjoint Nature of the gluon field that, for what
concerns color, can be seen (besides the singlet colorless configuration that is not present) as a quark-antiquark object.
For instance:
\begin{equation}
A_{\mu}^{(2,5)}\equiv\bar{C}_{2}C_{5}\text{ ,}%
\end{equation}
implying that this gluon configuration contains the color $C_5$ and the anticolor $\bar{C}_2$.
Indeed, this choice corresponds to a specific choice for the basis of the $t^{a}$ matrices:
\begin{equation}
A_{\mu}(x)\equiv\sum_{a=1}^{N_{c}}\sum_{b=1}^{N_{c}}A_{\mu}^{(a,b)}%
t^{(a,b\text{)}}%
\text{ ,}
\end{equation}
where the $N_{c}^{2}$ matrices $t^{(a,b\text{)}}$ are given by:
\begin{equation}
\left(  t^{(a,b\text{)}}\right)  _{c,d}=\delta_{ac}\delta_{bd}
\text{ .}
\label{trivial}
\end{equation}
The matrices of this basis are neither Hermitian nor traceless, so the `coefficients' $A_{\mu}^{(a,b)}$ are not real, nut need to fulfill the following requirements:
\begin{equation}
  A_{\mu}^{(a,b)} = \left(A_{\mu}^{(b,a)}  \right)^* 
  \text{  , }
  \sum_{a=1}^{N_{c}} A_{\mu}^{(a,a)} = 0
  \text{ .}
\end{equation}
In fact, the former equation guarantees that $A^{\mu}$ is hermitian, and the latter that it is traceless. Namely, the employed basis contains a matrix `too much', thus one needs to remove the color singlet (traceless) configuration. Yet,
for large-$N_{c}$ this additional contribution is negligible, so we will usually not 
`bother' to subtract it. Note, also the tensor $G_{\mu\nu}$ can be expressed in
this basis as
\begin{equation}
G_{\mu\nu}=\sum_{a=1}^{N_{c}}\sum_{b=1}^{N_{c}}G_{\mu\nu}^{(a,b)}%
t^{(a,b)}\text{ .}%
\end{equation}
Indeed, while this choice for the matrices $t^a$ may be useful to realize the double-line idea for a gluon, it is not what is usually employed for the specific cases of $N_c =2$ or $N_c =3$, see the next subsection.

\subsection{Brief recall of $SU(N)$}

Before we continue, it is important to recall some basic properties of the
groups\emph{ }$U(N)$\emph{ }and\emph{ }$SU(N)$ (see e.g. Ref. \cite{zeegroup}). An element of the group $U(N)$
is a complex $N\times N$ matrix such that:%
\begin{equation}
U^{\dagger}U=UU^{\dagger}=1_{N}\text{ ,} \label{un}%
\end{equation}
thus $U$ can be expressed as:%
\begin{equation}
U=e^{i\theta_{a}t^{a}}\text{, }a=0,1,...,N^{2}-1\text{ ,}%
\end{equation}
where the matrices $t^{a}$ are $N^{2}$ linearly independent $N\times N$
Hermitian matrices, implying that Eq. (\ref{un}) is fulfilled.Following the usual
convention, we set:
\begin{equation}
t^{0}=\frac{1}{\sqrt{2N}}1_{N}\text{ ,}%
\end{equation}
and for the other matrices, we choose:
\begin{equation}
\mathrm{Tr}\left[  t^{a}t^{b}\right]  =\frac{1}{2}\delta^{ab}\text{ with
}a,b=0,1,...,N^{2}-1\text{ , }%
\end{equation}
then%
\begin{equation}
\mathrm{Tr}\left[  t^{a}\right]  =0\text{ for }a=1,...,N^{2}-1\text{ .}%
\end{equation}
A $N\times N$ matrix $U$ belongs to the group $SU(N)$ if the following two equations are
fulfilled:
\begin{equation}
U^{\dagger}U=UU^{\dagger}=1_{N},\text{ }\det U=1\text{ .} \label{1csundef}%
\end{equation}
It is clear that a matrix belonging to $SU(N)$ can be written as
\begin{equation}
  U=e^{i\theta_{a}t^{a}} \text{ with }  a=1,...,N^{2}-1 \text{ ,} 
\end{equation}
(the identity matrix, which is
not traceless, is left out). 
Then:%
\begin{equation}
\det U=e^{Tr\left[  i\sum_{a=1}^{N^{2}-1}\theta_{a}t^{a}\right]  }=1\text{ .}%
\end{equation}
The matrices $t^{a}$ with $a=1,...,N^{2}-1$ are the generators of $SU(N)$ and fulfill the algebra:
\begin{equation}
\lbrack t^{a},t^{b}]=if^{abc}t^{c}\text{ with }a,b,c=1,...,N^{2}-1\text{ ,}%
\end{equation}
where $f^{abc}$ are the corresponding antisymmetric structure constants, see \cite{zeegroup} for their explicit form.
Namely, the commutator of two Hermitian matrices is anti-Hermitian and
traceless, therefore it must be expressed as a sum over $t^{a}$ for
$a=1,...,N^{2}-1$. By taking this choice, the matrix $A_{\mu}$ is Hermitian and traceless for arbitrary real coefficients $A_{\mu}^{a})$ in Eq. (\ref{amu}). 

For the color local case of QCD, the matrix the unitary matrix $U(x)$ can be expressed as 
\begin{equation}
U(x)=e^{i\theta_{a}(x)t^{a}}\text{ },\text{ }a=1,...,N_{c}^{2}-1\text{
}(=8\text{ for }N_{c}=3)\text{ ,}%
\end{equation}
where the quantities $\theta_{a}(x)$ are arbitrary functions of the spacetime variable $x$.

It is useful to briefly discuss some particular examples. For $N_{c}=2$, the generators are given by the matrices $t^a =\sigma^a/2$, where the $\sigma^a$ are the Pauli matrices. The structure constants read $f^{abc} = \epsilon^{abc}$. Finally,  the matrix $A_{\mu}$ takes the explicit form
\begin{align}
A_{\mu}  & =\sum_{a=1}^{N_{c}^{2}-1}A_{\mu}^{a}t^{a}=\sum_{a=1}^{N_{c}^{2}%
-1}A_{\mu}^{a}\frac{\tau^{a}}{2}\nonumber\\
& =\frac{1}{\sqrt{2}}\left(
\begin{array}
[c]{cc}%
\frac{A_{\mu}^{3}}{\sqrt{2}} & \frac{A_{\mu}^{1}-iA_{\mu}^{2}}{\sqrt{2}}\\
\frac{A_{\mu}^{1}-iA_{\mu}^{2}}{\sqrt{2}} & -\frac{A_{\mu}^{3}}{\sqrt{2}}%
\end{array}
\right)  \equiv\frac{1}{\sqrt{2}}\left(
\begin{array}
[c]{cc}%
\frac{R\bar{R}-G\bar{G}}{2} & R\bar{G}\\
G\bar{R} & \frac{-R\bar{R}+G\bar{G}}{2}%
\end{array}
\right)
\end{align}
where it is visible that $Tr[A_{\mu}]=0$ and $A_{\mu}^{\dagger}.$ The
off-diagonal components correspond to the expected double-line assignment.
Note, upon including the $0$-th component (a negligible error for large
$N_{c}$) we obtain
\begin{align}
A_{\mu}  & =\sum_{a=0}^{N_{c}^{2}-1}A_{\mu}^{a}t^{a}=\sum_{a=1}^{N_{c}^{2}%
-1}A_{\mu}^{a}\frac{\tau^{a}}{2}\nonumber\\
& =\frac{1}{\sqrt{2}}\left(
\begin{array}
[c]{cc}%
\frac{A_{\mu}^{0}+A_{\mu}^{3}}{\sqrt{2}} & \frac{A_{\mu}^{1}-iA_{\mu}^{2}%
}{\sqrt{2}}\\
\frac{A_{\mu}^{1}-iA_{\mu}^{2}}{\sqrt{2}} & \frac{A_{\mu}^{0}-A_{\mu}^{3}%
}{\sqrt{2}}%
\end{array}
\right)  \equiv\frac{1}{\sqrt{2}}\left(
\begin{array}
[c]{cc}%
R\bar{R} & R\bar{G}\\
G\bar{R} & G\bar{G}%
\end{array}
\right)
\end{align}
that correspond to the double-line notation also along the diagonal elements.
Note, in a classical view above $R$ and $G$ may be though as complex numbers
that characterize a quark component $\left(
\begin{array}
[c]{cc}%
R & G
\end{array}
\right)  ^{t},$ while the bar quantities are the complex conjugates. The
matrix $A_{\mu}$ (that includes the 0-th element, corresponds then to
\begin{equation}
A_{\mu}=\frac{1}{\sqrt{2}}\left(
\begin{array}
[c]{c}%
R\\
G
\end{array}
\right)  \left(
\begin{array}
[c]{cc}%
\bar{R} & \bar{G}%
\end{array}
\right)
\end{equation}
which clearly shows the origin of the double-line notation as well as the dual
nature of the gluon field.

It is interesting to notice that a similar decomposition holds for the pion triplet:
\begin{align}
\pi & =\sum_{a=1}^{N_{c}^{2}-1}\pi^{a}t^{a}=\frac{1}{\sqrt{2}}\left(
\begin{array}
[c]{cc}%
\frac{\pi^{3}}{\sqrt{2}} & \frac{\pi^{1}-i\pi^{2}}{\sqrt{2}}\\
\frac{\pi^{1}+i\pi^{2}}{\sqrt{2}} & -\frac{\pi^{3}}{\sqrt{2}}%
\end{array}
\right)  \nonumber\\
& =\frac{1}{\sqrt{2}}\left(
\begin{array}
[c]{cc}%
\frac{\pi^{0}}{\sqrt{2}} & \pi^{+}\\
\pi^{-} & -\frac{\pi^{0}}{\sqrt{2}}%
\end{array}
\right)  =\frac{1}{\sqrt{2}}\left(
\begin{array}
[c]{cc}%
\frac{u\bar{u}-d\bar{d}}{2} & u\bar{d}\\
d\bar{u} & \frac{-u\bar{u}+d\bar{d}}{2}%
\end{array}
\right)
\end{align}
while, upon introducing $P^{0}=\eta_{N}=\sqrt{1/2}(u\bar{u}+d\bar{d})$, the full multiplet (quartet) of pseudoscalar states can be described:
\begin{align}
P  & =\sum_{a=0}^{N_{c}^{2}-1}P^{a}t^{a}=\frac{1}{\sqrt{2}}\left(
\begin{array}
[c]{cc}%
\frac{\eta_{N}+\pi^{3}}{\sqrt{2}} & \frac{\pi^{1}-i\pi^{2}}{\sqrt{2}}\\
\frac{\pi^{1}+i\pi^{2}}{\sqrt{2}} & \frac{\eta_{N}-\pi^{3}}{\sqrt{2}}%
\end{array}
\right)  \nonumber\\
& =\frac{1}{\sqrt{2}}\left(
\begin{array}
[c]{cc}%
\frac{\eta_{N}+\pi^{0}}{\sqrt{2}} & \pi^{+}\\
\pi^{-} & \frac{\eta_{N}-\pi^{0}}{\sqrt{2}}%
\end{array}
\right)  =\frac{1}{\sqrt{2}}\left(
\begin{array}
[c]{cc}%
u\bar{u} & u\bar{d}\\
d\bar{u} & d\bar{d}%
\end{array}
\right)  \text{ .}%
\end{align}

The case $N_c =3$ can be treated in a similar way. The matrices are $t^a = \lambda^a/2$, with the $\lambda^a$ being the Gell-Mann matrices for $a=1,...,8$.

Finally, we recall also that there is discrete a subgroup of $SU(N)$, denoted as the
center $Z(N)$, whose $N$ elements are given by \cite{ratti}:
\begin{equation}
Z=Z_{n}=e^{i\frac{2\pi n}{N}}1_{N},\text{ }n=0,1,2,...,N-1\text{ .}
\label{zentrum}%
\end{equation}
Each $Z_{n}$ corresponds to a proper choice of the parameters $\theta_{a}$
(the case $Z_{0}=1_{N}$ corresponds to the simple case $\theta_{a}=0,$ the
other elements to more complicated choices).
This group plays an important role at nonzero temperature as an indicator of confinement, since the $Z(N)$-symmetry is realized in the QCD vacuum and in the confined phase (at small $T$), but is broken in the deconfined one (at large $T$) \cite{ratti,fuku}.

\subsection{Running coupling of QCD}

The coupling `constant' $g_{0}$ entering in the classical QCD Lagrangian of Eq.
(\ref{lqcd}) turns into a running coupling when QCD is quantized. At one-loop
level (which is enough for our illustrative purposes here) one has
\cite{weisebook}: 

\begin{equation}
\mu\frac{dg}{d\mu}=-bg^{3}\text{ ,}%
\end{equation}
with
\begin{equation}
b=\frac{1}{2}\frac{1}{8\pi^{2}}\left(  \frac{11}{3}N_{c}-\frac{2}{3}%
N_{f}\right)
\end{equation}
and where
\begin{equation}
g_{QCD}\equiv g \equiv g(\mu)
\end{equation}
refers to the QCD running coupling. In this work, whenever $g$ will be
presented, it always refers to the fundamental QCD coupling. Other coupling
constants shall carry an appropriate subscript specifying to what they refer. For a detailed description of the QCD running coupling for $N_c=3$ we refer to Ref. \cite{deur} and refs. therein.

In Nature $N_{c}=3$ and $N_{f}$ ranges from $2$ to $6,$ in dependence on the
number of considered quark flavors; in any case, $b>0.$ Note, $b>0$ is
definitely also true in the large-$N_{c}$ limit upon keeping $N_{f}$ fixed, as
we shall do here. This is the so-called 't Hooft large-$N_{c}$ scheme \cite{thooft}.

Upon fixing $g_{0}$ at a certain (large, or ultraviolet (UV)) energy scale $\Lambda_{UV}$ and by
integrating:
\begin{align}
\int_{g}^{g_{0}}dg^{\prime}\frac{dg^{\prime}}{g^{\prime3}}  &  =\left[
\frac{g^{\prime-2}}{-2}\right]  _{g}^{g_{0}}=-\int_{\mu
}^{\Lambda_{UV}}dg^{\prime}b\frac{d\mu^{\prime}}{\mu^{\prime}}=-b\ln\frac{\Lambda_{UV}}{\mu
}=b\ln\frac{\mu}{\Lambda_{UV}}\text{ };\\
\frac{1}{g_{0}^{2}}-\frac{1}{g^{2}}  &  =-2b\ln\frac{\mu}{\Lambda_{UV}}\text{
;}\\
\frac{1}{g^{2}}  &  =\frac{1}{g_{0}^{2}}+2b\ln\frac{\mu}{\Lambda_{UV}}%
=\frac{1+2bg_{0}^{2}\ln\frac{\mu}{\Lambda_{UV}}}{g_{0}^{2}}\text{ .}%
\end{align}
Hence:%
\begin{equation}
g^{2}=\frac{g_{0}^{2}}{1+2bg_{0}^{2}\ln\frac{\mu}{\Lambda_{UV}}} \text{ .}
\end{equation}
Then:
\begin{equation}
g^{2}(\mu)=\frac{g_{0}^{2}}{1+2bg_{0}^{2}\ln\frac{\mu}{\Lambda_{UV}}}%
=\frac{g_{0}^{2}}{1+\frac{g_{0}^{2}}{8\pi^{2}}\left(  \frac{11}{3}N_{c}%
-\frac{2}{3}N_{f}\right)  \ln\frac{\mu}{\Lambda_{UV}}} \text{ .}%
\end{equation}
One has (by construction):
\begin{equation}
g^{2}(\mu=\Lambda_{UV})=g_{0}^{2} \text{ .}
\end{equation}
Upon setting the denominator to zero%
\begin{equation}
1+\frac{g_{0}^{2}}{8\pi^{2}}\left(  \frac{11}{3}N_{c}-\frac{2}{3}N_{f}\right)
\ln\frac{\mu}{\Lambda_{UV}}=0\text{  ,}%
\end{equation}
we obtain the the so-called $\Lambda_{QCD}$ scale as a Landau pole of the
running coupling:%
\begin{equation}
\Lambda_{QCD}=\Lambda_{UV}\exp\left[  -\frac{8\pi^{2}}{g_{0}^{2}\left(
\frac{11}{3}N_{c}-\frac{2}{3}N_{f}\right)  }\right] \ll |Lambda_{UV} \text{ .}%
\end{equation}
The existence of a pole of the coupling at the low energy $\Lambda_{QCD} \ll |Lambda_{UV}$ is an artifact of the one-loop perturbative approach, but it
signalizes that the running coupling
becomes large. In Ref. \cite{gies}, using the FRG approach, it is shown that no infinity of the QCD running coupling takes place. 

The value of the bare coupling can be expressed as:%
\begin{equation}
g_{0}^{2}=-\frac{1}{\left(  \frac{11}{3}N_{c}-\frac{2}{3}N_{f}\right)  }%
\frac{8\pi^{2}}{\ln\frac{\Lambda_{QCD}}{\Lambda_{UV}}} \text{ .}
\end{equation}
Here, we intend to study the limit in which the low-energy scale $\Lambda_{QCD}$
is independent on $N_{c}$ ('t Hooft limit). We require that
\begin{equation}
g_{0}^{2}\propto\frac{1}{N_{c}}%
\text{ ,}
\end{equation}
then%
\begin{equation}
g_{0}\propto\frac{1}{\sqrt{N_{c}}}
\text{ .}
\end{equation}
Next, upon eliminating the UV scale $\Lambda_{UV}$ we find:
\begin{align}
g^{2}(\mu)  &  =\frac{g_{0}^{2}}{1+\frac{g_{0}^{2}}{8\pi^{2}}\left(  \frac
{11}{3}N_{c}-\frac{2}{3}N_{f}\right)  \ln\frac{\mu}{\Lambda_{UV}}} \nonumber \\
&  =-\frac{-\frac{1}{\left(  \frac{11}{3}N_{c}-\frac{2}{3}N_{f}\right)  }%
\frac{8\pi^{2}}{\ln\frac{\Lambda_{QCD}}{\Lambda_{UV}}}}{1+\left(  -\frac
{1}{\left(  \frac{11}{3}N_{c}-\frac{2}{3}N_{f}\right)  }\frac{8\pi^{2}}%
{\ln\frac{\Lambda_{QCD}}{\Lambda_{UV}}}\right)  \frac{1}{8\pi^{2}}\left(
\frac{11}{3}N_{c}-\frac{2}{3}N_{f}\right)  \ln\frac{\mu}{\Lambda_{UV}}} \nonumber \\
&  =\frac{8\pi^{2}}{\left(  \frac{11}{3}N_{c}-\frac{2}{3}N_{f}\right)  }%
\frac{1}{\ln\frac{\mu}{\Lambda_{QCD}}} \text{ .}
\end{align}

\begin{figure}[h]
        \centering       
        \includegraphics[scale=0.70]{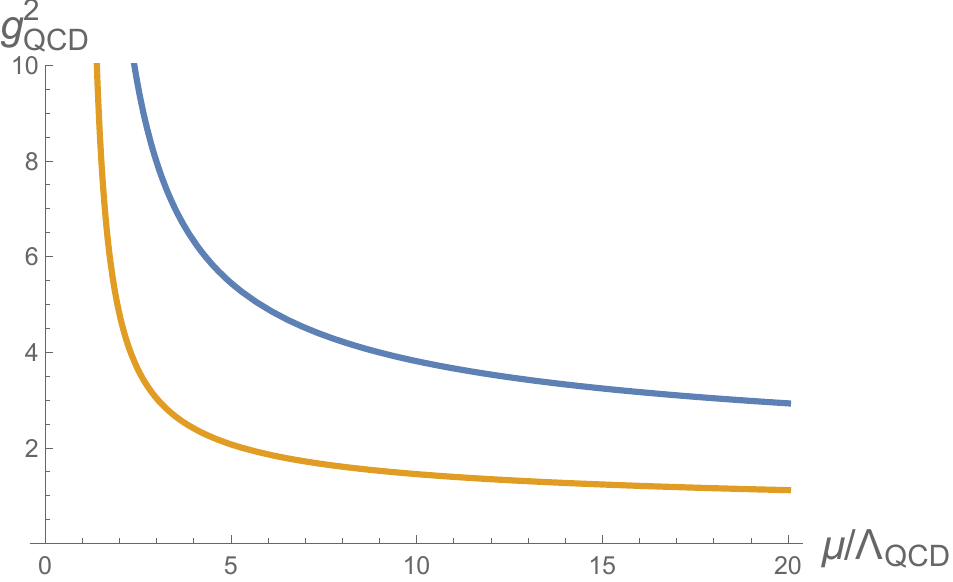}\\
        \caption{Running coupling of QCD of Eq. (\ref{gqcd}) for $N_c=3$ (upper, blue curve) and for $N_c=7$ (lower, yellow curve). The Landau pole is the same in both cases, in agreement with the 't Hooft large-$N_c$ limit.}
        \label{running}
   \end{figure}
   
Thus, in terms of $\mu$ and $\Lambda_{QCD}$ the one-loop running coupling can be expressed as:
\begin{equation}
g^{2}(\mu)=\frac{8\pi^{2}}{\left(  \frac{11}{3}N_{c}-\frac{2}{3}N_{f}\right)
}\frac{1}{\ln\frac{\mu}{\Lambda_{QCD}}}\text{ .}%
\end{equation}
For large-$N_{c}$we get:%
\begin{equation}
g_{QCD}^{2}\equiv g^{2}(\mu)=\frac{8\pi^{2}}{\left(  \frac{11}{3}N_{c}\right)
}\frac{1}{\ln\frac{\mu}{\Lambda_{QCD}}}\propto\frac{1}{N_{c}}\text{ .}%
\label{gqcd}
\end{equation}
As for the bare coupling, also the running coupling scales as $1/\sqrt{N_{c}}$
if $\Lambda_{QCD}$ and $N_{f}$ are kept $N_{c}$-independent. In Fig. \ref{running} the coupling $g(\mu)$ is plotted for $N_c=3$ and $N_c = 7$.

The fact that the coupling $g$ is a function of $\mu$ is also at the
basis of the so-called trace anomaly: the original classical invariance under dilatation symmetry (which is exact in the chiral limit) is broken by quantum fluctuations that lead the emergence of the low-energy scale $\Lambda_{QCD}$.

The behavior of the running coupling is in agreement with two crucial properties of QCD: asymptotic freedom and confinement. The former is due to the fact that the coupling becomes smaller at large energies, at which quarks and gluons interact perturbatively. The latter implies that quarks and gluons are
confined into hadrons. This is not directly provable, but it fits well with the
fact that the coupling becomes large at small energies.

The coupling constant becomes also small in the large-$N_{c}$ limit, but at
the same time the number of colors grows. So, at first it is hard to say what
it will happen in this regime. It is assumed (and one finds no
contradiction) that many of the properties of QCD are still valid in the
large-$N_{c}$ limit, among which confinement, asymptotic freedom, and
spontaneous symmetry breaking (SSB).

For completeness, we summarize below also additional symmetries (besides `local' color symmetry) of QCD for $m_{i}=0$ (more details in Ref. \cite{beyond}). To this end, the quark field $q_i$ is split into 
\begin{equation}
   q_i = q_{i,R} + q_{i,L} 
\end{equation}
with 
\begin{equation}
  q_{i,L}=\frac{1-\gamma^5}{2}q_i \text{  and  } q_{i,R}=\frac{1+\gamma^5}{2}q_i \text{ .}
  \end{equation}

(i) The dilatation transformation ($x^{\mu} \rightarrow \lambda^{-1} x^{\mu}$, together with $A_{\mu} \rightarrow \lambda A_{\mu}$ and $q_{i} \rightarrow \lambda^{3/2} q_{i}$ ) is a classical symmetry of QCD in the chiral limit, which is broken by quantum fluctuations (trace anomaly), in turn implying the emergence of the energy scale $\Lambda_{QCD}$ outlined above.

(ii) Chiral symmetry is expressed as
\begin{equation}
 U(N_{f})_{R}\times U(N_{f})_{L}\equiv U(1)_{V}\times
SU(N_{f})_{V}\times U(1)_{A}\times SU(N_{f})_{A}   \text{ .}
\nonumber 
\end{equation}
 According to it, quark fields transform as  
\begin{equation}
  q_i =q_{i,L}+q_{i,R}  \rightarrow U_{L,i,j}q_{j,L} + U_{R,i,j}q_{j,R}   \text{ ,}
\end{equation}
where $U_L$ and $U_R$ are $3 \times 3$ unitary matrices that mix flavor (but no color!) degrees of freedom. This symmetry undergoes SSB:  
\begin{equation}
    SU(N_{f})_{V}\times SU(N_{f}%
)_{A}\rightarrow SU(N_{f})_{V} \text{ .}
\end{equation}

(iii) The axial symmetry $U(1)_{A}$ corresponds to the choice 
$U_L = U_R^{\dagger}  = 
\exp{(-i \alpha/2)}$. This symmetry is also broken by quantum fluctuations and the corresponding anomaly is called axial or chiral anomaly.

(iv) There is also an explicit breaking of $U(1)_{A}$ and $SU(N_{f})_{A}$ through nonzero bare
quark masses. An explicit breaking of $SU(N_{f})_{V}$ occurs if the quark masses are different.

\bigskip

In the following, in order to keep the discussion as simple as possible, we will omit -if not stated differently- the flavor index $i=u,d,...$. Yet, the main features of QCD at large-$N_c$ are not dependent on the number of flavors. Yet, whenever needed, we will explicitly mention the quark flavors as well. 

\subsection{Quark and gluon propagators}

Here we shall have a quick look at the quark and the gluon propagators. The
main message is simple: their main features are retained in
the large-$N_{c}$ limit. It means that the can be \textquotedblleft
naively\textquotedblright\ regarded as large-$N_{c}$ invariant objects.

\begin{figure}[h]
        \centering       
        \includegraphics[scale=0.35]{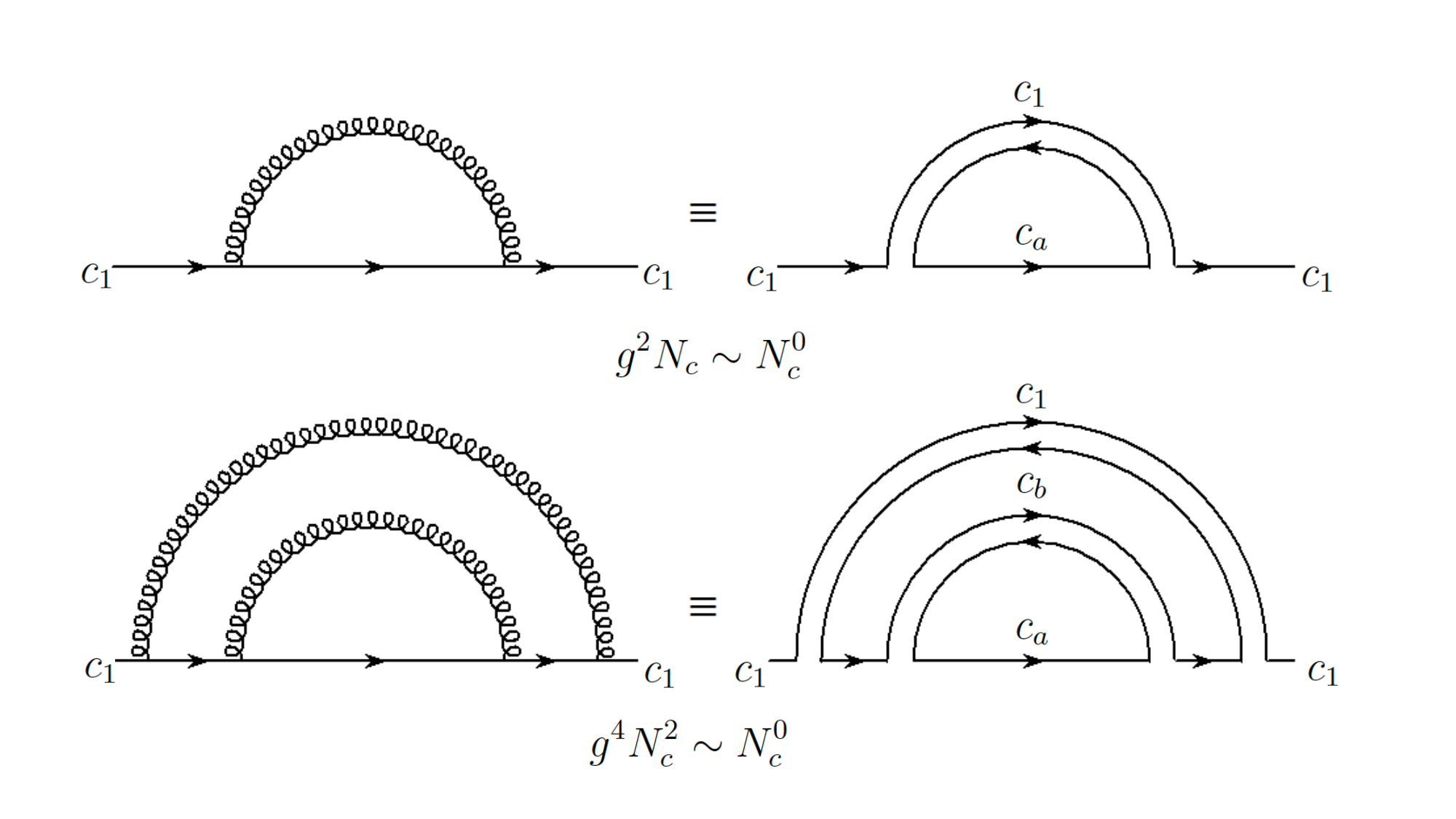}\\
        \caption{Two planar diagrams describing the self-energy of the quark (which is taken to have a specific color $C_1$ for definiteness). Both diagrams are `mass contributions' and scale as $N_c^0$. There is an infinity of such planar diagrams.}
        \label{qprop1}
   \end{figure}

Let us be more specific. For the quark propagator, there is an infinity class
of diagrams which are large-$N_{c}$ independent. They are depicted in Fig. \ref{qprop1}. These are the famous `planar diagrams' since they can be drawn on
a plane without intersection \cite{thooft,witten}. 

\begin{figure}[h]
        \centering       
        \includegraphics[scale=0.40]{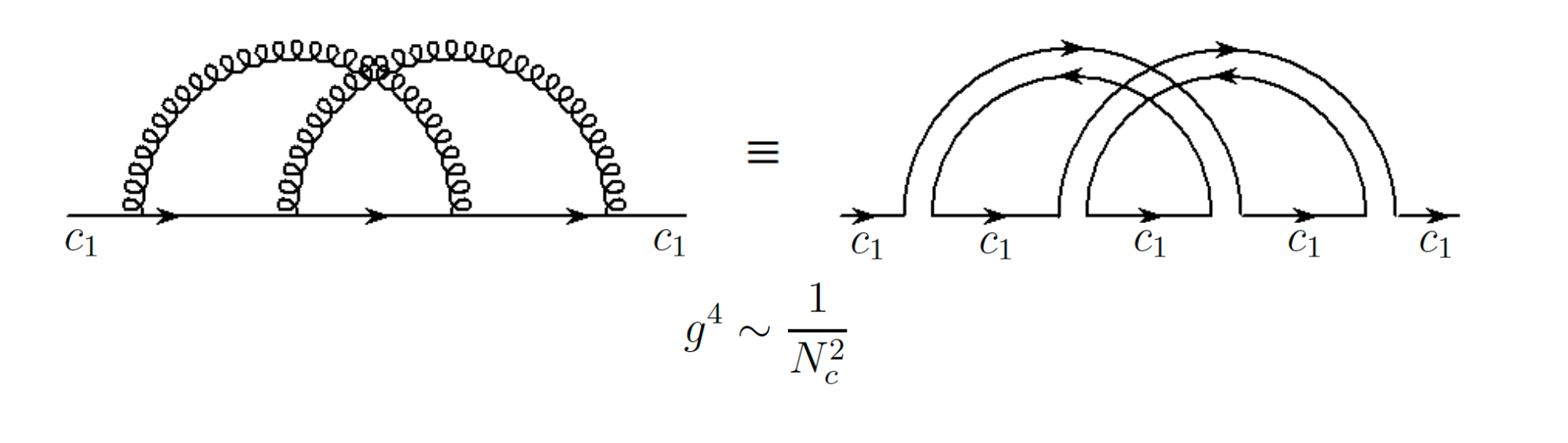}\\
        \caption{An example of a non-planar self-energy  diagram for the quark propagator. It scales as $N_c^{-2}$, thus suppressed.}
        \label{qprop3}
   \end{figure}

Admittedly, there are also non-planar diagrams which disappear in the
large-$N_{c}$ limit, thus the large-$N_{c}$ world is slightly simpler than the
one for $N_{c}=3$. An example of a non-planar (and therefore large-$N_c$ suppressed) diagram for the quark propagator can be found in Fig. \ref{qprop3}.  Yet, the main features are expected to be contained in the
large-$N_{c}$ dominant terms, which survive the limit $N_{c}\rightarrow\infty$.

An important consequence is that the most important properties related to low-energy QCD,
among which SSB \cite{coleman1}, are still valid for large values of $N_c$. For instance, the quark $u$
develops a constituent mass due to SSB as \cite{njl}:
\begin{equation}
m_{u}^{\text{bare}}\simeq3\text{ MeV}\rightarrow m_{u}^{\ast}\simeq300\text{
MeV,}%
\end{equation}
which holds also for arbitrary large values of $N_{c}.$ Since the dominant contributions
to the quark propagator are large-$N_{c}$ independent, the constituent mass
$m_{u}^{\ast}\simeq300$ MeV scales as $N_{c}^{0}.$ Then, as a consequence the
whole low-energy mesonic phenomenology is quite similar: the pions and kaons are still
(quasi-)Goldstone bosons, the vector particles carry a mass of about $2m_{u}^{\ast},$
see later on for more details on mesons. An important exception regards the
chiral anomaly, which indeed goes away for $N_{c}\rightarrow\infty$. In the
chiral limit, the mass of the singlet $\eta_{0}$ scales as $m_{\eta_{0}}%
^{2}\propto N_{c}^{-1}$ \cite{witteneta}, thus for $N_f =3$ a full nonet of Goldstone bosons is actually
realized in the chiral limit for large $N_c$.

The gluon propagator is dressed by large-$N_{c}$ independent planar diagrams,
that are expected to be responsible for its major properties. In this sense,
this feature is analogous to the quark propagator. One may roughly speak about
an effective gluon mass of about $m_{g}^{\ast}\simeq800$ MeV \cite{gluonmass}, even though the
term `mass' should be used with extreme care: one should better refer to an
energy scale entering into the propagator without breaking local color gauge invariance \cite{gluonmass2}.
This effective energy scale is $N_c$-independent. As a consequence, glueballs carry a mass
(starting at) about $2m_{g}^{\ast}$, which is also $N_c$ independent, as we shall
discuss in more detail in Sec. 3.

\begin{figure}[h]
        \centering       
        \includegraphics[scale=0.38]{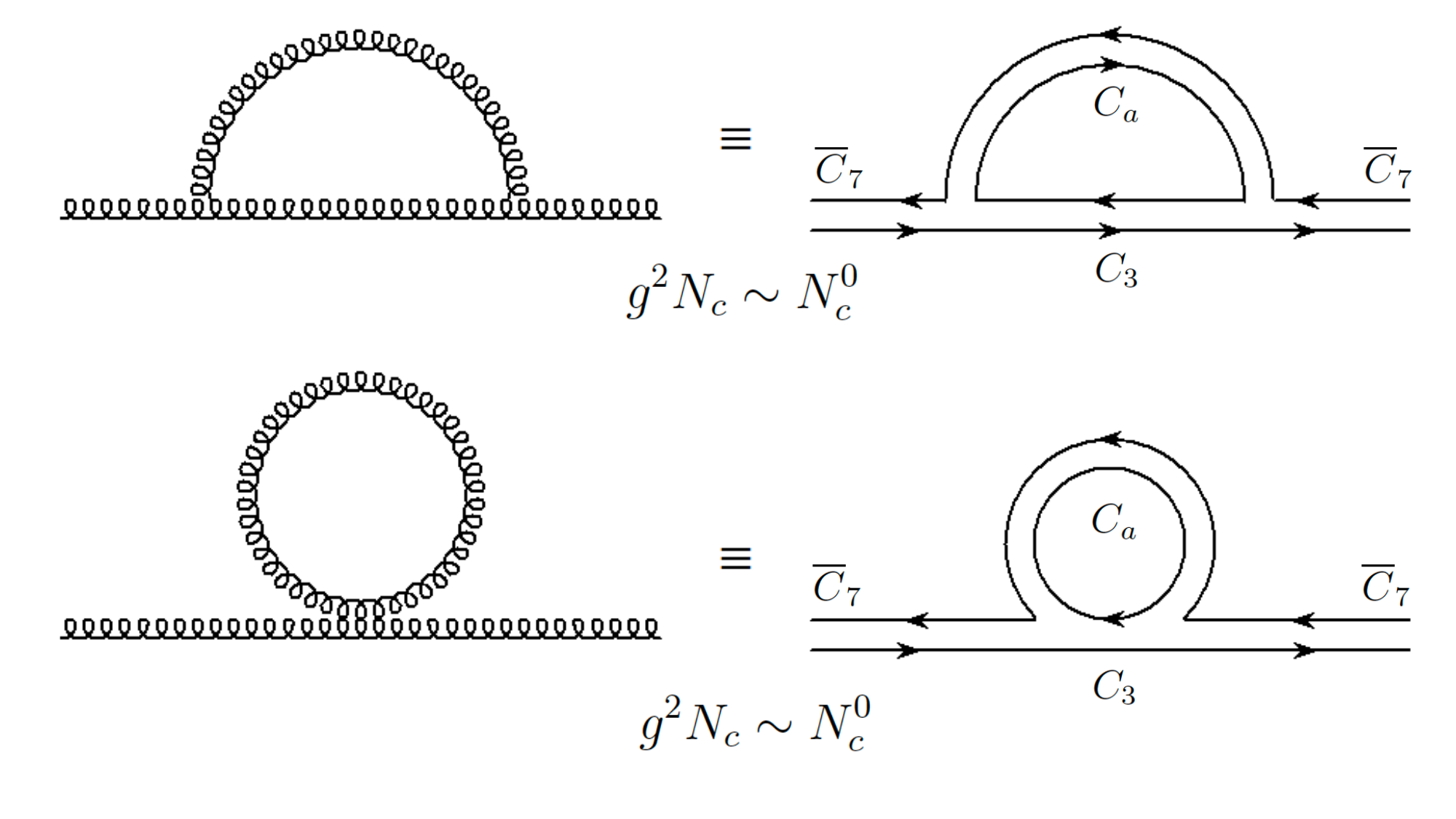}\\
        \caption{Two planar diagrams describing the self-energy of the gluon (which is taken to have a specific color $C_3 \bar{C}_7$ as an explicit example). Both diagrams are `mass contributions' and scale as $N_c^0$.}
        \label{gprop}
   \end{figure}

Summarizing, the quark and the gluon propagators contain a class of dominant
$N_c$-independent ($\propto N_c^0$) contributions. For our purposes, these propagators can be seen as
independent from the number of colors.

\subsection{Brief recall of mesons and baryons at large-$N_c$}

Quarks and gluons are not the physical states that hit our detectors. They are
confined into hadrons, i.e. mesons (integer spin) and baryons (semi-integer spin).

A \textit{conventional meson} is a meson constructed out of a quark and an
antiquark. Although it represents only one of (actually infinitely many)
possibilities to build a meson, the vast majority of mesons of the PDG \cite{pdg} can be
consistently (and successfully) interpreted as belonging to a quark-antiquark multiplet
(see also the results of the quark model \cite{isgur}). Mesons are classified
according to their total spin $J$, parity $P$, and charge conjugation $C$, forming
multiplets denoted with $J^{PC}$.

Moreover, in the quark model one may express a quark-antiquark state $Q$ using
the radial quantum number $n=1,2,3,...,$ the angular quantum number
$L=0,1,2,3,...\equiv S,P,D,F,...,$ the spin part $S=0,1,$ the flavor
composition, and finally the color wave function (that is crucial for this
work). 

Following this spectroscopic notation a meson is classified as 
\begin{equation}
    n ^{2S+1}L_{J}  \text{ ,}
\end{equation}
where $J$ is the total spin arising from the proper
combination of $L$ and $S.$ We remind that $P$ and $C$ are calculated
as:
\begin{equation}
    P=(-1)^{L+1} \text{   ,  }  C=(-1)^{L+S} \text{ .}
\end{equation}

In general, the wave function of a quarkonium state $Q$ takes the form%
\begin{gather}
\left\vert Q\text{ with }n\text{ }^{2S+1}L_{J}\text{ and }J^{PC}\right\rangle
\nonumber\\
=\left\vert \text{radial part }n\right\rangle \left\vert \text{angular part
}L\right\rangle \left\vert \text{spin part }S\right\rangle \left\vert
\text{flavor}\right\rangle \left\vert \text{color }\right\rangle \text{ .}%
\end{gather}

As an example, the wave function of the vector meson 
$\rho^{+} \equiv u \bar{d}$ reads%
\begin{equation}
\left\vert \rho^{+}\right\rangle =N\left\vert n=1\right\rangle \left\vert
L=0\right\rangle \left\vert S=1(\uparrow\downarrow+\downarrow\uparrow
)\right\rangle \left\vert u\bar{d}\right\rangle \left\vert \bar{R}R+\bar
{G}G+\bar{B}B\text{ }\right\rangle \text{ ,} \label{rho}%
\end{equation}
where $N$ is an overall normalization. The properly normalized color part is
\begin{equation}
\left\vert Q\text{-color }\right\rangle =\frac{1}{\sqrt{3}}\left\vert \bar
{R}R+\bar{G}G+\bar{B}B\text{ }\right\rangle \text{ .}%
\end{equation}
Interestingly, this is the color wave function of any quarkonium, independently on
the other quantum numbers. This combination is colorless, in the sense that
it is invariant under any (local) $SU(N_{c}=3)$ color transformations.

As we shall prove explicitly later, in the large-$N_{c}$ limit quark-antiquark
mesons retain their mass but become very narrow. For a generic $N_{c}$ the
color wave function takes the form:
\begin{equation}
\left\vert Q\text{-color }\right\rangle =\frac{1}{\sqrt{N_{c}}}\left\vert
\bar{C}_{1}C_{1}+\bar{C}_{2}C_{2}+...+\bar{C}_{N_{c}}C_{N_{c}}\text{
}\right\rangle \text{ ,} \label{qcolor}%
\end{equation}
which is invariant under (local) $SU(N_{c})$ color transformations. This fact
can be easily seen by considering%
\begin{equation}
\left\vert Q\text{-color }\right\rangle \simeq\frac{1}{\sqrt{N_{c}}}\sum
_{a=1}^{N_{c}}\bar{q}_{a}q_{a}\left\vert 0\right\rangle
\text{ .}
\end{equation}
A generic color transformation implies: 
\begin{equation}
q_{a}\rightarrow U_{ab}q_{b} 
\text{    ,   }
q_{a}^{\dagger}\rightarrow\left(  U_{ac}q_{c}\right)  ^{\dagger}%
=q_{c}^{\dagger}U_{ac}^{\ast}=q_{c}^{\dagger}\left(  U^{\dagger}\right)
_{ca}  \text{ ,}
\end{equation}
thus
\begin{equation}
\bar{q}_{a}q_{a}\rightarrow\bar{q}_{c}\left(  U^{\dagger}\right)  _{ca}%
U_{ab}q_{b}=\bar{q}_{a}q_{a}
\end{equation}
where $\left(  U^{\dagger}\right)  _{ca}U_{ab}=\delta_{bc}$ follows from
$U^{\dagger}U=1$ (sum over indices understood).

Other mesons (such as glueballs, hybrids,...) have more complicated wave
functions, see later for their detailed study.

The same procedure above can be carried out for conventional baryon states,
where a conventional baryon is a three-quark state. Even if the remaining part
of their w.f. is more complicated, the normalized color part is pretty simple:
\begin{equation}
\left\vert B\text{-color }\right\rangle =\frac{1}{\sqrt{6}}\left\vert
RGB+BRG+GBR-GRB-BGR-RBG\right\rangle \text{ ,}%
\end{equation}
which is invariant under $SU(N_{c}=3)$ color transformations. The extension to
$a$ generic number of colors gives:
\begin{equation}
\left\vert B\text{-color }\right\rangle =\frac{1}{\sqrt{N_{c}!}}%
\varepsilon_{a_1a_2...a_{N_c}}\left\vert C_{a_1}C_{a_2}...C_{a_{N_c}}\right\rangle \text{ }%
\end{equation}
or, by using quark fields:
\begin{equation}
\left\vert B\text{-color }\right\rangle \simeq\frac{1}{\sqrt{N_{c}!}%
}\varepsilon_{a_1a_2...a_{N_c}}q_{a_1}q_{a_2}...q_{a_{Nc}}\left\vert 0\right\rangle
\text{ .}
\end{equation}
In fact, under $SU(N_c)$ color transformations, one has:
\begin{align}
\varepsilon_{a_{1}a_{2}...a_{N_{c}}}q_{a_{1}}q_{a_{2}}...q_{N_{c}}  &
\rightarrow\varepsilon_{a_{1}a_{2}...a_{N_{c}}}U_{a_{1}a_{1}^{\prime}}%
U_{a_{2}a_{2}^{\prime}}...U_{a_{N_{c}}a_{N_{c}}^{\prime}}q_{a_{1}^{\prime}%
}q_{a_{2}^{\prime}}...q_{a_{N_{c}}^{\prime}}\nonumber\\
& =\varepsilon_{a_{1}^{\prime}a_{2}^{\prime}...a_{N_{c}}^{\prime}}%
q_{a_{1}^{\prime}}q_{a_{2}^{\prime}}...q_{a_{N_{c}}^{\prime}}%
\end{align}
where we have used that
\begin{equation}
\varepsilon_{a_{1}a_{2}...a_{N_{c}}}U_{a_{1}a_{1}^{\prime}}U_{a_{2}%
a_{2}^{\prime}}...U_{a_{N_{c}}a_{N_{c}}^{\prime}}=\varepsilon_{a_{1}^{\prime
}a_{2}^{\prime}...a_{N_{c}}^{\prime}}\text{ ,}%
\end{equation}
being a consequence of $\det U=1,$ namely:%
\begin{equation}
N!\det U=N!=\varepsilon_{a_{1}a_{2}...a_{N_{c}}}\varepsilon_{a_{1}^{\prime
}a_{2}^{\prime}...a_{N_{c}}^{\prime}}U_{a_{1}a_{1}^{\prime}}U_{a_{2}%
a_{2}^{\prime}}...U_{a_{N_{c}}a_{N_{c}}^{\prime}}\text{ .}%
\end{equation}

\subsection{Large-$N_{c}$: recall of basic results}

We present here a short summary of known large-$N_{c}$ rules
\cite{witten,lebedlect}. In the next sections we will re-derive them following (relatively simple) bound-state equations forming these states.

\begin{enumerate}
\item The masses of quark-antiquark states $Q\equiv\overline{q}q$, glueballs
$G\equiv gg$, and hybrids mesons $H\equiv\overline{q}qg$ are constant for
$N_{c}\rightarrow\infty$:
\begin{equation}
M_{Q}\propto N_{c}^{0}\text{ , }M_{G}\propto N_{c}^{0}\text{ , }M_{H}\propto
N_{c}^{0} \text{ .}
\end{equation}

\item The interaction between $n_{Q}$ quark-antiquark states $Q\equiv
\left\vert \overline{q}q\right\rangle $ scales as
\begin{equation}
A_{n_{Q}Q}\propto\frac{N_{c}}{N_{c}^{n_{Q}/2}}\text{ for }n_{Q}\geq1\text{ .}%
\end{equation}
This implies that the amplitude for a $n_Q$-meson scattering process becomes
smaller and smaller for increasing $N_{c}$. In particular the decay amplitude
is realized for $n_Q=3$, ergo $A_{\text{decay}}\propto N_{c}^{-1/2}$, implying
that the width scales as $\Gamma\propto1/N_{c}$. Conventional quarkonia become
very narrow for large $N_{c}.$

\item The interaction amplitude between $n_{G}$ glueballs is
\begin{equation}
A_{n_{G}G}\propto\frac{N_{c}^{2}}{N_{c}^{n_{G}}}\text{ for }n_{G}\geq1\text{ ,}%
\end{equation}
which is smaller than in the quarkonium case.

\item The interaction amplitude between $n_{Q}$ quarkonia and $n_{G}$
glueballs behaves as
\begin{equation}
A_{\left(  n_{Q}Q\right)  \left(  n_{G}G\right)  }\propto\frac{N_{c}}%
{N_{c}^{n_{Q}/2}N_{c}^{n_{G}}}\text{ for }n_{Q}\geq1\text{ ,}%
\end{equation}
thus the mixing ($n_{G}=n_{Q}=1$) scales as $A_{\text{mixing}}\propto
N_{c}^{-1/2}.$ Then, also the glueball-quarkonium mixing is suppressed for
$N_{c}\gg1$. Note, for $n_{G}=0$ one finds the correct interaction for
$n_{Q}$ quarkonia.

\item The amplitude for $n_Q$ quarkonia and $n_H$ hybrids scales as
\begin{equation}
A_{\left(  n_{Q}Q\right)  \left(  n_{H}H\right)  }\propto\frac{N_{c}}%
{N_{c}^{n_{Q}/2}N_{c}^{n_{H}/2}}\text{ for }n_{Q}+n_{H}\geq1\text{ .}%
\end{equation}
For $n_Q = n_H =1$ one recovers that the quarkonium-hybrid mixing scales as $N_c^0$, implying that quarkonia and hybrids behave in the same way at large-$N_c$.

\item For the general case of $n_{H}$ quarkonia, $n_{G}$ glueballs, and
$n_{H}$ hybrids one has:%
\begin{equation}
A_{\left(  n_{Q}Q\right)  \left(  n_{G}G\right)  \left(  n_{H}H\right)
}\propto\frac{N_{c}}{N_{c}^{n_{Q}/2}N_{c}^{n_{G}}N_{c}^{n_{H}/2}}\text{ for
}n_{Q}+n_{H}\geq1\text{ .}%
\end{equation}

\item Four-quark states (both as molecular objects and diquark-anti-diquark
objects) tend to fade away at large $N_c$. In fact, a part from (eventually existing, but not yet proven) peculiar tetraquarks
\cite{weinberg}, these objects typically do not survive in the large-$N_{c}$ limit.

\item Baryons are made of $N_{c}$ quarks for an arbitrary $N_{c}.$ As a
consequence
\begin{equation}
M_{B}\propto N_{c}\text{ .}%
\end{equation}

\item Interactions involving baryons: the baryon-baryon-meson interaction
scales as $N_{c}^{1/2},$ while baryon-baryon scattering goes as $N_{c}$.
In particular, for an arbitrary number of $\bar{B}B$ pairs, as well as for $n_Q$ quarkonia, $n_G$ glueballs, and $n_H$ hybrids, one has:
\begin{equation}
A_{ \left( \bar{B}B ... \right)\left(  n_{Q}Q\right)  \left(  n_{G}G\right)  \left(  n_{H}H\right)
}\propto\frac{N_{c}}{N_{c}^{n_{Q}/2}N_{c}^{n_{G}}N_{c}^{n_{H}/2}} \text{ .}%
\end{equation}
\end{enumerate}

Summing up, the large-$N_{c}$ limit is a firm theoretical method which explains
why the quark model works. In fact, a decay channel for a certain meson causes
quantum fluctuations: the propagator of the meson is dressed by loops of other
mesons. For instance, the state $\rho^{+}$ decays into $\pi^{+}\pi^{0},$ thus
the $\rho$-meson is dressed by loops of pions. In the end, one
finds that the wave function of the $\rho$-meson is given by:%
\begin{equation}
\left\vert \rho^{+}\right\rangle =a\left\vert u\bar{d}\right\rangle
+b\left\vert \pi^{+}\pi^{0}\right\rangle +...\text{ ,}%
\end{equation}
where the full expression of $\left\vert u\bar{d}\right\rangle $ is given in
Eq. (\ref{rho}). Being $a\propto N_{c}^{0}$ and $b\propto N_{c}^{-1/2},$ we
understand why the quark-antiquark configuration dominates. Dots refer to
further contributions which are even more suppressed.

Yet, for $N_{c}=3$ there are some mesons for which the meson-meson component
dominates. These are for instance, the light scalar mesons such as $a_0(980)$, see Sec. 3.5.

\section{Large-$N_{c}$ results for mesons}
In this Section we deal with mesons. First, we discuss conventional quarkonia states, then various exotic configurations: glueballs, hybrids, and (briefly) four-quark states.

\subsection{Quark-antiquark mesons}
A quark-antiquark meson has a rather simple color wave function, see Eq.
(\ref{qcolor}), that we rewrite here for convenience:
\begin{equation}
\left\vert Q\text{-color }\right\rangle =\frac{1}{\sqrt{N_{c}}}\left\vert
\bar{C}_{1}C_{1}+\bar{C}_{2}C_{2}+...+\bar{C}_{N_{c}}C_{N_{c}}\text{
}\right\rangle \text{ .} \label{qcolor2}%
\end{equation}
How does such a bound state emerge? For simplicity, let
us consider the processes that mix two elements of the wave function, for instance:%

\begin{equation}
\bar{C}_{1}C_{1}\rightarrow\bar{C}_{2}C_{2} \text{ .}
\label{c1c2}
\end{equation}
This particular transition implies that an initial state with color 1 and
anticolor 1 transforms into color 2 and anticolor 2. (For $N_{c}=3$ that would
correspond to e.g. $\bar{R}R$ into $\bar{G}G$.) Of course, any other example, such
as $\bar{C}_{3}C_{3}\rightarrow\bar{C}_{7}C_{7}$, is equally good. 
The important point is that
we start from a possible component of a quarkonium state and end up in another
component of its wave function. This is so because an eventual bound state would couple to any color combination $\bar{C}_a C_a$ with the same strength, and would appear as an intermediate state for processes of the type (\ref{c1c2}). In particular, close to the mass of the quarkonium, the $s$-channel becomes dominant.
Note, for the moment we do not `care' about
the normalization $1/\sqrt{N_{c}}$ entering in eq. (\ref{qcolor}), but we
simply study the amplitude for the transition of Eq. (\ref{c1c2}).

\begin{figure}[h]
        \centering       \includegraphics[scale=0.45]{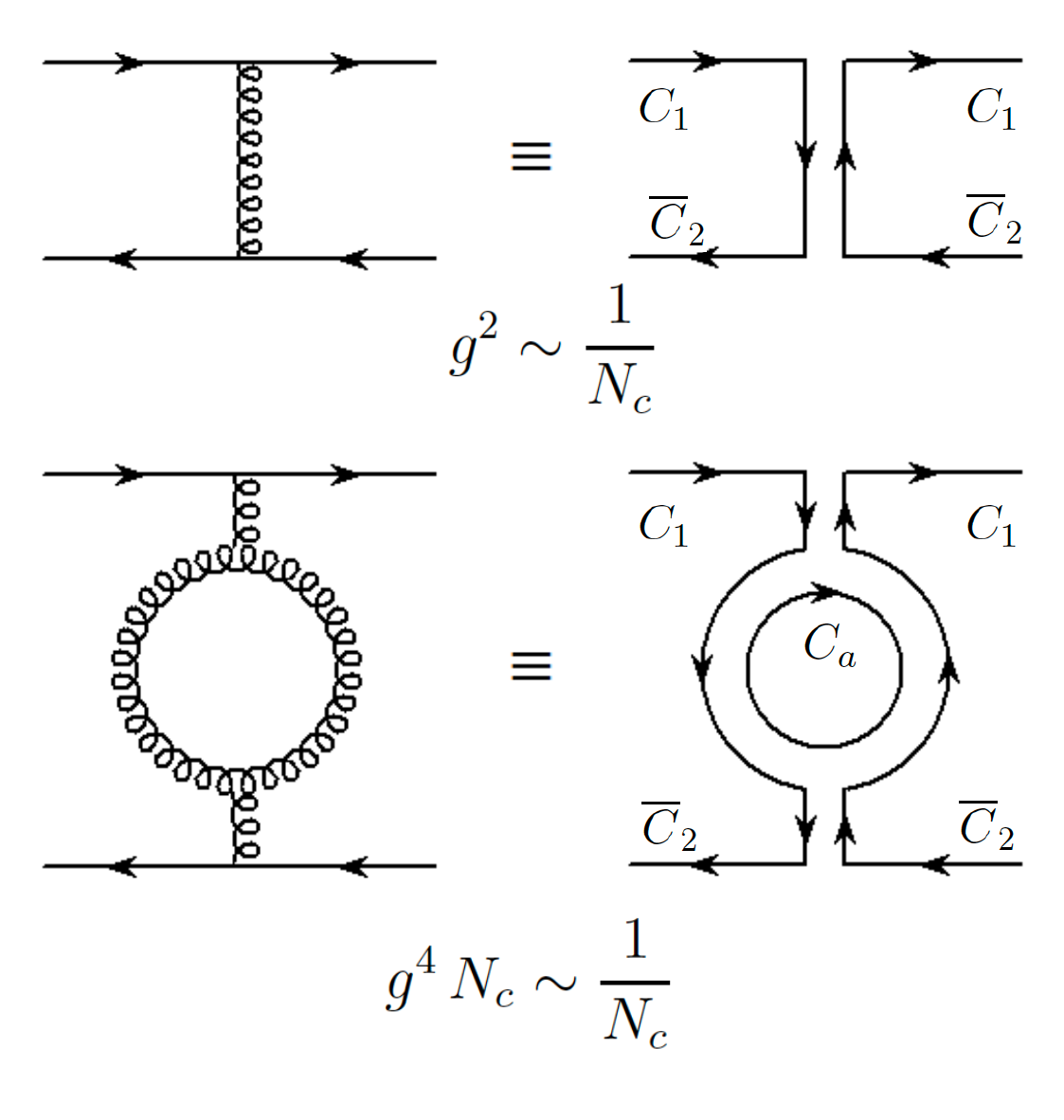}\\
        \caption{Examples of dominating diagrams for the scattering process $\bar{C}_1C_1 \rightarrow \bar{C}_2C_2$. These diagrams scale as $N_c^{-1}$. Of course, one could take any other color combination, such as   $\bar{C}_7C_7\rightarrow \bar{C}_{11}C_{11}$.}
        \label{qq1}
   \end{figure}

\begin{figure}[h]
        \centering       \includegraphics[scale=0.45]{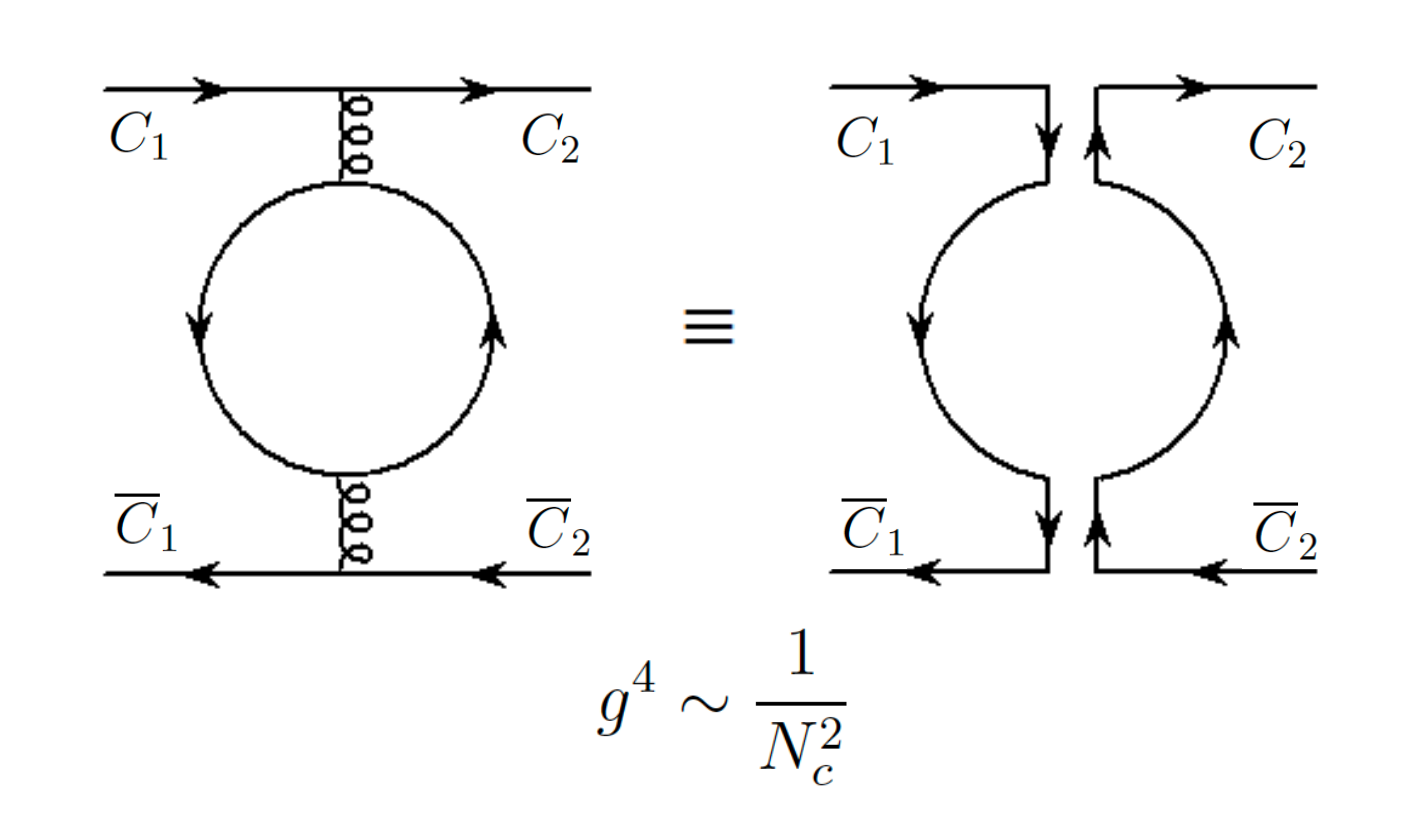}\\
        \caption{Example of a subleading diagram for the scattering process $\bar{C}_1C_1 \rightarrow \bar{C}_2C_2$. These types of diagram scale as $N_c^{-2}$.}
        \label{qq2}
   \end{figure}

\begin{figure}[h]
        \centering       \includegraphics[scale=0.40]{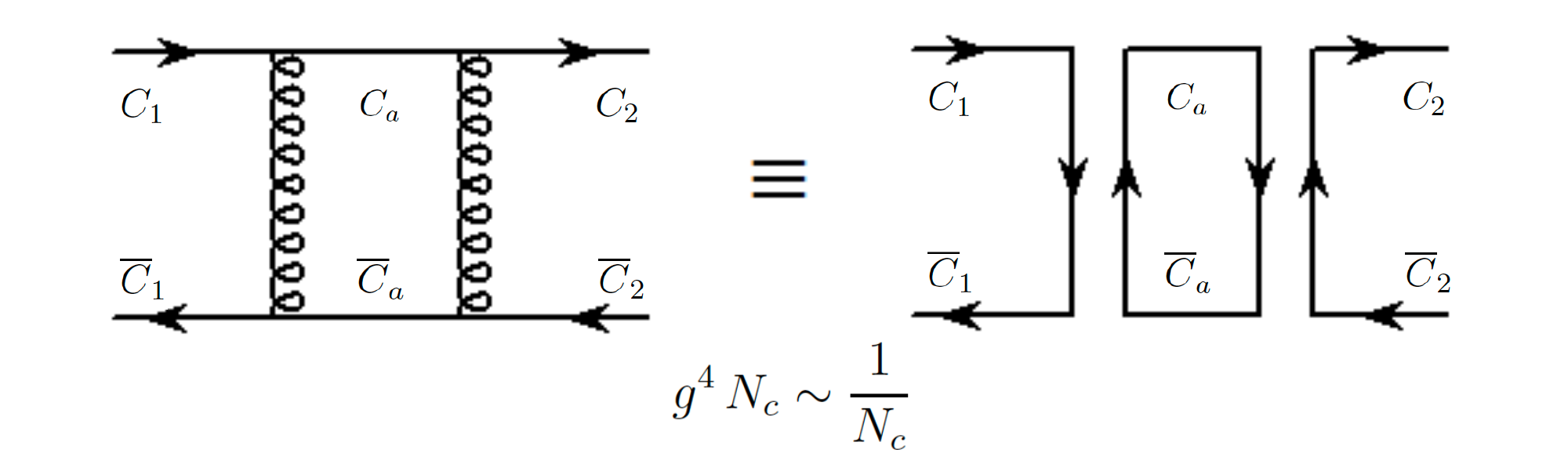}\\
        \caption{Another dominating diagram for the scattering process $\bar{C}_1C_1 \rightarrow \bar{C}_2C_2$, that involves an intermediate state with an arbitrary color $\bar{C}_aC_a$ for $a=1,...,N_c$. It scales as $N_c^{-1}$.}
        \label{qq3}
   \end{figure}

The simplest process of this type is depicted in Fig. \ref{qq1} (upper part), where it is evident that the dominant
amplitude for $\bar{C}_{1}C_{1}\rightarrow\bar{C}_{2}C_{2}$ scales as
$1/N_{c}.$ Interestingly, loop processes of the type of Fig. \ref{qq1} (lower part) also scale
in this way. Of course, there are also subleading terms that scale as
$1/N_{c}^{2}$, see Fig. \ref{qq2}, which can be dismissed in the large-$N_{c}$limit. In Fig. \ref{qq3} we show another type of diagrams, which scales also as $N_c^{-1}$ and displays an intermediate loop with any possible color combination. 

How to study the emergence of bound states in this context? Of course, the
full problem is complicated and one would need a Bethe-Salpeter approach. Yet, a
simple and in many respects successful approach makes use of a quartic separable
interaction, such as in the NJL model \cite{njl,volkov}. As previously mentioned, the large-$N_{c}$
counting is independent on the details of the employed approach, thus the results that
we will present are general.

\begin{figure}[h]
        \centering       \includegraphics[scale=0.45]{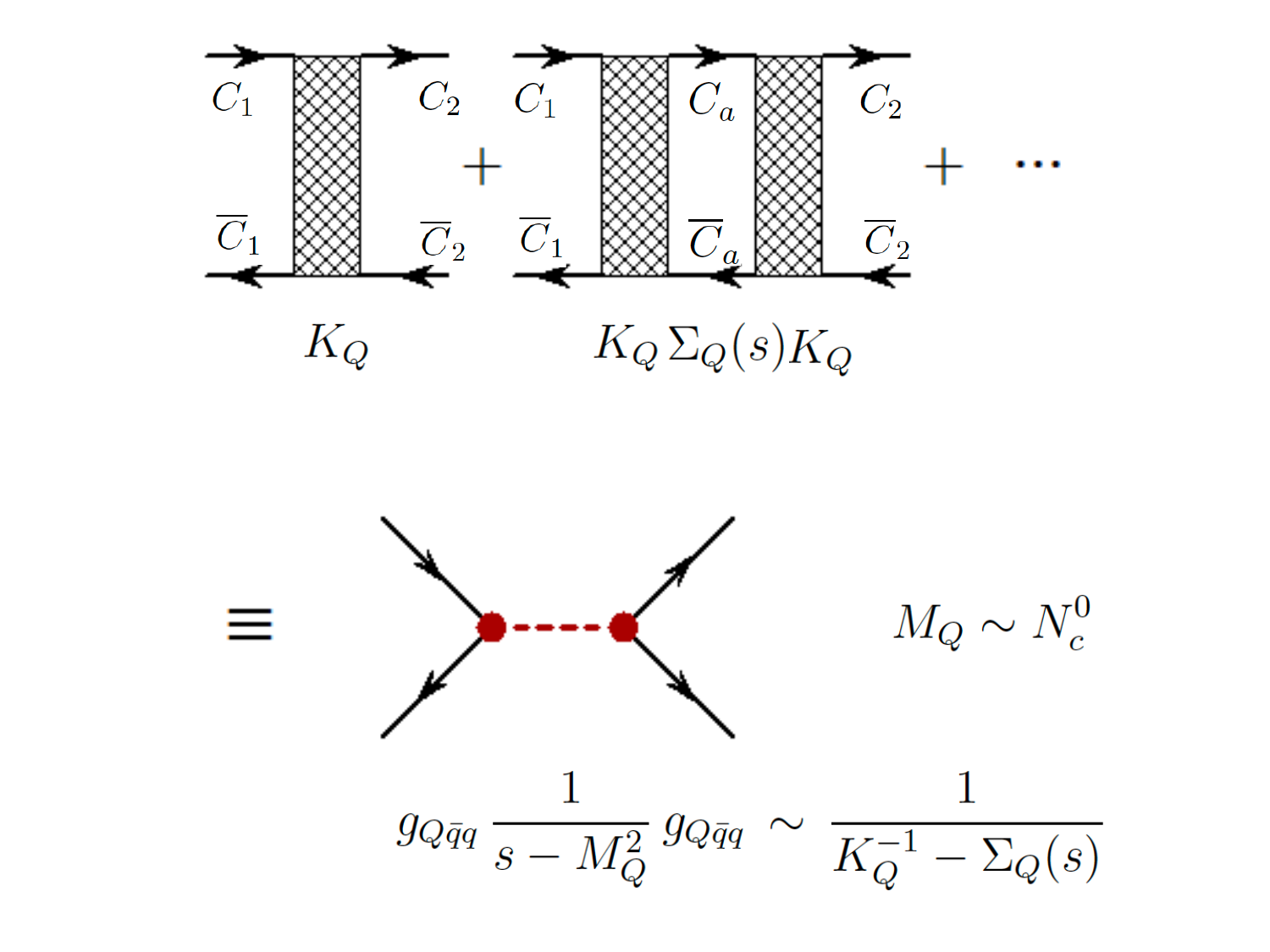}\\
        \caption{Generation of a quark-antiquark meson (red (dashed) line)  upon resummation of diagrams for the illustrative process $\bar{C}_1C_1 \rightarrow \bar{C}_2C_2$. }
        \label{qqJ}
   \end{figure}

\begin{figure}[h]
        \centering       \includegraphics[scale=0.45]{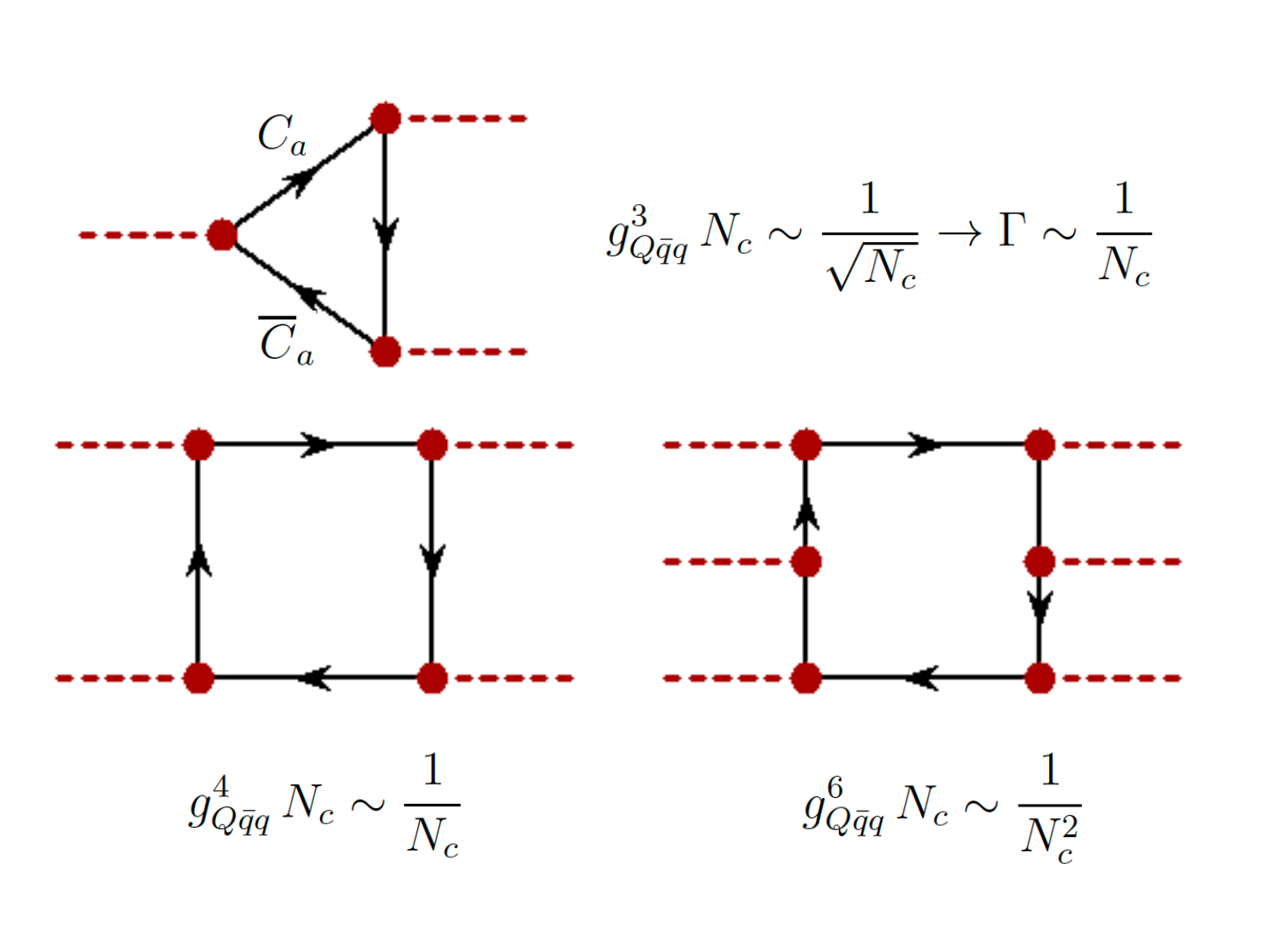}\\
        \caption{Conventional $\bar{q}q$ mesons $Q$ are in red (dashed), quarks in black (solid). Up: decay of a conventional $\bar{q}q$ meson into two conventional $\bar{q}q$ mesons via a loop of quarks. The leading amplitude scales as $N_c^{-1/2}$, hence the decay width scales as $N_c^{-1}$. Down: two-body scattering process of conventional $Q$ mesons, whose leading order is $N_c^{-1}$, thus the cross-section goes as $N_c^{-2}$, and three-body scattering process, , whose leading order is $N_c^{-2}$, thus the cross-section goes as $N_c^{-4}$. }
        \label{Qphen}
   \end{figure}
   
Let us consider a generic `colorless' current for an arbitrary quarkonium
meson $Q$:
\begin{equation}
J_{Q}(x)=\sum_{a=1}^{N_{c}}\bar{q}^{(a)}\Gamma q^{(a)}\equiv C_{1}\bar{C}%
_{1}+C_{2}\bar{C}_{2}+...+C_{N_{c}}\bar{C}_{N_{c}} \text{ ,}
\end{equation}
where no normalization is considered. (Note, the quantity $\Gamma$ is an appropriate combination of Dirac matrices and derivatives, that varies case by case in dependence of the mesonic quantum numbers, see Ref. \cite{anomaly1}; for instance, for pseudoscalar mesons one has $\Gamma = i\gamma^5$.)
The separable interaction term is proportional to $J_{Q}^{2}.$ The corresponding Lagrangian
takes the effective form
\begin{equation}
\mathcal{L}_{Q}=K_{Q}J_{Q}^{2} \text{ ,}\label{lQ}%
\end{equation}
where $K_{Q}$ is a coupling constant. In order to determine the scaling of
$K_{Q},$ one may consider the illustrative transition $C_{1}\bar{C}%
_{1}\rightarrow C_{2}\bar{C}_{2}$ or any other of that type, finding that
\begin{equation}
K_{Q}\propto g_{QCD}^{2} = g^2 \propto N_{c}^{-1}.
\end{equation}
We then introduce a useful notation: we define $\bar{K}_{Q}$ as a $N_{c}$-independent constant, thus:
\begin{equation}
K_{Q}=\frac{\bar{K}_{Q}}{N_{c}} \text{ .}
\end{equation}
In the following, any quantity with `bar' shall be regarded as $N_c$-independent.

The corresponding $T$-matrix $T_Q$ is obtained by properly resumming the interactions originated by the Lagrangian of Eq. (\ref{lQ}), see Fig. \ref{qqJ} for its pictorial representation. It then takes the form:
\begin{align}
iT_{Q}(s)  =iK_{Q}+iK_{Q}(-i\Sigma_{Q}(s))iK_{Q} 
+iK_{Q}(-i\Sigma
_{Q}(s))iK_{Q}(-i\Sigma_{Q}(s))iK_{Q}+...
\nonumber
\end{align}
leading to
\begin{equation}
    T_{Q}(s)   =K_{Q}+K_{Q}\Sigma_{Q}(s)K_{Q}+...=\frac{K_{Q}}{1-\Sigma
_{Q}(s)K_{Q}}=\frac{1}{K_{Q}^{-1}-\Sigma_{Q}(s)}%
\text{ ,}
\end{equation}
where $\Sigma_{Q}(s)$ is the quark-antiquark loop contribution, which scales
as:%
\begin{equation}
\Sigma_{Q}(s)=N_{c}\bar{\Sigma}_{Q}(s) \text{ .}
\end{equation}
 This result is a simple consequence of the $N_{c}$ possible loops when the
quark carries $N_{c}$ colors. The quantity $\bar{\Sigma}_{Q}(s)$ is, according to the adopted convention, independent on
$N_{c}$.

Next, the amplitude $T_{Q}(s)$ scales as
$1/N_{c}$ (indeed, each terms in the expansion is of order $1/N_{c},$ as one
can easily check). In this specific model, one can also write the explicit form of the loop function
$\Sigma_{Q}(s)$ as%
\begin{equation}
\Sigma_{Q}(s=p^{2})= -iN_{c}\int\frac{d^{4}k}{(2\pi)^{4}}Tr\left[  S_{q}\left(
p/2+k\right)  \Gamma S_{q}\left(  -p/2+k\right)  \Gamma\right]  f_{\Lambda}(k)
\text{ .}
\end{equation}
In this sense, the factor $N_{c}$ is simply a trace over color d.o.f.: this is
indeed a general result that does not depend on the model details. The
function $f_{\Lambda}(k)$ stays for a regulator (in turn, this function may arise from a nonlocal current \cite{gutsche,pionsigma}), but its specification is not needed since no explicit calculation will be performed.

The mass of the quark-antiquark bound states $M_{Q}$ corresponds to a pole of
the resummed amplitude $T_Q(s)$, hence to a zero of its denominator. As such, it is a solution of the equation
\begin{equation}
T_Q(s)^{-1} =0 \rightarrow K_{Q}^{-1}-\Sigma_{Q}(s=M_{Q}^{2})=0
\text{ .}
\label{poleq}
\end{equation}
Then, upon using $K_{Q}=\bar{K}_{Q}/N_{c}$ and $\Sigma_{Q}(s)=N_{c}\bar
{\Sigma}_{Q}(s)$, the previous equation takes the form%
\begin{equation}
\frac{N_{c}}{\bar{K}_{Q}}-N_{c}\bar{\Sigma}_{Q}(s=M_{Q}^{2})=0\rightarrow
\frac{1}{\bar{K}_{Q}}-\bar{\Sigma}_{Q}(s=M_{Q}^{2})=0 \text{ ,}
\end{equation}
which is $N_{c}$ \textit{independent}. Thus, the mass of the mesonic
quarkonium state $Q \equiv \bar{q}q$ scales as
\begin{equation}
M_{Q}\propto N_{c}^{0}\text{ .}%
\end{equation}
We were able to reproduce this very well known and general result of the large-$N_{c}$
phenomenology.

Next, let us expand the denominator $T_{Q}$ around
$s=M_{Q}^{2}$, finding:
\begin{align}
K_{Q}^{-1}-\Sigma_{Q}(s)  &  \simeq\underset{=0}{\underbrace{K_{Q}^{-1}%
-\Sigma_{Q}(M_{Q}^{2})}}-\left(  \frac{\partial\Sigma_{Q}(s)}{\partial
s}\right)  _{s=M_{Q}^{2}}(s-M_{Q}^{2})+...\nonumber\\
&  \simeq - N_{c}\left(  \frac{\partial\bar{\Sigma}_{Q}(s)}{\partial s}\right)
_{s=M_{Q}^{2}}(s-M_{Q}^{2})+..
\end{align}
Hence, the amplitude becomes:
\begin{equation}
T_{Q}(s)=\frac{1}{K_{Q}^{-1}-\Sigma_{Q}(s)}\simeq\frac{1}{-N_{c}\left(
\frac{\partial\bar{\Sigma}_{Q}(s)}{\partial s}\right)  _{s=M_{Q}^{2}}%
(s-M_{Q}^{2})}=\frac{(ig_{Q\bar{q}q})^{2}}{s-M_{Q}^{2}}%
\text{ ,}
\end{equation}
where we identify the coupling of the quarkonium  to its constituents, a quark and an
antiquark, as:
\begin{equation}
g_{Q\bar{q}q}=\frac{1}{\sqrt{N_{c}\left(  \frac{\partial\bar{\Sigma}_{Q}%
(s)}{\partial s}\right)  _{s=M_{Q}^{2}}}}=\frac{1}{\sqrt{N_{c}}}\bar{g}%
_{Q\bar{q}q}\text{ .}%
\end{equation}
Again, $\bar{g}_{Q\bar{q}q}$ is $N_{c}$ independent. Thus, the  coupling of a conventional
meson to a quark-antiquark pair $g_{Q\bar{q}q}$ scales as
$1/\sqrt{N_{c}}.$ In terms of the composite field $Q$ and the constituent
quark fields, one can write an effective interaction Lagrangian
\begin{equation}
\mathcal{L}_{Q\bar{q}q}=g_{Q\bar{q}q}QJ_{Q}\text{ .}%
\end{equation}
Such interactions enter, for example, in meson-quark chiral models
\cite{kovacscep,tripolt}, in approaches using the Weinberg compositeness
condition \cite{compcond,compo} (sometimes within nonlocal Lagrangians
\cite{gutsche,pionsigma}), as well as at intermediate stages of the
hadronization process of quark models such as the NJL one \cite{njl}.

Many results can be obtained from the previous outcomes. Let us first look at
3-leg meson interactions (see Fig. \ref{Qphen}), which is proportional to:%
\begin{equation}
A_{3Q}\propto g_{Q\bar{q}q}^{3}N_{c}\propto\frac{1}{\sqrt{N_{c}}}\text{ .}%
\end{equation}
As a consequence, the decay width of a conventional mesons into two
conventional mesons $Q\rightarrow Q_{1}Q_{2}$ scales as:%
\begin{equation}
\Gamma_{Q\rightarrow Q_{1}Q_{2}}\propto\left\vert A_{3Q}\right\vert
^{2}\propto\frac{1}{N_{c}}.
\end{equation}
This is also a very well known result of large-$N_{c}$ phenomenology.
Conventional quark-antiquark mesons become stable for $N_{c}\rightarrow\infty$
with a scaling of the type $N_{c}^{-1}.$

Similarly, the four-leg conventional meson interaction goes as (again Fig. \ref{Qphen}):
\begin{equation}
A_{4Q}\propto g_{Q\bar{q}q}^{4}N_{c}\propto\frac{1}{N_{c}}.
\end{equation}
For instance, the four-pion interaction term in an effective Lagrangian should
scale as $1/N_{c}$.

In general, the $n_{Q}$-th leg meson interaction among convectional mesons
scales as $(n_Q\geq1)$
\begin{equation}
A_{n_{Q}Q}\propto g_{Q\bar{q}q}^{n_{Q}}N_{c}\text{ }\propto N_{c}^{-n_{Q}%
/2}N_{c}=\frac{N_{c}}{N_{c}^{\frac{n_{Q}}{2}}}%
\text{ ,}
\end{equation}
in agreement with the general result quoted in Sec. 2.6 (point 2). Note, the case $n_{Q}=1$
generates $A_{Q}\sim N_{c}^{1/2},$ which coincides with the vacuum production
amplitude (and also with the weak decay constant), see below. The case
$n_{Q}=0$ implies $A_{0}\sim N_{c}$ which can be interpreted as the vacuum
contribution of quarks. In turn, the pressure generated by quarks scales as
$N_{c}$, see Sec. 5.

\bigskip

Next, we examine various additional consequences of the obtained large-$N_{c}$
scaling behavior.

\bigskip

1) \textbf{The example of the $\chi_{c,0}$ meson.}

\begin{figure}[h]
        \centering       \includegraphics[scale=0.45]{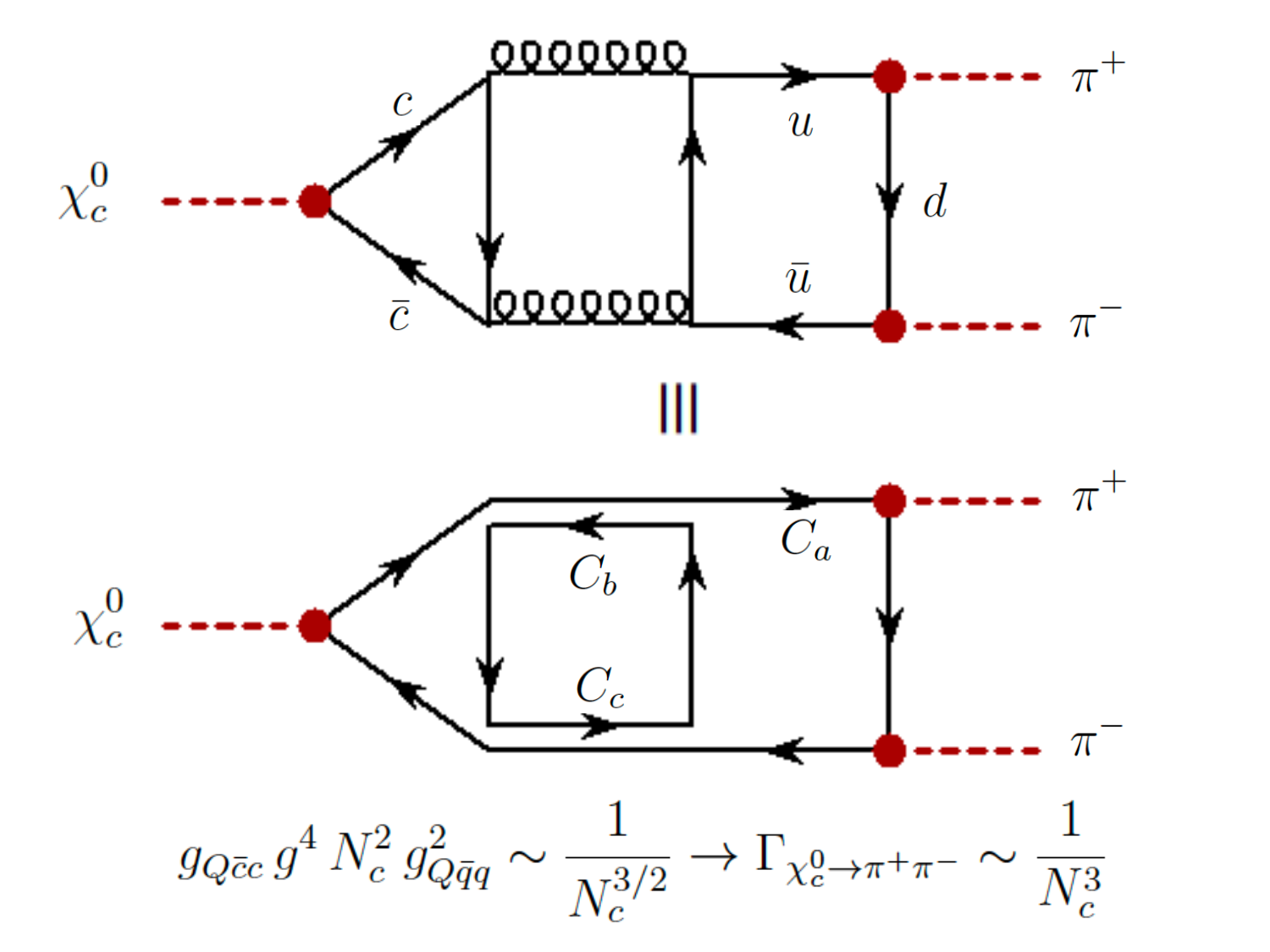}\\
        \caption{Decay of the charm-anticharm $\chi_{c,0}$ meson (red, dashed) into two $\pi$ mesons (also red, dashed). In this case, the leading decay into $\bar{D}D$ mesons as in Fig. \ref{Qphen} cannot take place because kinematically forbidden. The subleading diagram scales as $N_c^{-3/2}$, thus the decay width goes as $N_c^{-3}$, explaining why this charmonium meson is so narrow. Note, the  diagrams above also show that the flavor lines (upper part) behave differently than the color ones (lower part).}
        \label{chic0}
   \end{figure}

The $\chi_{c,0}$ meson is the ground-state scalar $\bar{c}c$ state. Its decay width is
very small \cite{pdg}. Can large-$N_{c}$ help us to understand why? Indeed, it does. The
dominant decay of the $\chi_{c,0}$ would be decays of the type $\bar{D} D$ or similar ones. The corresponding partial decay widths would be of the order of $N_{c}^{-1}$, but
cannot take place because it is kinematically forbidden. Schematically:
\begin{equation}
\Gamma_{\chi_{c,0} \rightarrow \bar{D} D}\propto g_{\chi_{c,0} \bar{D} D}^{2}\cdot
\frac{k_{D}}{M_{\chi_{c,0}}^{2}}\theta(M_{\chi_{c,0}}-2M_{D})=0%
\end{equation}
where $g_{\chi_{c,0} \bar{D} D}$ goes as $N_{c}^{-1/2}$ and $k_{D}$ is the modulus
of the three-momentum of an outgoing $D$-particle. This quantity is $N_{c}%
$-independent but is imaginary for $M_{\chi_{c,0}}<2M_{D}$. The step function
assures that in these cases the decay simply vanishes.

The $\chi_{c,0}$ can decay into light hadrons, e.g. into $\pi$-mesons. Formally, an analogous expression holds:
\begin{equation}
\Gamma_{\chi_{c,0}\rightarrow\pi\pi}\propto g_{\chi_{c,0}\pi\pi}^{2}\cdot\frac{k_{\pi}}{M_{\chi_{c,0}}^{2}}\theta(M_{\chi_{c,0}}-2M_{\pi})=g_{\chi_{c,0}\pi\pi}^{2}%
\cdot\frac{k_{\pi}}{M_{\chi_{c,0}}^{2}}\neq0\text{ .}%
\end{equation}
where $k_{\pi} = \sqrt{M^2_{\chi_{c,0}}/4-M_{\pi}^2} \sim N_c^{0}$. How does $g_{\chi_{c,0} \pi \pi}$ scale with $N_{c}?$ A simple diagrammatic analysis, see Fig. \ref{chic0}, 
shows that
\begin{equation}
g_{g_{\chi_{c,0} \pi \pi}}\propto N_{c}^{-3/2} \text{ ,}
\end{equation}
thus $\Gamma_{\chi_{c,0}\rightarrow\pi \pi}\propto N_{c}^{-3}.$ This result applies
to any similar mesonic channel. 
We thus find that:
\begin{equation}
\Gamma_{\chi_{c,0}\rightarrow\text{mesons}}\propto N_{c}^{-3}%
\text{ ,}
\end{equation}
or even smaller.
This explains why these decays are so suppressed. This result holds for any mesons whose (would be large-$N_c$) dominant decays are kinematically forbidden, most notably for the famous charm-anticharm $j/\psi$ state (where, however, because of $C$-parity, three intermediate gluons occur).
Indeed, this general outcome is a  realization of the so-called OZI (Okubo, Zweig, and Iizuka) rule, e.g. \cite{ozi}, according to which diagrams in which the quark lines are disconnected are suppressed. In this respect, the OZI rule can be understood as a consequence of the large-$N_c$ results.

\bigskip

\begin{figure}[h]
        \centering       \includegraphics[scale=0.45]{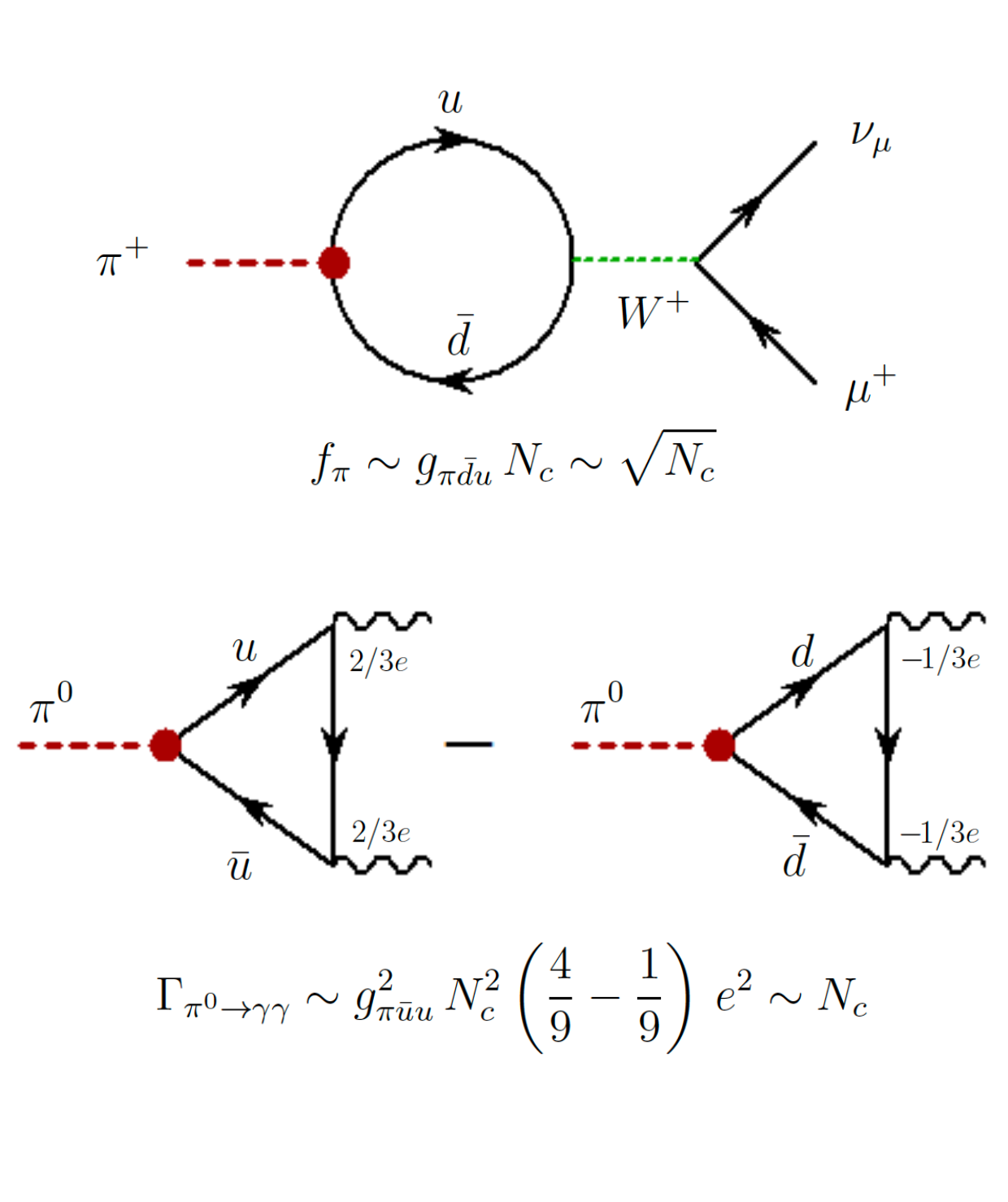}\\
        \caption{Schematic diagrams leading to the large-$N_c$ scaling of the weak decay constant $f_{\pi}$ and $\pi^0 \rightarrow \gamma \gamma$.}
        \label{Cons-Page-2}
   \end{figure}

\bigskip

2) \textbf{Pion decay constant and }$\pi^{0}\rightarrow\gamma\gamma$.

The pion decay constant $f_{\pi}$ refers to the quark-antiquark annihilation that forms it. It enters as a part of the amplitude
of the weak decay of $\pi^{+}$, for which the chain $\pi^{+}\rightarrow
W^{+}\rightarrow\mu^{+}\nu_{\mu}$ takes place, see Fig. \ref{Cons-Page-2}. It turns out hat%
\begin{equation}
f_{\pi}\propto N_{c}^{1/2} 
\text{ .}
\end{equation}
Indeed, this is the same scaling of the chiral condensate, see below.
The correct scaling can be also seen by writing the formula for $f_{\pi}$ as
\begin{equation}
f_{\pi}\sim g_{\pi^{+}u\bar{d}}\Sigma_{\pi}(s=M_{\pi}^{2})=\frac{\bar{g}%
_{\pi^{+}u\bar{d}}}{\sqrt{N_{c}}}N_{c}\bar{\Sigma}_{\pi}(s=M_{\pi}^{2}%
)\sim\sqrt{N_{c}}%
\text{ .}
\end{equation}
This result is indeed independent on the chosen quark-antiquark meson: the
weak decay constant of a generic conventional meson scales as $\sqrt{N_{c}}$.
[Note, while the scaling is correct, the expression $g_{\pi^{+}u\bar{d}}%
\Sigma_{\pi}(s=M_{\pi}^{2})$ has dimension Energy$^{2}$, but $f_{\pi}$ has dimension energy. This is due to the fact that the expression $g_{\pi^{+}u\bar{d}}%
\Sigma_{\pi}(s=M_{\pi}^{2})$ is indeed sufficient to determine the
large-$N_{c}$ scaling, but is not enough for an actual calculation of the decay constant. A closer inspection shows that $g_{\pi^{+}u\bar{d}}%
\Sigma_{\pi}(s=M_{\pi}^{2}) \propto M_{\pi}f_{\pi}$. For a detailed calculation that includes the large-$N_c$ discussion within a qualitatively similar approach, see Ref. \cite{volkovnc}. We also refer to point 3 below for a connection of this quantity to chiral models.

For what concerns the decay of $\pi^{0}$ into $\gamma\gamma,$ one obtains the
amplitude:%
\begin{equation}
A_{\pi^{0}\gamma\gamma}\propto\left[  g_{\pi^{0}\bar{u}u}\left(  \frac{2e}%
{3}\right)  ^{2}+g_{\pi^{0}\bar{d}d}\left(  -\frac{e}{3}\right)  ^{2}\right]
N_{c}\simeq g_{\pi^{0}\bar{u}u}N_{c}\left(  \frac{4}{9}-\frac{1}{9}\right) \text{ ,}
\end{equation}
where $g_{\pi^{0} \bar{d}d}\simeq-g_{\pi^{0}\bar{u}u}$ has been used (this
comes from $\left\vert \pi^{0}\right\rangle =\frac{1}{\sqrt{2}}\left\vert
\bar{u}u-\bar{d}d\right\rangle $.) 
Since $g_{\pi^{0}\bar{u}u}$ scales as
$N_{c}^{-1/2},$ one finds:
\begin{equation}
\Gamma_{\pi^{0}\rightarrow\gamma\gamma}\propto g_{\pi^{0}\bar{u}u}^{2}%
N_{c}^{2}\propto N_{c}\text{ .}%
\end{equation}
(If one would neglect the scaling of the coupling, a $N_{c}^{2}$ dependence
would emerge, but that is not the correct scaling).

The result above is valid as long as the electric charges of the quarks are
left untouched. This hypothesis is meaningful if we consider QCD only, yet
things change when other interactions are taken into account. In fact, following the scaling rules above, the large-$N_{c}$ analogous of the proton would not have charge 1, but
rather the charge (for $N_{c}$ odd and for the number of quarks  $u$ exceeding
that of quarks $d$ of one unit):
\begin{equation}
\left(  \frac{N_{c}-1}{2}+1\right)  \frac{2}{3}-\left(  \frac{N_{c}-1}%
{2}\right)  \frac{1}{3}=\frac{N_{c}+3}{6}\text{ .}%
\end{equation}
In general, baryons would not have integer charges. For instance, the
large-$N-c$ analogous of the $\Delta^{++}$ baryon would carry charge $2N_{c}/3,$ which is also not
necessarily an integer.

Following the discussion in Refs. \cite{gerard,wiese,bickert}, we impose the subsequent scaling behavior for the
charges of the $u$ and $d$ quarks:%
\begin{equation}
q_{u}=\frac{1+N_{c}}{2N_{c}}\text{, }q_{d}=\frac{1-N_{c}}{2N_{c}}\text{ ,}%
\end{equation}
out of which any baryon has an integer charge. In particular, the proton carries the charge
\begin{equation}
\left(  \frac{N_{c}-1}{2}+1\right)  q_{u}-\left(  \frac{N_{c}-1}{2}\right)
q_{d}=1\text{ .}%
\end{equation}
The $\pi^{+}$ has still the charge $q_{u}+q_{\bar{d}}=q_{u}-q_{d}=1,$ as
expected. Then, the amplitude for the $\pi^{0}$ decay reads%

\begin{align}
A_{\pi^{0}\gamma\gamma} &  =g_{\pi^{0}\bar{u}u}N_{c}\left(  q_{u}^{2}%
-q_{d}^{2}\right)  =g_{\pi^{0}\bar{u}u}N_{c}\left(  q_{u}^{2}-q_{d}%
^{2}\right)  \nonumber\\
&  =g_{\pi^{0}\bar{u}u}N_{c}\left(  \left(  \frac{1+N_{c}}{2N_{c}}\right)
^{2}-\left(  \frac{1-N_{c}}{2N_{c}}\right)  ^{2}\right)  =g_{\pi^{0}\bar{u}%
u}N_{c}\frac{4N_{c}}{4N_{c}^{2}}=g_{\pi^{0}\bar{u}u}\text{ .}%
\end{align}
It follows that with this scheme $\Gamma_{\pi^{0}\rightarrow\gamma\gamma}\sim g_{\pi
^{0}\bar{u}u}^{2}\sim N_{c}^{-1}$, just as a regular mesonic decay.

\bigskip

3)\textbf{ Three-body decay: direct process vs decay chain.}

\begin{figure}[h]
        \centering       \includegraphics[scale=0.35]{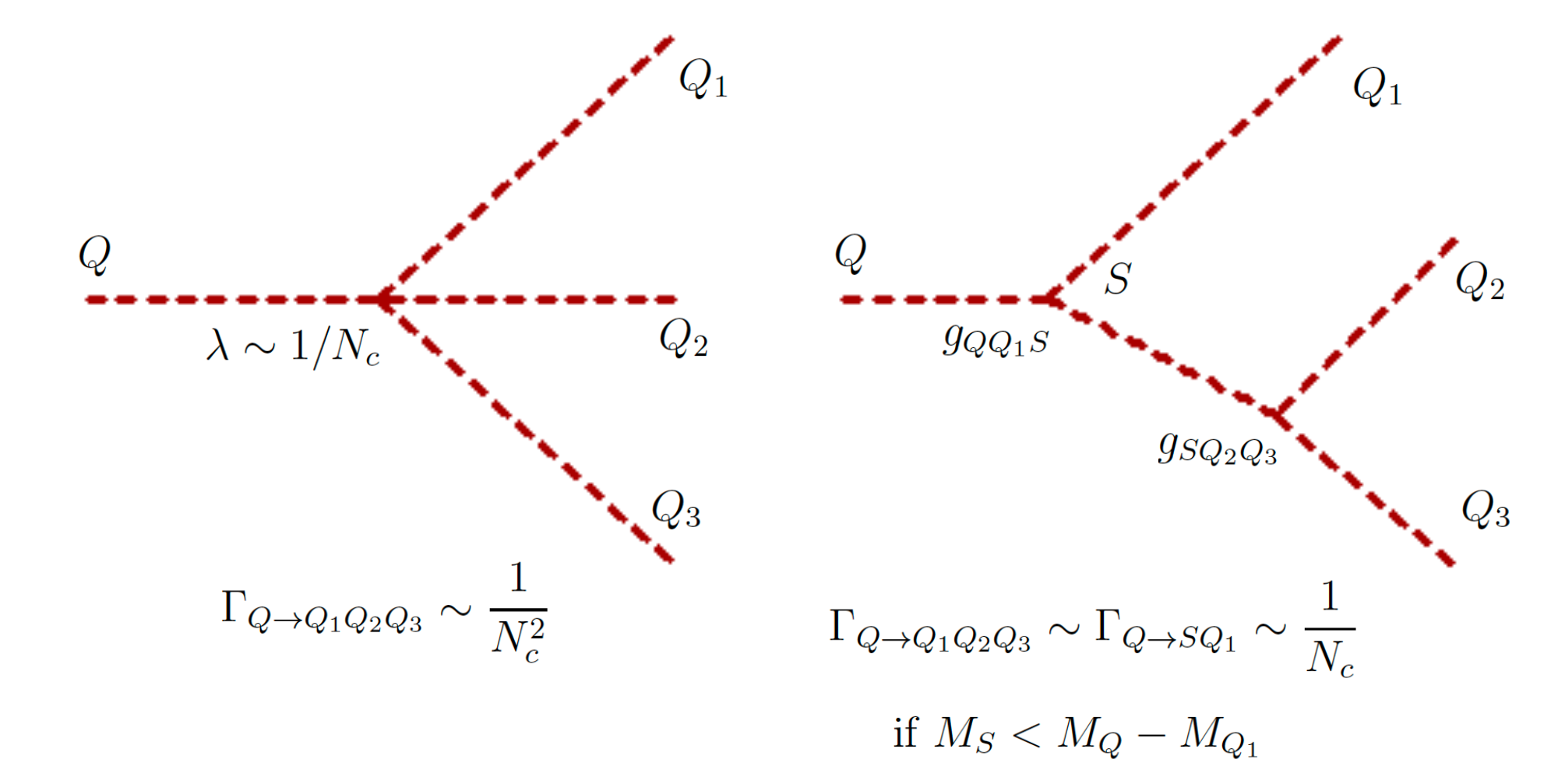}\\
        \caption{Direct three-body decay vs decay chain quarkonium $Q$ (red, dashed) into 3 quarkonia states. The decay chain takes place via an intermediate quarkonium state $S$. Under appropriate conditions, the decay chain dominates in the large-$N_c$ limit. }
        \label{Cons-Page-7}
   \end{figure}

We intend to study the three-body decay process
\begin{equation}
Q\rightarrow Q_{1}Q_{2}Q_{3}\text{.}%
\end{equation}
Let us consider two possible models for this decay. For the direct decay, a Lagrangian of the type (we call it `model A') reads (see Fig. \ref{Cons-Page-7}, left part)
\begin{equation}
\mathcal{L}_{A}=\lambda QQ_{1}Q_{2}Q_{3} \text{ ,}
\end{equation}
where $\lambda=\bar{\lambda}/N_{c}$, for which the three-body decay goes as:
\begin{equation}
\Gamma_{Q\rightarrow Q_{1}Q_{2}Q_{3}}\propto\left\vert A_{Q\rightarrow
Q_{1}Q_{2}Q_{3}}\right\vert ^{2}\propto\lambda^{2}\propto\frac{1}{N_{c}^{2}}%
\text{ .} 
\end{equation}
Next, let us consider the possibility that the decay takes place via an
additional intermediate quark-antiquark state $S$ via the Lagrangian (that we shall denote as `model B', see Fig. \ref{Cons-Page-7}, right part): 
\begin{equation}
\mathcal{L}_{B}=g_{QQ_{1}S}QQ_{1}S+g_{SQ_{2}Q_{3}}SQ_{2}Q_{3} \text{ ,}
\end{equation}
where both constants $g_{QQ_{1}S}$ and $g_{SQ_{2}Q_{3}}$ scale as
$1/\sqrt{N_{c}}$. Thus, one has the decay chain %
\begin{equation}
Q\rightarrow Q_{1}S\rightarrow Q_{1}Q_{2}Q_{3}
\text{ ,}
\end{equation}
where in the second step $S\rightarrow Q_{2}Q_{3}$ has taken place. We assume,
for simplicity, that $S\rightarrow Q_{2}Q_{3}$ is the only available decay channel for
$S$ (the main result holds also when this is not the case).

In scenario B, the decay amplitude takes the form:%
\begin{equation}
A_{Q\rightarrow Q_{1}Q_{2}Q_{3}}\propto g_{QQ_{1}S}\frac{1}{p_{S}^{2}%
-M_{S}^{2}+i\Gamma_{S}M_{S}}g_{SQ_{2}Q_{3}}
\text{ .}
\end{equation}
Since $M_{S}$ is not dependent on $N_{c}$ and $\Gamma_{S}$ is suppressed as
$1/N_{c},$ at first sight $A_{Q\rightarrow Q_{1}Q_{2}Q_{3}}$ scales also as
$1/N_{c},$ thus leading to the same result of model `A' (direct decay). Yet, a more careful analysis leads
to a different result. Following Ref. \cite{lupo,sill}, the decay chain leads
to the integration over final momenta leading to
\begin{equation}
\Gamma_{Q\rightarrow Q_{1}Q_{2}Q_{3}}=\int_{0}^{M_{Q}-M_{1}}dx\Gamma
_{Q\rightarrow Q_{1}S}(x)d_{S}(x)
\text{ ,}
\end{equation}
where $\Gamma_{Q\rightarrow Q_{1}S}(x)$ is the decay width for $Q\rightarrow
Q_{1}S$. The quantity $d_{S}(x)$ is the mass distribution of the
$S$ state with%
\begin{equation}
d_{S}(x)=\frac{\Gamma_{S}}{2\pi}\frac{1}{(x-M_{S})^{2}+\Gamma_{S}^{2}/4}
\text{ ,}
\end{equation}
where $x$ is the running mass of the intermediate state $S$; only for  $x <M_Q-M_1$ a nonzero contribution to $Q \rightarrow Q_1Q_2Q_3$ is possible. Above, the nonrelativistic Breit-Wigner approximation has been used. This approximation
is surely valid for narrow states. Anyway, the result is more general than
that (one could use for instance the relativistic Sill distribution of Ref.
\cite{sill} that takes into account threshold effects, getting the same
outcome). In the large-$N_{c}$ limit one obtains:%
\begin{equation}
d_{S}(x)=\delta(x-M_{S}) \text{ ,}
\end{equation}
thus
\begin{equation}
\Gamma_{Q\rightarrow Q_{1}Q_{2}Q_{3}}=\Gamma_{Q\rightarrow Q_{1}S}%
(x=M_{S})\propto\frac{1}{N_{c}}%
\text{ ,}
\end{equation}
instead of $N_{c}^{-2}.\ $It is then evident that in this case the decay chain dominates.
Yet, this term is nonzero only if $M_{S}\leq M_{Q}-M_{1}$. If this is not the case, one should consider the next to leading term for $d_{S}(x)$
that scales as $1/N_{c}$ which again would lead to an overall $1/N_{c}^{2}$
decay of $Q.$ This can be easily seen in the case in which $M_{S}$ is much
larger than $M_{Q},$ thus (for $x$ in the range $(0,M_{Q}-M_1)$)
\begin{equation}
d_{S}(x)=\frac{\Gamma_{S}}{2\pi}\frac{1}{M_{S}^{2}} + ... \propto\frac{1}{N_{c}}%
\text{ ,}
\end{equation}
out of which
\begin{gather}
\Gamma_{Q\rightarrow Q_{1}Q_{2}Q_{3}}=\int_{0}^{M_{Q}-M_1}dx\Gamma_{Q\rightarrow
Q_{1}S}(x)d_{S}(x)\nonumber \\
=\frac{\Gamma_{S}}{2\pi}\frac{1}{M_{S}^{2}}\int_{0}^{M_{Q}%
-M_1}dx\Gamma_{Q\rightarrow Q_{1}S}(x)\propto\frac{1}{N_{c}^{2}}%
\text{ .}
\end{gather}
(Note, this decay is additionally  also suppressed by the assumed large mass $M_{S}$).

As anticipated, the outcome is unchanged if $S$ has more than a single decay
channel. Namely, in this case the spectral function refers to the specific decay channel \cite{plb} with:
\begin{equation}
d_{S}(x)=\frac{\Gamma_{S\rightarrow Q_{2}Q_{3}}}{2\pi}\frac{1}{(x-M_{S}%
)^{2}+\Gamma_{S}^{2}/4}%
\end{equation}
which reduces to%
\begin{equation}
d_{S}(x)=\frac{\Gamma_{S\rightarrow Q_{2}Q_{3}}}{\Gamma_{S}}\delta(x-M_{S})
\end{equation}
in the large-$N_{c}$ limit. Then%

\begin{equation}
\Gamma_{Q\rightarrow Q_{1}Q_{2}Q_{3}}=\frac{\Gamma_{S\rightarrow Q_{2}Q_{3}}%
}{\Gamma_{S}}\Gamma_{Q\rightarrow Q_{1}S}(x=M_{S})\propto\frac{1}{N_{c}}%
\end{equation}
if, of course, $M_{S}\leq M_{Q}-M_{1}$.

In conclusion, the decay chain is dominant, if appropriate kinematic conditions
are met.

\bigskip

3) \textbf{Chiral models. }

Let us study the large-$N_c$ scaling in a chiral model. For simplicity, we consider one
scalar $\sigma$ particle and one pseudoscalar $\pi$ corresponding to the case
of a single flavor ($N_{f}=1)$. (Note, we neglect at first the chiral anomaly, thus
$\pi$ emerges as a Goldstone boson: in this respect, it is more
pion-like than $\eta'$-like).

\begin{figure}[h]
        \centering       \includegraphics[scale=0.55]{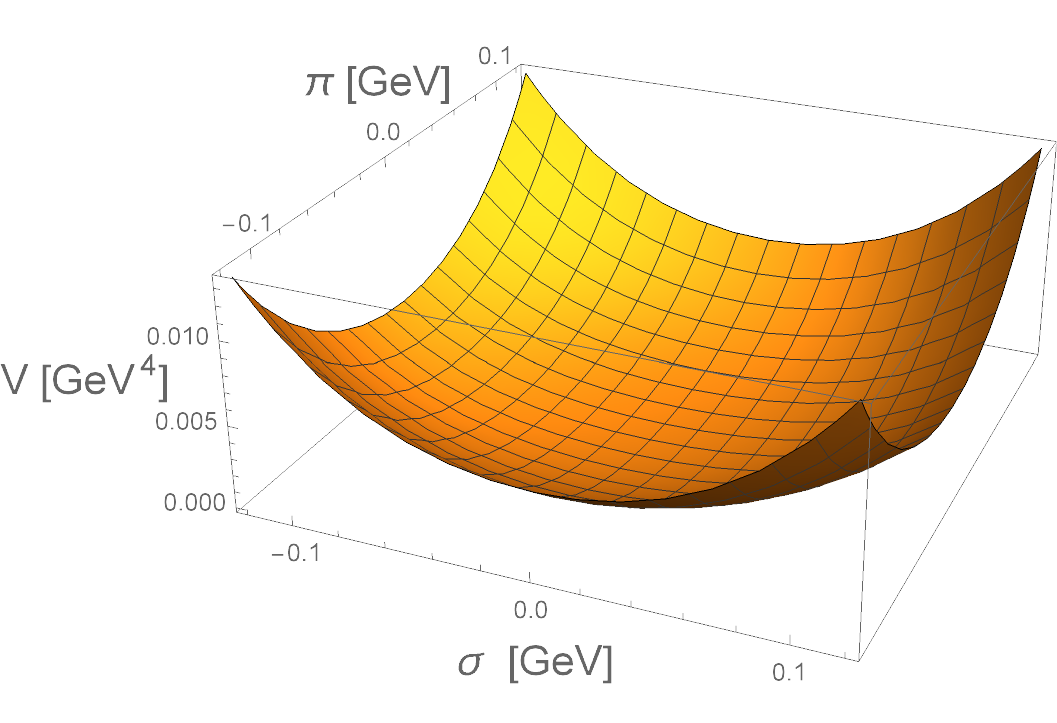}\\
        \caption{Form of the chiral potential of Eq. (\ref{pot}) for $m_0^2 >0$. The (unique and globnal) minimum sits at the origin. }
        \label{nomh}
   \end{figure}
   
The basic chiral `multiplet' reads
\begin{equation}
\Phi=\sigma+i\pi \text{ .}
\end{equation}
A chiral transformation amounts to $\Phi\rightarrow e^{i\alpha}\Phi$ (it is an
$O(2)$ rotation in the place spanned by $(\sigma,\pi)),$ thus the quantity
\begin{equation}
  \Phi^{\dag}\Phi=\Phi^{\ast}\Phi = \sigma^{2}+\pi^{2}  
\end{equation}
is a chirally invariant object. For a
generic $N_{f}$ the quantity $\Phi$ is a $N_{f}\times N_{f}$ matrix, e.g.
\cite{dick,ko,urban,nf2,fariborz}, but for $N_{f}=1$ it is a scalar, thus $\Phi^{\dag
}=\Phi^{\ast}$.  Interestingly, the main large-$N_{c}$ outcomes that we shall discuss 
are independent on $N_{f}$ (with an important exception, the chiral anomaly).

For $N_f=1$ the chirally invariant potential takes the simple form:%
\begin{equation}
V(\sigma,\pi)=\frac{m_{0}^{2}}{2}\Phi^{\ast}\Phi+\frac{\lambda}{4}\left(
\Phi^{\ast}\Phi\right)  ^{2}=\frac{m_{0}^{2}}{2}\left(  \pi^{2}+\sigma
^{2}\right)  +\frac{\lambda}{4}\left(  \pi^{2}+\sigma^{2}\right)  ^{2}
\text{ .}
\end{equation}
The large-$N_{c}$ scaling is an immediate consequence of our previous discussion:
\begin{equation}
m_{0}\sim N_{c}^{0}\text{ , }\lambda\sim N_{c}^{-1}%
\text{ .}
\label{pot}
\end{equation}

For $m_{0}^{2}>0$ the potential is plotted in Fig. \ref{nomh} for $m_0^2 = 0.6^2$ GeV$^2$ and $\lambda = 40$: it has a single minimum for $P_{\min}=(\sigma
=\pi=0)$. The masses of the particles correspond to the second derivatives
evaluated at the minimum:
\begin{align}
M_{\pi}^{2}  &  =\left.  \frac{\partial^{2}V}{\partial\pi^{2}}\right\vert
_{P=P_{\min}}=m_{0}^{2}\sim N_{c}^{0}\text{ ,}\\
M_{\sigma}^{2}  &  =\left.  \frac{\partial^{2}V}{\partial\sigma^{2}%
}\right\vert _{P=P_{\min}}=m_{0}^{2}\sim N_{c}^{0}\text{ .}%
\end{align}
Both particles have the same mass $m_{0}.$ This is a manifest realization of
chiral symmetry. Yet, this is not how Nature works. Chiral partners do not
have the same mass.

\begin{figure}[h]
        \centering       \includegraphics[scale=0.75]{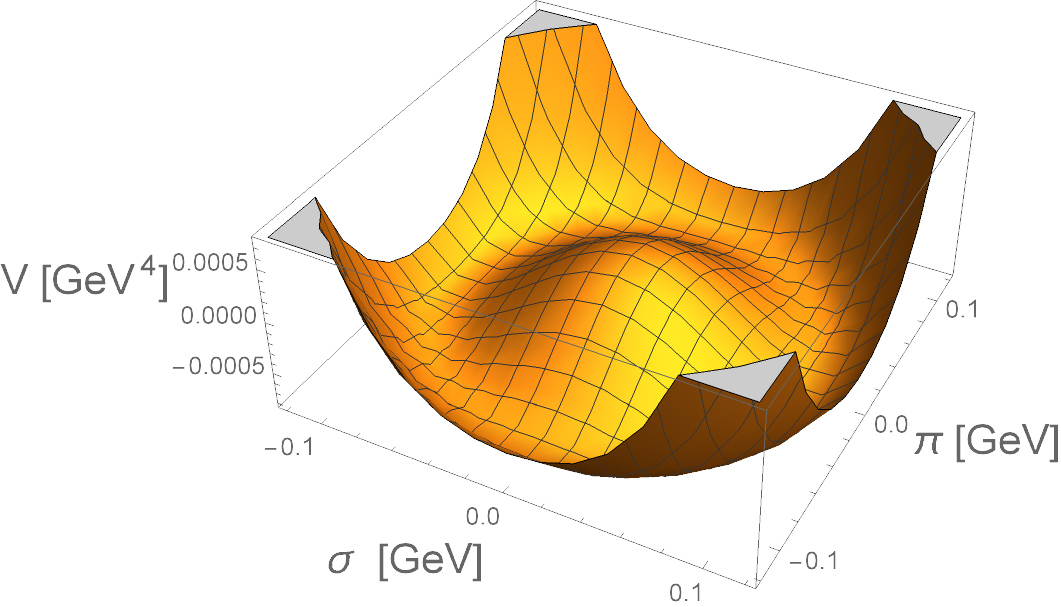}\\
        \caption{Form of the chiral potential of Eq. (\ref{pot}) for $m_0^2 <0$. There is a circle of minima for $\sigma^2+\pi^2 = F^2 = -m_0^2/\lambda$. The originan is a local maximum.}
        \label{mh1}
   \end{figure}

The splitting of masses is possible without explicit breaking of chiral symmetry by
considering $m_{0}^{2}<0$. The corresponding potential, plotted in Fig. \ref{mh1} for $m_0^2 = -0.6^2$ GeV$^2$ and $\lambda = 40$, can be rewritten as:
\begin{equation}
V(\sigma,\pi)=\frac{\lambda}{4}\left(  \pi^{2}+\sigma^{2}-F^{2}\right)
^{2} - \frac{m_0^4}{4 \lambda}\text{ with }F=\sqrt{\frac{-m_{0}^{2}}{\lambda}}\sim N_{c}^{1/2}>0
\text{ .}
\end{equation}
It has the typical shape of a Mexican hat, in which the origin (for
$\sigma=\pi=0$) is not a minimum but a maximum. In fact, upon calculating the
masses around the origin, they would turn out to be imaginary. There is
however a circle of equivalent minima for:
\begin{equation}
\pi^{2}+\sigma^{2}=F^{2}=-\frac{m_{0}^{2}}{\lambda}\sim N_{c}>0\text{ .}%
\end{equation}
Moreover, the radius of this circle goes as $N_{c}^{1/2}.$ SSB is realized when a specific minimum
is picked up. Following the usual convention we choose:

\begin{equation}
P_{\min}=\left(  \sigma_{\min}=\phi_{N}=\sqrt{-\frac{m_{0}^{2}}{\lambda}%
}=F,0\right)  \text{ .}%
\end{equation}
The quantity $\phi_{N}=F$ is also denoted as the chiral condensate. 
Quite remarkably, a closer inspection of chiral
models show that the previously studied pion decay constant is proportional to
$\phi_{N}$%
\begin{equation}
\phi_{N}\sim f_{\pi}\sim N_{c}^{1/2}.
\end{equation}
Namely, the coupling to the weak boson $W^{\pm}$ emerges from an interaction term of the
type
\begin{equation}
 g_{\text{weak}} \cos{\theta_C}W_{\mu}^{\pm}\sigma\partial^{\mu}\pi^{\mp} \text{ ,} 
\end{equation}
($\theta_C$ is the Cabibbo angle), hence when
$\sigma$ condenses to $\phi_{N}\sim f_{\pi}$, the direct 
$W$-$\pi$ mixing 
\begin{equation}
  W_{\mu}^{\pm
}\longleftrightarrow\pi^{\pm}  
\end{equation}
arises.

Within this model, the masses are:%
\begin{align}
M_{\pi}^{2}  &  =\left.  \frac{\partial^{2}V}{\partial\pi^{2}}\right\vert
_{P=P_{\min}}=0\sim N_{c}^{0}\text{ ,}\\
M_{\sigma}^{2}  &  =\left.  \frac{\partial^{2}V}{\partial\sigma^{2}%
}\right\vert _{P=P_{\min}}=m_{0}^{2}+3\lambda\phi_{N}^{2}=-2m_{0}^{2}%
=2\lambda\phi_{N}^{2}\sim N_{c}^{0}>0\text{ ,}%
\end{align}
which are not degenerate: the pion has a vanishing mass (it is a Goldstone
boson), while the $\sigma$ is massive. One then realizes how spontaneous chiral
symmetry breaking generates different masses for chiral partners. Note,
$M_{\sigma}$ is proportional to the chiral condensate $\phi_{N}.$ Both masses
are still $\sim N_{c}^{0}.$ Yet, as shown in \cite{coleman1}, SSB
is expected to occur for large $N_{c}$, just as it does for $N_c = 3$.

\begin{figure}[h]
        \centering       \includegraphics[scale=0.75]{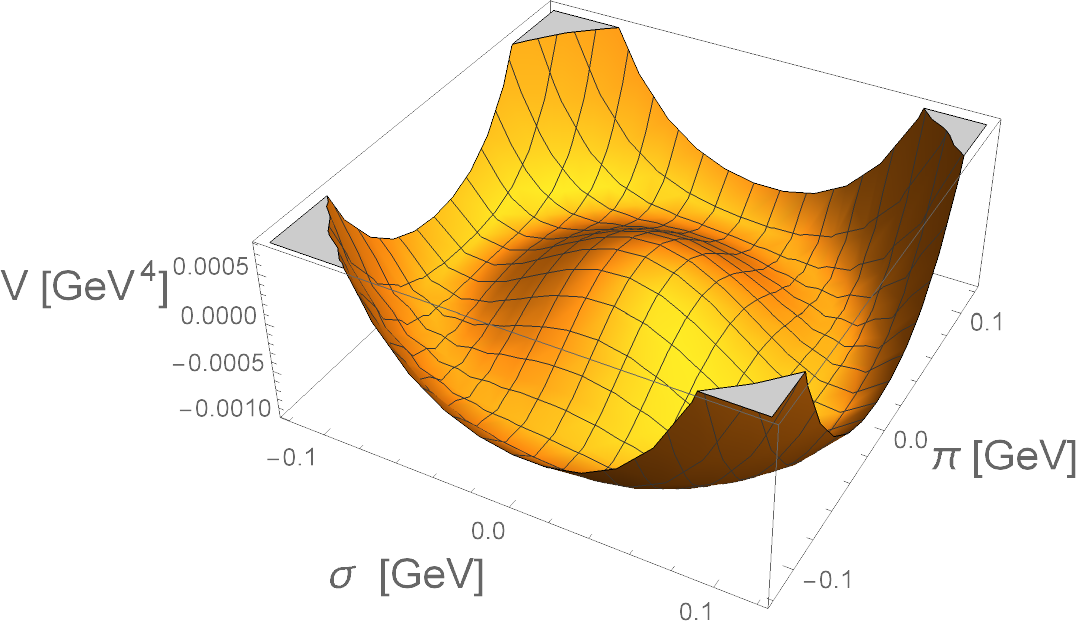}\\
        \caption{Same as in Fig. \ref{mh1} but with the explicit symmetry breaking of Eq. (\ref{poth}). There is now an absolute minimum for $\sigma = \phi_N>0$ and $\pi = 0$.}
        \label{mh2}
   \end{figure}
   
In Nature, the pion has a small but nonzero mass. In order to take this fact
into account, the potential is modified as:%
\begin{equation}
V(\sigma,\pi)=\frac{m_{0}^{2}}{2}\left(  \pi^{2}+\sigma^{2}\right)
+\frac{\lambda}{4}\left(  \pi^{2}+\sigma^{2}\right)  ^{2}-h\sigma\text{ ,}%
\label{poth}
\end{equation}
where $-h\sigma$ breaks chiral symmetry explicitly. This term follows directly
from the mass term $-m\bar{q}q$ in the QCD Lagrangian. We thus expect that
$h\propto m_{n}$, where $m_{n}$ is the bare quark mass (e.g. the average
$(m_{u}+m_{d})/2$). The potential, plotted in Fig. \ref{mh2} (same parameters as before and $h=m_{\pi}^2f_{\pi}$ with 
$f_{\pi} = 92$ MeV and $m_{\pi}= 135$ MeV) has now a unique minimum for
\begin{equation}
P_{\min}=\left(  \sigma_{\min}=\phi_{N},0\right)
\end{equation}
with%
\begin{equation}
\left.  \frac{\partial V(\sigma,0)}{\partial\sigma}\right\vert _{\sigma
=\sigma_{\min}=\phi_{N}}=m_{0}^{2}\phi_{N}+\lambda\phi_{N}^{3}-h\text{
}=0\text{ ,}\nonumber
\end{equation}
which is of third order. (Only one of the three solutions corresponds
to an absolute minimum). The pion mass is now nonzero:
\begin{equation}
M_{\pi}^{2}=\left.  \frac{\partial^{2}V}{\partial\pi^{2}}\right\vert
_{P=P_{\min}}=m_{0}^{2}+\lambda\phi_{N}^{2}=\frac{h}{\phi_{N}}>0\text{ .}%
\end{equation}
We realize that the pion mass scale as 
\begin{equation}
M_{\pi}\propto\sqrt{h}\propto
\sqrt{m_{n}} \text{ .}
\end{equation}
This is indeed a nontrivial dependence, since one would naively expect that $M_{\pi}\propto h \propto m|n$, i.e. to the mass of its constituents. This is not the case, signalizing that the tilted Mexican hat form of the potential is actually realized in Nature. This peculiar feature is also confirmed by lattice QCD studies, e.g. \cite{pionbox}.

In order to fulfill the
large-$N_{c}$ expectation, one must require that:
\begin{equation}
h\sim N_{c}^{1/2}\text{ }\rightarrow M_{\pi}^{2}\sim N_{c}^{0}\text{ .}%
\end{equation}
This is in agreement with the scaling $f_{\pi} \propto N_c^{1/2}$. The term
$h\sigma$ acts as a source term for $\sigma$, thus $h$ scales as $N_{c}%
^{1/2}.$ Yet, a closer look reveals a kind of problem: $h$ has dimension
energy$^{3}$ and is proportional to $m_{n},$ then one would naively write
$h\sim m_{n}\Lambda_{QCD}^{2},$ but  $\Lambda_{QCD}$ is $N_{c}$ independent. How
to reconciliate that with the required scaling 
$h\sim N_{c}^{1/2}$? This point
will be clarified in Sec. 3.2 after discussing the dilaton at large $N_c$.

\begin{figure}[h]
        \centering       \includegraphics[scale=0.75]{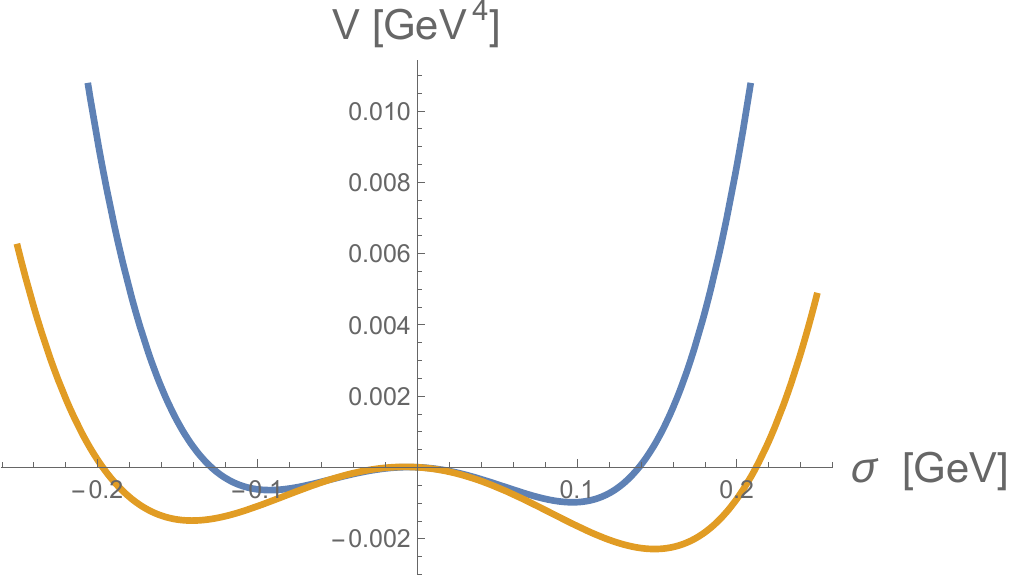}\\
        \caption{Potential of Eq. (\ref{poth}) along the $\sigma$ direction for $N_c =3$ (upper blue curve) and for $N_c =7$ (lower yellow curve). The minimum gets deeper on the vertical axis ($V_{min} \sim -N_c$) and its location on the horizontal axis moves to the right ($\phi_N \sim N_c^{1/2}$).  }
        \label{sigmadir}
   \end{figure}

The mass of the $\sigma$-particle is:%
\begin{equation}
M_{\sigma}^{2}=\left.  \frac{\partial^{2}V}{\partial\sigma^{2}}\right\vert
_{P=P_{\min}}=m_{0}^{2}+3\lambda\phi_{N}^{2}=M_{\pi}^{2}+2\lambda\phi_{N}%
^{2}\sim N_{c}^{0}\text{ .}%
\end{equation}
The mass difference $M_{\sigma}^{2}-M_{\pi}^{2}=2\lambda\phi_{N}^{2}\sim
N_{c}^{0}>0$ does not depend on $h.$
The plot of the potential along the $\sigma$-direction is shown in Fig. \ref{sigmadir} for two values of $N_c$.

The chiral condensate $\phi_N \sim f_{\pi}$ can be related to the 
quark condensate 
\begin{equation}
    \left\langle 0_{QCD}\left\vert \bar{q}q\right\vert
0_{QCD}\right\rangle < 0 
\end{equation}
via the GOR relation (e.g. \cite{gor}):
\begin{equation}
    M_{\pi}^2f_{\pi}^2 = -2m_n\left\langle 0_{QCD}\left\vert \bar{q}q\right\vert
0_{QCD}\right\rangle \text{ ,}
\end{equation}
where both members of this equation scale with $N_c$ (the r.h.s. is such because it comes from quark loops, see also Sec. 5). Note, this equation implies a nontrivial link between the chiral condensate and the quark condensate, $\phi_N^2 $ scales as $\langle 
 \bar{q}q \rangle$, Moreover, it is also in agreement with $M_{\pi}^2 \propto m_n$.

The chiral condensate $\phi_{N}$ enters also in decays, such as $\sigma
\rightarrow\pi\pi$. This is determined by performing the shift $\sigma\rightarrow
\sigma+\phi_{N}$ and then isolating the term $\lambda\phi_{N}\sigma\pi^{2},$
out of which:
\begin{equation}
\Gamma_{\sigma\rightarrow\pi\pi}=2\frac{k_{\pi}}{8\pi M_{\sigma}^{2}}\left[
\lambda\phi_{N}\right]  ^{2}\sim N_{c}^{-1}\text{ ,}%
\end{equation}
where
$k_{\pi}=\sqrt{\frac{M_{\sigma}^{2}}{4}-M_{\pi}^{2}} \propto N_c^0$
is the modulus of the three-momentum of one of the outgoing particle. Also in
this case, the expected scaling is recovered.

In the full $N_{f}=3$ version of the model, see e.g.  Refs. \cite{dick,stani}, the masses and decays are
calculated by following the very same steps. Obviously, there are many more
fields and decay channels, but the principles and the basic ideas are exactly
the same as those discussed here.

As stated above, in this $N_{f}=1$ example, the anomaly has been disregarded.
There are however some interesting large-$N_{c}$ considerations that can be
done. The Lagrangian describing the anomaly for any $N_f$ takes the form \cite{kovacsanomaly}:
\begin{align}
\mathcal{L}_{A}  &  =-c_{1}(\det\Phi+\det\Phi^{\dag})-c_{2}(\det\Phi-\det
\Phi^{\dag})^{2}-c_{3}(\det\Phi+\det\Phi^{\dag})^{2}
\end{align}
which for $N_f=1$ reduces to
\begin{equation}
  \mathcal{L}_{A}  =-2c_{1}\sigma+4c_{2}\pi^{2}-4c_{3}\sigma^{2}%
\text{ .}
\end{equation}

The mass arising from the anomaly scales as $M_{\pi}^{2}\sim N_{c}^{-1}.$
(Actually, the name $M_{\eta_{0}}^{2}\sim N_{c}^{-1}$ \cite{witteneta} with
$\eta_{0}$ being the flavor singlet would be more appropriate, but for
simplicity we stick to $\pi$). 
For $N_{f}=1$, the first terms is analogous to $h\sigma$ seen before, but the
scaling is different, $c_{1}\sim N_{c}^{-1/2}$so that $M_{\pi}^{2}\sim
N_{c}^{-1}$ follows. For the same reason $c_{2}\sim N_{c}^{-1}$ , $c_{3}\sim
N_{c}^{-1}.$ The former is evident, for the latter one needs to calculate the
pion mass that turns out to be (for $h=c_{1}=c_{2}=0)$ $M_{\pi}^{2}=8c_{3}.$ Thus:%
\begin{equation}
c_{1}\sim N_{c}^{-1/2}\text{ , }c_{2}\sim N_{c}^{-1}\text{ , }c_{3}\sim
N_{c}^{-1} \text{ .}
\end{equation}
When changing $N_{f},$ these scaling behaviors are modified in such a way to
preserve $M_{\eta_{0}}^{2}\sim N_{c}^{-1}.$ By properly counting the
condensates that scale as $N_{c}^{1/2},$ one finds:
\begin{equation}
c_{1}\sim N_{c}^{-N_{f}/2}\text{ , }c_{2}\sim N_{c}^{-N_{f}}\text{ , }%
c_{3}\sim N_{c}^{-N_{f}}\text{ .}%
\end{equation}
For recent applications and extensions of the axial anomaly, see Refs.
\cite{anomaly1,anomaly2}.

\bigskip

4) \textbf{Different interaction types: a single trace `wins'.}

\begin{figure}[h]
        \centering       \includegraphics[scale=0.55]{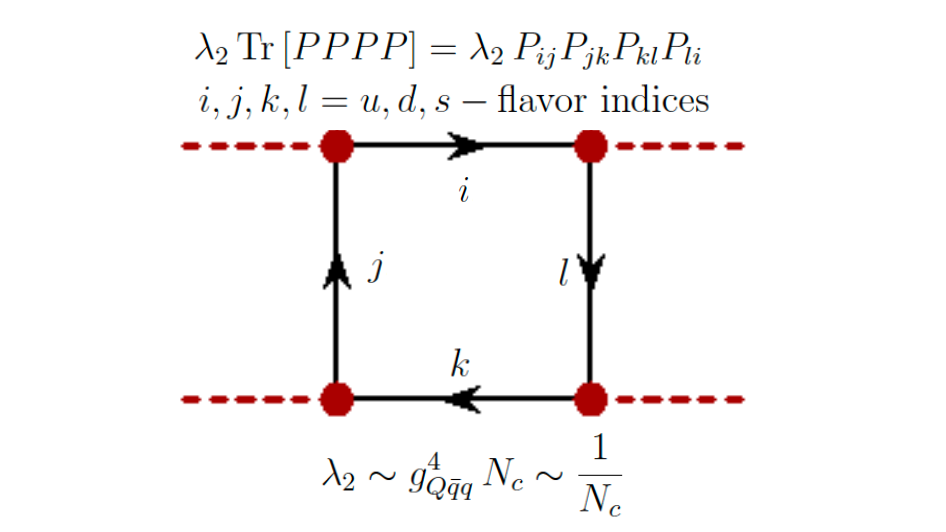}\\
        \caption{Large-$N_c$ scaling of the term proportional to $\lambda_2$ in the Lagrangian of Eq. (\ref{lambda12}). Note, the quark loops involves distinct flavors but the usual color factor $N_c$. }
        \label{Cons-Page-3}
   \end{figure}

\begin{figure}[h]
        \centering       \includegraphics[scale=0.45]{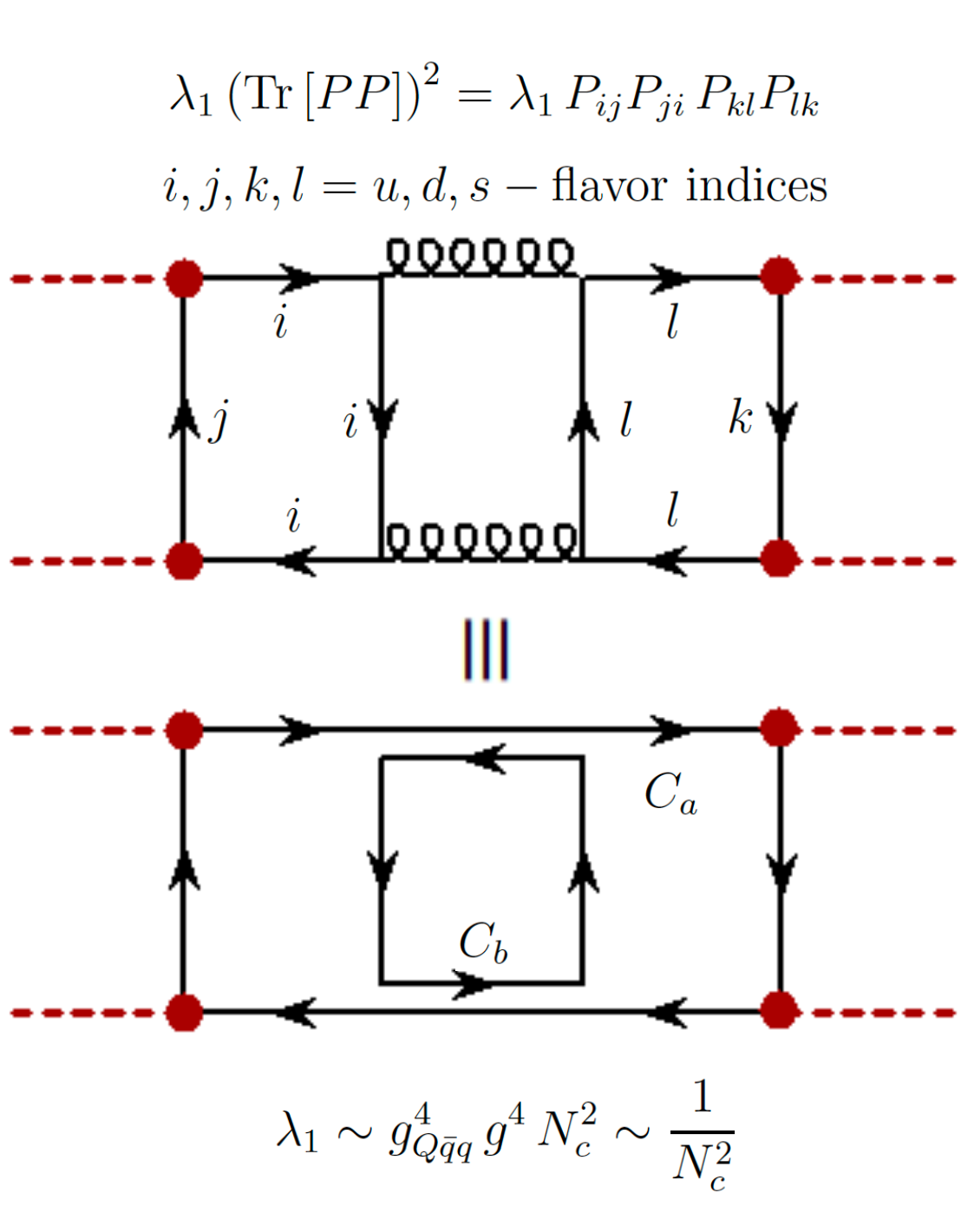}\\
        \caption{Large-$N_c$ scaling of the four-leg term proportional to $\lambda_1$ in the Lagrangian of Eq. (\ref{lambda12}).}
        \label{Cons-Page-4}
   \end{figure}

The study of the full $N_{f}=3$ chiral model is not within the scope of this lecture, but there is indeed an interesting point related to large-$N_{c}$
that is worth to be discussed. To this end, let us introduce the nonet of
pseudoscalar states \cite{dick,chpt} (see aslo Sec. 2.2 for $N_f=2$): 
\begin{equation}
P=\left(
\begin{array}
[c]{ccc}%
\frac{\eta_{N}+\pi^{0}}{\sqrt{2}} & \pi^{+} & K^{+}\\
\pi^{-} & \frac{\eta_{N}-\pi^{0}}{\sqrt{2}} & K^{0}\\
K^{-} & \bar{K}^{0} & \eta_{S}%
\end{array}
\right)  \equiv\left(
\begin{array}
[c]{ccc}%
u\bar{u} & u\bar{d} & u\bar{s}\\
d\bar{u} & d\bar{d} & d\bar{s}\\
s\bar{u} & s\bar{d} & s\bar{s}%
\end{array}
\right)
\end{equation}
with $\pi^{0}=\sqrt{1/2}\left(  u\bar{u}-d\bar{d}\right)  ,$ and where
$\eta(547)$ and $\eta^{\prime}(958)$ emerge as a mixing of $\eta_{N}%
=\sqrt{1/2}\left(  u\bar{u}+d\bar{d}\right)  $ and $\eta_{S}=s\bar{s}.$ In
chiral models such as \cite{dick,fariborz}, there are typically two types of
quartic interactions that emerge:
\begin{equation}
\mathcal{L}_{P}=-\lambda_{2}\mathrm{Tr}\left[  P^{4}\right]  -\lambda
_{1}(\mathrm{Tr}[P^{2}])^{2}\text{ .}%
\label{lambda12}
\end{equation}
Their scaling is depicted in Figs. \ref{Cons-Page-3} and \ref{Cons-Page-4} respectively showing that $\lambda_2 \sim N_c^{-1}$ and $\lambda_1 \sim N_c^{-2}$.

\begin{figure}[h]
        \centering       \includegraphics[scale=0.45]{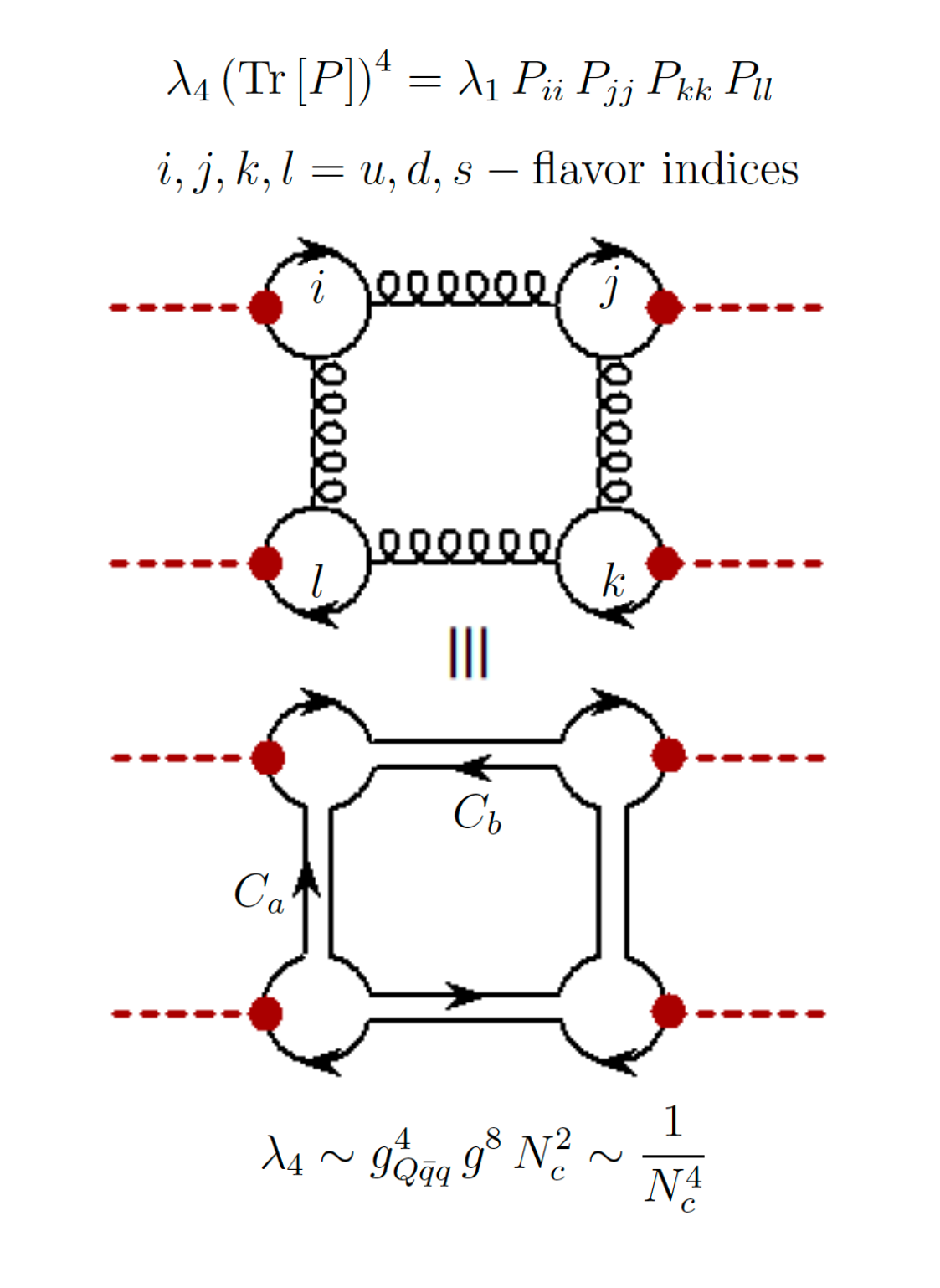}\\
        \caption{Large-$N_c$ scaling of the four-leg term proportional to $\lambda_4$  in the Lagrangian of Eq. (\ref{lambda34}).}
        \label{Cons-Page-5}
   \end{figure}

Note, other terms are possible, as:
\begin{equation}
\mathcal{L}_{P}^{\prime}=-\lambda_{3}\mathrm{Tr}\left[  P\right]
\mathrm{Tr}\left[  P^{3}\right]  -\lambda_{4}(\mathrm{Tr}[P])^{4} \label{lambda34} \text{ .}%
\end{equation}
The scaling for $\lambda_4$ can be determined by drawing the corresponding diagrams of Fig. \ref{Cons-Page-5}, leading to $\lambda_4 \sim N_c^{-4}$. A similar study for $\lambda_3$ leads to $\lambda_3 \sim N_c^{-2}$.

In conclusion, even if all the terms proportionals to $\lambda_{1,2,3,4}$ are quartic terms in the mesonic fields, the large-$N_c$ results show that only one dominates, the one that contains a single trace (the $\lambda_2$-term). Interestingly, this result deals with an interplay of flavor and color d.o.f.. It is also relevant for models, since it makes clear which terms should be at first kept and which can be disregarded, see e.g. \cite{koenig1,koenig2}.

\bigskip

5)\textbf{\ Connection to correlations.}

\begin{figure}[h]
        \centering       \includegraphics[scale=0.45]{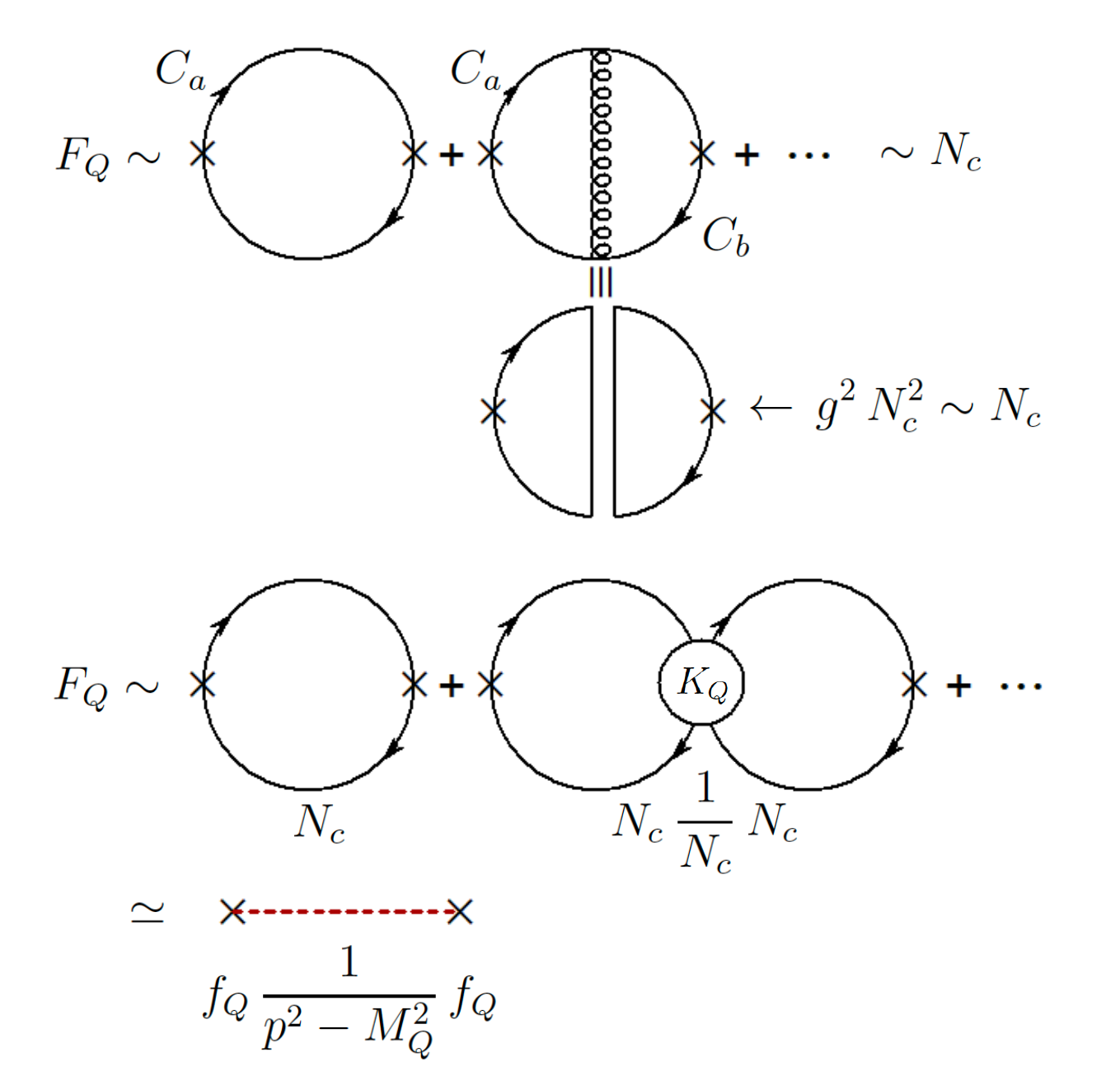}\\
        \caption{Large-$N_c$ scaling of the correlator $F_Q(s) \propto N_c$. }
        \label{Cons-Page-6}
   \end{figure}

In \cite{witten,lebed} as well as other works on QCD at large-$N_{c},$ the starting
point is the correlation function
\begin{equation}
\left\langle J_{Q}(x_{2})J_{Q}(x_{1})\right\rangle =-i\int\frac{d^{4}p}%
{(2\pi)^{2}}F_{Q}(p^2)e^{ip(x_{1}-x_{2})}%
\text{ ,}
\end{equation}
where the quantity $F_{Q}(p)$ is the loop contribution with total momentum
$p.$ Within our framework, at lowest order this is just the loop function $-\Sigma_{Q}(s=p^{2}).$
Expanding $F_{Q}(s)$ we get (see Fig. \ref{Cons-Page-6} for an illustration of these processes):
\begin{gather}
F_{Q}(s = p^2)=-\Sigma_{Q}(s=p^{2})\left(  1+\Sigma_{Q}(s)K_{Q}+...\right) = 
\nonumber \\ -\frac{\Sigma_{Q}(s)}{1-\Sigma_{Q}(s)K_{Q}}=-\frac{\Sigma_{Q}(s)}{K_{Q}}%
\frac{1}{K_{Q}^{-1}-\Sigma_{Q}(s)}
\text{ .}
\end{gather}
The pole is realized for the quarkonium mass $M_{Q}^{2}$ with the already encountered equation 
$K_{Q}^{-1}-\Sigma_{Q}(s=M_{Q}^{2})=0.$
Upon expanding close to the pole, we
find:%
\begin{equation}
F_{Q}(s)=\frac{\Sigma_{Q}(s)\Sigma_{Q}(M_{Q}^{2})}{\Sigma_{Q}^{^{\prime}%
}(M_{Q}^{2})(s-M_{Q}^{2})}\simeq\frac{\Sigma_{Q}^{2}(M_{Q}^{2})g_{Q\bar{q}%
q}^{2}}{s-M_{Q}^{2}}\simeq\frac{f_{Q}^{2}}{s-M_{Q}^{2}}
\text{ ,}
\end{equation}
where
\begin{equation}
f_{Q}=\Sigma_{Q}(M_{Q}^{2})g_{Q\bar{q}q}=\frac{\Sigma_{Q}(M_{Q}^{2})}%
{\sqrt{\Sigma_{Q}^{\prime}(s=M_{Q}^{2})}}\sim N_{c}^{1/2}%
\end{equation}
is the amplitude for the vacuum creation of the meson $Q.$ Indeed, this
expression has the same large-$N_c$ behavior of the weak decay constant of this meson, see the previous
discussion about $f_{\pi}.$ (This equivalence does not hold for glueballs or hybrids).

Since for any given set of quantum numbers an infinity of conventional
mesons exists \cite{witten}, the previous equation may be generalized as:%
\begin{equation}
F_{Q}(s)\simeq\sum_{n=1}^{\infty}\frac{f_{Q,n}^{2}}{s-M_{Q,n}^{2}}\sim N_{c}%
\end{equation}
since $f_{Q,n}^{2}\sim N_{c}$ and $M_{Q,n}^{2}\sim N_{c}^{0}.$ This equation
can be found in Ref. \cite{witten}.

\bigskip

6)\textbf{\ Connection to the Weinberg compositeness condition.}

Finally, we study the Weinberg compositeness
condition \cite{compcond,compo} along the
large-$N_{c}$ direction. The starting point is the Lagrangian
\begin{equation}
\mathcal{L}_{Qq}=\mathcal{L}_{q}+g_{Q\bar{q}q}Q(x)J_{Q}(x)-\frac{\alpha}%
{2}Q^{2}\text{ ,}%
\label{cc1}
\end{equation}
where $Q$ has no kinetic term and $\mathcal{L}_{q}$ contains the kinetic part
for the quark field as well as eventual other interactions not relevant here.
Note, the parameter $\alpha$ is \textit{not} the physical mass squared, see below. The basic
assumption is that $g_{Q\bar{q}q}\sim N_{c}^{-1/2}.$

By using the e.o.m. for the field $Q$ one gets:
\begin{equation}
\frac{\partial\mathcal{L}_{Qq}}{\partial Q}=g_{Q\bar{q}q}J_{Q}(x)-\alpha
Q(x)=0\rightarrow Q(x)=\frac{g_{Q\bar{q}q}}{\alpha}J_{Q}(x)
\text{ .}
\end{equation}
Plugging it back into Eq. \ref{cc1} one obtains a Lagrangian that depends on the quark field $q$ only:%
\begin{equation}
\mathcal{L}_{Qq}\equiv\mathcal{L}_{q}+\frac{g_{Q\bar{q}q}^{2}}{2\alpha}%
J_{Q}^{2}\text{. }%
\end{equation}
This is not a surprise, since $Q$ had no kinetic term from the very beginning. 
One then gets the
correspondence
\begin{equation}
K_{Q}=\frac{g_{Q\bar{q}q}^{2}}{\alpha}=\frac{1}{\Sigma_{Q}(M_{Q}^{2}%
)} \text{,}
\end{equation}
where Eq. (\ref{poleq}) has been used. Then:
\begin{equation}
    \alpha=\frac{\Sigma_{Q}(M_{Q}^{2})}{\Sigma_{Q}^{\prime}(M_{Q}%
^{2})}\sim N_{c}^{0}\text{ ,}
\end{equation}
out of which:%
\begin{equation}
J_{Q}(x)=\frac{\alpha}{g_{Q\bar{q}q}}Q(x)=\Sigma_{Q}(M_{Q}^{2})g_{Q\bar{q}%
q}Q(x)=f_{Q}Q(x) \label{JQ} \text{ ,}%
\end{equation}
which shows that the microscopic quark current $J_Q(x)$ is proportional to the composite meson field $Q(x)$, and the
constant of proportionality is the amplitude for production of this field 
in the QCD vacuum. In this way, the correlator $\left\langle J_{Q}(x_{2}%
)J_{Q}(x_{1})\right\rangle $ takes the form%
\begin{equation}
\left\langle J_{Q}(x_{2})J_{Q}(x_{1})\right\rangle =f_{Q}^{2}\left\langle
Q(x_{2})Q(x_{1})\right\rangle =-if_{Q}^{2}\int\frac{d^{4}p}{(2\pi)^{2}}%
\frac{1}{p^{2}-M_{Q}^{2}}e^{ip(x_{1}-x_{2})}
\end{equation}
hence%
\begin{equation}
F_{Q}(p)=\frac{f_{Q}^{2}}{p^{2}-M_{Q}^{2}}%
\end{equation}
follows consistently.

The formal way to show the result above (especially Eq. (\ref{JQ})) starts from the Lagrangian containing the
bare mesonic coupling, mass and field:%
\begin{equation}
\mathcal{L}_{Qq}=\mathcal{L}_{q}+g_{Q_0\bar{q}q}Q_{0}(x)J_{Q}(x)-\frac
{M_{Q_0}^{2}}{2}Q_{0}^{2}+\frac{1}{2}\left(  \partial_{\mu}Q_{0}\right)
^{2}\text{ ,}%
\end{equation}
with%
\begin{equation}
g_{Q_0\bar{q}q}\sim N_{c}^{-1/2}\text{ and }M_{0}\sim N_{c}^{0}%
\text{ .}
\end{equation}
The propagator of the bare field $Q_{0}$ reads:
\begin{equation}
\frac{1}{p^{2}-M_{Q_0}^{2}+g_{Q_0\bar{q}q}^{2}\Sigma_{Q}(p^{2})}\sim N_{c}%
^{0}\text{ .}%
\end{equation}
The first natural condition is to impose that the pole is realized for the
physical mass $M_{Q}^{2}$:%
\begin{equation}
M_{Q}^{2}-M_{Q_0}^{2}+g_{Q_0\bar{q}q}^{2}\Sigma_{Q}(M_{Q}^{2})=0 \text{ .}
\end{equation}
By expanding the denominator, the propagator of $Q_{0}$ reads:%
\begin{align}
&  \frac{1}{\left(  p^{2}-M_{Q}^{2}\right)  \left(  1+g_{Q_0\bar{q}q}%
^{2}\Sigma_{Q}^{\prime}(M_{Q}^{2})\right)  +g_{Q_0\bar{q}q}^{2}\tilde{\Sigma
}_{Q}(p^{2})}\nonumber\\
&  \simeq\frac{1}{\left(  p^{2}-M_{Q}^{2}\right)  \left(  1+g_{0,Q\bar{q}%
q}^{2}\Sigma_{Q}^{\prime}(M_{Q}^{2})\right)  }=\frac{Z_{2}}{p^{2}-M_{Q}^{2}%
}\text{ ,}%
\end{align}
where $\tilde{\Sigma}_{Q}(p^{2})$ contains terms of the type $(p^{2}-M_{Q}%
^{2})^{2}$ and higher powers, which are negligible close to the pole.

In order to obtain a correctly normalized propagator, the field
\begin{equation}
Q=\frac{1}{\sqrt{Z_{2}}}Q_{0}\Leftrightarrow Q_{0}=\sqrt{Z_{2}}Q
\end{equation}
with the renormalization constant
\begin{equation}
Z_{2}=\frac{1}{1+g_{Q_0\bar{q}q}^{2}\Sigma_{Q}^{\prime}(M_{Q}^{2})}\sim
N_{c}^{0}%
\end{equation}
is introduced.

We impose here $g_{Q_0\bar{q}q}\rightarrow\infty$, out of which $Z_{2}%
\rightarrow0.$ Intuitively, it means that the dressed field $Q$ is not fundamental,
realizing the main idea behind the compositeness condition. The Lagrangian takes
the form:%
\begin{align}
\mathcal{L}_{Qq}  &  =\mathcal{L}_{q}+g_{Q_0\bar{q}q}\sqrt{Z_{2}}%
Q(x)J_{Q}(x)-\frac{M_{Q_0}^{2}}{2}Z_{2}Q_{0}^{2}+\frac{1}{2}Z_{2}\left(
\partial_{\mu}Q\right)  ^{2}\nonumber \\
&  =\mathcal{L}_{q}+g_{Q_0\bar{q}q}\sqrt{Z_{2}}Q(x)J_{Q}(x)-\frac{M_{Q_0}^{2}%
}{2}Z_{2}Q^{2}\text{ ,}%
\end{align}
where the dynamical term for the field $Q(x)$ has disappeared. Next:%
\begin{equation}
g_{Q\bar{q}q}=g_{Q_0\bar{q}q}\sqrt{Z_{2}}=\frac{g_{Q_0\bar{q}q}}%
{\sqrt{1+g_{Q_0\bar{q}q}^{2}\Sigma_{Q}^{\prime}(M_{Q}^{2})}}\overset
{g_{Q_0\bar{q}q}\rightarrow\infty}{=}\frac{1}{\sqrt{\Sigma_{Q}^{\prime}%
(M_{Q}^{2})}}\sim N_{c}^{-1/2}%
\end{equation}
and%
\begin{equation}
\alpha=M_{Q_0}^{2}Z_{2}=\frac{g_{Q_0\bar{q}q}^{2}\Sigma_{Q}(M_{Q}^{2}%
)}{1+g_{Q_0\bar{q}q}^{2}\Sigma_{Q}^{\prime}(M_{Q}^{2})}\overset{g_{0,Q\bar
{q}q}\rightarrow\infty}{=}\frac{\Sigma_{Q}(M_{Q}^{2})}{\Sigma_{Q}^{\prime
}(M_{Q}^{2})}\sim N_{c}^{0}\text{ ,}%
\end{equation}
which coincide with the previous derivation.

\subsection{Glueballs}

According to lattice QCD many glueballs with various quantum numbers should
exist, see e.g. the original predictions within the bag model \cite{bag}, the review of Ref. \cite{vento} and the lattice works of e.g. \cite{mainlattice,at} (for a recent compilation and comparison of lattice results, see Ref. \cite{grg}). The glueball wave
function must be colorless. Considering for definiteness the case of the
scalar glueball, the corresponding (local and gauge invariant) current reads
\begin{equation}
J_{G} = \sum_{a=1}^{N_{c}^{2}-1}G_{\mu\nu}^{a}G^{\mu\nu,a}\text{ .}%
\end{equation}
(For other currents, see e.g. \cite{bag,updatedrev}). Using the double-line
index notation, we may rewrite it as:
\begin{equation}
J_{G} \simeq \sum_{a=1}^{N_{c}}\sum_{b=1}^{N_{c}}G_{\mu\nu}^{(a,b)}G^{\mu\nu,(b,a)} \text{ .}
\end{equation}

   \begin{figure}[h]
        \centering       \includegraphics[scale=0.38]{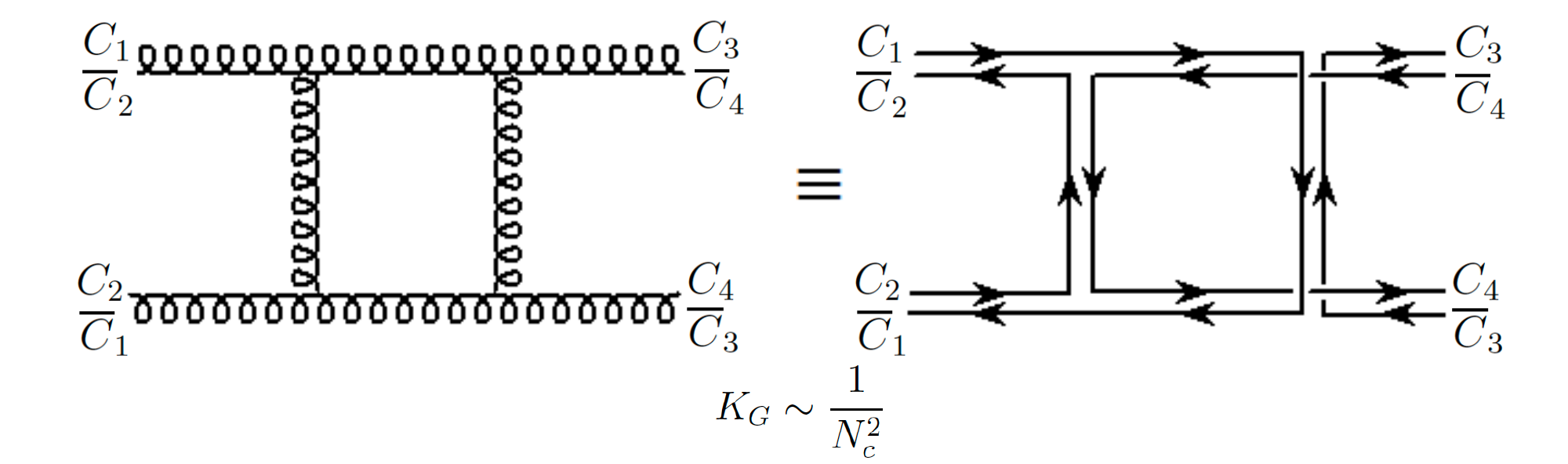}\\
        \caption{Example of a leading diagram for the gluon-gluon scattering of the type $C_1\bar{C}_2C_2\bar{C}_1\rightarrow C_3\bar{C}_4C_4\bar{C}_3$. Note, all gluonic colors have switched. The amplitude scales as $N_c^{-2}$ and models the coupling in Eq. (\ref{lagG}).}
        \label{GG-Page-1}
   \end{figure}

      \begin{figure}[h]
        \centering       \includegraphics[scale=0.38]{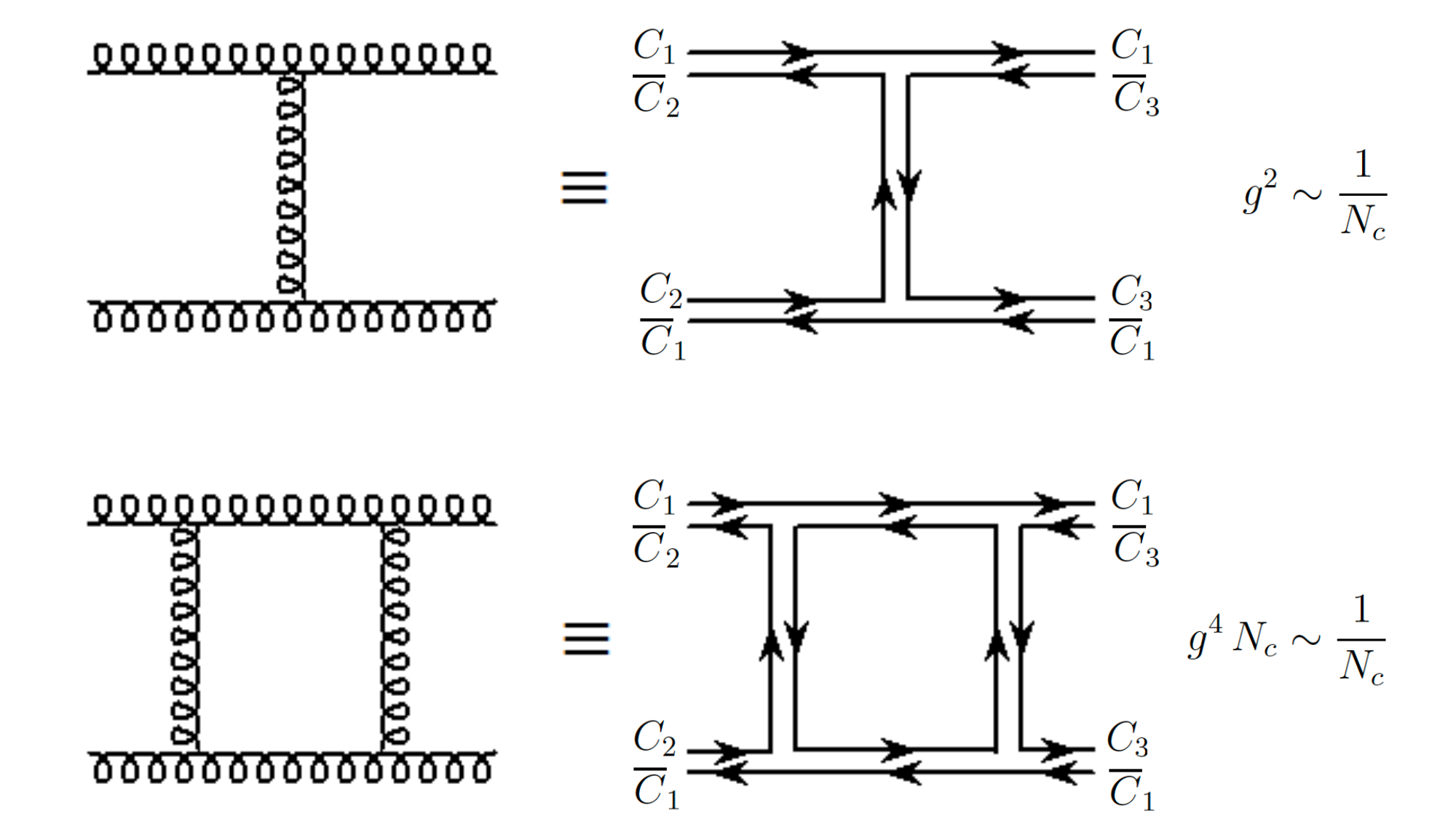}\\
        \caption{Example of a leading diagram for the gluon-gluon scattering of the type $C_1\bar{C}_2C_2\bar{C}_1\rightarrow C_1\bar{C}_4C_4\bar{C}_3$. Note, not all colors have switched ($C_1$ is in the beginning and in the end). This term, whose amplitude goes as $N_c^{-1}$, does \textbf{not} model the constant $K_G$ in Eq. (\ref{lagG}).}
        \label{GG-Page-2}
   \end{figure}
   
\begin{figure}[h]
        \centering       \includegraphics[scale=0.35]{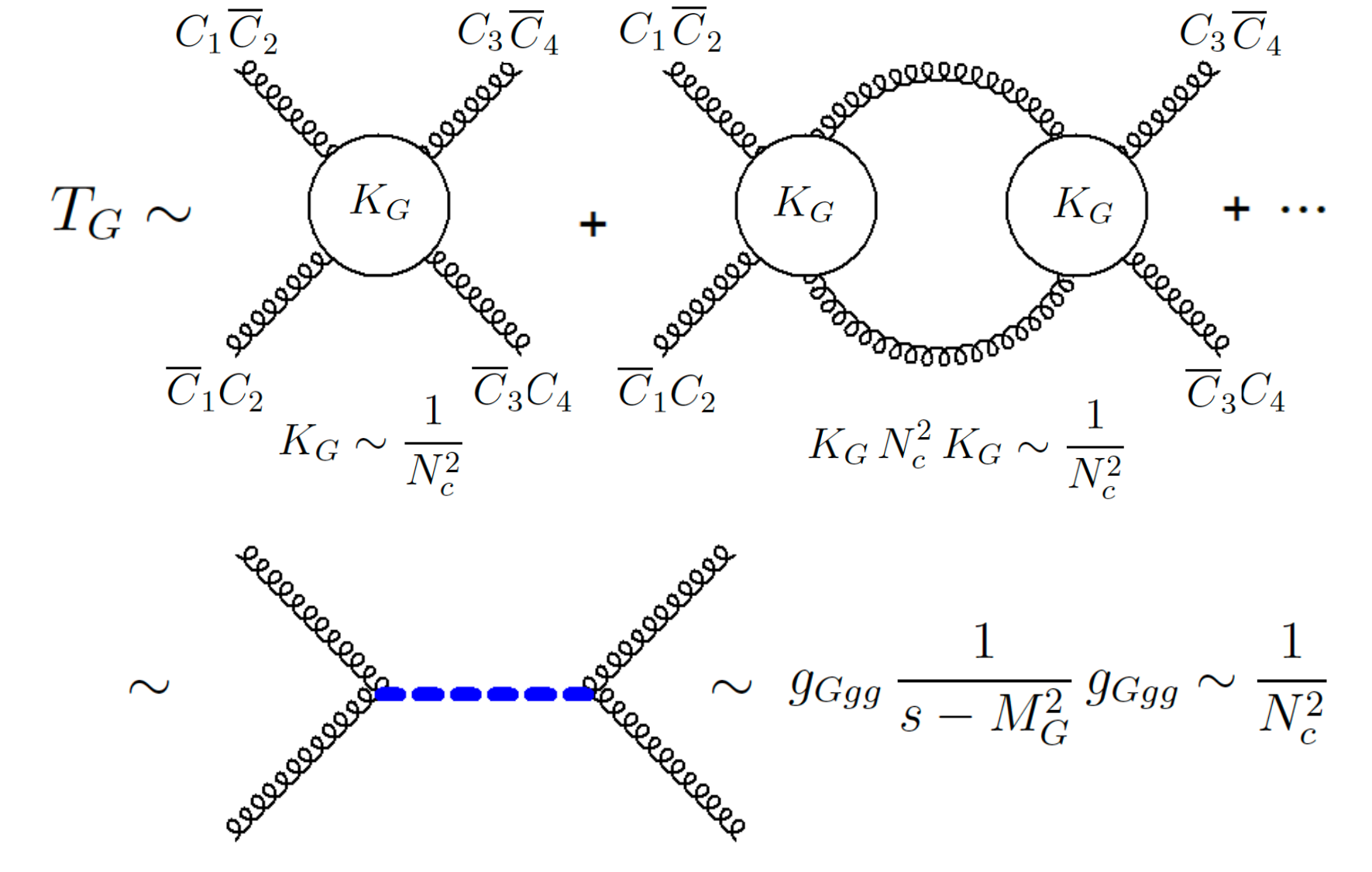}\\
        \caption{Resummation of diagrams for the gluon-gluon scattering of the type $C_1\bar{C}_2C_2\bar{C}_1\rightarrow C_3\bar{C}_4C_4\bar{C}_3$ with consequent formation of an intermediate glueball state (blue, thick-dashed line).}
        \label{GG-Page-3}
   \end{figure}
Then, the color wave function of this glueball can be written as%
\begin{equation}
\left\vert G\text{-color }\right\rangle \simeq\frac{1}{\sqrt{N_{c}^{2}-1}%
}J_{G}\left\vert 0\right\rangle \text{ .}
\end{equation}
Explicitly: 
\begin{equation}
\left\vert G\text{-color }\right\rangle \simeq\frac{1}{N_{c}}\left\vert
\bar{C}_{1}C_{1}\bar{C}_{2}C_{2}+\bar{C}_{1}C_{1}\bar{C}_{3}C_{3}%
+...\right\rangle \simeq\frac{1}{N_{c}}\sum_{a=1}^{N_{c}}\sum_{b=1}^{N_{c}%
}\left\vert \bar{C}_{a}C_{a}\bar{C}_{b}C_{b}\right\rangle \text{ ,}%
\end{equation}
where, for simplicity, on the r.h.s. all the combinations are taken into
account. Again, there is one combination (the colorless one) that should be subtracted, but this is unimportant for large $N_{c}$. Besides that, the previous equation is fully general,
and applies to any two-gluon glueball.

Following the same procedure as for quark-antiquark states, let us
consider the processes leading to the illustrative transition:
\begin{equation}
\bar{C}_{1}C_{1}\bar{C}_{2}C_{2}\rightarrow\bar{C}_{3}C_{3}\bar{C}_{4}C_{4}%
\end{equation}
in which all the colors have changed. It is easy to see that the dominant processes leading to this type of transitions scales as $N_{c}^{-2}$, see Fig. \ref{GG-Page-1}. 
In fact,
a single gluon exchange or the quartic interaction, proportional to $g^{2}$,
are not sufficient for a switch of all colors, see Fig.  \ref{GG-Page-2}.

We write down an effective
Lagrangian
\begin{equation}
\mathcal{L}_{G}=K_{G}J_{G}^{2}%
\label{lagG}
\end{equation}
where
\begin{equation}
K_{G}\sim g^{4}\sim N_{c}^{-4}\text{ , thus }K_{G}=\frac{\bar{K}_{G}}%
{N_{c}^{2}}%
\text{ ,}
\end{equation}
with $\bar{K}_{G}$ being $N_{c}$ independent. Note, a (nonlocal version of
this) Lagrangian was implemented in Ref. \cite{gutsche} to study the mixing of
glueballs with quarkonia.

The gluon-gluon scattering matrix for a given selected process such as
$\bar{C}_{1}C_{1}\bar{C}_{2}C_{2}\rightarrow\bar{C}_{3}C_{3}\bar{C}_{4}C_{4}$
is given by%
\begin{equation}
T_{G}(s)=\frac{1}{K_{G}^{-1}-\Sigma_{G}(s)}
\end{equation}
see Fig. \ref{GG-Page-3}  for its pictorial representation. 
Now, the loop $\Sigma_{G}(s)$ scales as \cite{gutsche,gounaris}:%
\begin{equation}
\Sigma_{G}(s)=N_{c}^{2}\bar{\Sigma}_{G}(s) \text{ .}
\end{equation}
Then:
\begin{equation}
T_{G}=\frac{1}{\frac{N_{c}^{2}}{\bar{K}_{G}}-N_{c}^{2}\bar{\Sigma}_{G}(s)}%
\text{ .}
\end{equation}
Just as for the quarkonium, the glueball mass is $N_{c}$-independent and
solves the equation
\begin{equation}
\frac{1}{\bar{K}_{G}}-\bar{\Sigma}_{G}(s=M_{G}^{2})=0\rightarrow M_{G}\sim
N_{c}^{0}
\text{ .}
\end{equation}
Following the same steps, the amplitude can be written as%
\begin{equation}
T_{G}\simeq\frac{(ig_{Ggg})^{2}}{s-M_{G}^{2}}
\end{equation}
where the coupling of the glueball to its gluonic constituents is
\begin{equation}
g_{Ggg}=\frac{1}{\sqrt{N_{c}^{2}\left(  \frac{\partial\bar{\Sigma}_{G}%
(s)}{\partial s}\right)  _{s=M_{G}^{2}}}}=\frac{\bar{g}_{Ggg}}{N_{c}}\text{ .}%
\end{equation}

\begin{figure}[h]
        \centering       \includegraphics[scale=0.45]{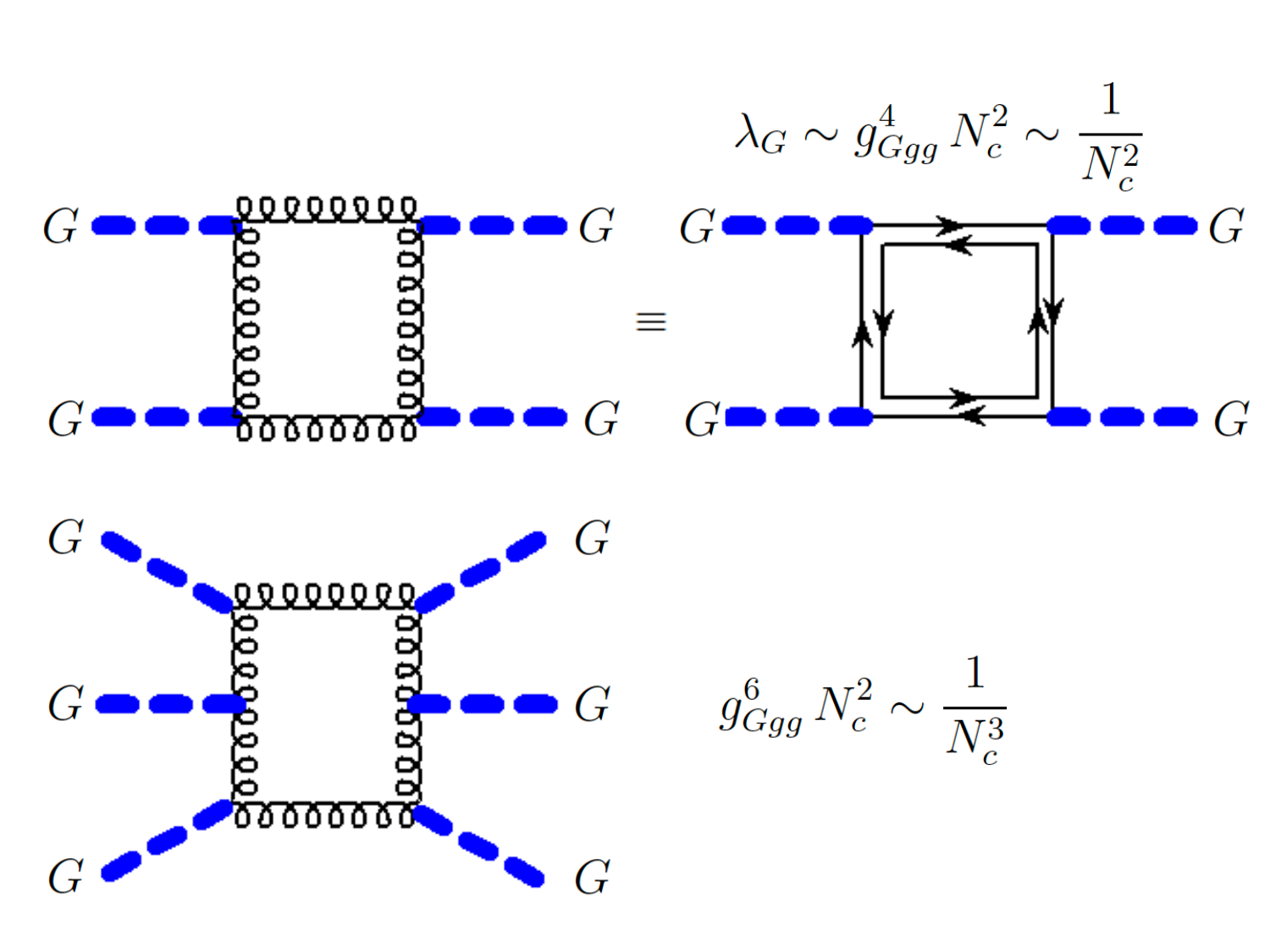}\\
        \caption{Examples of processes involving only glueballs as initial and final states. Up: the amplitude $GG \rightarrow GG$ goes as $N_c^{-2}$. Down: the amplitude $GGG \rightarrow GGG$ goes as $N_c^{-3}$. }
        \label{GGcons-Page-5}
   \end{figure}

 \begin{figure}[h]
        \centering       \includegraphics[scale=0.45]{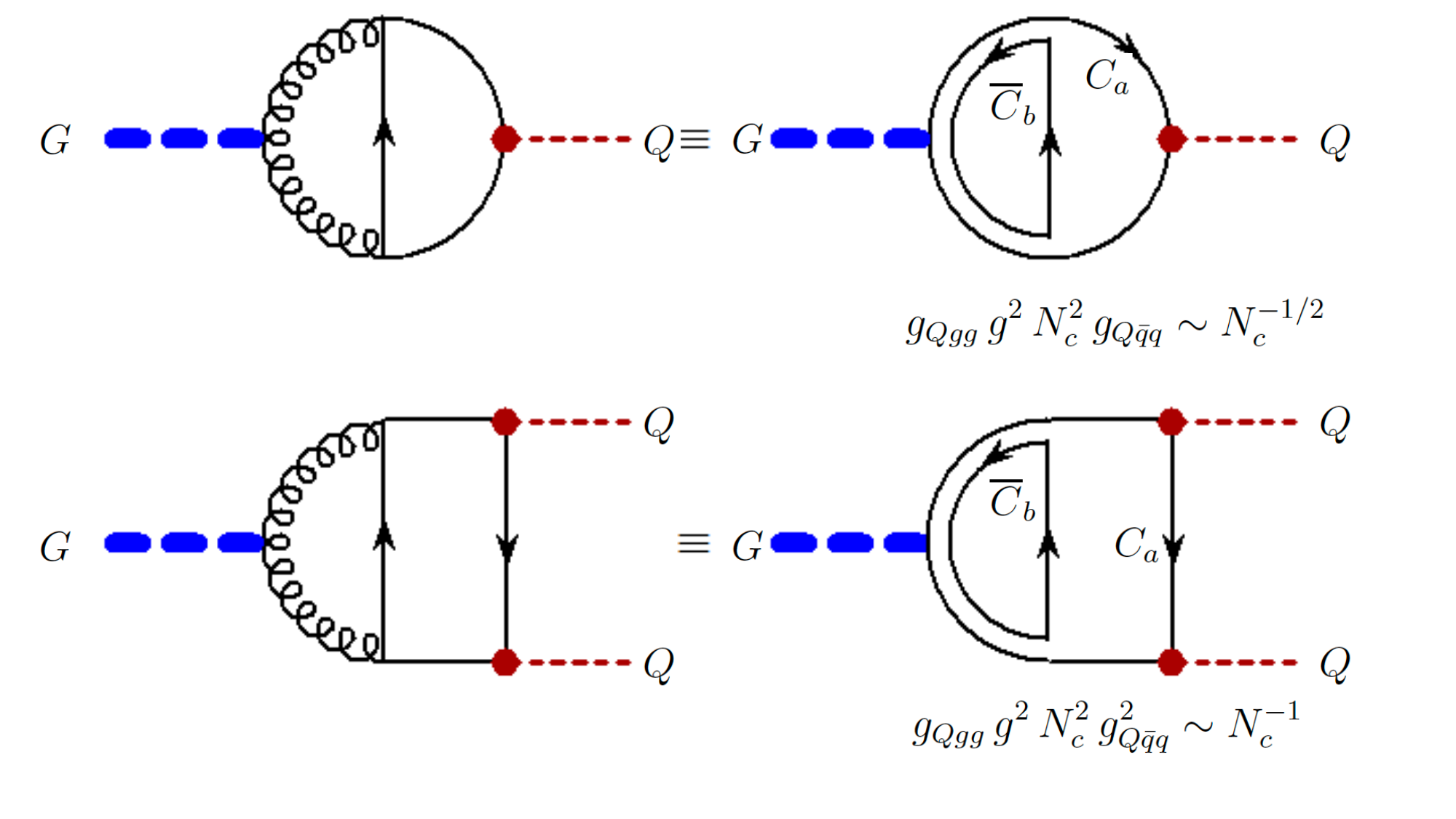}\\
        \caption{Up: Mixing of a glueball (blue) and a quarkonium (red) via quarks and gluons (black straight and springy lines). The amplitude scales as $N_c^{-1/2}$. Down: Decay of a glueball (blue) into two quarkonia (red) via quarks and gluons (black). This diagram scales as $N_c^{-1}$, thus the glueball decay width goes as $N_c^{-2}$.}
        \label{GGcons-Page-1}
   \end{figure}

    \begin{figure}[h]
        \centering       \includegraphics[scale=0.35]{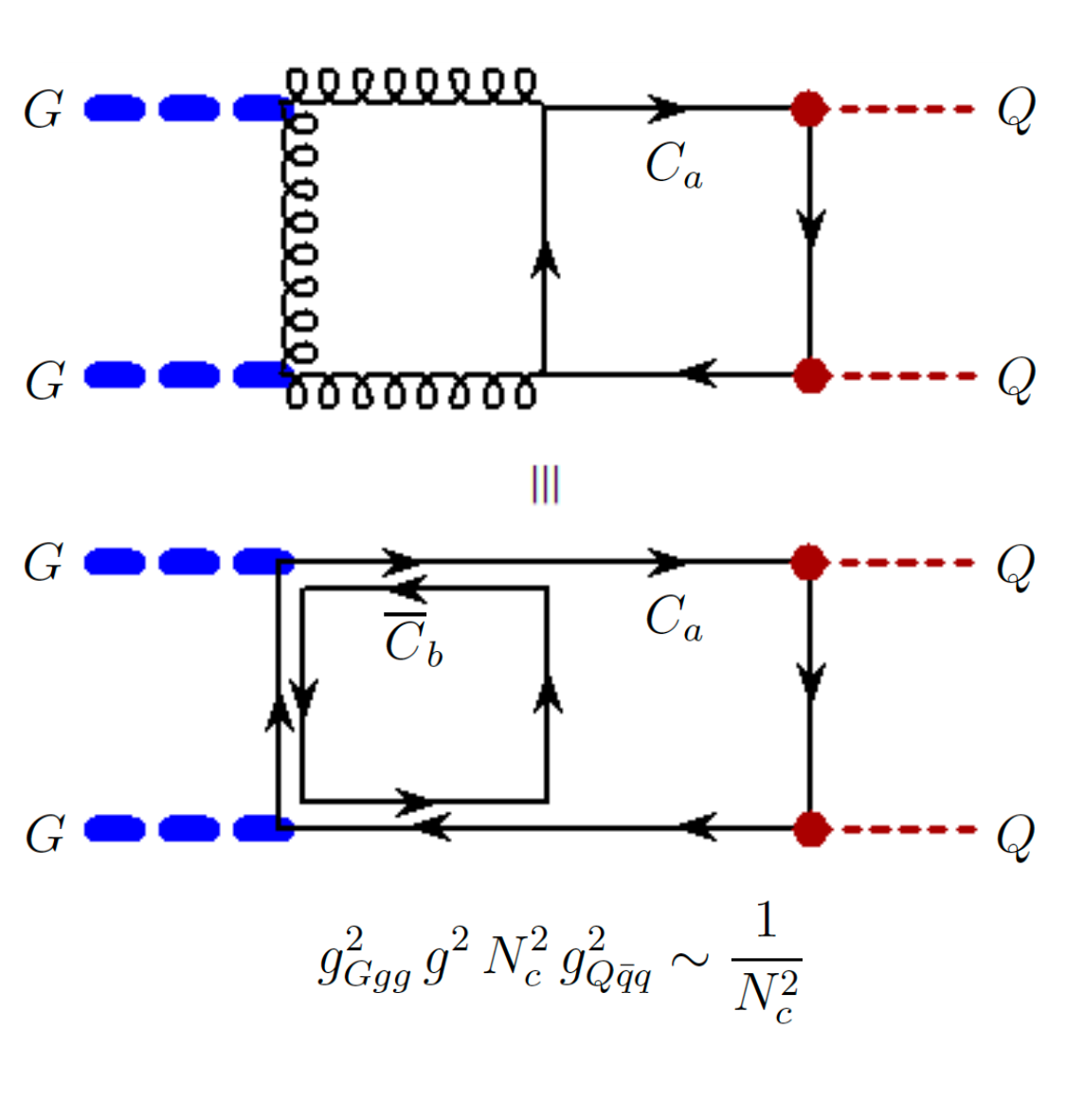}\\
        \caption{Leading amplitude for the process $GG \rightarrow QQ$, that scales as $N_c^{-2}$. As usual, glueballs are blue, quarkonia red, quarks and gluons black. }
        \label{GGcons-Page-2}
   \end{figure}
   
From the results above, we can easily derive the phenomenology of glueballs at large-$N_c$.

First, we describe the four- and six-leg purely glueball couplings, which scale as $N_c^{-2}$ and $N_c^{-3}$, respectively, see Fig. \ref{GGcons-Page-5}.
This is in agreement with the general amplitude for $n_G$ glueballs being
$A_{n_G G}\propto\frac{N_{c}^{2}}{N_{c}^{n_{G}}}$, see Sec. 2.6.

Next, we calculate the interaction of a glueball with mesons. The basic mixing
goes as $A_{GQ}\sim N_{c}^{-1/2}$, while the decay amplitude scales as
$A_{GQQ}\sim N_{c}^{-1}$ (see Fig. \ref{GGcons-Page-1}). 
It then follows that the glueball decay into two
standard mesons is suppressed as
\begin{equation}
\Gamma_{G\rightarrow QQ}\sim N_{c}^{-2} \text{ ,}
\end{equation}
thus even more suppressed than the quarkonium decay.

\begin{figure}[h]
        \centering       \includegraphics[scale=0.55]{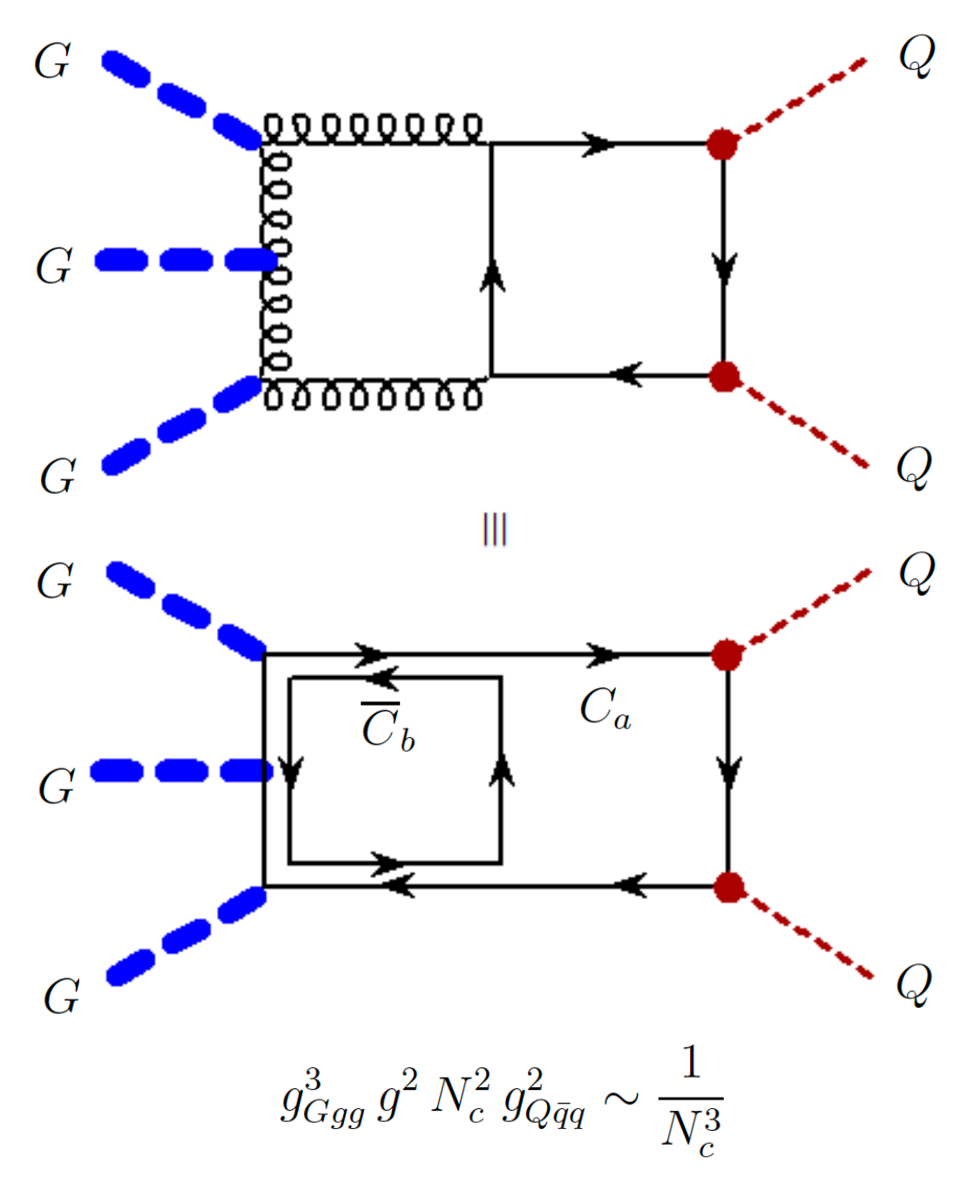}\\
        \caption{Leading amplitude for the process $GGG \rightarrow QQ$, that scales as $N_c^{-3}$.}
        \label{GGcons-Page-4}
   \end{figure}

As additional examples, in Fig. \ref{GGcons-Page-4} we present the scattering $GG \rightarrow QQ$  (two glueballs into two quarkonia), that behaves as $N_c^{-2}$, and in Fig. \ref{GGcons-Page-5} the scattering $GGG \rightarrow QQ$ scaling with $N_c^{-3}$.

The general amplitude for $n_{Q}$ quarkonia and $n_{G}$ glueballs as $A_{\left(  n_{Q}Q\right)
\left(  n_{G}G\right)  }\propto\frac{N_{c}}{N_{c}^{n_{Q}/2}N_{c}^{n_{G}}}.$ In
this way we confirm the previously quoted results (Sec. 2.6).

An important remark is in order: glueballs with thee gluons work just as above. The scaling laws are left unchanged. 

Next, we describe additional consequences concerning glueballs. 

\bigskip

1) \textbf{The dilaton Lagrangian in the large-}$\mathbf{N}_{c}$\textbf{
limit.}

The scalar glueball $G$ can be described as a dilaton field, which is an
important element of many chiral models (among which the extended linear sigma model \cite{dick}).

\begin{figure}[h]
        \centering       \includegraphics[scale=0.75]{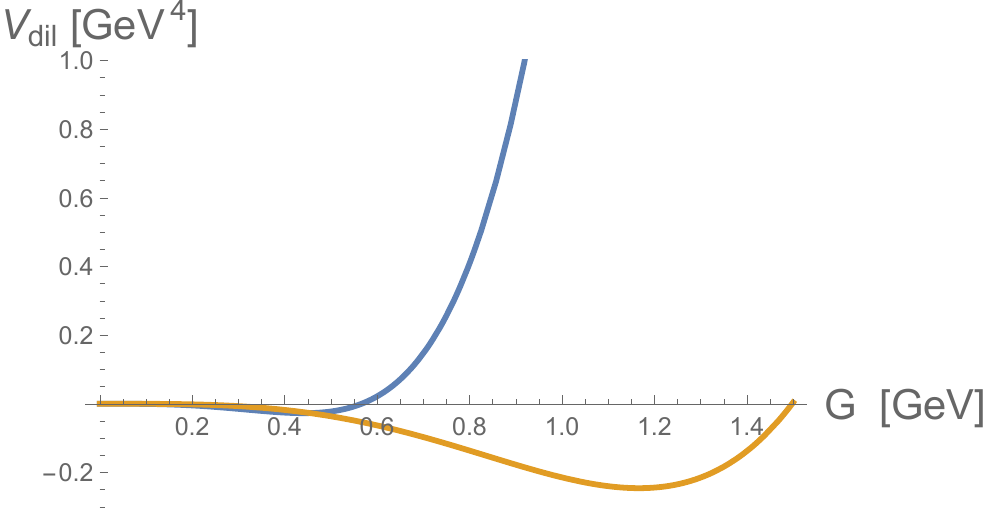}\\
        \caption{Function $V_{dil}(G)$ expressed in Eq. (\ref{vdil}) for $N_c=3$ (blue line) and for $N_c=7$ (yellow line). The value of the minimum $G_0$ increases with $N_c$, while its depth scales with $-N_c^2$. The numerical value for $N_c=3$ are: $\Lambda_G = 0.5$ GeV, and $\lambda_G=1.7^2/\Lambda_G^2$, corresponding to a scalar glueball mass of $M_G=1.7$ GeV. }
        \label{dildir}
   \end{figure}

The dilaton Lagrangian reads
\cite{migdal,salo,ellis,stani}
\begin{equation}
\mathcal{L}_{dil}=\frac{1}{2}(\partial_{\mu}G)^{2}-V_{dil}(G)\text{ , }%
\end{equation}
with%
\begin{equation}
V_{dil}(G)=\frac{1}{4}\lambda_{G}\left[  G^{4}\ln\left(  \frac{G}{\Lambda_{G}%
}\right)  -\frac{G^{4}}{4}\right]  \text{ ,}
\label{vdil}%
\end{equation}
which contains the dimensionless constant $\lambda_{G}$ and the dimensionful
constant $\Lambda_{G}.$ The scaling laws, to be explained below, are given by:%
\begin{equation}
\lambda_{G}\sim N_{c}^{-2}\text{ , }\Lambda_{G}\sim N_{c}\text{ .}%
\end{equation}
The potential is shown in Fig. \ref{dildir} for two different values of $N_c$ (3 and 7, respectively). For $N_c=3$, the numerical values are given by $\lambda_G = 1.7^2/0.5^2$ and  $\Lambda_G = 0.5^2$ GeV$^2$, corresponding to a glueball mass of $1.7$ GeV, in agreement with lattice estimates \cite{mainlattice,at}.

The logarithm and the dimensional parameter $\Lambda_{G}$ are required for
describing the breaking of dilatation symmetry
\[
x^{\mu}\rightarrow\lambda^{-1}x^{\mu}\text{ and }G(x)\rightarrow G^{\prime
}(x^{\prime})=\lambda G(x)\text{ }%
\]
in the following way:
\begin{equation}
\partial_{\mu}J^{\mu}=T_{\mu}^{\mu}= -G\partial_{G}V_{dil}(G) +4V_{dil}(G)=-\frac{1}%
{4}\lambda_{G}G^{4}\text{ .}%
\end{equation}
This equation resembles the QCD result \cite{weisebook,ellis}
\begin{equation}
\left(  T_{\mu}^{\mu}\right)  _{QCD}=-\frac{\alpha_{s}}{16\pi}\left(
\frac{11}{3}N_{c}-\frac{2}{3}N_{f}\right)  G_{\mu\nu}^{a}G^{a,\mu\nu} \text{ .}
\end{equation}
Taking the expectation value of the former equation we get:%
\begin{equation}
\left\langle \left(  T_{\mu}^{\mu}\right)  _{QCD}\right\rangle =-\frac
{\alpha_{s}}{16\pi}\left(  \frac{11}{3}N_{c}-\frac{2}{3}N_{f}\right)
\left\langle G_{\mu\nu}^{a}G^{a,\mu\nu}\right\rangle \text{ ,} 
\label{gluoncond}
\end{equation}
which scales as $N_{c}^{2}$ because the gluon condensate $\left\langle G_{\mu\nu
}^{a}G^{a,\mu\nu}\right\rangle \sim N_{c}^{2}$ and $\alpha_{s}\sim N_{c}^{-1}$
(for a numerical estimate of the gluon condensate for $N_c=3$, see Ref. \cite{gluoncondensat}). Does the dilaton
potential reproduce this scaling? In order to see that, let us expand the
dilaton potential around the minimum, which is realized for
$G_{0}=\Lambda
_{G}.$ Upon performing the shift $G\rightarrow G_{0}+\Lambda_{G},$ we obtain \cite{glueballonium}:
\begin{equation}
V_{dil}(G)=-\lambda_{G}\Lambda_{G}^{4}+\frac{1}{2}\lambda_{G}\Lambda_{G}%
^{2}G^{2}+\frac{5\lambda_{G}\Lambda_{G}}{3!}G^{3}+\frac{11\lambda_{G}}%
{4!}G^{4}+...
\end{equation}
where $V_{dil}(G=\Lambda_{G})=-\lambda_{G}\Lambda_{G}^{4} \sim N_c^2$ has been used. From
the term proportional to $G^{4}$ it follows that
\begin{equation}
\lambda_{G}\sim N_{c}^{-2}
\text{ .}
\end{equation}
The glueball mass reads
\begin{equation}
M_{G}^{2}=\lambda_{G}\Lambda_{G}^{2}\sim N_{c}^{0}\rightarrow\Lambda_{G}\sim
N_{c}\text{ .}%
\end{equation}
Note, the $G^{3}$ terms scales as $\lambda_{G}\Lambda_{G}\sim N_{c}^{-1},$
which is in agreement with the previous general rules. Going to higher order
in the expansions would also generate terms in agreement with those rules
(for instance, $G^{5}$ goes as $\lambda_{G}/\Lambda_{G}\sim N_{c}^{-3}$ as expected, etc.).

The fact that the energy parameter $\Lambda_{G}$ scales as $N_{c}$ implies that it can be
intuitively expressed as
\begin{equation}
\Lambda_{G}\sim N_{c}\Lambda_{QCD}\text{ .}%
\end{equation}

Finally, let us have a look at the condensate of the dilaton field:%
\begin{equation}
\left\langle T_{\mu}^{\mu}\right\rangle =-\frac{1}{4}\lambda_{G}\left\langle
G^{4}\right\rangle =-\frac{1}{4}\lambda_{G}\Lambda_{G}^{4}\sim - N_{c}^{2}%
\end{equation}
in agreement with the QCD scaling of Eq. (\ref{gluoncond}).

The scalar glueball is the lightest gluonic state predicted by lattice QCD
and is a natural element of the chiral models with dilatation invariance  \cite{dick,stani}. Presently, the resonance $f_0(1710)$ is a good candidate for being predominantly the scalar glueball, see e.g. \cite{stani,weingarten,long,cheng,rebhan,jpsilatt} and refs. therein. 

2) \textbf{Coupling the dilaton to other glueballs.}

The lightest scalar glueball is special since it is related to dilatation
symmetry and its breaking, but other fields can be easily introduced. For illustrative
purposes, let us couple the dilaton to the pseudoscalar glueball $\tilde{G}$ and the tensor
glueball $T_{\mu\nu}$:
\begin{equation}
\mathcal{L}=\mathcal{L}_{dil}+\mathcal{L}_{kin}-\frac{\lambda_{\tilde{G}G}}%
{2}\tilde{G}^{2}G^{2}-\lambda_{\tilde{G}}\tilde{G}^{4}+\frac{\lambda_{TG}}{2}T_{\mu\nu}T^{\mu\nu}G^{2}+\frac
{\lambda_{T}}{2}\left(  T_{\mu\nu}T^{\mu\nu}\right)  ^{2}+...
\label{glueballgen}%
\end{equation}
where $\Lambda_{G}$ is the only dimensionful parameter entering in
$\mathcal{L}_{dil}.$ All the $\lambda$ parameters scale as $N_{c}^{-2},$ since
each of them describes a four-leg interaction between glueballs.

When considering the shift $G\rightarrow\Lambda_{G}+G$, the other glueballs
get a mass: $m_{\tilde{G}}^{2}=\lambda_{\tilde{G}G}\Lambda_{G}^{2}\sim
N_{c}^{0}$ and $m_{T}^{2}=\lambda_{TG}\Lambda_{G}^{2}\sim N_{c}^{0}.$ For an
explicit study of the scattering of tensor glueballs using the Lagrangian
above see Ref. \cite{grg}. 
\bigskip

3) \textbf{Coupling the dilaton to the LSM.}

We consider, again for simplicity, the case $N_{f}=1$.
The potential for the chiral model containing
both the dilaton as well as the chiral multiplet $\Phi=\sigma+i\pi$
is given by:
\begin{equation}
V(G,\sigma,\pi)=V_{dil}(G)+\frac{a}{2}G^{2}\Phi^{\ast}\Phi+\frac{\lambda}%
{4}\left(  \Phi^{\ast}\Phi\right)  ^{2}
\text{ .}
\end{equation}
Again, $\Lambda_{G}$ is the only dimensionful parameter. Above, the constant $a$ scales as
\begin{equation}
a\sim N_{c}^{-2}\text{ ,}%
\end{equation}
since it parameterizes a vertex with 2 glueballs and two quarkonia. The
realistic $N_{f}=3$ treatment of this model can be found in Ref. \cite{stani}.

The search for the minimum of the model is more complicated than in the LSM case, since now two scalar fields are present and can condense. Namely, setting the pion field to zero ($\pi = 0$), one has:
\begin{equation}
V(G,\sigma,0)=\frac{1}{4}\lambda_{G}\left[  G^{4}\ln\left(  \frac{G}%
{\Lambda_{G}}\right)  -\frac{G^{4}}{4}\right]  +\frac{a}{2}G^{2}\sigma
^{2}+\frac{\lambda}{4}\sigma^{4} \text{ ,}%
\end{equation}
with%
\begin{equation}
\lambda_{G}\sim N_{c}^{-2}\text{ , }\Lambda_{G}\sim N_{c}\text{, }a\sim
N_{c}^{-2}\text{ },\lambda\sim N_{c}^{-1}\text{ .}%
\end{equation}
The minimum is searched for:%
\begin{align}
\partial_{G}V(G,\sigma,0)  &  =\lambda_{G}G^{3}\ln\left(  \frac{G}{\Lambda
_{G}}\right)  +aG\sigma^{2}=0 \text{ ,}\\
\partial_{\sigma}V(G,\sigma,0)  &  =aG^{2}\sigma+\lambda\sigma^{3}=0.
\end{align}

For $a>0,$ the minimum is realized for $G_{0}=\Lambda_{G}\neq0,$ $\sigma
_{0}=0$, but as explained above this is not what we have in Nature.

For $a<0$ , the minimum is realized for $G_{0}\neq0,$ $\sigma_{0}\neq0$:
spontaneous breaking of chiral symmetry (on top of the breaking of dilatation symmetry) is realized. In particular, one has%

\begin{equation}
\sigma_{0}^{2}=-\frac{aG_{0}^{2}}{\lambda}\rightarrow\sigma_{0}=\sqrt
{\frac{-m_{0}^{2}}{\lambda}}\sim N_{c}^{1/2}\text{ with }m_{0}^{2}=aG_{0}%
^{2}<0 \text{ .}
\end{equation}
The equation for $G_{0}$ reads:
\begin{equation}
\lambda_{G}G_{0}^{3}\ln\left(  \frac{G_{0}}{\Lambda_{G}}\right)
=-aG_{0}\sigma_{0}^{2}=\frac{a^{2}G_{0}^{3}}{\lambda}%
\text{ ,}
\end{equation}
or%
\begin{equation}
\ln\left(  \frac{G_{0}}{\Lambda_{G}}\right)  =\frac{a^{2}}{\lambda\lambda_{G}} \text{ ,}
\end{equation}
hence:%
\begin{equation}
G_{0}=\Lambda_{G}e^{\frac{a^{2}}{\lambda\lambda_{G}}}\geqslant\Lambda_{G}%
\text{ .}
\end{equation}
In terms of large-$N_{c}$, we have:%
\begin{equation}
G_{0}\sim N_{c}e^{1/N_{c}}\sim N_{c}\left(  1+\frac{1}{N_{c}}+...\right)
\text{ .}%
\end{equation}
Thus, in the large-$N_{c}$ limit%
\begin{equation}
G_{0}=\Lambda_{G} \propto N_c \text{ .}
\end{equation}
How far is $G_{0}$ from $\Lambda_{G}?$ For $N_{c}=3$ that depends on the
numerical values, but typically $G_{0}\simeq\Lambda_{G}$ is well fulfilled \cite{glueballonium}.

Finally, let us briefly discuss the explicit symmetry breaking. That can be
achieved by a term of the type
\begin{equation}
\lambda_{n}m_{n}G^{2}\sigma
\end{equation}
where the coupling constant $\lambda_{n}$ scales as
\begin{equation}
\lambda_{n}\sim N_{c}^{-3/2}%
\end{equation}
since it represents the coupling to two glueballs to an ordinary meson.

Then, the parameter $h$ in Eq. (\ref{poth}) turns out to be (upon dilaton condensation ($G = G_0$):
\begin{equation}
h=\lambda_{n}m_{n}G_{0}^{2}\simeq\lambda_{n}m_{n}\Lambda_{G}^{2}\sim
N_{c}^{1/2} \text{ ,}
\end{equation}
as required for getting a pion mass that does not depend on $N_{c}$. This result shows that this is the appropriate way to model the explicit chiral symmetry breaking.

Another interesting consequence of this toy model is the decay of $G$ into
pions \cite{glueballonium}:
\begin{align}
\Gamma_{G\rightarrow\pi\pi}  &  =2\frac{k_{\pi}}{8\pi M_{\sigma}^{2}}\left[
aG_{0}\right]  ^{2}=2\frac{k_{\pi}}{8\pi M_{\sigma}^{2}}\left[  \frac
{-m_{0}^{2}}{G_{0}}\right]  ^{2}\\
&  \simeq2\frac{k_{\pi}}{8\pi M_{\sigma}^{2}}\left[  \frac{M_{\sigma}^{2}%
}{2\Lambda_{G}}\right]  ^{2}\sim N_{c}^{-2}\text{ ,}%
\end{align}
where $k_{\pi}=\sqrt{\frac{M_{G}^{2}}{4}-M_{\pi}^{2}}$. 
The scaling laws, that
follow from $\Lambda_{G}\sim N_{c}$, are in agreement with the expected results.

All in all, a fully consistent picture, that is correctly embedded in chiral models with the dilaton, is obtained in the large-$N_{c}$ limit.

\newpage

\textbf{4) Decays of other glueballs.}

Other glueballs also decays into conventional quark-antiquark mesons. A special case is given by the pseudoscalar glueball, whose coupling to mesons may be written down as \cite{psg}
\begin{equation}
    \mathcal{L}_{\tilde{G}} = ic_{\tilde{G}}\tilde{G}(\det\Phi - \det\Phi^{\dagger})
\end{equation}
where 
\begin{equation}
    c_{\tilde{G}} \propto N_c^{-1/2 -N_f/2} \text{ .}
\end{equation}
A numerical evaluation of $c_{\tilde{G}}$ via instantons can be found in Ref. \cite{anomaly2}. The recently discovered resonance $X(2600)$ by the BES III collaboration is a promising candidate for being the pseudoscalar glueball \cite{bes3}. 

For the study of various glueball masses and decays we refer to chiral models of Refs. \cite{psg,sammet,vereijken} and to the 
Witten-Sakai-Sugimoto approach (which also makes use of large-$N_{c}$ arguments) of Refs. \cite{rebhan,rebhan2,rebhan3} (for additional holographic considerations related to the spectrum, see Ref. \cite{misra}).

\bigskip

\textbf{5) Connection to correlations}

\begin{figure}[h]
        \centering       \includegraphics[scale=0.45]{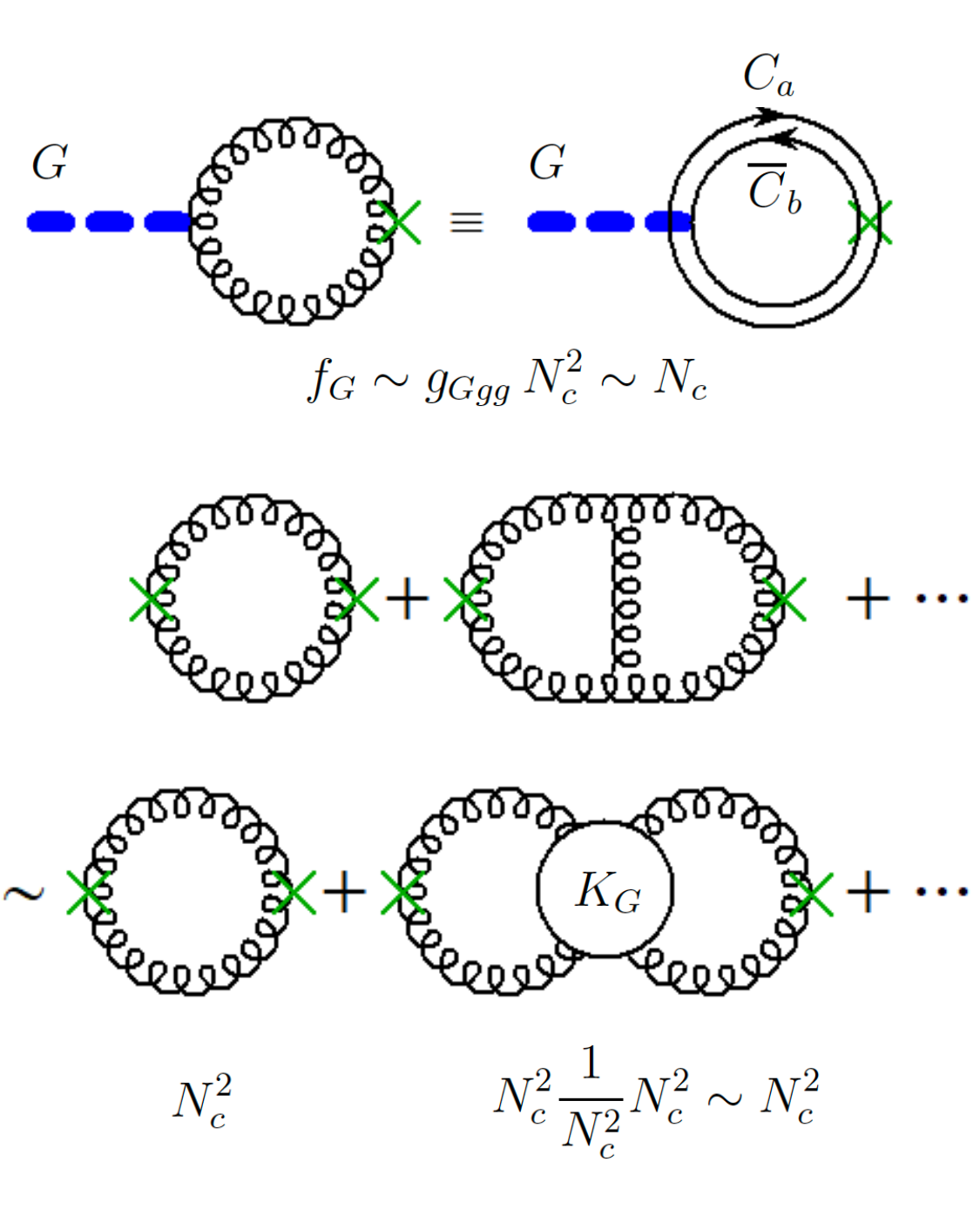}\\
        \caption{Up: Amplitude for the production of the glueball, denoted as $f_G$, scales with $N_c$. Down: Correlator due to colorless sources of gluon-gluon states (glueballs): the bubbles scale as $N_c^2$. One may understand this process with a (tower of) glueball(s) as intermediate states, see Eq. (\ref{FG}). }
        \label{GGcons-Page-3}
   \end{figure}
   
When considering the correlation involving glueball currents
\begin{equation}
\left\langle J_{G}(x_{2})J_{G}(x_{1})\right\rangle = -i\int d^{4}pF_{G}%
(s = p^2)e^{ip(x_{1}-x_{2})}\text{ ,}%
\end{equation}
$F_{G}(s)$ is the loop contribution with total momentum $p.$ At lowest order
this is just the loop function $-\Sigma_{G}(s=p^{2}).$ Evaluating $F_{G}(p)$ according to Fig. \ref{GGcons-Page-3} we
get:
\begin{gather}
F_{G}(p)=-\Sigma_{G}(s=p^{2})\left(  1+\Sigma_{G}(s)K_{G}+...\right)
= \nonumber \\ -\frac{\Sigma_{G}(s)}{1-\Sigma_{G}(s)K_{G}}=-\frac{\Sigma_{G}(s)}{K_{G}}%
\frac{1}{K_{G}^{-1}-\Sigma_{G}(s)}
\text{ .}
\end{gather}
The pole takes place (just as previously) for $K_{G}^{-1}-\Sigma_{G}(s=M_{G}^{2})=0.$ Upon
expanding close to the pole, we find:%
\begin{equation}
F_{G}(p)=\frac{\Sigma_{G}(s)\Sigma_{G}(M_{G}^{2})}{\Sigma_{G}^{^{\prime}%
}(M_{G}^{2})(s-M_{G}^{2})}\simeq\frac{\Sigma_{G}^{2}(M_{Q}^{2})g_{Ggg}^{2}%
}{s-M_{G}^{2}}\simeq\frac{f_{G}^{2}}{s-M_{Q}^{2}}%
\end{equation}
\label{FG}
where
\begin{equation}
f_{G}=g_{Ggg}\Sigma_{G}(M_{G}^{2})\sim N_{c}^{-1}\cdot N_{c}^{2}\sim N_{c}%
\end{equation}
is the vacuum production/annihilation amplitude of the glueball $G$. In some
cases, $f_{G}$ may also be referred as a `decay constant', yet it should be
stressed that this is not the weak decay constant. This is so because $W^{\pm
}$ and $Z^{0}$ couple directly to quarks and not to gluons. In order to obtain
the weak decay constant of glueballs, an additional suppression of $N_{c}$
enters, leading to
\begin{equation}
f_{G}^{weak}\sim N_{c}^{0}.
\end{equation}
Indeed, the same result can be obtained starting with an external glueball $G$, which transforms to a $Q$ (mixing proportional $N_{c}^{-1/2}$), which then
annihilates weakly (process proportional to $f_{Q}\sim N_{c}^{1/2}$). The
scaling goes as $f_{G}^{weak}\sim N_{c}^{-1/2}\cdot f_{Q}\sim N_{c}^{0}.$

\subsection{Hybrids}

Hybrids are bound states containing a quark-antiquark pair and a gluon, see the review of Ref. 
\cite{hybrid} and the lattice results in Ref. \cite{dudek}. As an example of an hybrid current, we consider the lightest
$1^{-+}$ hybrid case:
\begin{equation}
J_{H}^{\mu}=\sum_{a=1}^{N_{c}^{2}-1}\bar{q}G^{a,\mu\nu}t^{a}\gamma^{5}%
\gamma_{\nu}q \text{ .}
\end{equation}
The quantity $\bar{q}G^{a,\mu\nu}t^{a}\gamma^{5}\gamma
_{\nu}q$ is a scalar in color space. Schematically (and neglecting Lorentz
indices and Dirac matrices), the hybrid current in the double-line notation reads
\begin{equation}
J_{H}=\sum_{a=1}^{N_{c}}\sum_{b=1}^{N_{c}}\bar{q}^{(a)}A^{(a,b)}q^{(b)}%
=C_{1}\left(  \bar{C}_{1}C_{2}\right)  \bar{C}_{2}+C_{3}\left(  \bar{C}%
_{3}C_{4}\right)  \bar{C}_{4}+...
\end{equation}
The corresponding interaction Lagrangian for the hybrid formation is expressed as:%
\begin{equation}
\mathcal{L}_{H}=K_{H}J_{H}^{2}%
\label{LH}
\end{equation}

 \begin{figure}[h]
        \centering       \includegraphics[scale=0.39]{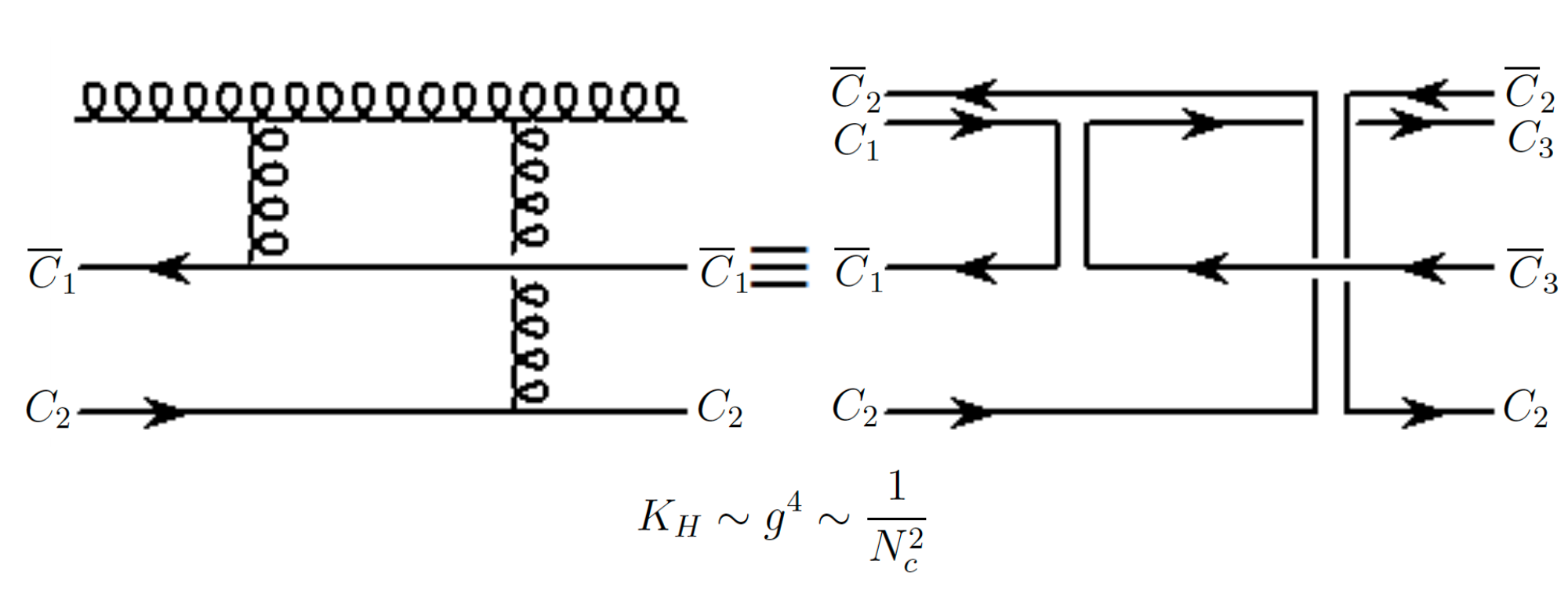}\\
        \caption{Amplitude for gluon-quark-antiquark (hybrid) scattering of the type $(C_1\bar{C}_2)C_2\bar{C}_1 \rightarrow (C_4\bar{C}_3)C_3\bar{C}_4$. Then initial gluon is $(C_1\bar{C}_2)$ and the final one $(C_4\bar{C}_3)$. Note, all colors have switched. The corresponding amplitude scales as $N_c^{-2}$ and models the constant $K_H$ appearing in the interaction Lagrangian of Eq. (\ref{LH}).}
        \label{HH-Page-1}
   \end{figure}

    \begin{figure}[h]
        \centering       \includegraphics[scale=0.35]{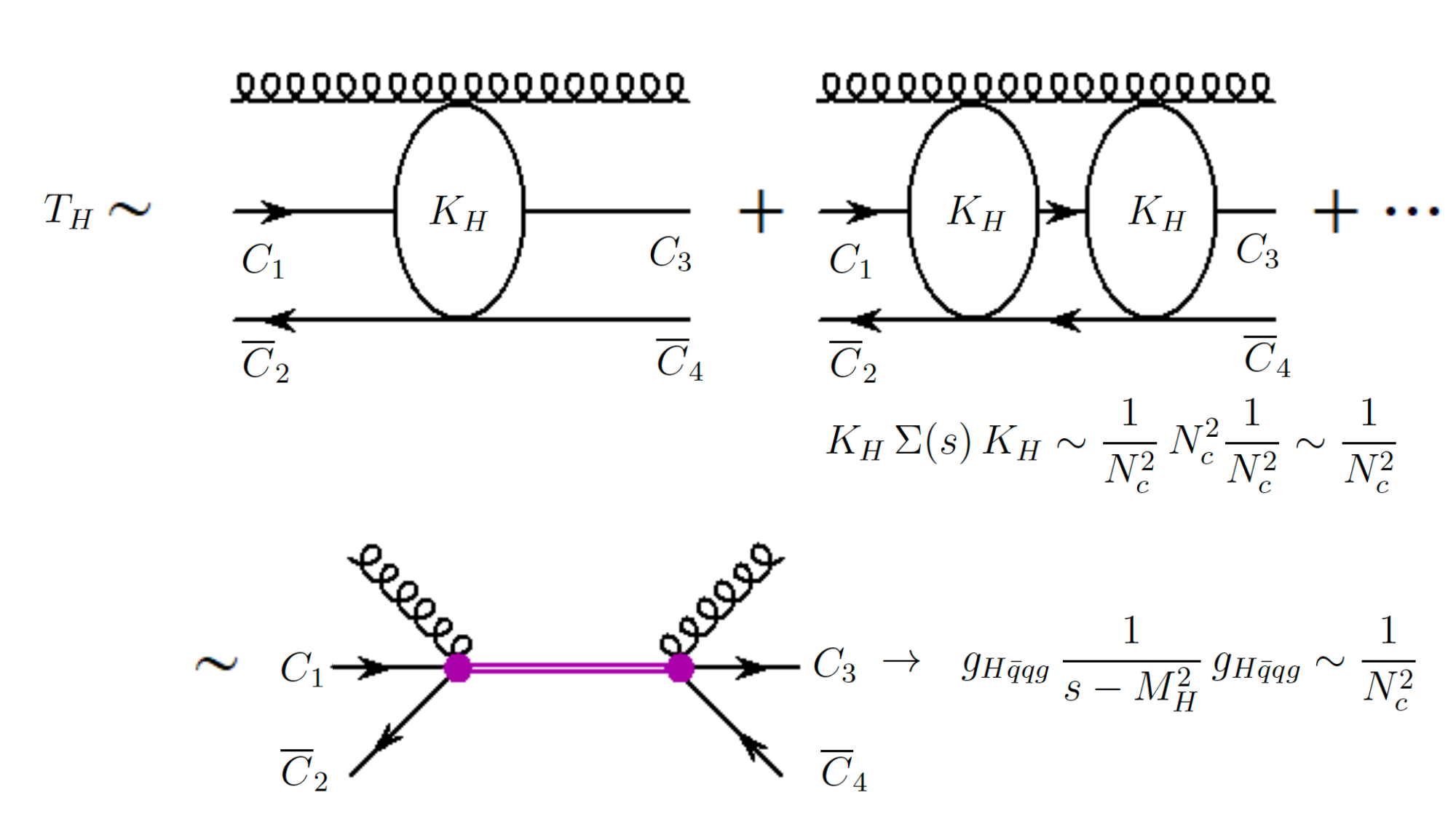}\\
        \caption{Resummed diagrams leading to the formation of an hybrid meson (purple, double-solid line). }
        \label{HH-Page-2}
   \end{figure}

In line with the previous cases, let us consider a specific transition (see Fig. \ref{HH-Page-1}):
\begin{equation}
C_{1}\left(  \bar{C}_{1}C_{2}\right)  \bar{C}_{2}\rightarrow C_{3}\left(
\bar{C}_{3}C_{4}\right)  \bar{C}_{4}%
\end{equation}
in which all colors have been switched. 

The basic (connected) interaction turns out to be of the order of (see Fig. \ref{HH-Page-1}):%
\begin{equation}
K_{H}\propto g^{4}\propto N_{c}^{-2}\rightarrow K_{H}=\frac{\bar{K}_{H}}%
{N_{c}^{2}}\text{ .}%
\end{equation}
We then proceed as before by resumming over loop diagrams, see Fig. \ref{HH-Page-2}, finding:%
\begin{equation}
T_{H}(s)=\frac{1}{K_{H}^{-1}-\Sigma_{H}(s)}%
\text{ .}
\end{equation}
Now, the loop $\Sigma_{H}(s)$ scales as:%
\begin{equation}
\Sigma_{H}(s)=N_{c}^{2}\bar{\Sigma}_{H}(s)
\text{ .}
\end{equation}
Then:
\begin{equation}
T_{H}=\frac{1}{\frac{N_{c}^{2}}{\bar{K}_{H}}-N_{c}^{2}\bar{\Sigma}_{H}(s)}%
\text{ .}
\end{equation}
Just as for quarkonia and glueballs, the hybrid mass is $N_{c}$-independent:
\begin{equation}
\frac{1}{\bar{K}_{H}}-\bar{\Sigma}_{H}(s=M_{H}^{2})=0\rightarrow M_{H}\sim
N_{c}^{0} \text{ .}
\end{equation}
Following analogous steps, the $T$-amplitude can be written as%
\begin{equation}
T_{H}\simeq\frac{(ig_{H\bar{q}qg})^{2}}{s-M_{H}^{2}}
\text{ ,}
\end{equation}
where the coupling of an hybrid meson to its constituents is
\begin{equation}
g_{H\bar{q}qg}=\frac{1}{\sqrt{N_{c}^{2}\left(  \frac{\partial\bar{\Sigma}%
_{G}(s)}{\partial s}\right)  _{s=M_{G}^{2}}}}=\frac{\bar{g}_{H\bar{q}qg}%
}{N_{c}}\text{ .}%
\end{equation}

 \begin{figure}[h]
        \centering       \includegraphics[scale=0.40]{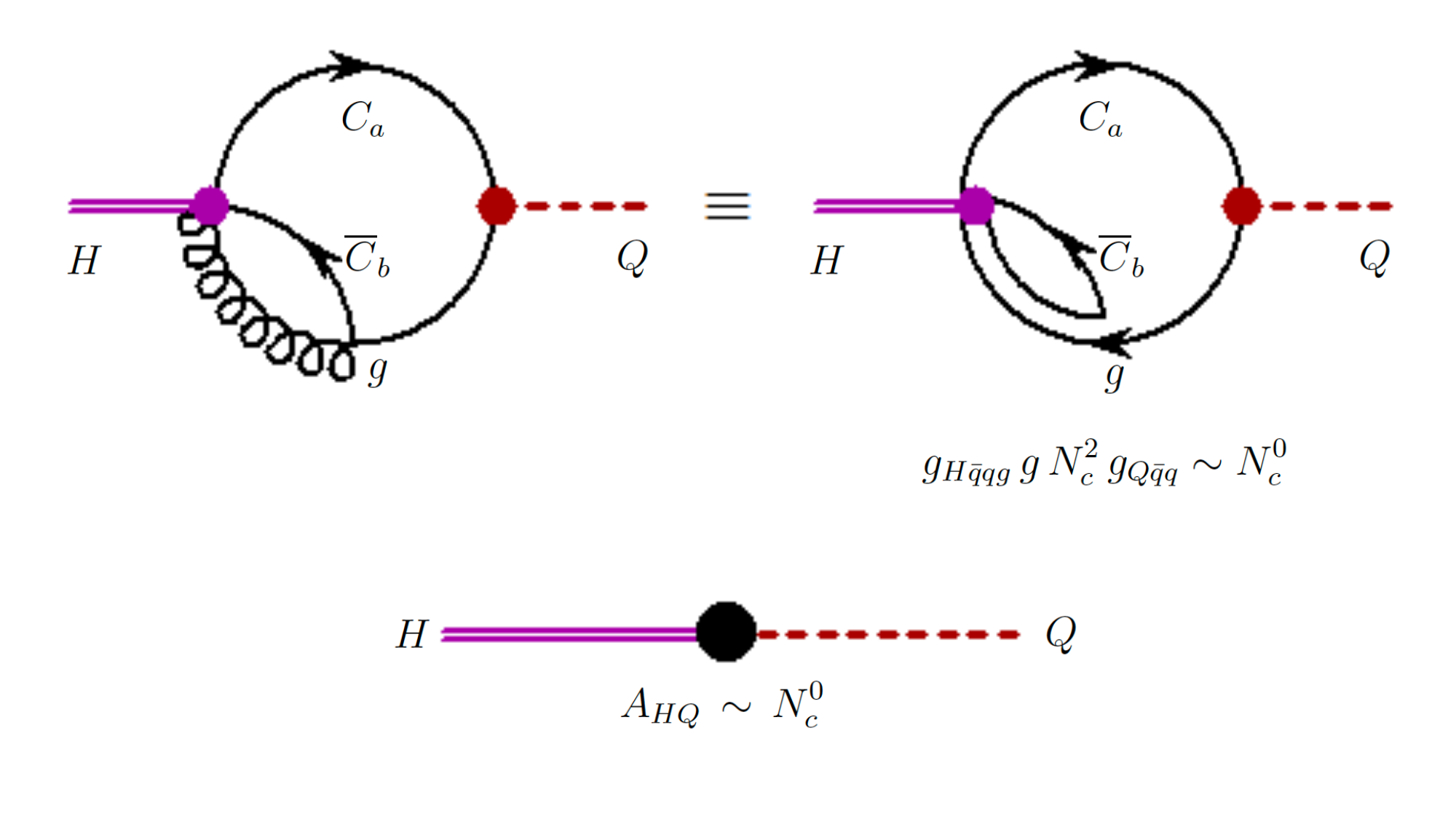}\\
        \caption{The mixing of an hybrid meson $H$ (purple, double-solid line) with a conventional quarkonium $Q$ (red, dashed line) scales as $N_c^0$, implying that hybrids and quarkonia can freely mix at large-$N_c$ and therefore behave (mostly) in a similar way.}
        \label{HHcons-Page-1}
   \end{figure}

 \begin{figure}[h]
        \centering       \includegraphics[scale=0.45]{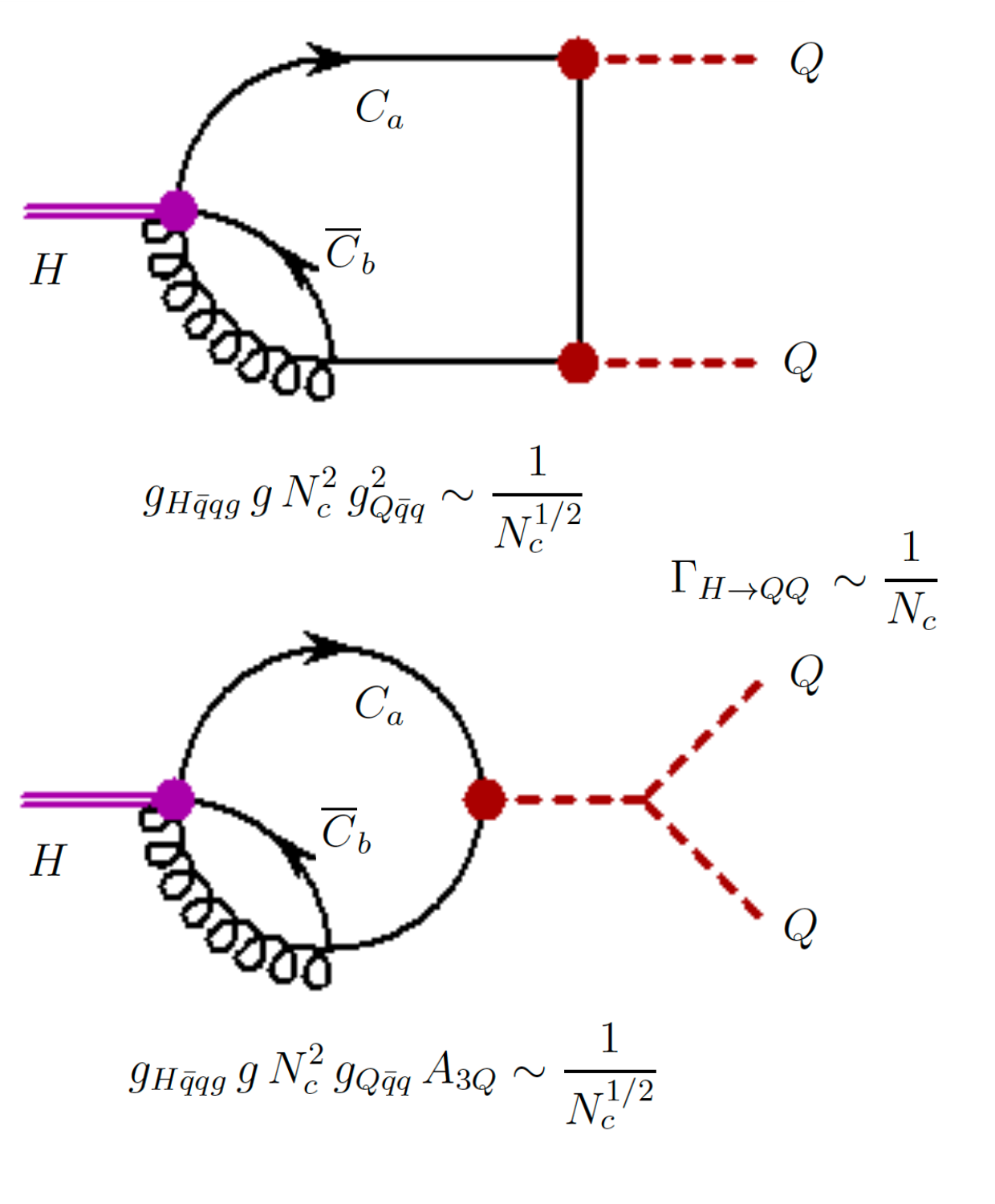}\\
        \caption{Upper part: Decay of an hybrid meson $H$ (purple double-solid  line) into two quarkonia mesons $Q$ (red dashed lines) via quarks and gluons (black lines). The amplitude scales as $N_c^{-1/2}$, just as for the standard decay of a quarkonium. Lower part: the same result is obtained by converting an hybrid $H$ into a quarkonium $Q$, which then decays into two quarkonia. Even simpler, one could just use the previous result concerning mixing and study the chain $H\rightarrow Q \rightarrow QQ$, that goes as $N_c^{-1/2}$. }
        \label{HHcons-Page-2}
   \end{figure}
   
We list some phenomenological consequences of hybrids at large-$N_{c}$. The most important one is
that the mixing of a hybrid state $H$ with a quarkonium state $Q$ (with the same quantum numbers, of course) scales as:%
\begin{equation}
A_{HQ}\sim\frac{1}{N_{c}}N_{c}^{2}\frac{1}{\sqrt{N_{c}}}\frac{1}{\sqrt{N_{c}}%
}\sim N_{c}^{0}  \text{ ,}
\end{equation}
which is $N_c$ independent! (This case is depicted in Fig. \ref{HHcons-Page-1}). This result means that hybrids behave as quarkonia in the
large-$N_{c}$ limit.

For example, the decay of a hybrid into two standard quarkonia mesons (see Fig. \ref{HHcons-Page-2}) scales
as%
\begin{equation}
A_{HQQ}\sim\frac{1}{N_{c}}N_{c}^{2}\frac{1}{\sqrt{N_{c}}}\left(  \frac
{1}{\sqrt{N_{c}}}\right)  ^{2}=\frac{1}{\sqrt{N_{c}}}\text{ ,}%
\end{equation}
just as a regular quark-antiquark mesonic decay.

 \begin{figure}[h]
        \centering       \includegraphics[scale=0.40]{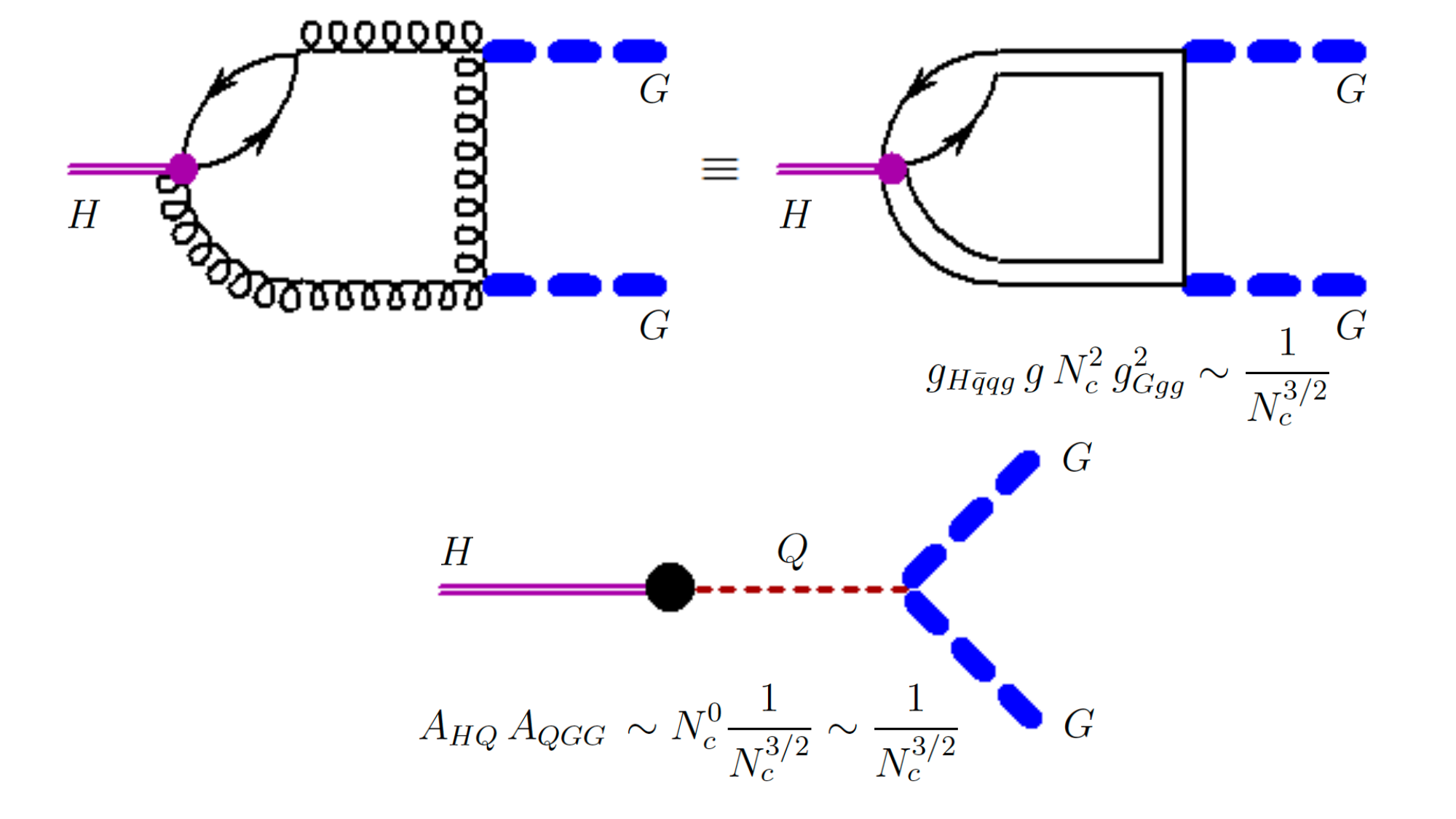}\\
        \caption{Up: Decay of an hybrid meson $H$ (purple double-solid line) into two glueballs $G$ (blue thisk-dashed lines) via quarks and gluons (black lines). The amplitude scales as $N_c^{-3/2}$. Lower part: the same result is obtained by converting an hybrid into a quarkonium which then converts into two glueballs. }
        \label{HHcons-Page-3}
   \end{figure}
   
The same applies to interactions with an arbitrary number of hybrids, that
scales as $N_{c}/N_{c}^{n_{H}/2},$ as well as of hybrids and quarkonia, that
goes as $N_{c}/\left(  N_{c}^{n_{Q}/2}N_{c}^{n_{H}/2}\right)$.
In the case
of $n_{Q}=0$ and $n_{H}=1,$ one obtains $N_{c}^{1/2},$ which corresponds to
the weak decay constant of an hybrid meson. Namely, the hybrid
production/annihilation amplitude goes as
\begin{equation}
f_{H}\sim g_{Hqqg}\Sigma_{H}(M_{H}^{2})\sim N_{c}^{-1}\cdot N_{c}^{2}\sim
N_{c}\text{ ,}%
\end{equation}
but the weak decay goes with an additional suppression of $N_{c}^{1/2}$ (the
gluon needs to disappear):%
\begin{equation}
f_{H}^{weak}\sim N_{c}^{1/2}\text{ .}%
\end{equation}

This result is also obtained by taking an external $H$, which converts to $Q$
(amplitude $N_{c}^{0})$, which subsequently annihilates (amplitude
$N_{c}^{1/2})$, resulting in $N_{c}^{1/2}$.

Indeed, hybrids can form nonets just as regular mesons, and thus can be
embedded into chiral approaches \cite{shastry,elsmhyb}.

The interaction of hybrids and glueballs can also be studied, see Fig.
\ref{HHcons-Page-3} for the explicit case of the decay of an hybrid meson into two glueballs, that scales as $N_c^{-3/2}$. 

\subsection{Summary of the scaling for an arbitrary number of $Q,G,H$ states}

Putting all the results together, we recover the general scaling law for the amplitude with $n_Q$ quarkonia, $n_G$ glueballs, and $n_H$ hybrids (see Sec. 2.6) as:
\begin{equation}
A_{\left(  n_{Q}Q\right)  \left(  n_{G}G\right)  \left(  n_{H}H\right)
}\propto\frac{N_{c}\cdot N_{c}^{2(1-sign(n_{Q}+n_{H}))}}{N_{c}^{n_{Q}/2}%
N_{c}^{n_{G}}N_{c}^{n_{H}/2}}%
\end{equation}
where $sign(x)$ is the sign function with $sign(0)=1/2.$ If $n_{Q}+n_{H}>0,$
thus at least one quarkonium or an hybrid is present, $N_{c}\cdot
N_{c}^{2(1-sign(n_{Q}+n_{H}))}=N_{c},$ while for $n_{Q}=n_{H}=0,$ the purely
gluonic case is recovered: $N_{c}\cdot N_{c}^{2(1-sign(n_{Q}+n_{H}))}%
=N_{c}^{2}.$

\bigskip

\subsection{Four-quark states}

The treatment of four-quark states in the large $N_c$ limit is subject to an
ongoing debate, see \cite{lucha} for a review. Some of the questions related
to it are unsettled yet. The first question is what one can understand under
four-quark states, since different possibilities are available.

As a specific example, we shall consider the meson $a_{0}(980)$, for which
different interpretations have been proposed in the literature, that we briefly review in the following.

\textbf{(1) Molecular states, such as the binding of two colorless quark-antiquark
mesons.}

For the illustrative state $a_{0}^{+}(980),$ this amounts to consider a bound
state of $K^{+}$ and $\bar{K}^{0}$ \cite{barua0,branz} resulting into the state:
\begin{equation}
\left\vert a_{0}^{+}(980)\right\rangle =\left\vert K^{+}\bar{K}^{0}%
\right\rangle
\text{ .}
\end{equation}

In general, such a molecule is of the type $\left\vert QQ\right\rangle $ and
its current can be expressed as $J_{QQ}(x)=Q^{2}(x).$ The basic interaction
takes the form
\begin{equation}
\mathcal{L}_{QQ}=K_{QQ}J_{QQ}^{2}(x)
\text{ ,}
\end{equation}
where
\begin{equation}
K_{QQ}\sim N_{c}^{-1}\rightarrow K_{QQ}\simeq\frac{\bar{K}_{QQ}}{N_{c}}
\text{ ,}
\end{equation}
being a quartic interaction between conventional mesonic fields.

\begin{figure}[h]
        \centering       \includegraphics[scale=0.35]{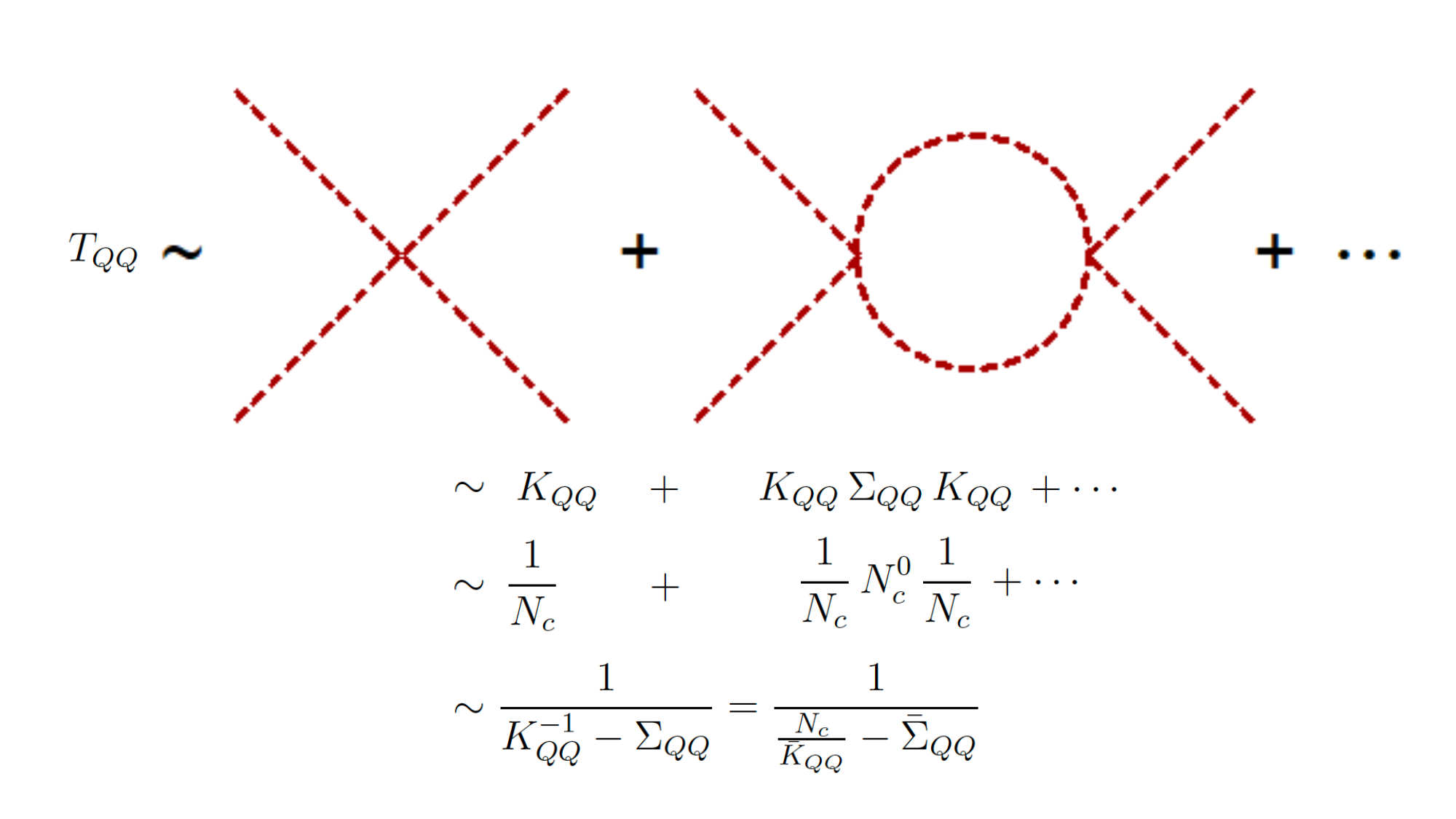}\\
        \caption{Resummation of regular meson-meson scattering diagrams needed to investigated the eventual emergence of a molecular bound state. While for $N_c=3$ these states can form, this is not the case for large-$N_c$. Namely, the attraction decreases as $N_c^{-1}$ but the intermediate states in the bubble are colorless, thus the loop function cannot compensate for the decrease of attraction. See text for more details.  }
        \label{Four-Page-1}
   \end{figure}

Upon repeating the previous steps, the resummed $T$-matrix for the eventual
formation of a $QQ$ bound states (see Fig. \ref{Four-Page-1}) reads:
\begin{equation}
T_{QQ}(s)=\frac{1}{K_{QQ}^{-1}-\Sigma_{QQ}(s)}%
\end{equation}
with $\Sigma_{QQ}(s)=\bar{\Sigma}_{QQ}(s)$ being large-$N_{c}$ independent (it
is the loop of two colorless states). The mass of the molecular states corresponds to a solution of the equation
\begin{equation}
\frac{N_{c}}{\bar{K}_{QQ}}-\bar{\Sigma}_{QQ}(s=M_{QQ}^{2})=0 \text{ .}
\end{equation}
This equation might have a solution for $N_{c}=3,$ but this is not the case of
large $N_{c}$. 

Let us first consider the case of a genuine molecular state
whose mass $M_{QQ}$ is below the threshold $2M_{Q}$ for $N_c = 3$. The function $\Sigma
_{QQ}(s)$ is real below threshold and has a maximum (cusp) just at it. If the bound
state exists for $N_{c}=3$ for a value $M_{QQ}<2M_{Q}$ below threshold, there
is a maximal value $N_{c}^{\max}$ for which the molecular state forms just
at threshold, $M_{QQ}=2M_{Q}.$ Yet, upon increasing $N_{c}$ further, the state
ceases to form. Indeed, this is very intuitive: by increasing $N_{c},$ the
attraction decreases and there is no $N_{c}$ factor to compensate it.
Molecular states of this type fade away for large $N_{c}$.

Indeed, the previous argumentation may be extended also to molecular resonances with
mass above $2M_{Q}$, since $\Sigma_{QQ}(s)$ is a non-diverging function. In any case, for $N_{c}$
large enough one has $T_{QQ}(s)\simeq K_{QQ}\simeq\bar{K}_{QQ}/N_{c}$, thus no bound state is possible.

Molecular states may also emerge from glueballs or hybrids. The case of
glueball molecular states, so-called glueballonia, has been recently studied
in Ref. \cite{glueballonium}. Quite remarkably, the bound state of two scalar glueballs may be
stable in pure Yang-Mills and be a resonance in full QCD for $N_{c}=3$.
Yet, for large $N_{c}$ it fades away even faster then $QQ$ molecular states. 
In fact, one has:
\begin{equation}
\mathcal{L}_{GG}=K_{GG}J_{GG}^{2}(x)
\end{equation}
where $J_{GG} = G^2(x)$ and 
\begin{equation}
K_{GG}\sim N_{c}^{-2}\rightarrow K_{GG}\simeq\frac{\bar{K}_{GG}}{N_{c}^{2}}%
\text{ ,}
\end{equation}
as it follows from being a quartic interaction between glueballs. Then:
\begin{equation}
T_{GG}(s)=\frac{1}{K_{GG}^{-1}-\Sigma_{GG}(s)}=\frac{1}{\frac{N_c^2}{\bar{K}_{GG}}-\bar{\Sigma}_{GG}(s)}%
\end{equation}
with $\Sigma_{GG}(s)=\bar{\Sigma}_{GG}(s)$ being $N_{c}$-independent. This result shows that
no glueballonium can form at large-$N_{c}.$ It is important to note that the
same large-$N_{c}$ scaling for the glueballonium formation is obtained in the
more advanced approach of Ref. \cite{glueballonium}, where two unitarization methods have
been used to study its formation. This fact shows again that the rather simple
separable interaction considered above is fully consistent with general large-$N_{c}$ results
and is therefore suitable to study the large-$N_{c}$ scaling. Additional
bound states of regular mesons with glueballs and/or hybrids can be studied
\cite{petrov}, but for the very same reason they shall also not survive in the large-$N_{c}$ domain.

\bigskip

\textbf{(2) Dynamically generated states: the example of companion poles.}

A specific example of a dynamically generated state is the so-called emergence of a companion pole, as it was presented for the case of the
meson $a_{0}(980)$ in Refs. \cite{boglione,wolkaa0}.
The starting point is a Lagrangian
which contains a single conventional scalar quark-antiquark bare state, roughly
correspondent to $a_{0}\equiv a_{0}(1450)$. 
One then writes the interaction term as:
\begin{equation}
\mathcal{L}_{int}=g_{a_{0}KK}a_{0}^{+}K^{-}K^{0}+g_{a_{0}\pi\eta}a_{0}^{+}%
\pi^{-}\eta+... \text{ ,}
\end{equation}
with the standard scaling
\begin{equation}
g_{a_{0}KK}\sim N_{c}^{-1/2}\text{ , }g_{a_{0}\pi\eta}\sim N_{c}^{-1/2} \text{ .}
\end{equation}
Then, upon studying mesonic loops, the full dressed propagator of $a_{0}(1450)$ arising from the decays into $\bar{K}K$, $\pi \eta$, etc., 
takes the form
\begin{equation}
\frac{1}{p^{2}-M_{a_{0}}^{2}+g_{a_{0}KK}^{2}\Sigma_{KK}(p^{2})+g_{a_{0}\pi
\eta}^{2}\Sigma_{\pi\eta}(p^{2})+...}%
\end{equation}
with $M_{a_{0}}\simeq1.4$ GeV$\sim N_{c}^{0}$. The coupling constants are set to
reproduce the physical results for the resonance $a_0(1450)$ for the physical case $N_{c}=3$.

Then, upon solving the pole equation in the complex plane%
\begin{equation}
p^{2}-M_{a_{0}}^{2}+g_{a_{0}KK}^{2}\Sigma_{KK}(p^{2})+g_{a_{0}\pi\eta}%
^{2}\Sigma_{\pi\eta}(p^{2})+...=0
\end{equation}
two poles are found (in this specific case on the third and second Riemann
sheets, respectively, but this aspect is not relevant for our analysis).

One pole is close to the expected bare quarkonium result and
corresponds to $a_0(1450)$: when increasing $N_{c}$, this pole converges
toward the real axis, that is its imaginary part decreases as $N_{c}^{-1},$ as expected for a regular $\bar{q}q$ quarkonium state. 

The second pole appears  close the $\bar{K}K$ threshold and
corresponds to $a_{0}(980)$. In the large-$N_{c}$ limit it behaves
differently: its width increases instead of decreasing, showing that this
additional state is not an ordinary quark-antiquark object. Eventually it
disappears from the original (second)  Riemann sheet.

These features concerning dynamical generated companion poles are quite
general and apply, with minor changes, to other states as well, such as the scalar resonances $f_0(500)$ \cite{pelaeznc,sigmareview} and 
$K_{0}^{\ast}(700)$ \cite{rodas,wolkak}, or the famous $X(3872)$ (as a virtual pole)
\cite{x3872}.

It should be also stressed that companion poles are not the only possibility
for dynamically generated states, see e.g. \cite{sigmareview,lowscalars} but
it shows a quite general feature: these solutions fade away in the
large-$N_{c}$ limit\footnote{A word of caution is required; in some cases, one
may obtain certain mesons as solutions of bound-state equations out of
effective Lagrangians. Yet, these mesons can be ordinary quark-antiquark
states, but the way they have been obtained would make them look like
molecular states that do not survive the large-$N_{c}$ limit. We refer to Ref.
\cite{dynrec} (see also \cite{lesson}) for this subtlety and for the related notion of `dynamical
reconstruction'.}.

\bigskip
\begin{figure}[h]
        \centering       \includegraphics[scale=0.40]{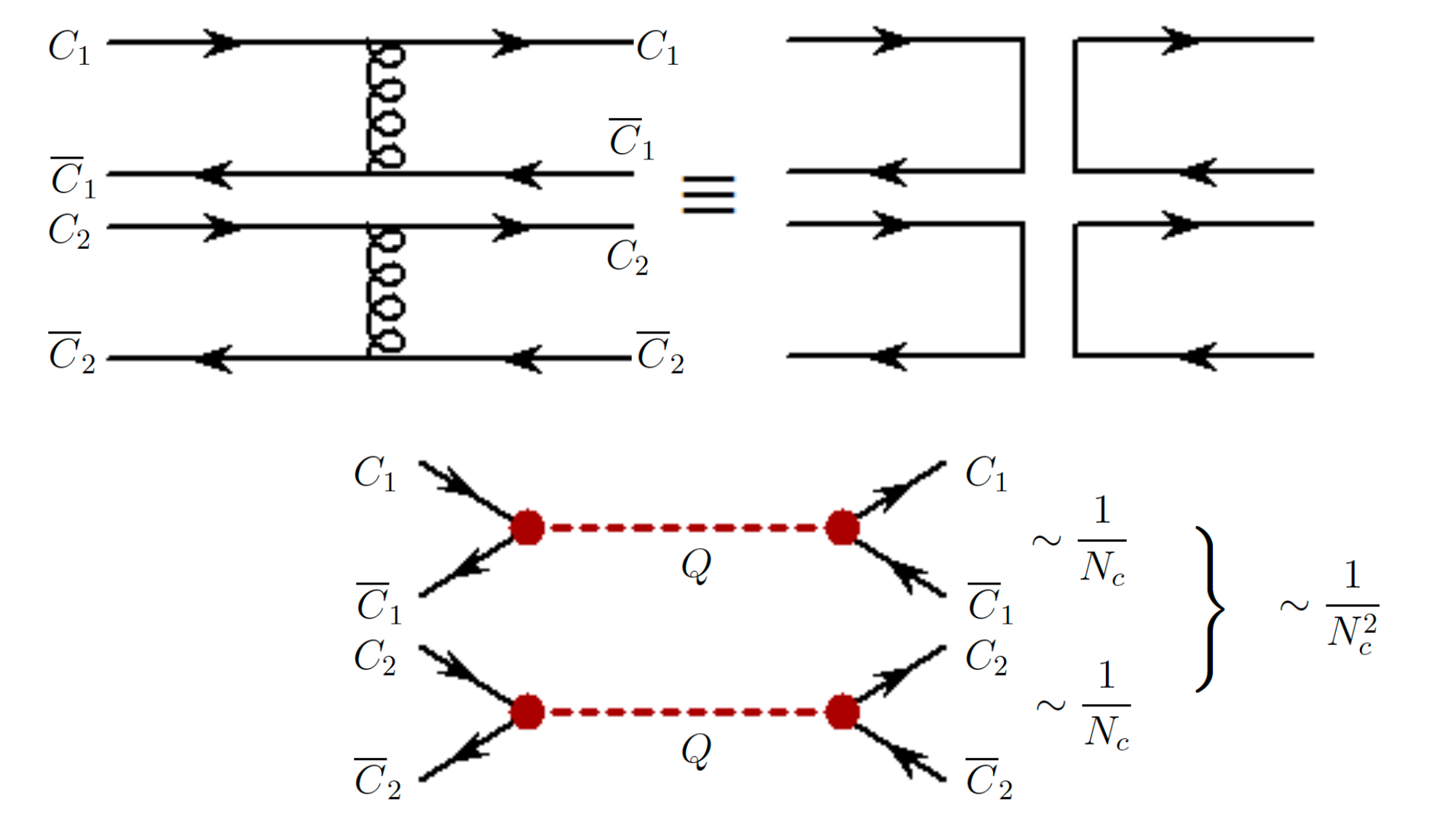}\\
        \caption{Disconnected diagram for the reaction $C_1\bar{C}_1C_2\bar{C}_2 \rightarrow C_3\bar{C}_3C_4\bar{C}_4$. All the colors have switched, but these diagrams (properly resummed) eventually generate two intermediate conventional $Q$ states. Indeed, the scaling $N_c^{-2}$ is in agreement with this interpretation.}
        \label{Four-Page-2}
   \end{figure}

\textbf{(3) Genuine tetraquark state as a bound state of two diquarks. }

Referring to $a_{0}(980)$ as our example, we may interpret it as a bound state
of a good diquark and a good anti-diquark \cite{jaffeorig,exotica}, where a good diquark state is antisymmetric in both color and flavors, e.g.:
\begin{equation}
\left\vert us,\text{good}\right\rangle =\left\vert \text{space: }%
L=0\right\rangle \left\vert \text{spin: }S=0\right\rangle \left\vert
\text{color: }RG-GR\right\rangle \left\vert \text{flavor: }us-su\right\rangle.
\nonumber
\end{equation}
Then%
\begin{equation}
\left\vert a_{0}^{+}(980)\right\rangle =\left\vert us,\text{good}\right\rangle
\left\vert \bar{d}\bar{s},\text{good}\right\rangle \text{ .}%
\end{equation}
Indeed, one can build nonets of states and describe these objects in a
chiral context \cite{fariborz,maianilow,tq,tqchiral}.

What about the large-$N_{c}$ scaling of these configurations?  The
issue is that the straightforward generalization of the good diquark is an
object that contains $N_{c}-1$ quarks, e.g. \cite{estrada}. This object may be used to construct
baryons in the large-$N_{c}$ limit (see next Section for its explicit
implementation), but is not useful for building tetraquarks states (with `tetra' in the sense of four).

\begin{figure}[h]
        \centering       \includegraphics[scale=0.35]{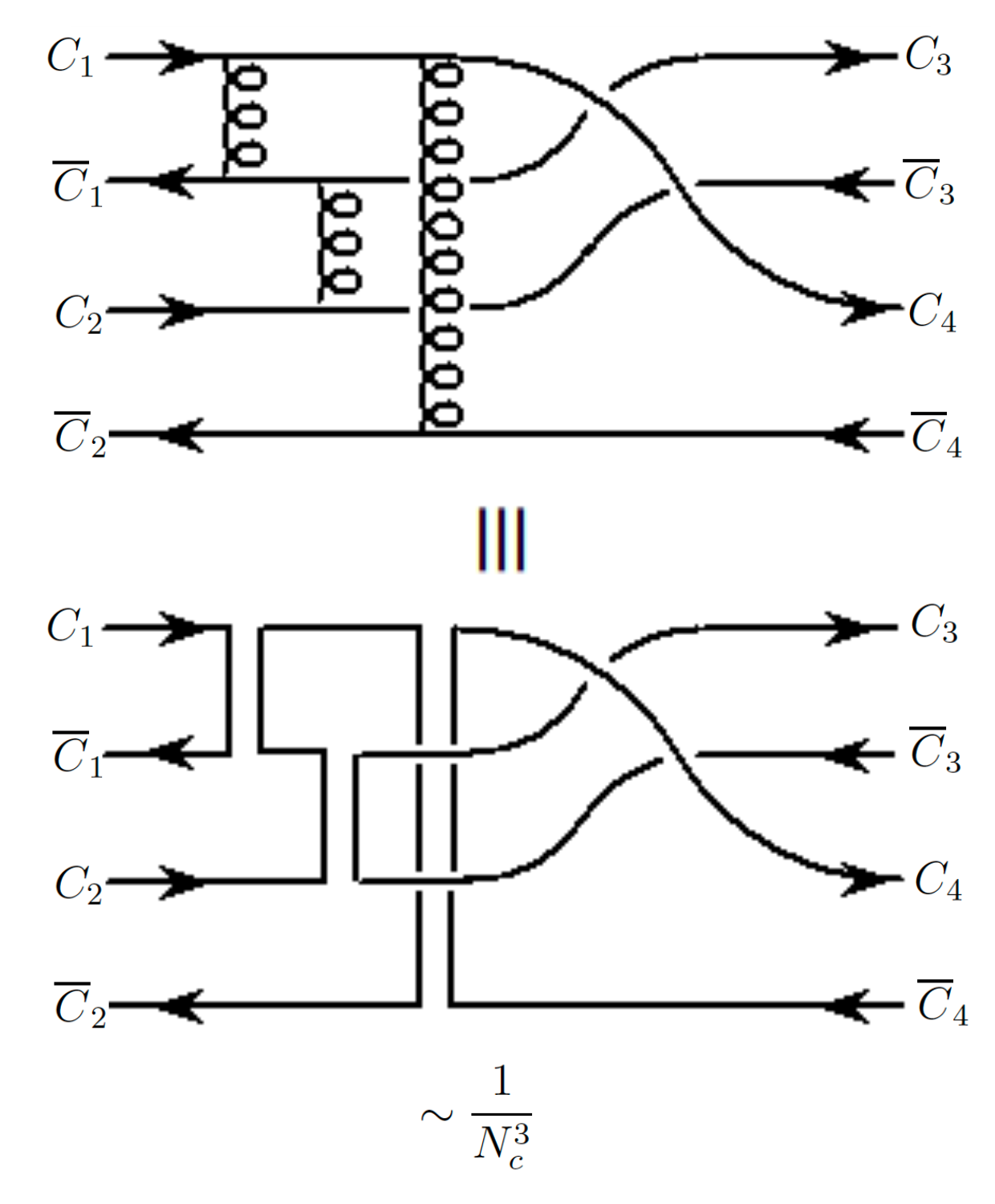}\\
        \caption{Connected diagram for the reaction $C_1\bar{C}_1C_2\bar{C}_2 \rightarrow C_3\bar{C}_3C_4\bar{C}_4$. All the colors have been switched. It scales as $N_c^{-3}$. }
        \label{Four-Page-3}
   \end{figure}

    \begin{figure}[h]
        \centering       \includegraphics[scale=0.35]{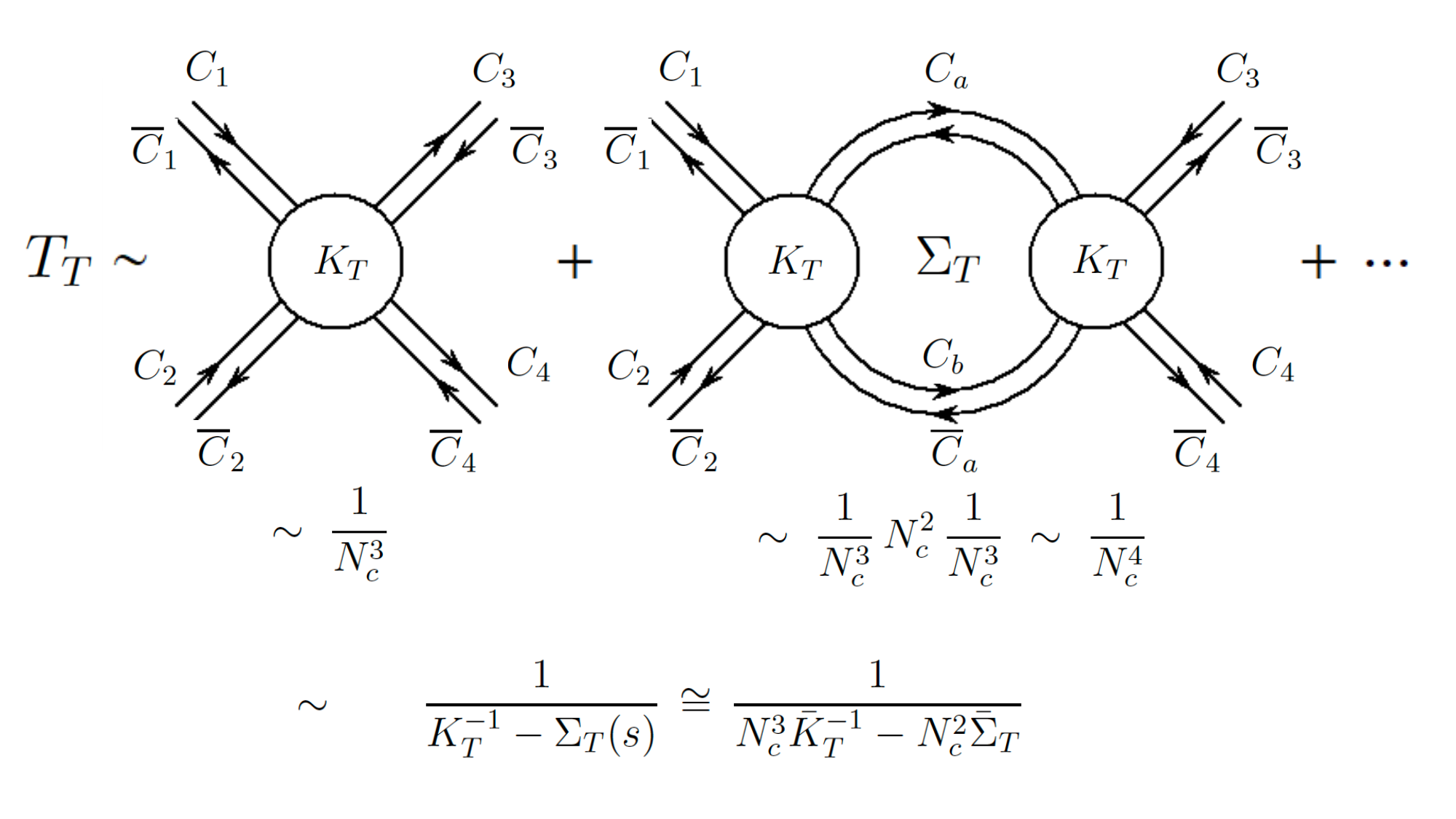}\\
        \caption{Tentative resummation of connected diagrams with a colorless four-quark configuration in the initial and in the final states. The attraction seems to decrease too fast to allow for a bound-state (a genuine tetraquark state) formation.   }
        \label{Four-Page-4}
   \end{figure}

In the classic lecture of Coleman \cite{colemanlect}, it is stated that
tetraquarks (of whatever type) do not exist in the large-$N_{c}$ limit because four-quark states
preferably arrange themselves into two free mesons, see Fig. \ref{Four-Page-2}. 

Weinberg realized years later
\cite{weinberg} that one should rather look at connected diagrams, hence certain
tetraquarks states might exist and show scaling laws similar to regular mesons.
A debate has followed
\cite{lucha,lebed,cohenlebed1,cohenlebed2,estrada,knecht}, basically
confirming Weinberg point of view but always stressing that it is not clear if
such tetraquark states do form in the large-$N_{c}$ limit.\ Indeed, in
\cite{cohenlebed2} it is argued that they eventually do not.

Does the bound-state approach discussed in these lectures help? Here, we just make some basic considerations that are not conclusive, but may be the starting
point for future investigations. 
To this end, let us consider the
most general four-quark current \cite{estrada}:
\begin{equation}
J_{T}=C_{1}\delta^{ac}\delta^{bd}q^{a}q^{b}\bar{q}^{c}\bar{q}^{d}+C_{2}%
\delta^{ad}\delta^{bc}q^{a}q^{b}\bar{q}^{c}\bar{q}^{d}%
\text{ .}
\end{equation}
Then, the separable
interaction takes the form%
\begin{equation}
\mathcal{L}_{T}=K_{T}J_{T}^{2}(x) 
\text{ ,}
\end{equation}
where we need to discuss the large-$N_c$ scaling of $K_{T}.$ If one considers a
connected four-quark diagram, one obtains (see Fig. \ref{Four-Page-3}):
\begin{equation}
K_{T}\sim N_{c}^{-3}\rightarrow K_{T}\simeq\frac{\bar{K}_{T}}{N_{c}^{3}}
\text{ ,}
\end{equation}
since it is a quartic interaction between conventional mesonic fields. Namely, a
disconnected diagram, in which the two quark-antiquark parts interact separately, 
would scale as $N_{c}^{-2}$, see again Fig. \ref{Four-Page-2}, but this is not what we search.

Then, the resummed $T$-matrix for the eventual formation of a tetraquark state state, see Fig. \ref{Four-Page-4}, reads:
\begin{equation}
T_{T}(s)=\frac{1}{K_{T}^{-1}-\Sigma_{T}(s)}
\end{equation}
where $\Sigma_{T}(s)$ scales as $N_{c}^{2}$. 
Namely, Whatever is the specific
tetraquark configuration, the order is always $N_{c}^{2}$.
(If, for example,
we consider only antisymmetric diquark color configurations, there are $N_{c}%
(N_{c}-1)/2\sim$ $N_{c}^{2}$ choices for the diquark color.) Hence, $\Sigma
_{T}(s)\simeq N_{c}^{2}\bar{\Sigma}_{T}(s),$ leading to
\begin{equation}
T_{T}(s)=\frac{1}{\frac{N_{c}^{3}}{K_{T}}-N_{c}^{2}\bar{\Sigma}_{T}(s)}
\text { .}
\end{equation}
This result suggests that no tetraquark bind for large-$N_{c}$, since the
interaction strength decreases too fast, just as mesonic molecular states.

Note, if we would study the tetraquark correlator $\left\langle J_{T}(x_{2})J_{T}(x_{1})\right\rangle$, one should indeed remove the
disconnected part $\Sigma_{T}(s)$ $\sim N_{c}^{2},$ hence the lowest order
contribution $\Sigma_{T}(s)K_{T}^{-1}\Sigma_{T}(s)$ scales as $N_{c},$ as
expected. Future studies are needed to check if the present heuristic
arguments against the emergence of tetraquarks can be made rigorous.

\section{Brief study of baryons at large-$N_{c}$}

The topic of baryons at large-$N_c$ cannot be fully covered in these lectures. Here, our aim is to show that an approach similar to the one applied to mesons (bound state formation) is also consistent with baryonic
large-$N_{c}$ scaling properties.\ To this end, let us introduce the
generalized `diquark'\ $D^{a_{1}}$ for $a_1 =1,...,N_c$ as a $N_{c}-1$ quark object with the
structure%
\begin{equation}
D^{a_{1}}=\frac{1}{\sqrt{(N_{c}-1)!}}\sum_{a_{2},a_{3},...,a_{N_{c}}=1}%
^{N_{c}}\varepsilon_{a_{1}a_{2}....a_{N_{c}}}q^{a_{2}}q^{a_{3}}...q^{a_{N_{c}%
}}\text{ .}%
\end{equation}

 \begin{figure}[h]
        \centering       \includegraphics[scale=0.35]{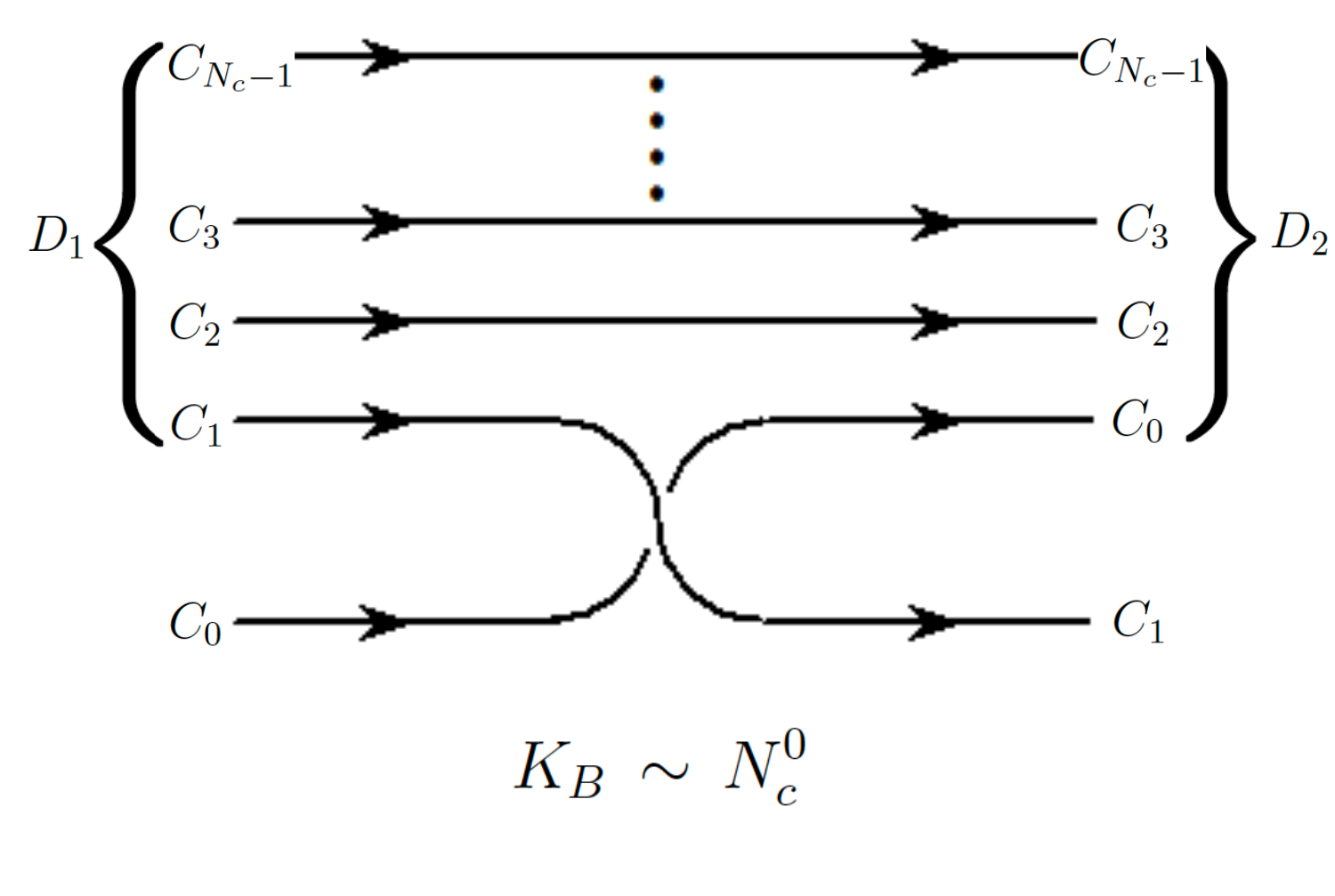}\\
        \caption{Scattering of a generalized diquark $D_1$ and a quark with color $C_1$ into $D_2$ and $C_2$ (thus color `changed'). A simple switch of quarks does the job, thus at leading order no gluon is present and the amplitude goes with $N_c^0$. }
        \label{B-Page-1}
   \end{figure}

   \begin{figure}[h]
        \centering       \includegraphics[scale=0.35]{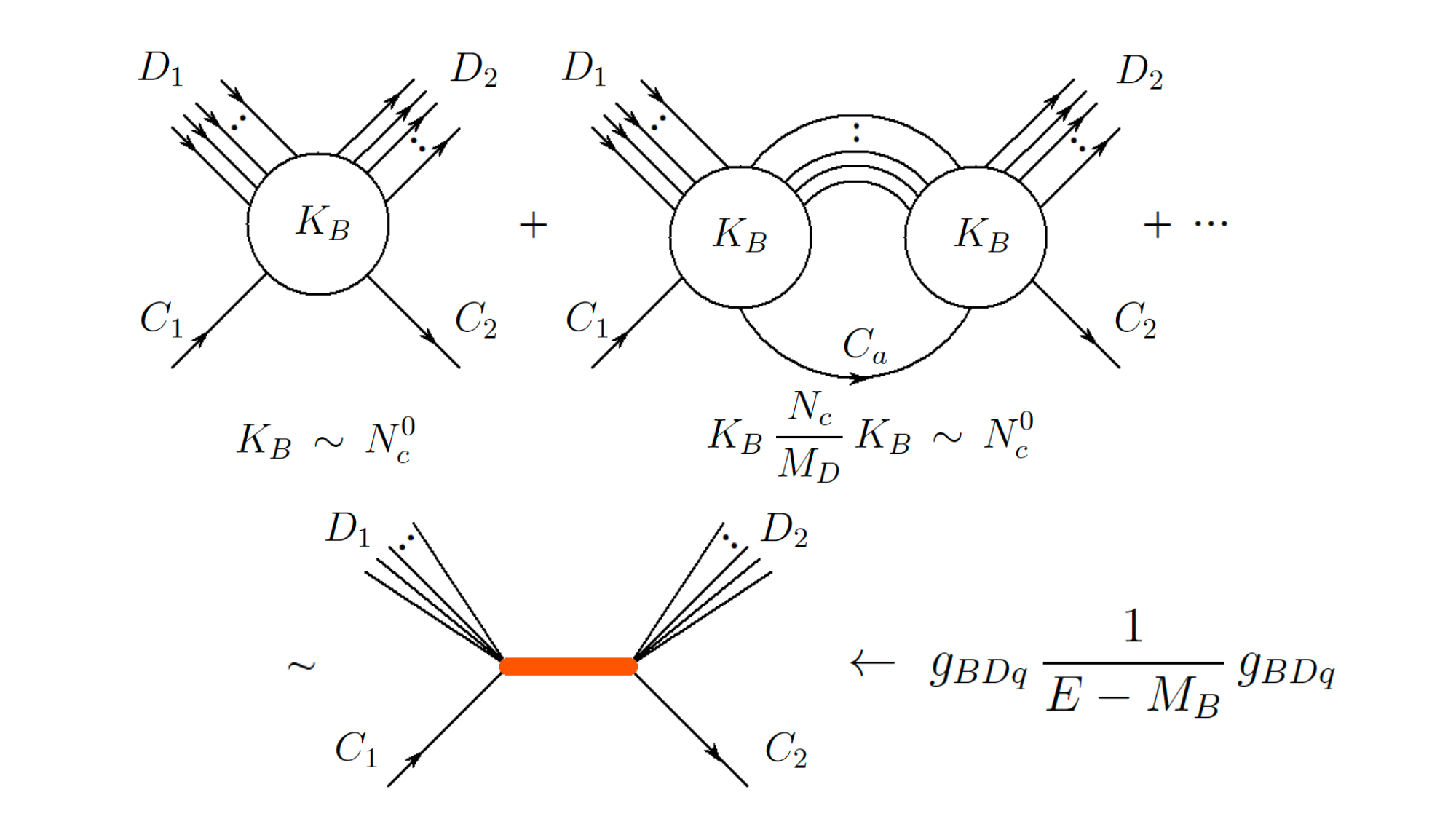}\\
        \caption{Resummation of the scattering $D_1C_1 \rightarrow D_2C_2$ with consequent formation of one baryon (as brown thick line) as intermediate state.}
        \label{B-Page-2}
   \end{figure}
There are $N_{c}$ generalized diquarks, just as there are $N_{c}$ antiquarks. Indeed, under color transformations $D^{a_1}$ transforms as an antiquark. Then, one
may interpret the baryon as a bound state of such a generalized diquark and a
quark. The current is given by:
\begin{equation}
J_{B}=\sum_{a_{1}=1}^{N_{c}}D^{a_{1}}q^{a_{1}}\text{ .}%
\end{equation}
Following the mesonic case, we write down the interaction Lagrangian as
\begin{equation}
\mathcal{L}_{B}=K_{B}J_{B}^{2}
\text{ .}
\end{equation}
The determination of the scaling of $K_{B}$ can be deduced from Fig. \ref{B-Page-1}.
Since at lowest order no gluon is involved because a simple switch of quarks suffices to
change the color of both $D$ and $q$ (this is due to the fact that $D$
contains already $N_c-1$ colors). It then follows that:%
\begin{equation}
K_{B}\sim N_{c}^{0}\text{ ,}%
\end{equation}
hence $K_{B}\simeq\bar{K}_{B}.$ On the other hand, the loop $\Sigma_{B}(E)$
involving $D$-$q$ goes with $N_{c}$ for what concerns the color circulating in
it, but contains also a dependence on the mass of the generalized diquark $D$
with $m_{D}\propto N_{c}-1$. Since $m_{D}$ is very large, we resort to non-relativistic propagators and use the energy
$E$ (and not $s = E^2$)  as an argument of the loop $\Sigma_{B}(E)$. This function can be expanded in $E$ finding:
\begin{equation}
\Sigma_{B}(E)\simeq N_{c}\left(  \frac{c_{1}%
}{m_{D}}+c_{2}\frac{E}{m_{D}^{2}}+...\right)
\text{ ,}
\end{equation}
which scales as $N_c^0 +N_c^{-1}+...$.
The $T$-matrix for the illustrative scattering process $D^{1}%
q^{1}\rightarrow D^{2}q^{2}$ reads (Fig. \ref{B-Page-2}):
\begin{equation}
T_{B}=\frac{1}{K_{B}^{-1}-\Sigma_{B}(E)} \text{ .}
\end{equation}
The pole equation reads%
\begin{equation}
\bar{K}_{B}^{-1}-N_{c}\left(  \frac{c_{1}}{m_{D}}+c_{2}\frac{E}{m_{D}^{2}%
}+...\right)  =0 \text{ ,}
\end{equation}
leading to%
\begin{equation}
\frac{N_{c}}{m_{D}}c_{1}+c_{2}\frac{N_{c}}{m_{D}}\frac{E}{m_{D}}+...=\bar
{K}_{B}^{-1}
\text{ ,}
\end{equation}
thus%
\begin{equation}
c_{2}\frac{N_{c}}{m_{D}}\frac{E}{m_{D}}=\bar{K}_{B}^{-1}-\frac{N_{c}}{m_{D}%
}c_{1}\sim N_{c}^{0} \text{ .}
\end{equation}
It then follows that
\begin{equation}
E\equiv M_{B}\sim m_{D}\frac{m_{D}}{N_{c}}\sim m_{D}\sim N_{c}
\text{ .}
\end{equation}
In other words, we find that, if $m_{D}\sim N_{c},$ then $M_{B}\sim m_{D}\sim
N_{c}$ as well, being a consistent (and expected) result.

\begin{figure}[h]
        \centering       \includegraphics[scale=0.35]{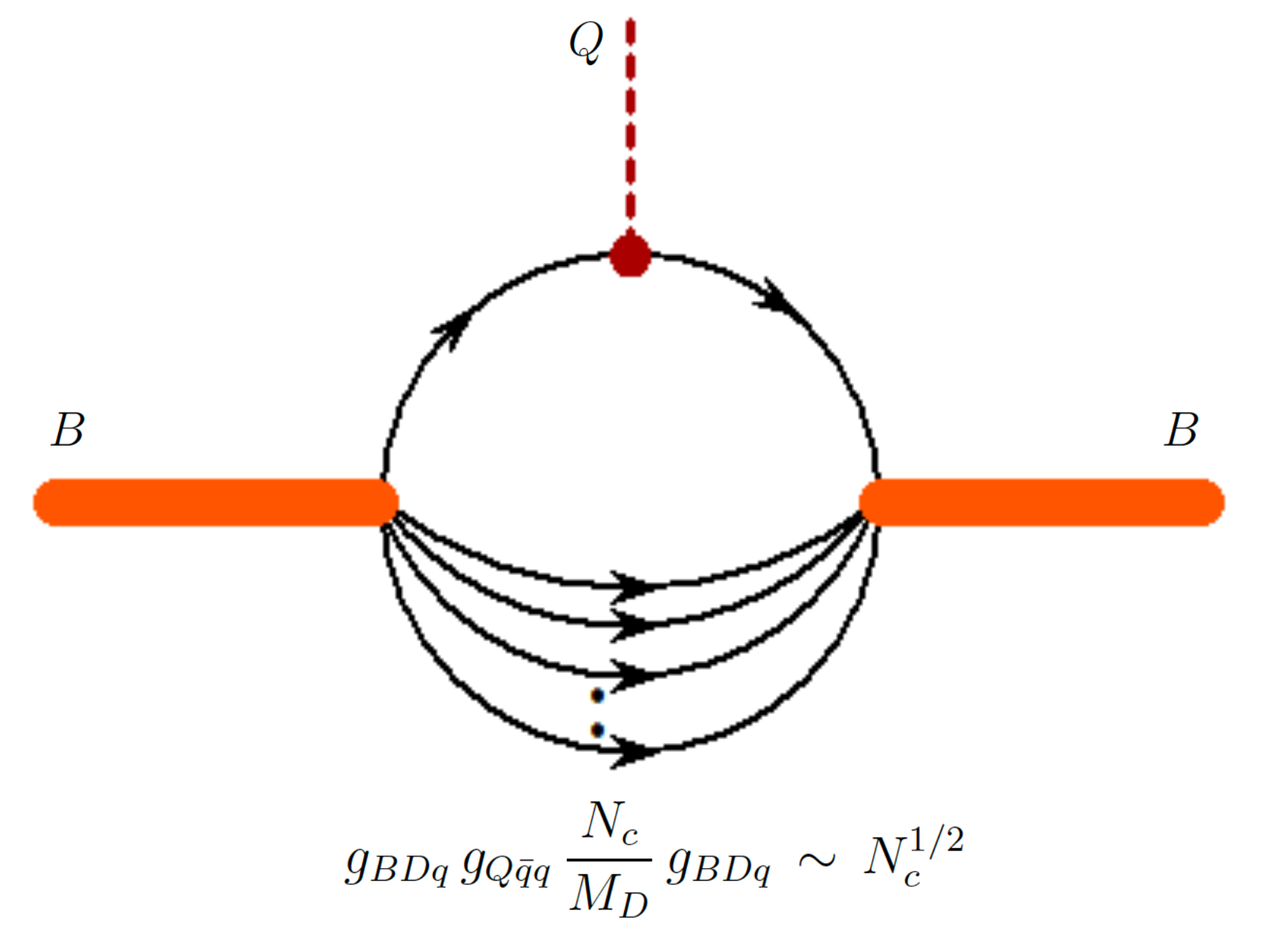}\\
        \caption{Vertex of two baryons $B$ (brown thick lines) and one conventional meson $Q$ (red dashed line). The amplitude, that corresponds to the generic meson-baryon-baryon coupling scales with $N_c^{1/2}$. }
        \label{B-Page-3}
   \end{figure}

   \begin{figure}[h]
        \centering       \includegraphics[scale=0.35]{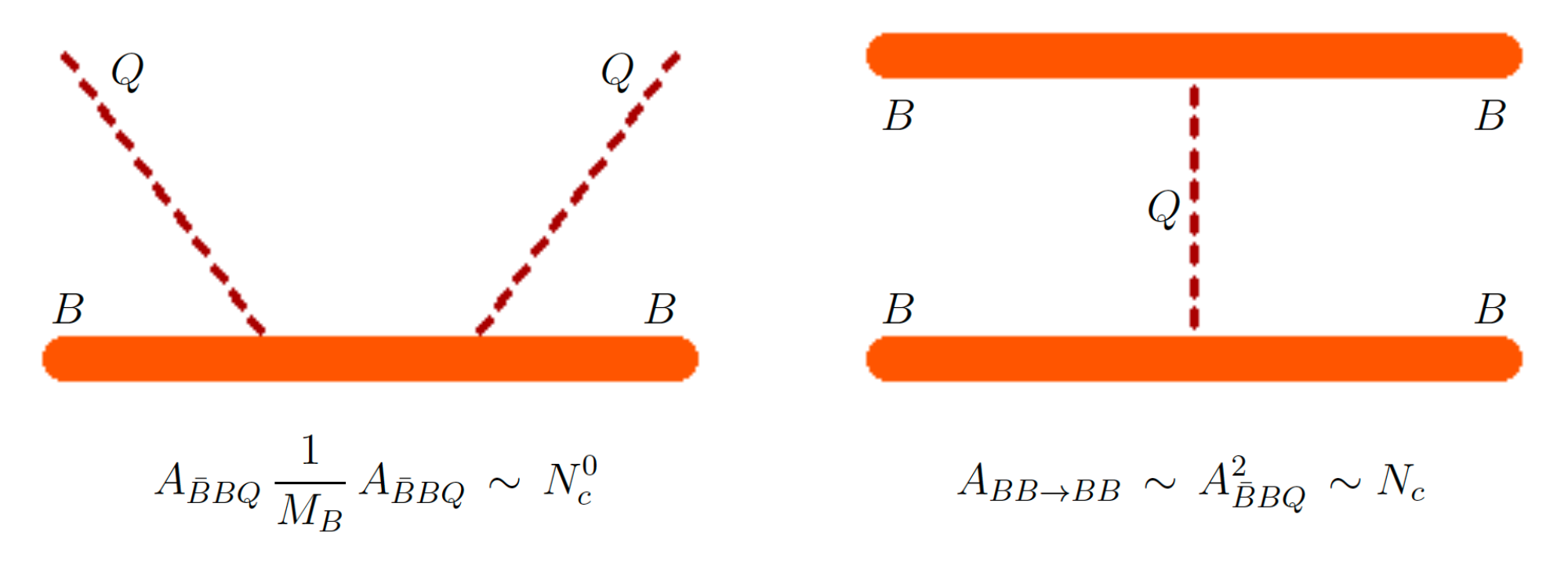}\\
        \caption{Left: baryon-meson scattering, which goes with $N_c^0$ (note, the intermediate baryon gives a $N_c^{-1}$ contribution. Right: baryon-baryon scattering that scales as $N_c$. Baryons are brown thick lines, quarkonia are red dashed lines.}
        \label{B-Page-4}
   \end{figure}
   
Upon expanding around the pole, we find:%
\begin{equation}
T_{B}\simeq\frac{(ig_{BDq})^{2}}{E-M_{B}}%
\end{equation}
with
\begin{equation}
g_{BDq}\simeq\sqrt{\frac{1}{\Sigma_{B}^{^{\prime}}(E=M_{B})}}\simeq\sqrt
{\frac{1}{\frac{N_{c}c_{2}}{m_{D}^{2}}+...}}\sim\sqrt{\frac{1}{\frac{1}{N_{c}%
}}}\sim N_{c}^{1/2}
\text{ ,}
\end{equation}
thus the baryon coupling to its generalized diquark $D$ and quark $q$
increases as $N_{c}^{1/2}.$

From these scaling laws, one can determine all the others. The quarkonium-baryon coupling goes as (see Fig. \ref{B-Page-3} as well as Ref. \cite{olbrich2}):
\begin{equation}
g_{\bar{B}BQ}\sim N_{c}^{1/2}
\text{ .}
\end{equation}
The scaling is the same for any number of $\bar{B}B$ pairs. 

\begin{figure}[h]
        \centering       \includegraphics[scale=0.35]{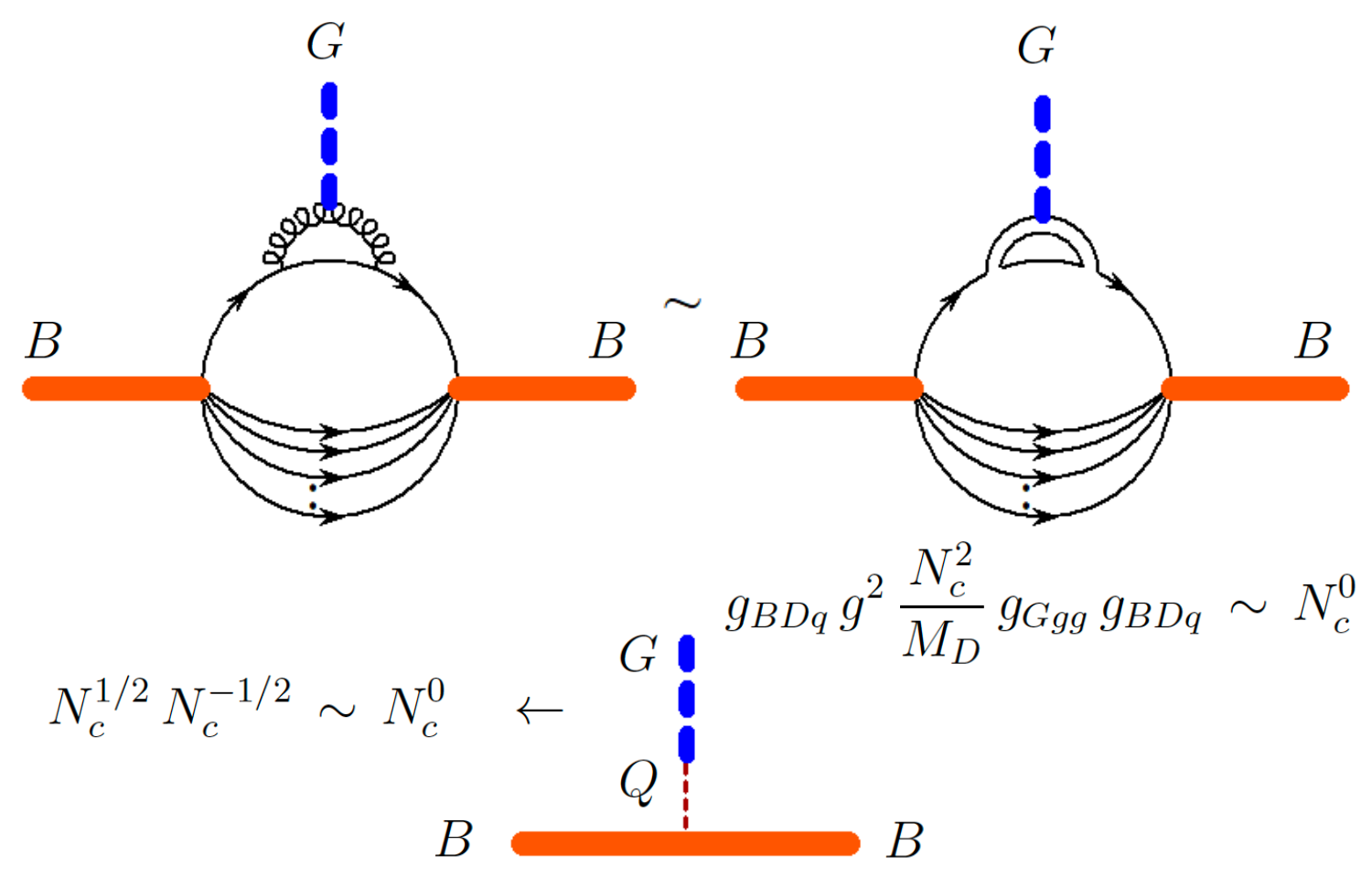}\\
        \caption{Coupling of a baryon (brown thick lines) to a glueball (blue thick-dashed line). This is shown both at the level of intermediate quarks and gluons (upper left) and color double-line notation (upper right). This interaction scales with $N_c^0$. At the bottom, the very same result can be obtained by coupling  the baryon to a quarkonium $Q$ (red dashed line) and subsequently $Q$ to $G$. }
        \label{B-Page-5}
   \end{figure}

By increasing the number of quarkonia to $n_{Q}$, we find (see Fig. \ref{B-Page-4} for two examples) that the generic
amplitude
\begin{equation}
A_{(\bar{B}B\bar{B}B\cdot\cdot\cdot)(n_{Q}Q)}\sim\frac{N_{c}}{N_{c}^{n_{Q}/2}} \text{ .}
\end{equation}
The coupling to a single glueball goes as (see Fig. \ref{B-Page-5}):
\begin{equation}
g_{\bar{B}BG}\sim N_{c}^{0} \text{ ,}
\end{equation}
and then the one to $n_{G}$ glueballs as%
\begin{equation}
A_{(\bar{B}B\bar{B}B\cdot\cdot\cdot)(n_{G}G)}\sim\frac{N_{c}}{N_{c}^{n_{G}}}%
\text{ .}
\end{equation}
The coupling to hybrid meson is identical to quark-antiquark ones:
\begin{equation}
g_{\bar{B}BH}\sim N_{c}^{1/2}
\text{ ,}
\end{equation}%
\begin{equation}
A_{(\bar{B}B\bar{B}B\cdot\cdot\cdot)(n_{H}H)}\sim\frac{N_{c}}{N_{c}^{n_{H}/2}}
\text{ .}
\end{equation}
The final coupling of an arbitrary number  of baryon-antibaryon pairs to $n_{Q}$ quarkonia, $n_{G}$ glueballs, and $n_{H}$ hybrid
mesons is:
\begin{equation}
A_{(\bar{B}B\bar{B}B\cdot\cdot\cdot)(n_{Q}Q)(n_{G}G)(n_{H}H)}\sim\frac{N_{c}%
}{N_{c}^{n_{Q}/2}N_{c}^{n_{G}}N_{c}^{n_{H}/2}}%
\label{genbar} 
\text{ .}
\end{equation}
What about the scattering of baryons? Following the same `visual' approach (Fig. \ref{B-Page-4}), the
amplitude for the process $BB \rightarrow BB$ goes as
\begin{equation}
A_{BB \rightarrow BB}\sim N_{c}
\text{ .}
\end{equation}
Indeed, it does not change for any arbitrary number of baryons, provided that the initial number of baryons is equal to the final one (baryon number conservation). In turn, it is equal to the amplitude with $n_B$ baryons and $n_B$ antibaryons, and can be formally recovered from the previous case of Eq. (\ref{genbar}) upon setting $n_{Q}=n_{G}=n_{H}=0$.

   \begin{figure}[h]
        \centering       \includegraphics[scale=0.35]{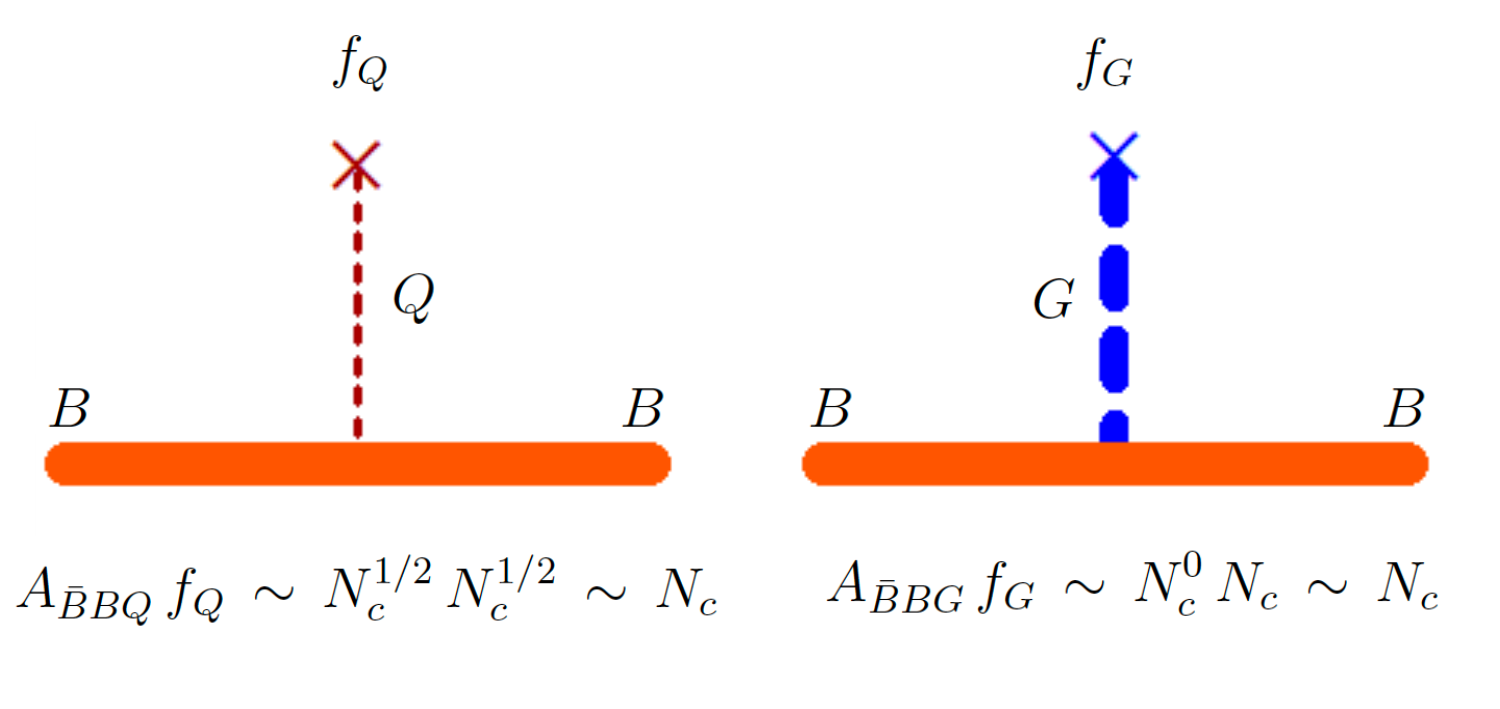}\\
        \caption{Contribution to the mass of the baryon emerging from its interaction to a scalar quarkonium field $Q$ and to a scalar glueball $G$. One obtains that, in both cases, when the scalar quarkonium field condenses or the scalar glueball field (the dilaton) condenses, the mass of the baryon scales with $N_c$, as expected. These results are implemented in the context of chiral models for the nucleon: the upper one corresponds to the standard LSM, the second one to the so-called mirror assignment (see text). }
        \label{B-Page-6}
   \end{figure}

   \begin{figure}[h]
        \centering       \includegraphics[scale=0.75]{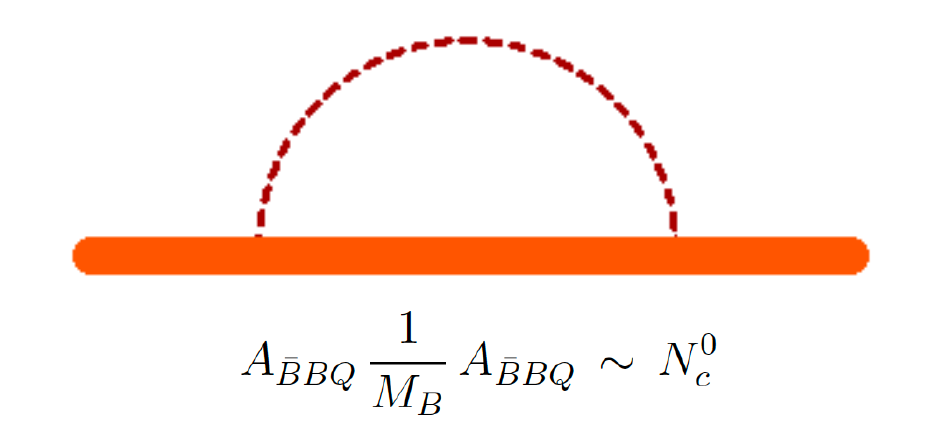}\\
        \caption{The contribution to the mass of a baryon generated by a meson emission and absorption is suppressed in the large-$N_c$ limit. }
        \label{B-Page-7}
   \end{figure}

\bigskip

How to implement baryons in a chiral model? In line with previous simplified
treatments, we consider a single flavor (and disregard the chiral anomaly). We
introduce a nucleon field $\Psi_{1},$ which is as usual split into
right-handed and left-handed parts as:
\begin{equation}
\Psi_{1,R}=\frac{1+\gamma^{5}}{2}\Psi_{1}\text{ , }\Psi_{1,L}=\frac
{1-\gamma^{5}}{2}\Psi_{1}%
\text{ .}
\end{equation}
A chiral transformation at the level of the nucleon amounts to:%
\begin{equation}
\Psi_{1,R}\rightarrow e^{i\alpha/2}\Psi_{1,R}\text{ , }\Psi_{1,L}\rightarrow
e^{-i\alpha/2}\Psi_{1,L}%
\text{ ,}
\end{equation}
where the the right and left pieces transform with different sign of the phase (for equal
sign, one has a simple $U(1)$ baryon-number transformation). Because of this chiral
transformation, a mass term of the type
\begin{equation}
\bar{\Psi}_{1}\Psi_{1}=\bar{\Psi}_{1,R}\Psi_{1,L}+\bar{\Psi}_{1,L}\Psi_{1,R}%
\end{equation}
is \textit{not} chirally invariant! Thus, it seems that the nucleon needs to be -at
first- massless.

As indeed well known, one can generate a massive nucleon by fulfilling chiral
symmetry upon coupling the nucleon field to the mesonic field $\Phi,$ which
transforms as $\Phi\rightarrow e^{i\alpha}\Phi.\ $Then, an invariant term
that generates a mass (via SSB) is obtained as (e.g. \cite{koch}):%
\begin{equation}
\mathcal{L}_{\Psi_{1}\Phi}=-g_{\Psi_{1}\Phi}\left(  \bar{\Psi}_{1,R}\Phi
\Psi_{1,L}+\bar{\Psi}_{1,L}\Phi^{\dagger}\Psi_{1,R}\right)
\end{equation}
where%
\begin{equation}
g_{\Psi_{1}\Phi}\sim N_{c}^{1/2}
\text{ .}
\end{equation}
When $\Phi$ condenses via SSB to $\phi_{N}\sim N_{c}^{1/2}$, a nucleon mass
proportional to the chiral condensate $\phi_{N}$ is generated as:%
\begin{equation}
M_{N}\sim g_{\Psi_{1}\Phi}\phi_{N}\sim N_{c}^{1/2}\cdot N_{c}^{1/2}\sim N_{c}%
\label{massc1}
\end{equation}
with the expected large-$N_{c}$ behavior, see Fig. \ref{B-Page-6} for a pictorial representation of this result. 

Note, the mesonic loop mass correction contributes with $N_{c}^{0}$ to the
nucleon mass and is thus suppressed, as shown in Fig. \ref{B-Page-7}. The `bulk' mass dominates the formation
of the nucleon mass.

If the only mass term is the one of Eq. (\ref{massc1}), it means that the nucleon mass disappears when 
the chiral condensate vanishes (as e.g. in a confined but chirally restored phase of matter, see next Section). 

Interestingly, there is a second chiral way to give
mass to the nucleon. To this end, a second nucleon field (the chiral partner
of the nucleon, with opposite parity w.r.t. the bare nucleon field $\Psi_{1}$)
is considered, see for example \cite{detar,tolos,gallas,gallas2,vonsmekal,laka,olbrich} and refs. therein. We consider its chiral
transformation as being mirror-like%
\begin{equation}
\Psi_{2,R}\rightarrow e^{-i\alpha/2}\Psi_{2,R}\text{ , }\Psi_{2,L}\rightarrow
e^{i\alpha/2}\Psi_{2,L}%
\text{ .}
\end{equation}
Besides a standard interaction%
\begin{equation}
\mathcal{L}_{\Psi_{2}\Phi}=g_{\Psi_{2}\Phi}\left(  \bar{\Psi}_{2,R}%
\Phi^{\dagger}\Psi_{2,L}+\bar{\Psi}_{2,L}\Phi\Psi_{2,R}\right)
\end{equation}
an invariant mass term is obtained as \cite{gallas2,laka}:%

\begin{align}
\mathcal{L}_{mirror,mass}  &  =-c_{N}G\left(  \bar{\Psi}_{1,L}\Psi_{2,R}%
-\bar{\Psi}_{1,R}\Psi_{2,L}-\bar{\Psi}_{2,L}\Psi_{1,R}-\bar{\Psi}_{2,R}%
\Psi_{1,L}\right) \nonumber\\
&  =-c_{N}G\left(  \bar{\Psi}_{1}\gamma^{5}\Psi_{2}-\bar{\Psi}_{2}\gamma
^{5}\Psi_{1}\right)  \text{ ,}%
\end{align}
where $c_{N}\sim N_{c}^{0}$ is a dimensionless constant independent on $N_c$ and $G$ is, as usual, the
dilaton/scalar glueball field. Hence, a chirally invariant mass terms
\begin{equation}
m_{N,0}=c_{N}G_{0}\simeq c\Lambda_{QCD}\sim N_{c}%
\end{equation}
emerges with the correct scaling, see Fig. \ref{B-Page-6}. 

The mass of the nucleon and its chiral
partners are obtained by properly diagonalizing the system, finding:%
\begin{equation}
\left(
\begin{array}
[c]{c}%
N\\
N^{\ast}%
\end{array}
\right)  =\frac{1}{\sqrt{2\cosh\delta}}\left(
\begin{array}
[c]{cc}%
e^{\delta/2} & \gamma^{5}e^{-\delta/2}\\
\gamma^{5}e^{-\delta/2} & e^{\delta/2}%
\end{array}
\right)  \left(
\begin{array}
[c]{c}%
\Psi_{1}\\
\Psi_{2}%
\end{array}
\right)
\end{equation}
with%
\begin{equation}
M_{N,N^{\ast}}=\sqrt{m_{N,0}^{2}+\frac{1}{4}\left(  g_{\Psi_{1}\Phi}%
+g_{\Psi_{2}\Phi}\right)  ^{2}\phi_{N}^{2}}\pm\frac{1}{2}\left(  g_{\Psi
_{1}\Phi}-g_{\Psi_{2}\Phi}\right)  \phi_{N}\sim N_{c}%
\end{equation}
and
\begin{equation}
\cosh\delta=\frac{M_{N}+M_{N^{\ast}}}{2m_{N,0}}\sim N_{c}^{0}\text{ .}%
\end{equation}
Note, for $m_{N,0}=0,$ one has $M_{N}=g_{\Psi_{1}\Phi}\phi_{N}$ and
$M_{N^{\ast}}=g_{\Psi_{2}\Phi}\phi_{N},$ as expected. On the other hand, if
$\phi_{N} = 0$, the chiral partners are degenerate with $M_{N}=M_{N^{\ast}}%
=m_{N,0}.$

There is another interesting aspect of chiral models, that regards the axial
coupling constant of the nucleon. In the model above, for $m_{N,0}=0$ the
axial coupling constant turns out to be one, $g_A^N =1$. If $m_{N,0}$ is nonzero, the
mixing sets in with%
\begin{equation}
g_{A}^{N}=\frac{1}{2\cosh\delta}\left(  e^{\delta/2}-e^{-\delta/2}\right)
\sim N_{c}^{0}\text{ .}%
\end{equation}
Yet, the Skyrme model predicts $g_{A}^{N}\sim N_{c}$ \cite{mclerranissues}. How
to reconcile these different results?

The key is to consider vector and axial-vector mesons. We introduce
\begin{equation}
R^{\mu}=\rho^{\mu}-a_{1}^{\mu}\text{ , }L^{\mu}=\rho^{\mu}+a_{1}^{\mu}%
\end{equation}
(recall that we are in the one-flavor case, so we could have as well used
$\omega^{\mu}$ and $f_{1}^{\mu}$ instead), which under a chiral transformation
are separately invariant: $R^{\mu}\rightarrow R^{\mu}$ , $L^{\mu}\rightarrow
L^{\mu}$. However, under parity: $R^{\mu}\Leftrightarrow L^{\mu}$.  Hence, the chirally
and parity invariant coupling to (axial-)vector states is:
\begin{align}
\mathcal{L}_{(axial-)vector}  &  =c_{\Psi_{1}}R_{\mu}\bar{\Psi}_{1,R}%
\gamma^{\mu}\Psi_{1,R}+c_{\Psi_{1}}L_{\mu}\bar{\Psi}_{1,L}\gamma^{\mu}%
\Psi_{1,L}\nonumber\\
&  +c_{\Psi_{2}}R_{\mu}\bar{\Psi}_{2,R}\gamma^{\mu}\Psi_{2,R}+c_{\Psi_{2}%
}L_{\mu}\bar{\Psi}_{2,L}\gamma^{\mu}\Psi_{2,L}%
\end{align}
where
\begin{equation}
c_{\Psi_{1}}\sim N_{c}^{1/2}\text{ , }c_{\Psi2}\sim N_{c}^{1/2}%
\text{ .}
\end{equation}
The exact calculation of the axial coupling constants goes beyond the scope of
these lectures, since it involves subtle issues, such as the mixing of the
$a_{1}$ and $\pi$ occurring in chiral models with (axial-)vector mesons \cite{dick,stani,ko,urban,nf2}. Yet, we can provide the final results of this calculation. The previous expression for the axial coupling $g_A$ is modified into:%
\begin{equation}
g_{A}^{N}=\frac{1}{2\cosh\delta}\left(  g_{A}^{(1)}e^{\delta/2}+g_{A}%
^{(2)}e^{-\delta/2}\right)
\end{equation}
with%
\begin{equation}
g_{A}^{(1)}=1-\frac{c_{\Psi_{1}}}{g_{1}}\left(  1-\frac{1}{Z^{2}}\right)
\text{ , }g_{A}^{(2)}=-1+\frac{c_{\Psi_{2}}}{g_{1}}\left(  1-\frac{1}{Z^{2}%
}\right)
\end{equation}
where $g_{1}\sim N_{c}^{-1/2}$ is the coupling of one vector meson to two
pseudoscalar ones (hence a standard $QQQ$ vertex that scales as $N_{c}^{-1/2}%
$), and 
\begin{equation}
Z=\left(  1-\frac{g_{1}^{2}\phi_{N}^{2}}{M_{a_{1}}^{2}}\right)
^{-1/2}\sim N_{c}^{0}  
\end{equation}
is a constant that appears when dealing with the mixing between the
pion and the $a_{1}$ meson  \cite{dick,stani,ko,urban,nf2}. 
It then follows that
\begin{equation}
g_{A}^{(1)}\sim N_{c}\text{ and }g_{A}^{(2)}\sim N_{c}%
\end{equation}
thus
\begin{equation}
g_{A}^{N}\sim N_{c}
\text{ ,}
\end{equation}
in agreement with the Skyrme approach. This result implies that, in hte present case, one cannot neglect axial-vector mesons, otherwise
basic large-$N_{c}$ properties might be lost.

\section{Brief description of QCD at nonzero temperature and densities at
large-$N_{c}$}

The QCD phase diagram at large-$N_c$ is a rich topic that would deserve a series of lectures on its own. Here, we present a summary of some basic facts and to some interesting recent developments.

\begin{figure}[h]
        \centering       \includegraphics[scale=0.75]{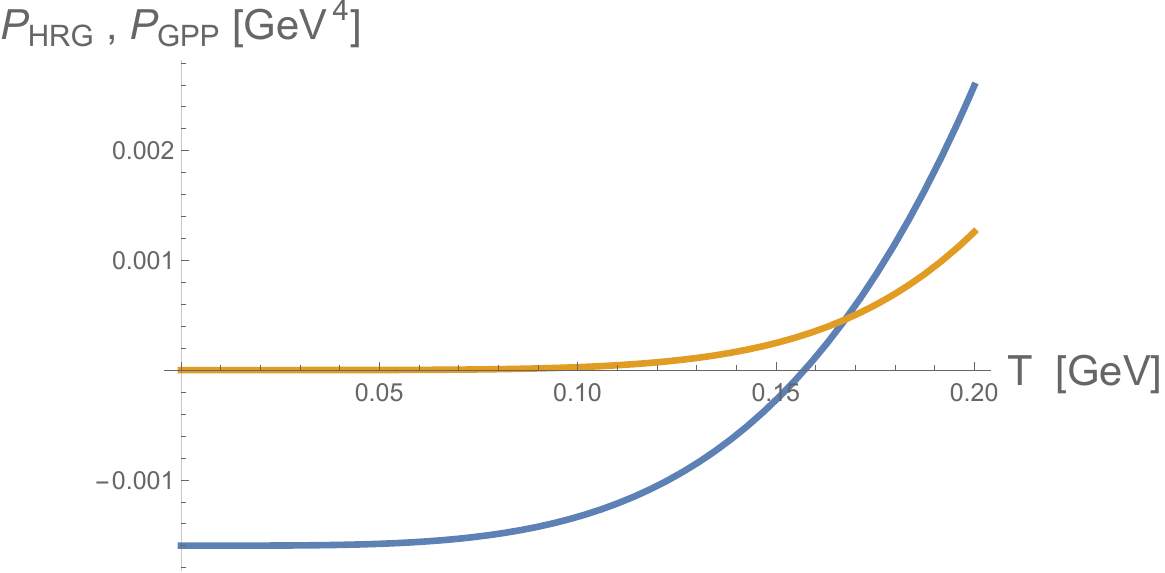}\\
        \caption{Schematic representation of the pressure in the confined and the deconfined phases. In the confined one, the pressure $P_HRG$ (Eq. \ref{prhrg}, yellow line, the upper one for low $T$) of the hadron resonance gas scales as $N_c^0$ and vanishes at $T=0$. Then, the quark-gluon pressure $P_QGP$ (Eq. \ref{pqgp}, blue line,, the lower one for low $T$) contains the free parts of quarks ans gluons, that scale as $N_c$ and $N_c^2$ respectively, as well  a vacuum part that has two similar terms, but with opposite sign. Thus, in order the QGP to be realized, the QGP pressure needs first to become positive: this occurs for a temperature of the order of $N_c^0$. Shortly after, the QGP pressure overcomes the HRG one, realizing deconfinement.}
        \label{Cep-Page-3}
   \end{figure}

As shown in \cite{chiralspirals1}, when $N_{c}$ is large, gluons dominate: a
first order phase transition between the deconfined and confined phases at $T_{dec}\simeq300 \propto N_c^0$ MeV (see \cite{grg} for a
compilation of results) is expected for any value of the chemical potential.
This is utterly different from the QCD phase diagram for $N_c =3 $ \cite{ratti}, which is
cross-over along the $T$ direction and first order along the $\mu$ one. 

In the
following, we first concentrate on the main expected properties of
confinement/deconfinement phase transition for varying $N_{c}$ and then on
four snapshots concerning large-$N_{c}$ properties in the medium.

Since relevant for our purposes, we write down the vacuum contribution of the dilaton-LSM confined matter, see Sec. 3.1 (consequence 3) and Sec. 3.2 (consequence 1): 
\begin{gather}
V_{vac}=V(G=G_{0},\sigma=\sigma_{0}=\phi_N,0)= \nonumber \\ 
\frac{1}{4}\lambda_{G}G_{0}^{4}\left[
\ln\left(  \frac{G_{0}}{\Lambda_{G}}\right)  -\frac{1}{4}\right]  +\frac{a}%
{2}G_{0}^{2}\sigma_{0}^{2}+\frac{\lambda}{4}\sigma_{0}^{4}\text{.}%
\end{gather}
with
\begin{equation}
G_{0}=\Lambda_{G}e^{\frac{a^{2}}{\lambda\lambda_{G}}}\sim N_{c}\text{ ,
}\sigma_{0}^{2}=-\frac{aG_{0}^{2}}{\lambda}\sim N_{c}%
\text{ .}
\end{equation}
Hence,%
\begin{equation}
V_{vac}=V_{G,vac}+V_{\sigma.vav}%
\text{ ,}
\end{equation}
with%
\begin{equation}
V_{G,vac}=\frac{1}{4}\lambda_{G}\Lambda_{G}^{4}e^{\frac{4a^{2}}{\lambda
\lambda_{G}}}\left[  \frac{a^{2}}{\lambda\lambda_{G}}-\frac{1}{4}\right]  \sim
N_{c}^2\left[  N_{c}^{-1}-\frac{1}{4}\right]  \sim-N_{c}^{2}<0
\text{ ,}
\end{equation}
and%
\begin{equation}
V_{\sigma.vac}=\frac{a}{2}G_{0}^{2}\sigma_{0}^{2}+\frac{\lambda}{4}\sigma
_{0}^{4}=-\frac{a^{2}}{4\lambda}G_{0}^{4}\sim-N_{c}<0
\text{ .}
\end{equation}

We may then express the QCD vacuum pressure as function of $N_{c}$ by two
terms:
\begin{equation}
P_{QCD,vac}=P_{G,vac}+P_{\sigma,vac}%
\end{equation}
with%
\begin{equation}
P_{G,vac}\simeq \bar{B}_{G}N_{c}^{2}>0\text{ },P_{\sigma,vac}\simeq \bar{B}_{\sigma}N_{c}>0
\end{equation}
Now, strictly speaking this pressure should be added (as a positive term) to the confined phase, as
it is derived from a confined model of QCD. 
We however follow the usual
convention of subtracting this term from the QCD vacuum, thus at zero temperature and
density the pressure of the confined phase vanishes. One has then to
subtract this term from the corresponding quark-gluon-plasma (QGP) phase.

Hence, the (schematic and simplified) pressure of the QGP can be expressed as \cite{satz}%
\begin{align}
P_{QGP}(T)  &  =2N_{c}^{2}\frac{\pi^{2}}{90}T^{4}+\frac{7}{4} N_{c}N_{f}\frac{\pi^{2}%
}{90}T^{4}-P_{G,vac}\nonumber\\
&  =2N_{c}^{2}\frac{\pi^{2}}{90}T^{4}+\frac{7}{4} N_{c}N_{f}\frac{\pi^{2}}{90}T^{4}%
-\bar{B}_{G}N_{c}^{2}-\bar{B}_{\sigma}N_{c} \text{ ,}
\label{prqgp}
\end{align}
where it is visible that gluons dominate for $N_c$ large enough.
The pressure for the confined phase (referred to as Hadron Resonance Gas (HRG) \cite{ratti}) reads
\begin{equation}
P_{HRG}(T)=\sum_{n}P_{n}(T)\text{  ,}%
\label{prhrg}
\end{equation}
with%
\begin{align}
\text{ }P_{n}(T) &  =-T\varsigma_{n}\int_{k}\ln\left[  1-\exp\left(
-\frac{\sqrt{k^{2}+M_{n}^{2}}}{T}\right)  \right]  \text{ for a meson ,}\\
P_{n}(T) &  =T\varsigma_{n}\int_{k}\ln\left[  1-\exp\left(  -\frac{\sqrt
{k^{2}+M_{n}^{2}}}{T}\right)  \right]  \text{ for a baryon ,}%
\end{align}
where $\varsigma_{n}$ is the appropriate degeneracy factor and $M_{n}$ is the
mass of the $n$-th hadron. Clearly, mesons dominate at large-$N_{c}$, since
baryons are very heavy in this limit. Hence:
\begin{equation}
P_{HRG}(T)\simeq P_{HRG}^{\text{mesons}}(T)\sim N_{c}^{0}.
\end{equation}
Note,
\begin{equation}
P_{HRG}(T=0)=0
\end{equation}
in agreement with the adopted normalization.

The confinement/deconfinement phase transition takes place for $T = T_{dec}$ given by (see Fig. \ref{Cep-Page-3}):
\begin{equation}
P_{HRG}(T_{dec})=P_{QGP}(T_{dec}) \text{ .}
\end{equation}
It is easy to understand that this is the case for
\begin{equation}
T_{dec}\sim N_{c}^{0} \text{ .}
\end{equation}
Namely, at large $N_c$ the gluons dominate and the transition takes place basically
just after the QGP pressure comes positive:
\begin{equation}
T_{dec}\gtrsim\frac{\bar{B}_{G}}{2\frac{\pi^{2}}{90}}\sim N_{c}^{0}
\text{ .}
\end{equation}
Note, the
present simple treatment is not able to correctly guess the order of the transition, which is by construction a first-order, but  in reality it is known to be a cross-over \cite{ratti}.

Moreover, it is also expected that the phase transition for chiral restoration takes
place for a similar temperature as the deconfined one, as supported by models, e.g. \cite{ratti,kovacscep,lena}. 

Next, let us consider zero temperature and finite density. To this end,
we introduce the quark chemical potential $\mu_{q}$ and the baryon chemical
potential $\mu_{B}=N_{c}\mu_{q}$. 
In the confined phase, one has only baryons. We follow here one possible choice
described in Ref. \cite{pagliara}, that corresponds to a confined stiff matter
with speed of sound equal to the speed of light. Intuitively, it corresponds
to an interaction dominated gas. In this case, the pressure in the baryonic
phase is (for large values of the baryonic chemical potential)
\begin{equation}
P_{B}=a_{B}\mu_{B}^{2}%
\end{equation}
where $a_{B}$ is a constant with dimension Energy$^{2}$, whose $N_{c}$
dependence must be established\footnote{In a more realistic treatment, one may
consider at high density $P_{B}=a_{B}\mu_{B}^{\alpha}-K$ (with $\alpha$ a free parameter) which needs
to be
matched to known equation of state of nuclear matter at about $2n_{0}$ \cite{pagliara}. The
large-$N_{c}$ behavior is not affected.}.

The baryon density follows as
\begin{equation}
n_{B}=dP_{B}/d\mu_{B}=2a_{B}\mu_{B}
\text{ ,}
\end{equation}
while the energy density as 
\begin{equation}
   \varepsilon_{B}=n_{B}P_{B} - P_{B} = P_{B} \text{ ,}
\end{equation}
hence the speed if sound is
\begin{equation}
v_{B}=\sqrt{dP_{B}/d\varepsilon_{B}}=1 \text{ .}    
\end{equation}
This result is indeed in agreement
with the high density limit of nuclear matter described in Ref.
\cite{glendenning}, which takes place when vector meson driven interaction dominates.
Within, since $a_{B}$ has energy$^{2}$ and since the appropriate dimension is
constructed by $m_{V}^{2}/g_{\bar{B}BV}^{2}$ with $g_{\bar{B}BV}\sim
N_{c}^{1/2}$ being the baryon-baryon-vector coupling and $m_V$ the mass of the vector meson (such as the $\omega$ meson) , one gets%
\begin{equation}
a_{B}\sim \frac{m_V^2}{g_{BBV}^2} \sim N_{c}^{-1}\rightarrow a_{B}=\frac{\bar{a}_{B}}{N_{c}} \text{ .}
\end{equation}
In fact, the pressure as function of the density can be written as
\begin{equation}
P_{B}=\frac{n_{B}^{2}}{4a_{B}}\simeq\frac{1}{2}\frac{g_{\bar{B}BV}^{2}}%
{m_{V}^{2}}n_{B}^{2}
\end{equation}
where the right hand side follows from Sec. 4.11 of Ref.
\cite{glendenning} (high-density limit). It thus shows the direct proportionality of $a_{B}$ to
$m_{V}^{2}/g_{\bar{B}BV}^{2}.$

Finally, in terms of the quark chemical potential, we may write the baryonic pressure
\begin{equation}
P_{B}(\mu_{q})=N_{c}\bar{a}_{B}\mu_{q}^{2} \label{pb}%
\end{equation}
implying that we have a confined phase whose pressure scales with $N_{c}$.
This is in agreement with the quarkyonic phase discussed in\ Refs.
\cite{chiralspirals1,chiralspirals2,dichotomous}, and also
with the result of the Walecka-type model at large $N_{c}$ and for high
densities, see Ref. \cite{bonanno}.

Next, let us consider the QGP phase, which contains only quarks as d.o.f. (gluons are not present at $T =0$) as well as the already discussed vacuum contribution:
\begin{equation}
P_{QGP}(\mu_{q}) = P_{q}(\mu_{q})=\frac{N_{c}N_{f}}{12\pi^{2}}\mu_{q}^{4}+P_{QCD,vac}=\frac
{N_{c}N_{f}}{12\pi^{2}}\mu_{q}^{4}-\bar{B}_{G}N_{c}^{2}-\bar{B}_{G}N_{c}\text{.}%
\end{equation}
The deconfinement phase transition takes place for $P_B(\mu_{q,dec}) = P_{q}(\mu_{q,dec})$, leading to
\begin{equation}
\mu_{q,dec}\sim N_{c}^{1/4} \text{ ,}
\end{equation}
thus the deconfinement phase transition takes place at higher and higher
densities when $N_{c}$ increases. Here, one expects a different behavior of
the chiral phase transition, which should take place for $\mu_{q,c}\sim
N_{c}^{0}$. There is therefore a wide range of $\mu_q$ for which matter is chirally restored but confined, the already mentioned quarkyonic phase \cite{chiralspirals1,chiralspirals2} (see also the large-$N_c$ considerations of \cite{lottini,mishustin}).

Additional topics related to large-$N_{c}$ in the medium are
briefly discussed below.

\textbf{1) Chiral mesonic models at nonzero }$T$\textbf{: a problem and how
to cure it.}
\begin{figure}[h]
        \centering       \includegraphics[scale=0.75]{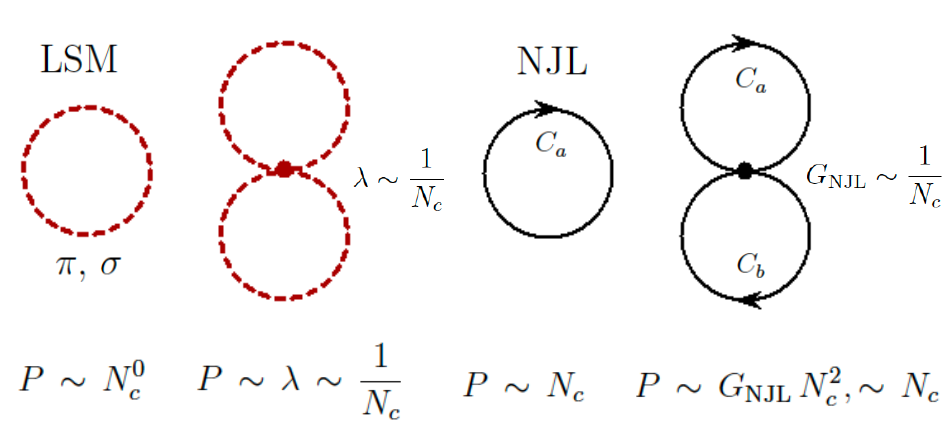}\\
        \caption{Vacuum diagrams can be used to establish the scaling of different contribution to the pressure. Upper part: in the LSM, the pressure of free mesons goes with $N_c^0$, while the interaction contribution goes as $N_c^{-1}$. This is why in the large-$N_c$ limit chiral restoration takes place at higher and higher $T$. Lower part: in the NJL model, the free quarks give rise to a pressure proportional to $N_c$, just as the interaction terms. Accordingly, the critical temperature for chiral restoration is $N_c$-independent, as it should. }
        \label{Cep-Page-1}
   \end{figure}

\begin{figure}[h]
        \centering       \includegraphics[scale=0.45]{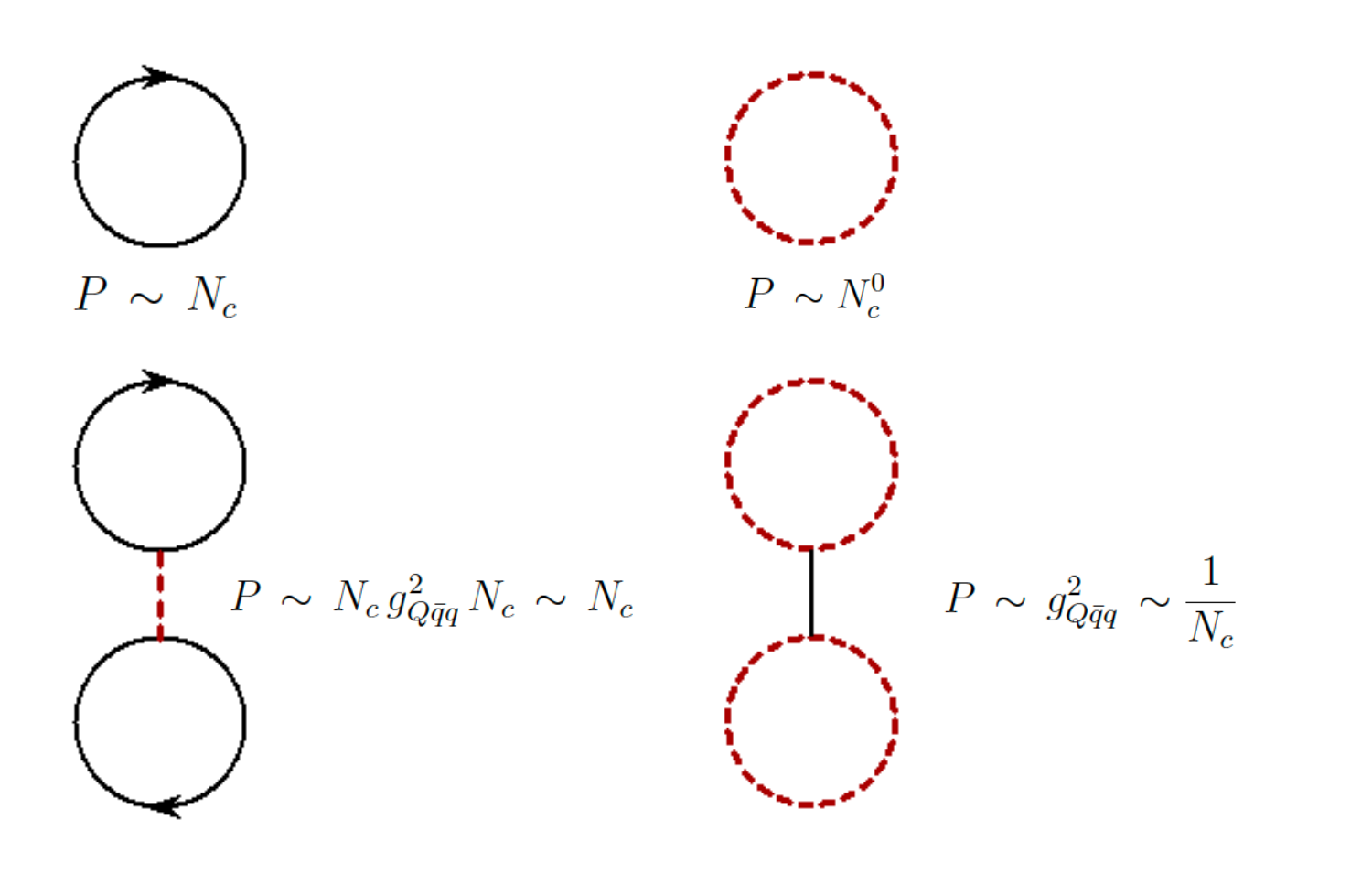}\\
        \caption{Quark-level LSM. The free part contains two contributions for the pressure, the meson one ($N_c^0$) and the quark one ($N_c$). Two interaction terms are outlined: one with quark loops, whose pressure contribution goes as $N_c$ (just as the free quark part) and one with mesonic loops, whose pressure contribution goes as $N_c^{-1}$. Quarks dominate in both cases, and the final outcome is similar to the NJL model: the critical temperature for chiral restoration scales as $N_c^0$.}
        \label{Cep-Page-2}
   \end{figure}
   
If a purely mesonic model as the one of Eq. (\ref{poth}) is considered, chiral restoration can be
studied by evaluating the chiral condensate as function of the temperature, $\phi_{N}\rightarrow\phi_{N}(T)$.
For large $T$,
$\phi_{N}(T)$ tends to zero and the way it does also specifies the
order of the chiral phase transition. For $N_{c}=3$ one expects a smooth cross-over \cite{ratti,lena,achimtq}.

What about the behavior of the chiral phase transition at large $N_c$? One indeed expects that the critical temperature for chiral restoration, denoted as $T_c$, should be $N_c$-independent:
\begin{equation}
T_{c}\sim T_{dec} \sim \Lambda_{QCD}\sim N_{c}^{0}
\text{ .}
\end{equation}

Yet, in purely mesonic models, the contribution of the interaction to the effective potential (or equivalently to the
pressure) scales as $1/N_{c}$ and is therefore suppressed. Correspondingly, one finds \cite{achimnc}: 
\begin{equation}
T_{c}\sim f_{\pi}\sim N_{c}^{1/2} \text{ ,}
\end{equation}
hence chiral restoration takes place at larger and larger $N_{c}$.
This result is depicted as in Fig. \ref{Cep-Page-1} (left), in which the vacuum diagrams give rise to the contribution for the pressure, see Ref. \cite{lena} and refs. therein. It seems therefore that such models cannot describe the expected large-$N_c$ results.

How to reconcile chiral models with the expected large-$N_{c}$ scaling? In
Ref. \cite{achimnc} some recipes were put forward. An intuitive heuristic approach
consists in considering the following potential with an explicit dependence on the temperature $T$:
\begin{align}
V(\sigma,\pi)  &  =\frac{m_{0}^{2}}{2}\left(  1-\frac{T^{2}}{T_{0}^{2}%
}\right)  \Phi^{\ast}\Phi+\frac{\lambda}{4}\left(  \Phi^{\ast}\Phi\right)
^{2}\nonumber\\
&  =\frac{m_{0}^{2}}{2}\left(  1-\frac{T^{2}}{T_{0}^{2}}\right)  \left(
\sigma^{2}+\pi^{2}\right)  +\frac{\lambda}{4}\left(  \sigma^{2}+\pi
^{2}\right)  ^{2}\text{,}%
\end{align}
where $T_{0}\sim\Lambda_{QCD}\sim N_{c}^{0}.$ In this way, the chiral
restoration is enforced by this modification. At $T=T_{0}$ the bare potential
is such that only the quartic interaction is present. With this `ad hoc'
modification, $T_{c}\sim N_{c}^{0}$ is obtained.

A more formal way of achieving this result is realized by coupling the chiral multiplet
$\Phi$ to the expectation value of the Polyakov loop (see e.g.
\cite{ratti,fuku}). This quantity, denoted as $l(T),$ describes effectively
the gluonic sector, more specifically the restoration of the symmetry under $Z(N_{c})$ transformations
in the vacuum. In particular, $\left\vert l(T)\right\vert =1$ at large
temperature (in the deconfined phase) while it vanishes at small $T$. One can couple the Polyakov loop to the LSM in the following way: 
\begin{equation}
V(\sigma,\pi,l)=\frac{m_{0}^{2}}{2}\left(  1-c_{l}T^{2}\left\vert
l(T)\right\vert ^{2}\right)  \Phi^{\ast}\Phi+\frac{\lambda}{4}\left(
\Phi^{\ast}\Phi\right)  ^{2}
\text{ ,}
\end{equation}
where $c_{l}\sim N_{c}^{0}$ is a dimensionless constant. Indeed, the proper
description of the Polyakov loop at large N$_{c}$ is not an easy task, but certain relatively simple choices are possible
\cite{cpnc,fuku}. Within LSMs with Polyakov loop, the critical temperature for chiral restoration scales as $T_c \propto N_c^0$, as expected.

\bigskip

The already mentioned famous NJL model \cite{njlorig,njl} contains only quarks
with a chiral interaction of the type
\begin{equation}
V_{NJL}(\sigma,\pi,l)=G_{NJL} [\left(  \bar{\psi}\psi\right)  ^{2}+\left(  \bar{\psi
}i\gamma^{5}\psi\right)  ^{2}]
\end{equation}
where the chiral transformation (in this one-flavor case) is $\psi\rightarrow
e^{i\alpha\gamma^{5}/2}\psi,$ and where $G_{NJL}\sim N_{c}^{-1}$ (this is just as the
constant $K_{Q}$ studied in the case of the quarkonium formation in Sec. 3.1; indeed, the NJL model has been widely used to study $\bar{q}q$ bound states \cite{njl,volkov}). Here, the
interaction type is of the same order of the free quark ones. The SSB takes
place if $G_{NJL}$ is large enough and chiral restoration takes place at nonzero
$T$, with $T_{c}\sim N_{c}^{0}$, see Fig. \ref{Cep-Page-1} (right part).

The last case that we mention is the one that involves quark-meson type model
\cite{cpnc,kovacscep,tripolt}, in which both mesons and quarks are present:
\begin{equation}
V_{LSM,quarks}(\sigma,\pi,l)=g_{\sigma}\sigma\left(  \bar{\psi}\psi\right)  +g_{\pi
}\pi\left(  \bar{\psi}i\gamma^{5}\psi\right)
\text{ .}
\end{equation}
In this case, the interaction contribution to the pressure is also of the order of $N_{c}$ just
as the quarks, thus $T_{c}\sim N_{c}^{0}$, see Fig. \ref{Cep-Page-2}.

\bigskip

\textbf{2) Critical endpoint CEP at large-}$N_{c}$

It is well known that the confinement/deconfinement as well as the chiral
phase transition(s) are of cross-over type along the $T$ direction, and first
order along the $\mu_{q}$ one. At least one critical point is therefore
expected, whose search is important for both theoretical and experimental
investigations, e.g. \cite{ratti,kovacscep,gazd,hara}.

Yet, at large $N_{c}$, as shown in details in \cite{cpnc} using an extended
linear sigma model with quarks and Polyakov loop, the phase diagram turns out
to be utterly different: one has a first-order transtion along $T$ and a cross-over one
along $\mu_{q}.$ A critical point is present at about $(T_{CEP},\mu_{q,CEP})$
where $T_{CEP} \propto \Lambda_{QCD} \sim N_c^0$ while $\mu_{q,CEP}$ increases for increasing $N_{c})$. 
Quite remarkably,
for intermediate $N_{c}$ (from $4$ up to about $50)$, only cross-over phases are present in the whole phase diagram \cite{cpnc}. 

\begin{figure}[h]
        \centering       \includegraphics[scale=0.90]{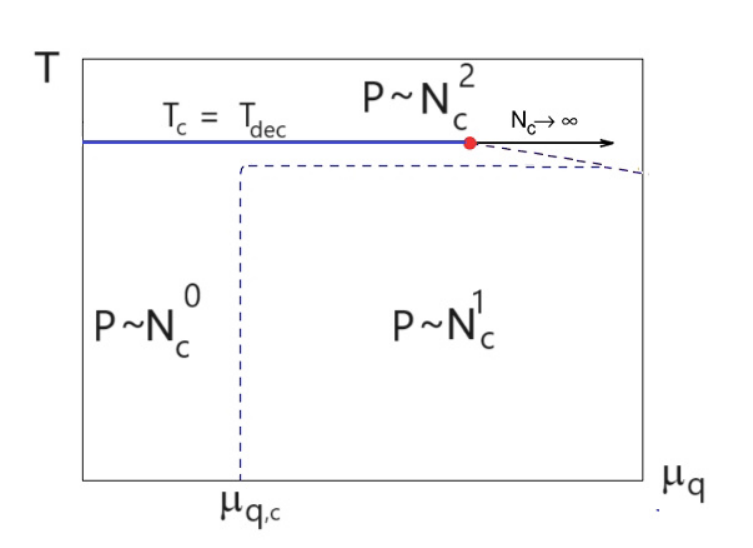}\\
        \caption{Schematic and simplified representation of the QCD phase diagram for large-$N_c$. For low $T$ and $\mu$, a confined and chirally broken (SSB) phase with pressure proportional to $N_c^0$ is present. If the temperature is above $T_c=T_{dec}$, for any chemical potential $\mu$, a deconfined phase with pressure proportional to $N_c^2$ is realized. Here, chiral symmetry is also restored. For large $\mu$ but $T<T_{dec}=T_c$, the system is still confined but chiral symmetry is restored: the pressure is proportional to $N_c$ (as a gas of quarks would have). Important aspects: the type of the chiral phase transition is a first-order one along $T$ (solid line) and a cross-over along $\mu$ (dashed line). Along $T$, the chiral and the deconfinement transitions coincide. Along $\mu$, they do not: the chiral one takes place for $\mu_{q,c} \sim N_c^0$, the deconfinement one ( moves toward infinity. The critical end point (red dot in the figure) also tends to infinity along the $\mu$ direction. For more details, see Ref. \cite{cpnc}.  }
        \label{Cep-Page-4}
   \end{figure}

The pressure at very large $N_{c}$ resembles the expected behavior, in particular we have the following areas, see Fig. \ref{Cep-Page-4} as well as the detailed explanations in ref. \cite{cpnc}:

(i) $P\sim N_{c}^{0}$ for low-$T$ and low-$\mu_{q}$ within the confined and
chirally broken phase.

(ii) $P\sim N_{c}$ for low-$T$ and high-$\mu_{q}$ within the confined and chirally
restored (quarkyonic) phase. Note, in this  phase the nucleons do not not need to be massless, see the discussion in Sec. 4. 

(iii) $P\sim N_{c}^{2}$ for high $T$ and high-$\mu_{q}$ within the deconfined QGP phase.

Finally, along the $T$ line and for small $\mu_{q}$ the chiral and the
deconfinement phase transition coincide ($T_{dec} = T_c$), while along the $\mu_{q}$ line and
for small $T$ the chiral transition occurs for $\mu_{q,c}\sim N_{c}^{0}$ and the
deconfinement one for $\mu_{q,dec}$ increasing $N_{c}$.

In the recent work of Ref. \cite{pisarskilast} the effect of the chiral anomaly on the phase diagram is discussed. Yet, anomaly terms  decrease fast for increasing $N_c$, thus they shall not modify the overall large-$N_c$ picture outlined above. 

\bigskip

\textbf{3) Nuclear matter at large-}$N_{c}$

Does nuclear matter bind at large $N_{c}?$ This question was already posed in
the introduction, and at first put aside since quite (too?) philosophical. Yet, is
$N_{c}=3$ somewhat special?

Indeed, the issue is quite subtle. In the easiest scenario, one considers
a standard $\sigma$-$\omega$ model coupled to the nucleon
\begin{equation}
V=g_{\sigma N}\bar{\Psi}_{1}\Psi_{1}+g_{\omega N} \omega_{\mu}\bar{\Psi}_{1}\gamma^{\mu
}\Psi_{1}
\text{ ,}
\end{equation}
with%
\begin{equation}
g_{\sigma N}\sim N_{c}^{1/2}\text{ and }g_{\omega N}\sim N_{c}^{1/2}
\text{ ,}
\end{equation}
where $\sigma$ corresponds to $f_{0}(500)$ and $\omega$ to $\omega(782).$
If the masses of these fields behave as $N_{c}^{0}$ (as quark-antiquark regular
states), then the mean field equations show that nuclear matter forms for any
$N_{c}$ (and becomes more and more bound).

Yet, it is well known that $f_{0}(500)$ is \textit{not} predominantly a $\bar{q}q$ \cite{sigmareview}.
Considering this fact, changes the picture completely: nuclear matter
does not form, already for $N_{c}\geq4$.

Indeed, for very large $N_{c}$ the pion cannot be neglected and being `de facto'
massless should generate a kind of loosely bound nuclear matter similar to atomic
matter (basically an attractive almost Coulomb force would act among nucleons in this
limit). The detailed study of this hypothetical state of matter is a task for
the future.

\bigskip

\textbf{4) Neutron stars at large-}$N_{c}$

One may apply the previous discussion about baryonic and quark
matter  to neutron stars, see Ref. \cite{pagliara} for details.

Using the baryonic equation of state of Eq. (\ref{pb}) that corresponds to
stiff matter (speed of sound equal $1$), for $N_{c}=3$ the maximal mass of a
neutron star turns out to be about
\begin{equation}
M_{NS}^{\max}\simeq2.2M_{\odot}\text{ .}%
\end{equation}
Namely, for higher masses a phase transition to deconfined quark phase takes
place inside the neutron star, which is however unstable because quark matter cannot sustain the gravitational pressure \cite{bielich}. (The maximal value quoted above can be increased if the vacuum pressure is increased).

When increasing $N_{c}$, the phase transition to quark matter takes place at
higher and higher density. Already for $N_{c}\gtrsim5.5,$ quark matter plays no role, and
the maximum mass of a neutron star is about $3M_{\odot}.$

\section{Conclusions}

In these lectures we have revisited the main features of QCD at large-$N_{c}$ for
both mesons and baryons.
To this end, we have used a bound-state approach in
which a simple separable Ansatz has been applied. This is indeed similar to
certain approaches of QCD, such as the NJL model. Yet, the large-$N_{c}$ scaling laws 
that can be derived are fully general and do not depend on this specific
Ansatz. In this way, we could recover all the large-$N_{c}$ scaling behavior
for regular mesons, for glueballs, and for hybrid mesons. Also their mutual
interactions, decays and mixing, could be properly described.

Many consequences of these results have been investigated, among which the
reason why certain mesons are narrow, how dominant and subdominant
interaction terms arise, and most importantly, how chiral models behave in the
large-$N_{c}$ limit. Also the behavior of the dilaton/scalar glueball field and its
coupling to chiral models has been reviewed in some detail. An overall nice
and consistent picture of large-$N_c$ QCD has emerged.

Four-quark states were briefly discussed. Molecular states and dynamically
generated states do not form in the large-$N_{c}$ limit, as our bound-state
approach easily shows. Yet, the case of tetraquark objects is more
complicated. Our present results suggest that all of them share the same fate:
they do not form ion the large-$N_{c}$ limit, but this last statement is not
yet conclusive.

Baryons were presented following the same line applied to mesons, upon
interpreting them as bound states of a generalized diquark built with $N_{c}-1$ quarks, and a quark. Upon taking the mass of the generalized diquark as increasing with $N_c$, it was possible to recover the known large-$N_c$ results for baryons. Chiral models
have been studied for baryons as well: the chirally invariant  mass generation via the
quark condensate and via the dilaton condensate (the latter in the so-called
mirror assignment) fulfill the expected large-$N_{c}$ properties.

Finally, we have discussed the main features of the phase diagram of QCD in the
large-$N_{c}$ domain. Simple scaling laws show that at large $N_c$ gluons
dominate if the temperature is high enough. The temperature for
confinement/deconfinement transition is $N_{c}$ independent, but the chemical
potential increases for increasing $N_c$. This means that in the large-$N_{c}$
limit matter is confined below a certain $T_{dec}$ and deconfined above, for
any chemical potential. Yet, for high density, one may have a confined and
chirally restored matter whose pressure is proportional to $N_c$, which well
fits with the concept of a  quarkyonic phase. These results also imply that the phase diagram
at large $N_c$ looks quite different than the one for the real $N_c = 3$ world.

Moreover, we have also discussed some related issues, such as the failure of
certain chiral models at large $N_c$ and how to improve them, the formation (or non-formation)
of nuclear matter, and implications for neutron stars.

In conclusion, large-$N_{c}$ QCD is an interesting theoretical framework,
often the only one available, to understand certain properties of QCD. It offers a
consistent picture, somewhat simplified from our physical one, but definitely
not trivial.

Coming back to our original question: is $N_c$=3 large? We have shown that in most cases it is, but not in all of them. There
is therefore not a simple and always valid answer to that seemingly naive question but one needs,
case by case, to study the consequences of increasing $N_c$ and see how much the outcomes
depart from the real world. Yet, in many cases the lesson gained from large $N_c$ is very useful for understanding our world with $N_c =3$.

\bigskip

\bigskip

\textbf{Acknowledgments: }The author thanks A. Pilloni, G. Pagliara, Gy.
Kovacs, P. Kovacs, C. Fischer, R. Pisarski, G. Torrieri, L. Bonanno, S. Jafarzade, E. Trotti, J. Pelaez, L. McLerran, D. Rischke, A. Koenigstein, and V. Shastry for very useful discussions that extended over the last 15 years. The author also acknowledges useful remarks to the first version of these lectures by M. Praszałowicz, J.~M.~Gerard, and S. Leupold.
Acta Phys. Pol. B is thanked for help in preparing a new version of the figures. 
Financial support
from the Polish National Science Centre (NCN) via the OPUS project
2019/33/B/ST2/00613 is acknowledged.


\bigskip

\bigskip

\bigskip

\newpage

\begin{thebibliography}{999}                                                                                              %


\bibitem {pdg}R.L. Workman et al. (Particle Data Group), Prog. Theor. Exp.
Phys. 2022, 083C01 (2022) and 2023 update.

\bibitem {weisebook}
A.~W.~Thomas and W.~Weise, The Structure of the
Nucleon,
Berlin, Germany: Wiley-VCH (2001) 389 p.

\bibitem {ratti}
C.~Ratti and R.~Bellwied,
The Deconfinement Transition of QCD: Theory Meets Experiment,
Lect. Notes Phys. \textbf{981} (2021), 1-216 2021, ISBN 978-3-030-67234-8,
978-3-030-67235-5 
doi:10.1007/978-3-030-67235-5

\bibitem {bonanno}
L.~Bonanno and F.~Giacosa,
Does nuclear matter bind at large $N_c$?,
Nucl. Phys. A \textbf{859} (2011), 49-62
[arXiv:1102.3367 [hep-ph]].


\bibitem {thooft}G. t Hooft,
`A PLANAR DIAGRAM THEORY FOR STRONG INTERACTIONS,
Nucl.\ Phys.\ B \textbf{72} (1974) 461.


\bibitem {witten}E.~Witten,
Nucl.\ Phys.\ B \textbf{160} (1979) 57.


\bibitem {thooftlect}
G.~'t Hooft, \textquotedblleft Large N,\textquotedblright\
doi:10.1142/9789812776914\_0001
[arXiv:hep-th/0204069 [hep-th]].


\bibitem {lebedlect}
R.~F.~Lebed, 
Phenomenology of large N(c) QCD,
Czech. J. Phys. \textbf{49}
(1999), 1273-1306
doi:10.1023/A:1022820227262
[arXiv:nucl-th/9810080 [nucl-th]].

\bibitem {colemanlect}
S.~R.~Coleman, \textquotedblleft1/N,\textquotedblright\ 17th International
School of Subnuclear Physics: Pointlike Structures Inside and Outside Hadrons
31 July-10 August 1979. Erice, Italy (C79-07-31) SLAC-PUB-2484.


\bibitem {lucinilect}
B.~Lucini and M.~Panero, Introductory lectures to large-{N}
QCD phenomenology and lattice results,
Prog. Part. Nucl.
Phys. \textbf{75} (2014), 1-40
doi:10.1016/j.ppnp.2014.01.001
[arXiv:1309.3638 [hep-th]].


\bibitem {luciniglueballsN}
C.~Bonanno, M.~D'Elia, B.~Lucini and D.~Vadacchino, 
Towards
glueball masses of large-N SU(N) pure-gauge theories without topological
freezing,Phys. Lett. B \textbf{833} (2022), 137281
doi:10.1016/j.physletb.2022.137281
[arXiv:2205.06190 [hep-lat]].


\bibitem {lucha}
W.~Lucha, D.~Melikhov and H.~Sazdjian,  Tetraquarks in
large-Nc QCD,
Prog. Part. Nucl. Phys. \textbf{120} (2021),
103867
doi:10.1016/j.ppnp.2021.103867
[arXiv:2102.02542 [hep-ph]].


\bibitem {njlorig}
Y.~Nambu and G.~Jona-Lasinio,
Dynamical Model of Elementary Particles Based on an Analogy with Superconductivity. 1.,
Phys. Rev. \textbf{122} (1961), 345-358 doi:10.1103/PhysRev.122.345

\bibitem {njl}
T.~Hatsuda and T.~Kunihiro,
QCD phenomenology based on a chiral effective Lagrangian,
Phys. Rept. \textbf{247} (1994), 221-367
doi:10.1016/0370-1573(94)90022-1
[arXiv:hep-ph/9401310 [hep-ph]];
S.~P.~Klevansky,
The Nambu-Jona-Lasinio model of quantum chromodynamics,Rev. Mod. Phys. \textbf{64} (1992), 649-708
doi:10.1103/RevModPhys.64.649

U.~Vogl and W.~Weise,
The Nambu and Jona Lasinio model: Its implications for hadrons and nuclei,
Prog. Part. Nucl. Phys. \textbf{27} (1991), 195-272
doi:10.1016/0146-6410(91)90005-9

\bibitem{volkov}
M.~K.~Volkov and A.~E.~Radzhabov,
The Nambu-Jona-Lasinio model and its development,
Phys. Usp. \textbf{49} (2006), 551-561
doi:10.1070/PU2006v049n06ABEH005905
[arXiv:hep-ph/0508263 [hep-ph]].




\bibitem {compcond}
S.~Weinberg,
Evidence That the Deuteron Is Not an Elementary Particle,
Phys. Rev. \textbf{137} (1965), B672-B678 
doi:10.1103/PhysRev.137.B672


\bibitem {compo}
K.~Hayashi, M.~Hirayama, T.~Muta, N.~Seto and T.~Shirafuji,
Compositeness criteria of particles in quantum field theory and S-matrix theory,''
Fortsch. Phys. \textbf{15} (1967) no.10, 625-660 doi:10.1002/prop.19670151002


\bibitem {gutsche}
F.~Giacosa, T.~Gutsche and A.~Faessler,  A Covariant
constituent quark/gluon model for the glueball-quarkonia content of scalar -
isoscalar mesons,
Phys. Rev. C \textbf{71} (2005), 025202
doi:10.1103/PhysRevC.71.025202
[arXiv:hep-ph/0408085 [hep-ph]].


\bibitem {pionsigma}
A.~Faessler, T.~Gutsche, M.~A.~Ivanov, V.~E.~Lyubovitskij and P.~Wang,
Pion and sigma meson properties in a relativistic quark
model,
Phys. Rev. D \textbf{68} (2003), 014011
doi:10.1103/PhysRevD.68.014011
[arXiv:hep-ph/0304031 [hep-ph]].


\bibitem {alkofer}
R.~Alkofer and L.~von Smekal,
The Infrared behavior of QCD Green's functions: Confinement dynamical symmetry breaking, and hadrons as relativistic bound states,
Phys. Rept. \textbf{353} (2001), 281 doi:10.1016/S0370-1573(01)00010-2
[arXiv:hep-ph/0007355 [hep-ph]].


\bibitem {fischer}
C.~S.~Fischer,
Infrared properties of QCD from Dyson-Schwinger equations,
J. Phys. G \textbf{32} (2006), R253-R291 doi:10.1088/0954-3899/32/8/R02
[arXiv:hep-ph/0605173 [hep-ph]].


\bibitem {eichmann}
G.~Eichmann, H.~Sanchis-Alepuz, R.~Williams, R.~Alkofer and C.~S.~Fischer,
Baryons as relativistic three-quark bound states,
Prog. Part. Nucl. Phys. \textbf{91} (2016), 1-100
doi:10.1016/j.ppnp.2016.07.001 [arXiv:1606.09602 [hep-ph]].


\bibitem{dick}
D.~Parganlija, P.~Kovacs, G.~Wolf, F.~Giacosa and D.~H.~Rischke,
Meson vacuum phenomenology in a three-flavor linear sigma model with (axial-)vector mesons,
Phys. Rev. D \textbf{87} (2013) no.1, 014011
doi:10.1103/PhysRevD.87.014011
[arXiv:1208.0585 [hep-ph]].


\bibitem {cpnc}
P.~Kov\'{a}cs, G.~Kov\'{a}cs and F.~Giacosa,
Fate of the critical endpoint at large Nc,
Phys. Rev. D \textbf{106} (2022) no.11, 116016
doi:10.1103/PhysRevD.106.116016
[arXiv:2209.09568 [hep-ph]].


\bibitem{migdal}
A.~A.~Migdal and M.~A.~Shifman,
Dilaton Effective Lagrangian in Gluodynamics,
Phys. Lett. B \textbf{114} (1982), 445-449
doi:10.1016/0370-2693(82)90089-2
\bibitem{salo}
A.~Salomone, J.~Schechter and T.~Tudron,
Properties of Scalar Gluonium,
Phys. Rev. D \textbf{23} (1981), 1143
doi:10.1103/PhysRevD.23.1143


\bibitem{ellis}
J.~R.~Ellis and J.~Lanik,
IS SCALAR GLUONIUM OBSERVABLE?,
Phys. Lett. B \textbf{150} (1985), 289-294
doi:10.1016/0370-2693(85)91013-5

\bibitem{stani}
S.~Janowski, F.~Giacosa and D.~H.~Rischke,
Is f0(1710) a glueball?,
Phys. Rev. D \textbf{90} (2014) no.11, 114005
doi:10.1103/PhysRevD.90.114005
[arXiv:1408.4921 [hep-ph]].


\bibitem {weinberg}
S.~Weinberg,
Tetraquark Mesons in Large $N$ Quantum Chromodynamics,
Phys.\ Rev.\ Lett.\ \textbf{110} (2013) 261601
doi:10.1103/PhysRevLett.110.261601
[arXiv:1303.0342 [hep-ph]].


\bibitem {lebed}
R.~F.~Lebed, 
 Large-N Structure of Tetraquark
Mesons,
Phys. Rev. D \textbf{88} (2013), 057901
doi:10.1103/PhysRevD.88.057901
[arXiv:1308.2657 [hep-ph]].


\bibitem {cohenlebed1}
T.~D.~Cohen and R.~F.~Lebed, 
Tetraquarks with exotic flavor
quantum numbers at large $N_{c}$ in QCD(AS),
Phys. Rev. D
\textbf{89} (2014) no.5, 054018
doi:10.1103/PhysRevD.89.054018
[arXiv:1401.1815 [hep-ph]].


\bibitem {cohenlebed2}
T.~D.~Cohen and R.~F.~Lebed, 
Are There Tetraquarks at Large $N_{c}$ in
QCD(F)?,
Phys. Rev. D \textbf{90} (2014) no.1, 016001
doi:10.1103/PhysRevD.90.016001
[arXiv:1403.8090 [hep-ph]].

\bibitem {estrada}
T.~Cohen, F.~J.~Llanes-Estrada, J.~R.~Pelaez and J.~Ruiz de Elvira,
Nonordinary light meson couplings and the $1/N_{c}$ expansion,
Phys. Rev.
D \textbf{90} (2014) no.3, 036003
doi:10.1103/PhysRevD.90.036003
[arXiv:1405.4831 [hep-ph]].


\bibitem {knecht}
M.~Knecht and S.~Peris, 
Narrow Tetraquarks at Large N,
Phys. Rev. D
\textbf{88} (2013), 036016
doi:10.1103/PhysRevD.88.036016
[arXiv:1307.1273 [hep-ph]].


\bibitem {chiralspirals1}
L.~McLerran and R.~D.~Pisarski, 
Phases of cold, dense quarks at large
N(c),
Nucl. Phys. A \textbf{796} (2007), 83-100
doi:10.1016/j.nuclphysa.2007.08.013
[arXiv:0706.2191 [hep-ph]].


\bibitem {chiralspirals2}
T.~Kojo, Y.~Hidaka, L.~McLerran and R.~D.~Pisarski,
Quarkyonic Chiral Spirals,
Nucl. Phys. A \textbf{843} (2010), 37-58
doi:10.1016/j.nuclphysa.2010.05.053
[arXiv:0912.3800 [hep-ph]].


\bibitem {redlich}
L.~McLerran, K.~Redlich and C.~Sasaki, Quarkyonic Matter and Chiral Symmetry
Breaking,
Nucl. Phys. A \textbf{824} (2009),
86-100
doi:10.1016/j.nuclphysa.2009.04.001 [arXiv:0812.3585 [hep-ph]].


\bibitem {mclerranissues}
L.~McLerran,
The Phase Diagram of QCD and Some Issues of Large N(c),
Nucl. Phys. B Proc. Suppl. \textbf{195} (2009), 275-280
doi:10.1016/j.nuclphysbps.2009.10.020
[arXiv:0906.2651 [hep-ph]].


\bibitem{achimnc}
A.~Heinz, F.~Giacosa and D.~H.~Rischke,
Restoration of chiral symmetry in the large-$N_c$ limit,
Phys. Rev. D \textbf{85} (2012), 056005
doi:10.1103/PhysRevD.85.056005
[arXiv:1110.1528 [hep-ph]].

\bibitem {dichotomous}
Y.~Hidaka, T.~Kojo, L.~McLerran and R.~D.~Pisarski, 
The
Dichotomous Nucleon: Some Radical Conjectures for the Large-$N_{c}$
Limit,
Nucl. Phys. A \textbf{852} (2011), 155-174
doi:10.1016/j.nuclphysa.2011.01.008
[arXiv:1004.2261 [hep-ph]].




\bibitem {quarkyonicneutronstars}
L.~McLerran and S.~Reddy, 
Quarkyonic Matter and Neutron
Stars
Phys. Rev. Lett. \textbf{122} (2019) no.12, 122701
doi:10.1103/PhysRevLett.122.122701
[arXiv:1811.12503 [nucl-th]].



\bibitem {pagliara}
F.~Giacosa and G.~Pagliara, 
Neutron stars in the
large-$N_{c}$ limit,
Nucl. Phys. A \textbf{968} (2017),
366-378
doi:10.1016/j.nuclphysa.2017.08.006
[arXiv:1707.02644 [nucl-th]].


\bibitem {beyond}
F.~Giacosa, ``Mesons beyond the quark-antiquark picture,'' Acta Phys. Polon. B
\textbf{47} (2016), 7 
doi:10.5506/APhysPolB.47.7 [arXiv:1511.04605 [hep-ph]].

\bibitem{mainlattice}
Y.~Chen, A.~Alexandru, S.~J.~Dong, T.~Draper, I.~Horvath, F.~X.~Lee, K.~F.~Liu, N.~Mathur, C.~Morningstar and M.~Peardon, \textit{et al.}
Glueball spectrum and matrix elements on anisotropic lattices,
Phys. Rev. D \textbf{73} (2006), 014516
doi:10.1103/PhysRevD.73.014516
[arXiv:hep-lat/0510074 [hep-lat]].

\bibitem {bag}
R.~L.~Jaffe, K.~Johnson and Z.~Ryzak,
Qualitative Features of the Glueball Spectrum,
Annals Phys. \textbf{168} (1986), 344 doi:10.1016/0003-4916(86)90035-7


\bibitem{vento}
V.~Mathieu, N.~Kochelev and V.~Vento,
The Physics of Glueballs,
Int. J. Mod. Phys. E \textbf{18} (2009), 1-49
doi:10.1142/S0218301309012124
[arXiv:0810.4453 [hep-ph]].

\bibitem{zeegroup}
A.~Zee,
Group Theory in a Nutshell for Physicists,
Princeton University Press, 2016,
ISBN 978-1-4008-8118-5, 978-0-691-16269-0

\bibitem {fuku}
K.~Fukushima and V.~Skokov,
Polyakov loop modeling for hot QCD,''
Prog. Part. Nucl. Phys. \textbf{96} (2017), 154-199
doi:10.1016/j.ppnp.2017.05.002 [arXiv:1705.00718 [hep-ph]].


\bibitem{deur}
A.~Deur, S.~J.~Brodsky and G.~F.~de Teramond,
The QCD Running Coupling,''
Nucl. Phys. \textbf{90} (2016), 1
doi:10.1016/j.ppnp.2016.04.003
[arXiv:1604.08082 [hep-ph]].

\bibitem{gies}
H.~Gies,
Phys. Rev. D \textbf{66} (2002), 025006
doi:10.1103/PhysRevD.66.025006
[arXiv:hep-th/0202207 [hep-th]].


\bibitem {coleman1}
S.~R.~Coleman and E.~Witten,
Chiral Symmetry Breakdown in Large N Chromodynamics,
Phys. Rev. Lett. \textbf{45} (1980), 100 doi:10.1103/PhysRevLett.45.100



\bibitem {witteneta}
E.~Witten,
Current Algebra Theorems for the U(1) Goldstone Boson,
Nucl. Phys. B \textbf{156} (1979), 269-283 doi:10.1016/0550-3213(79)90031-2

\bibitem{gluonmass}
J.~F.~Donoghue,
The Gluon 'Mass' in the Bag Model,
Phys. Rev. D \textbf{29} (1984), 2559
doi:10.1103/PhysRevD.29.2559

\bibitem{gluonmass2}
D.~Binosi, D.~Ibanez and J.~Papavassiliou,
The all-order equation of the effective gluon mass,
Phys. Rev. D \textbf{86} (2012), 085033
doi:10.1103/PhysRevD.86.085033
[arXiv:1208.1451 [hep-ph]].
S.~Strauss, C.~S.~Fischer and C.~Kellermann,
Analytic structure of the Landau gauge gluon propagator,
Phys. Rev. Lett. \textbf{109} (2012), 252001
doi:10.1103/PhysRevLett.109.252001
[arXiv:1208.6239 [hep-ph]].



\bibitem{isgur}
S.~Godfrey and N.~Isgur,
Mesons in a Relativized Quark Model with Chromodynamics,
Phys. Rev. D \textbf{32} (1985), 189-231
doi:10.1103/PhysRevD.32.189
See also the summary on `quark model' in\ Ref. \cite{pdg}.

\bibitem {anomaly1}
F.~Giacosa, A.~Koenigstein and R.~D.~Pisarski, 
How the axial anomaly
controls flavor mixing among mesons,
Phys. Rev. D \textbf{97} (2018) no.9, 091901
doi:10.1103/PhysRevD.97.091901
[arXiv:1709.07454 [hep-ph]].

\bibitem {kovacscep}
P.~Kov\'acs, Z.~Sz\'ep and G.~Wolf,
Existence of the critical endpoint in the vector meson extended linear sigma model,
Phys. Rev. D \textbf{93} (2016) no.11, 114014 doi:10.1103/PhysRevD.93.114014
[arXiv:1601.05291 [hep-ph]].

\bibitem {tripolt}
R.~A.~Tripolt, N.~Strodthoff, L.~von Smekal and J.~Wambach,
Spectral Functions for the Quark-Meson Model Phase Diagram from the Functional Renormalization Group,
Phys. Rev. D \textbf{89} (2014) no.3, 034010 doi:10.1103/PhysRevD.89.034010
[arXiv:1311.0630 [hep-ph]].


\bibitem{ozi}
H.~J.~Lipkin,
The {OZI} Rule in Charmonium Decays Above $D \bar{D}$ Threshold,
Phys. Lett. B \textbf{179} (1986), 278
doi:10.1016/0370-2693(86)90580-0





\bibitem{volkovnc}
M.~K.~Volkov, A.~A.~Osipov, A.~A.~Pivovarov and K.~Nurlan,
1/Nc approximation and universality of vector mesons,
Phys. Rev. D \textbf{104} (2021) no.3, 034021
doi:10.1103/PhysRevD.104.034021
[arXiv:2105.02160 [hep-ph]].

\bibitem{gerard}
J.~M.~Gerard and T.~Lahna,
The Asymptotic behavior of the pi0 gamma* gamma* vertex,
Phys. Lett. B \textbf{356} (1995), 381-385
doi:10.1016/0370-2693(95)00811-X
[arXiv:hep-ph/9506255 [hep-ph]].

\bibitem{wiese}
O.~Bar and U.~J.~Wiese,
Can one see the number of colors?,
Nucl. Phys. B \textbf{609} (2001), 225-246
doi:10.1016/S0550-3213(01)00288-7
[arXiv:hep-ph/0105258 [hep-ph]].


\bibitem{bickert}
P.~Bickert and S.~Scherer,
$\eta^{(')}\to\pi^+\pi^-\gamma^{(\ast)}$ in large-$N_c$ chiral perturbation theory,
Phys. Rev. D \textbf{104} (2021), 074021
doi:10.1103/PhysRevD.104.074021
[arXiv:2106.12482 [hep-ph]].


\bibitem{lupo}
F.~Giacosa and G.~Pagliara,
On the spectral functions of scalar mesons,
Phys. Rev. C \textbf{76} (2007), 065204
doi:10.1103/PhysRevC.76.065204
[arXiv:0707.3594 [hep-ph]].




\bibitem {sill}
F.~Giacosa, A.~Okopi\'{n}ska and V.~Shastry,
A simple alternative to the relativistic Breit\textendash{}Wigner distribution,
Eur. Phys. J. A \textbf{57} (2021) no.12, 336
doi:10.1140/epja/s10050-021-00641-2 [arXiv:2106.03749 [hep-ph]].

\bibitem{plb}
F.~Giacosa,
Multichannel decay law,
Phys. Lett. B \textbf{831} (2022), 137200
doi:10.1016/j.physletb.2022.137200
[arXiv:2108.07838 [quant-ph]].



\bibitem{ko}
P.~Ko and S.~Rudaz,
Phenomenology of scalar and vector mesons in the linear sigma model,
Phys. Rev. D \textbf{50} (1994), 6877-6894
doi:10.1103/PhysRevD.50.6877
J.~K.~Kim, P.~Ko, K.~Y.~Lee and S.~Rudaz,
A1 (1260) contribution to photon and dilepton productions from hot hadronic matter: Revisited,
Phys. Rev. D \textbf{53} (1996), 4787-4792
doi:10.1103/PhysRevD.53.4787
[arXiv:hep-ph/9602293 [hep-ph]].

\bibitem{urban}
M.~Urban, M.~Buballa and J.~Wambach,
Vector and axial vector correlators in a chirally symmetric model,
Nucl. Phys. A \textbf{697} (2002), 338-371
doi:10.1016/S0375-9474(01)01248-9
[arXiv:hep-ph/0102260 [hep-ph]].

\bibitem {nf2}
D.~Parganlija, F.~Giacosa and D.~H.~Rischke,
Vacuum Properties of Mesons in a Linear Sigma Model with Vector Mesons and Global Chiral Invariance,
Phys. Rev. D \textbf{82} (2010), 054024 doi:10.1103/PhysRevD.82.054024
[arXiv:1003.4934 [hep-ph]].


\bibitem {fariborz}
A.~H.~Fariborz, R.~Jora and J.~Schechter,
A.~H.~Fariborz, R.~Jora and J.~Schechter,
Toy model for two chiral nonets,
Phys. Rev. D \textbf{72} (2005), 034001
doi:10.1103/PhysRevD.72.034001
[arXiv:hep-ph/0506170 [hep-ph]].
A.~H.~Fariborz,
Isosinglet scalar mesons below 2-GeV and the scalar glueball mass,
Int. J. Mod. Phys. A \textbf{19} (2004), 2095-2112
doi:10.1142/S0217751X04018695
[arXiv:hep-ph/0302133 [hep-ph]].
M.~Napsuciale and S.~Rodriguez,
A Chiral model for anti-q q and anti-qq qq mesons,
Phys. Rev. D \textbf{70} (2004), 094043
doi:10.1103/PhysRevD.70.094043
[arXiv:hep-ph/0407037 [hep-ph]].



\bibitem{pionbox}
W.~Bietenholz \textit{et al.}
Pion in a Box,
Phys. Lett. B \textbf{687} (2010), 410-414
doi:10.1016/j.physletb.2010.03.063
[arXiv:1002.1696 [hep-lat]].


\bibitem{gor}
K.~Langfeld and C.~Kettner,
The Quark condensate in the GMOR relation,
Mod. Phys. Lett. A \textbf{11} (1996), 1331-1338
doi:10.1142/S0217732396001338
[arXiv:hep-ph/9601370 [hep-ph]].


\bibitem{kovacsanomaly}
P.~Kov\'{a}cs and G.~Wolf,
Meson Vacuum Phenomenology in a Three-flavor Linear Sigma Model with (Axial-)Vector Mesons: Investigation of the U$(1)_A$ Anomaly Term,
Acta Phys. Polon. Supp. \textbf{6} (2013) no.3, 853-858
doi:10.5506/APhysPolBSupp.6.853 [arXiv:1304.5362 [hep-ph]].





\bibitem {anomaly2}
F.~Giacosa, S.~Jafarzade and R.~Pisarski, ``Anomalous interactions for
heterochiral mesons with $J^{PC}=1^{+-}$ and $2^{-+}$,
[arXiv:2309.00086
[hep-ph]].



\bibitem{chpt}
J.~Gasser and H.~Leutwyler,
Chiral Perturbation Theory to One Loop,
Annals Phys. \textbf{158} (1984), 142
doi:10.1016/0003-4916(84)90242-2
See also S.~Scherer,
Introduction to chiral perturbation theory,
Adv.\ Nucl.\ Phys.\ \textbf{27} (2003) 277 [arXiv:hep-ph/0210398]
and refs. therein.


\bibitem{koenig1}
A.~Koenigstein and F.~Giacosa,
Phenomenology of pseudotensor mesons and the pseudotensor glueball,
Eur. Phys. J. A \textbf{52} (2016) no.12, 356
doi:10.1140/epja/i2016-16356-x
[arXiv:1608.08777 [hep-ph]].

\bibitem{koenig2}
S.~Jafarzade, A.~Koenigstein and F.~Giacosa,
Phenomenology of $J^{PC}$ = $3^{--}$ tensor mesons,
Phys. Rev. D \textbf{103} (2021) no.9, 096027
doi:10.1103/PhysRevD.103.096027
[arXiv:2101.03195 [hep-ph]].



\bibitem {at}
A.~Athenodorou and M.~Teper,
The glueball spectrum of SU(3) gauge theory in 3 + 1 dimensions,
JHEP \textbf{11} (2020), 172 doi:10.1007/JHEP11(2020)172 [arXiv:2007.06422
[hep-lat]].



\bibitem {grg}
E.~Trotti, S.~Jafarzade and F.~Giacosa,
Thermodynamics of the glueball resonance gas,
Eur. Phys. J. C \textbf{83} (2023) no.5, 390
doi:10.1140/epjc/s10052-023-11557-0
[arXiv:2212.03272 [hep-ph]].


\bibitem {updatedrev}
H.~X.~Chen, W.~Chen, X.~Liu, Y.~R.~Liu and S.~L.~Zhu,
An updated review of the new hadron states,
Rept. Prog. Phys. \textbf{86} (2023) no.2, 026201 doi:10.1088/1361-6633/aca3b6
[arXiv:2204.02649 [hep-ph]].



\bibitem {gounaris}
G.~J.~Gounaris, J.~E.~Paschalis and R.~Kogerler,
The 0 Spin Glueballs: A New Approach to Relativistic Bound States,
Z. Phys. C \textbf{31} (1986), 277 [erratum: Z. Phys. C \textbf{33} (1987),
474] doi:10.1007/BF01479537



\bibitem{gluoncondensat}
V.~A.~Novikov, L.~B.~Okun, M.~A.~Shifman, A.~I.~Vainshtein, M.~B.~Voloshin and V.~I.~Zakharov,
Charmonium and Gluons: Basic Experimental Facts and Theoretical Introduction,
Phys. Rept. \textbf{41} (1978), 1-133
doi:10.1016/0370-1573(78)90120-5
M.~Campostrini, A.~Di Giacomo and Y.~Gunduc,
Gluon Condensation in SU(3) Lattice Gauge Theory,
Phys. Lett. B \textbf{225} (1989), 393-397
doi:10.1016/0370-2693(89)90588-1


\bibitem {glueballonium}
F.~Giacosa, A.~Pilloni and E.~Trotti,  Glueball--{}glueball
scattering and the glueballonium,
Eur. Phys. J. C
\textbf{82} (2022) no.5, 487
doi:10.1140/epjc/s10052-022-10403-z
[arXiv:2110.05582 [hep-ph]].

\bibitem{weingarten}
W.~J.~Lee and D.~Weingarten,
Phys. Rev. D \textbf{61} (2000), 014015
doi:10.1103/PhysRevD.61.014015
[arXiv:hep-lat/9910008 [hep-lat]].


\bibitem{long}
F.~Giacosa, T.~Gutsche, V.~E.~Lyubovitskij and A.~Faessler,
Scalar nonet quarkonia and the scalar glueball: Mixing and decays in an effective chiral approach,
Phys. Rev. D \textbf{72} (2005), 094006
doi:10.1103/PhysRevD.72.094006
[arXiv:hep-ph/0509247 [hep-ph]].

\bibitem{cheng}
H.~Y.~Cheng, C.~K.~Chua and K.~F.~Liu,
Scalar glueball, scalar quarkonia, and their mixing,
Phys. Rev. D \textbf{74} (2006), 094005
doi:10.1103/PhysRevD.74.094005
[arXiv:hep-ph/0607206 [hep-ph]].

\bibitem{rebhan}
F.~Br\"unner, D.~Parganlija and A.~Rebhan,
Glueball Decay Rates in the Witten-Sakai-Sugimoto Model,
Phys. Rev. D \textbf{91} (2015) no.10, 106002
[erratum: Phys. Rev. D \textbf{93} (2016) no.10, 109903]
doi:10.1103/PhysRevD.91.106002
[arXiv:1501.07906 [hep-ph]].



\bibitem{jpsilatt}
L.~C.~Gui \textit{et al.} [CLQCD],
Scalar Glueball in Radiative $J/\psi$ Decay on the Lattice,
Phys. Rev. Lett. \textbf{110} (2013) no.2, 021601
doi:10.1103/PhysRevLett.110.021601
[arXiv:1206.0125 [hep-lat]].




\bibitem{psg}
W.~I.~Eshraim, S.~Janowski, F.~Giacosa and D.~H.~Rischke,
Decay of the pseudoscalar glueball into scalar and pseudoscalar mesons,
Phys. Rev. D \textbf{87} (2013) no.5, 054036
doi:10.1103/PhysRevD.87.054036
[arXiv:1208.6474 [hep-ph]].

\bibitem{bes3}
M.~Ablikim \textit{et al.} [(BESIII Collaboration)* and BESIII],
Observation of a State $X(2600)$ in the $\pi^{+}\pi^{-}\eta'$ System in the Process $J/\psi\rightarrow\gamma\pi^{+}\pi^{-}\eta'$,
Phys. Rev. Lett. \textbf{129} (2022) no.4, 042001
doi:10.1103/PhysRevLett.129.042001
[arXiv:2201.10796 [hep-ex]].

\bibitem{sammet}
F.~Giacosa, J.~Sammet and S.~Janowski,
Decays of the vector glueball,
Phys. Rev. D \textbf{95} (2017) no.11, 114004
doi:10.1103/PhysRevD.95.114004
[arXiv:1607.03640 [hep-ph]].

\bibitem{vereijken}
A.~Vereijken, S.~Jafarzade, M.~Piotrowska and F.~Giacosa,
Is f2(1950) the tensor glueball?,
Phys. Rev. D \textbf{108} (2023) no.1, 014023
doi:10.1103/PhysRevD.108.014023
[arXiv:2304.05225 [hep-ph]].


\bibitem {rebhan2}F.~Hechenberger, J.~Leutgeb and A.~Rebhan,
Radiative meson and glueball decays in the Witten-Sakai-Sugimoto model,
Phys. Rev. D \textbf{107} (2023) no.11, 114020 doi:10.1103/PhysRevD.107.114020
[arXiv:2302.13379 [hep-ph]].


\bibitem {rebhan3}
F.~Hechenberger, J.~Leutgeb and A.~Rebhan,
Spin-1 glueballs in the Witten-Sakai-Sugimoto model,
Phys. Rev. D \textbf{109} (2024) no.7, 074014
doi:10.1103/PhysRevD.109.074014
[arXiv:2401.17986 [hep-ph]].

\bibitem{misra}
K.~Sil, V.~Yadav and A.~Misra,
Top\textendash{}down holographic G-structure glueball spectroscopy at (N)LO in $N$ and finite coupling,
Eur. Phys. J. C \textbf{77} (2017) no.6, 381
doi:10.1140/epjc/s10052-017-4921-7
[arXiv:1703.01306 [hep-th]].







\bibitem{hybrid}
C.~A.~Meyer and E.~S.~Swanson,
Hybrid Mesons,
Prog. Part. Nucl. Phys. \textbf{82} (2015), 21-58
doi:10.1016/j.ppnp.2015.03.001
[arXiv:1502.07276 [hep-ph]].


\bibitem {dudek}
J.~J.~Dudek, R.~G.~Edwards, M.~J.~Peardon, D.~G.~Richards and C.~E.~Thomas,
Toward the excited meson spectrum of dynamical QCD,
Phys. Rev. D \textbf{82} (2010), 034508 doi:10.1103/PhysRevD.82.034508
[arXiv:1004.4930 [hep-ph]].
J.~J.~Dudek \textit{et al.} [Hadron Spectrum],
Toward the excited isoscalar meson spectrum from lattice QCD,
Phys. Rev. D \textbf{88} (2013) no.9, 094505 doi:10.1103/PhysRevD.88.094505
[arXiv:1309.2608 [hep-lat]].





\bibitem {shastry}
V.~Shastry, C.~S.~Fischer and F.~Giacosa,
The phenomenology of the exotic hybrid nonet with \ensuremath{\pi}1(1600) and \ensuremath{\eta}1(1855),
Phys. Lett. B \textbf{834} (2022), 137478
doi:10.1016/j.physletb.2022.137478
[arXiv:2203.04327 [hep-ph]];
V.~Shastry and F.~Giacosa,
Radiative production and decays of the exotic \ensuremath{\eta}1'(1855) and its siblings,
Nucl. Phys. A \textbf{1037} (2023), 122683
doi:10.1016/j.nuclphysa.2023.122683
[arXiv:2302.07687 [hep-ph]].

\bibitem {elsmhyb}
W.~I.~Eshraim, C.~S.~Fischer, F.~Giacosa and D.~Parganlija,
Hybrid phenomenology in a chiral approach,
Eur. Phys. J. Plus \textbf{135} (2020) no.12, 945
doi:10.1140/epjp/s13360-020-00900-z
[arXiv:2001.06106 [hep-ph]].


\bibitem {barua0}
V.~Baru, J.~Haidenbauer, C.~Hanhart, Y.~Kalashnikova and A.~E.~Kudryavtsev,
Evidence that the a(0)(980) and f(0)(980) are not elementary particles,
Phys. Lett. B \textbf{586} (2004), 53-61 doi:10.1016/j.physletb.2004.01.088
[arXiv:hep-ph/0308129 [hep-ph]].


\bibitem {branz}
T.~Branz, T.~Gutsche and V.~E.~Lyubovitskij,
Strong and radiative decays of the scalars f(0)(980) and a(0)(980) in a hadronic molecule approach,
Phys. Rev. D \textbf{78} (2008), 114004 doi:10.1103/PhysRevD.78.114004
[arXiv:0808.0705 [hep-ph]].

\bibitem {petrov}
A.~A.~Petrov,
Glueball-meson molecules,
Phys. Lett. B \textbf{843} (2023), 138030 doi:10.1016/j.physletb.2023.138030
[arXiv:2204.11269 [hep-ph]].



\bibitem{boglione}
M.~Boglione and M.~R.~Pennington,
Dynamical generation of scalar mesons,
Phys. Rev. D \textbf{65} (2002), 114010
doi:10.1103/PhysRevD.65.114010
[arXiv:hep-ph/0203149 [hep-ph]].

\bibitem {wolkaa0}
T.~Wolkanowski, F.~Giacosa and D.~H.~Rischke,
$a_{0}(980)$ revisited,
Phys. Rev. D \textbf{93} (2016) no.1, 014002 doi:10.1103/PhysRevD.93.014002
[arXiv:1508.00372 [hep-ph]].

\bibitem{pelaeznc}
J.~R.~Pelaez,
On the Nature of light scalar mesons from their large N(c) behavior,
Phys. Rev. Lett. \textbf{92} (2004), 102001
doi:10.1103/PhysRevLett.92.102001
[arXiv:hep-ph/0309292 [hep-ph]].





\bibitem {sigmareview}
J.~R.~Pelaez,
From controversy to precision on the sigma meson: a review on the status of the non-ordinary $f_0(500)$ resonance,
Phys. Rept. \textbf{658} (2016), 1 doi:10.1016/j.physrep.2016.09.001
[arXiv:1510.00653 [hep-ph]].

\bibitem{rodas}
J.~R.~Pel\'aez, A.~Rodas and J.~Ruiz de Elvira,
Eur. Phys. J. C \textbf{77} (2017) no.2, 91
doi:10.1140/epjc/s10052-017-4668-1
[arXiv:1612.07966 [hep-ph]].

\bibitem {wolkak}
T.~Wolkanowski, M.~So\l {}tysiak and F.~Giacosa,
$K_{0}^{\ast}(800)$ as a companion pole of $K_{0}^{\ast}(1430)$,
Nucl. Phys. B \textbf{909} (2016), 418-428 doi:10.1016/j.nuclphysb.2016.05.025
[arXiv:1512.01071 [hep-ph]].









\bibitem {x3872}
F.~Giacosa, M.~Piotrowska and S.~Coito,
$X(3872)$ as virtual companion pole of the charm\textendash{}anticharm state $\chi_{c1}(2P)$,
Int. J. Mod. Phys. A \textbf{34} (2019) no.29, 1950173
doi:10.1142/S0217751X19501732 [arXiv:1903.06926 [hep-ph]].






\bibitem {lowscalars}
E.~van Beveren, T.~A.~Rijken, K.~Metzger, C.~Dullemond, G.~Rupp and J.~E.~Ribeiro,
A Low Lying Scalar Meson Nonet in a Unitarized Meson Model,
Z. Phys. C \textbf{30} (1986), 615-620
doi:10.1007/BF01571811
[arXiv:0710.4067 [hep-ph]].

E.~van Beveren, D.~V.~Bugg, F.~Kleefeld and G.~Rupp,
Phys. Lett. B \textbf{641} (2006), 265-271
doi:10.1016/j.physletb.2006.08.051
[arXiv:hep-ph/0606022 [hep-ph]].


J.~A.~Oller and E.~Oset,
Chiral symmetry amplitudes in the S wave isoscalar and isovector channels and the $\sigma$, f$_0$(980), a$_0$(980) scalar mesons,
Nucl. Phys. A \textbf{620} (1997), 438-456
[erratum: Nucl. Phys. A \textbf{652} (1999), 407-409]
doi:10.1016/S0375-9474(97)00160-7
[arXiv:hep-ph/9702314 [hep-ph]].


J.~A.~Oller, E.~Oset and J.~R.~Pelaez,
Meson meson interaction in a nonperturbative chiral approach,
Phys. Rev. D \textbf{59} (1999), 074001
[erratum: Phys. Rev. D \textbf{60} (1999), 099906; erratum: Phys. Rev. D \textbf{75} (2007), 099903]
doi:10.1103/PhysRevD.59.074001
[arXiv:hep-ph/9804209 [hep-ph]].

\bibitem{dynrec}
F.~Giacosa,
Dynamical generation and dynamical reconstruction,
Phys. Rev. D \textbf{80} (2009), 074028
doi:10.1103/PhysRevD.80.074028
[arXiv:0903.4481 [hep-ph]].

\bibitem{lesson}
Z.~H.~Guo, L.~Y.~Xiao and H.~Q.~Zheng,
Is the f0(600) meson a dynamically generated resonance? A Lesson learned from the O(N) model and beyond,
Int. J. Mod. Phys. A \textbf{22} (2007), 4603-4616
doi:10.1142/S0217751X0703710X
[arXiv:hep-ph/0610434 [hep-ph]].


\bibitem{jaffeorig}
R.~L.~Jaffe,
Multi-Quark Hadrons. 1. The Phenomenology of (2 Quark 2 anti-Quark) Mesons,
Phys. Rev. D \textbf{15} (1977), 267
doi:10.1103/PhysRevD.15.267


\bibitem{exotica}
R.~L.~Jaffe,
Exotica,
Phys. Rept. \textbf{409} (2005), 1-45
doi:10.1016/j.physrep.2004.11.005
[arXiv:hep-ph/0409065 [hep-ph]].






\bibitem{maianilow}
L.~Maiani, F.~Piccinini, A.~D.~Polosa and V.~Riquer,
A New look at scalar mesons,
Phys. Rev. Lett. \textbf{93} (2004), 212002
doi:10.1103/PhysRevLett.93.212002
[arXiv:hep-ph/0407017 [hep-ph]].

\bibitem {tq}
F.~Giacosa,
Strong and electromagnetic decays of the light scalar mesons interpreted as tetraquark states,
Phys. Rev. D \textbf{74} (2006), 014028
doi:10.1103/PhysRevD.74.014028
[arXiv:hep-ph/0605191 [hep-ph]].

\bibitem{tqchiral}
F.~Giacosa,
Mixing of scalar tetraquark and quarkonia states in a chiral approach,
Phys. Rev. D \textbf{75} (2007), 054007
doi:10.1103/PhysRevD.75.054007
[arXiv:hep-ph/0611388 [hep-ph]].


\bibitem {olbrich2}
L.~Olbrich, M.~Z\'{e}t\'{e}nyi, F.~Giacosa and D.~H.~Rischke,
Influence of the axial anomaly on the decay $N(1535) \rightarrow N\eta $,
Phys. Rev. D \textbf{97} (2018) no.1, 014007
doi:10.1103/PhysRevD.97.014007
[arXiv:1708.01061 [hep-ph]].


\bibitem {koch}
V.~Koch,
Aspects of chiral symmetry,
Int. J. Mod. Phys. E \textbf{6} (1997), 
203-250 doi:10.1142/S0218301397000147
[arXiv:nucl-th/9706075 [nucl-th]].



\bibitem {detar}
C.~E.~Detar and T.~Kunihiro,
Linear $\sigma$ Model With Parity Doubling,
Phys. Rev. D \textbf{39} (1989), 2805 doi:10.1103/PhysRevD.39.2805

\bibitem {tolos}
D.~Zschiesche, L.~Tolos, J.~Schaffner-Bielich and R.~D.~Pisarski,
Cold, dense nuclear matter in a SU(2) parity doublet model,
Phys. Rev. C \textbf{75} (2007), 055202 doi:10.1103/PhysRevC.75.055202
[arXiv:nucl-th/0608044 [nucl-th]].


\bibitem {gallas}
S.~Gallas, F.~Giacosa and D.~H.~Rischke,
Vacuum phenomenology of the chiral partner of the nucleon in a linear sigma model with vector mesons,
Phys. Rev. D \textbf{82} (2010), 014004 doi:10.1103/PhysRevD.82.014004
[arXiv:0907.5084 [hep-ph]].

\bibitem {gallas2}
S.~Gallas, F.~Giacosa and G.~Pagliara,
Nuclear matter within a dilatation-invariant parity doublet model: the role of the tetraquark at nonzero density,
Nucl. Phys. A \textbf{872} (2011), 13-24 doi:10.1016/j.nuclphysa.2011.09.008
[arXiv:1105.5003 [hep-ph]].

\bibitem {vonsmekal}
R.~A.~Tripolt, C.~Jung, L.~von Smekal and J.~Wambach,
Vector and axial-vector mesons in nuclear matter,
Phys. Rev. D \textbf{104} (2021) no.5, 054005 doi:10.1103/PhysRevD.104.054005
[arXiv:2105.00861 [hep-ph]].


\bibitem {laka}
P.~Lakaschus, J.~L.~P.~Mauldin, F.~Giacosa and D.~H.~Rischke,
Phys. Rev. C \textbf{99} (2019) no.4, 045203 doi:10.1103/PhysRevC.99.045203
[arXiv:1807.03735 [hep-ph]].




\bibitem{olbrich}
L.~Olbrich, M.~Z\'et\'enyi, F.~Giacosa and D.~H.~Rischke,
Three-flavor chiral effective model with four baryonic multiplets within the mirror assignment,
Phys. Rev. D \textbf{93} (2016) no.3, 034021
doi:10.1103/PhysRevD.93.034021
[arXiv:1511.05035 [hep-ph]].

\bibitem{satz}
H.~Satz,
The Thermodynamics of Quarks and Gluons,
Lect. Notes Phys. \textbf{785} (2010), 1-21
doi:10.1007/978-3-642-02286-9\_1
[arXiv:0803.1611 [hep-ph]].
H.~Satz,
The Quark-Gluon Plasma: A Short Introduction,
Nucl. Phys. A \textbf{862-863} (2011), 4-12
doi:10.1016/j.nuclphysa.2011.05.014
[arXiv:1101.3937 [hep-ph]].


\bibitem {lena}
J.~T.~Lenaghan, D.~H.~Rischke and
J.~Schaffner-Bielich,
Chiral symmetry restoration at nonzero temperature in the SU(3)(r) x SU(3)(l) linear sigma model,
Phys. Rev. D \textbf{62} (2000), 085008 doi:10.1103/PhysRevD.62.085008
[arXiv:nucl-th/0004006 [nucl-th]].


\bibitem {glendenning}
N.~K.~Glendenning, Compact stars: Nuclear
physics, particle physics, and general relativity,
Springer Verlag, 2000.

\bibitem{lottini}
S.~Lottini and G.~Torrieri,
A percolation transition in Yang-Mills matter at finite number of colours,
Phys. Rev. Lett. \textbf{107} (2011), 152301
doi:10.1103/PhysRevLett.107.152301
[arXiv:1103.4824 [nucl-th]].

\bibitem{mishustin}
G.~Torrieri and I.~Mishustin,
The nuclear liquid-gas phase transition at large $N_c$ in the Van der Waals approximation,
Phys. Rev. C \textbf{82} (2010), 055202
doi:10.1103/PhysRevC.82.055202
[arXiv:1006.2471 [nucl-th]].

\bibitem{achimtq}
A.~Heinz, S.~Struber, F.~Giacosa and D.~H.~Rischke,
Role of the tetraquark in the chiral phase transition,
Phys. Rev. D \textbf{79} (2009), 037502
doi:10.1103/PhysRevD.79.037502
[arXiv:0805.1134 [hep-ph]].





\bibitem {pisarskilast}
R.~D.~Pisarski and F.~Rennecke,
The chiral phase transition and the axial anomaly,
[arXiv:2401.06130 [hep-ph]].



\bibitem {gazd}
M.~Gazdzicki, M.~Gorenstein and P.~Seyboth,
Onset of deconfinement in nucleus-nucleus collisions: Review for pedestrians and experts,
Acta Phys. Polon. B \textbf{42} (2011), 307-351 doi:10.5506/APhysPolB.42.307
[arXiv:1006.1765 [hep-ph]].




\bibitem {hara}
H.~Adhikary \textit{et al.} [NA61/SHINE],
Search for the critical point of strongly-interacting matter in $^{40}$Ar ~+~$^{45}$Sc collisions at 150A ~Ge V /c using scaled factorial moments of protons,
Eur. Phys. J. C \textbf{83} (2023) no.9, 881
doi:10.1140/epjc/s10052-023-11942-9 [arXiv:2305.07557 [nucl-ex]].
M.~I.~Gorenstein, M.~Gazdzicki and W.~Greiner,
Critical line of the deconfinement phase transitions,''
Phys. Rev. C \textbf{72} (2005), 024909 doi:10.1103/PhysRevC.72.024909
[arXiv:nucl-th/0505050 [nucl-th]].


\bibitem{bielich}
G.~Pagliara and J.~Schaffner-Bielich,
Stability of CFL cores in Hybrid Stars,
Phys. Rev. D \textbf{77} (2008), 063004
doi:10.1103/PhysRevD.77.063004
[arXiv:0711.1119 [astro-ph]].






\end{thebibliography}
\end{document}